\documentclass[fontsize=12pt]{scrartcl}
\usepackage[left=3.0cm,right=3.0cm,top=2.6cm,bottom=4cm]{geometry}
\usepackage{parskip} 
\usepackage{ucs} 
\usepackage{inputenc} 
\usepackage[ngerman]{babel} 
\usepackage{ulem} 
\usepackage{xcolor} 
\usepackage{wrapfig} 
\makeatletter 
\newcommand\wrapfill{\par
  \ifx\parshape\WF@fudgeparshape
  \nobreak
  \vskip-\baselineskip
  \vskip\c@WF@wrappedlines\baselineskip
  \allowbreak
  \WFclear
  \fi
}
\makeatother
\usepackage{graphicx}
\usepackage{setspace}
\usepackage{array,longtable}
\usepackage{fancyhdr} 
\usepackage{url} 

\usepackage[colorlinks=true,urlcolor=blue,citecolor=blue]{hyperref} 
\begin{document}

\begin{titlepage}

\begin{center}
\vspace*{0,2cm} 
\textsc{\large\textbf{Nicht-propositionales Wissen aus Literaturlekt\"ure\\ und Bedingungen seiner Darstellbarkeit\\ in Wikipedia-Eintr\"agen zu literarischen Werken}}\\
\vspace{1cm}
\textsc{\large\textbf{von Claudia Koltzenburg}}

\vspace{1cm}
Dieses Werk hat die Lizenz CC BY-SA 4.0\\ (\href{https://creativecommons.org/licenses/by-sa/4.0/}{Creative Commons Attribution-Share Alike 4.0 Unported})
\end{center}

\vspace{1cm}
Wo diese Arbeit in welchem Ausgabeformat und welcher Fassung zu finden ist:\\
\newline
\textbf{Belegfassung (Print-pdf)} (Fassung 1, time stamp: 3. September 2015, enth\"alt 7 Anh\"ange und tr\"agt einen l\"angeren Titel: \textit{"`The sum of all human knowledge"': Nicht-propositionales Wissen aus Literaturlekt\"ure und Bedingungen seiner Darstellbarkeit in Wikipedia-Eintr\"agen zu literarischen Werken}), Dissertation \href{http://hdl.handle.net/10900/64678}{Universit\"at T\"ubingen, Philo\-so\-phi\-sche Fakult\"at}, \href{https://de.wikipedia.org/wiki/Open_Access}{Open Access}, doi: 10.15496/publikation-6100\\
\newline
\textbf{als pdf mit Verlinkungen} (Fassung 2, \"uberarbeitet und gek\"urzt, time stamp: 7. September 2015)\\
1. bei \href{http://hdl.handle.net/10760/25717}{eprints.rclis.org} (pdf)\\
2. bei \href{https://commons.wikimedia.org/wiki/File:Koltzenburg_Sept2015_Nicht-propositionales-Wissen_Literarische-Werke_Wikipedia.pdf}{commons.wikimedia.org} (pdf)\\
3. bei \href{https://arxiv.org}{arxiv.org} (pdf)\\
Enth\"alt kleinere Korrekturen, die in Fassung 3 bis einschlie{\ss}lich 12. September 2015 vorgenommen worden sind, mit Dank an User:95.113.139.178, User:Chricho und User:Cethegus\\
\newline
bei \href{https://de.wikiversity.org/wiki/Nicht-propositionales_Wissen_aus_Literaturlektüre_und_Bedingungen_seiner_Darstellbarkeit_in_Wikipedia-Einträgen_zu_literarischen_Werken}{Wikiversity} \textbf{zum Kommentieren und Weiterschreiben} (Fassung 3, mit Erweiterungen in Kapitel \textit{Schreibweisen bei Ulrike Draesner, Hans Ulrich Gumbrecht und Ina Hartwig}, time stamp der ersten Version: 12:13:47‎ UTC am 1. September 2015, Antikriegstag)

\begin{center}
Mail: \flq koltzenburg@w4w.net\frq
\end{center}
\null
\newpage
\thispagestyle{empty}
\end{titlepage}

\newpage
\thispagestyle{empty}
\singlespacing
¸\newline
\newline
\newline
\newline
\newline
\newline
\newline
\textbf{Abstract}
(English title: \textbf{Wikipedia entries on fiction and non-propositional knowledge representation})

Given that Wikipedia entries are likely to extert a strong influence on how literary texts are perceived -- due to their preferential ranking in Google -- there is some demand that research dealing with the transfer of know\-ledge on li\-te\-ra\-tur\-e to the public be more concerned with looking into both the content that is available for free on the Web and any aspect that may come with writing about li\-te\-ra\-tur\-e for free. This contribution argues from within Wi\-ki\-pe\-dia's multidisciplinary consensus-driven space in which propositional know\-ledge is given priority that it would be essential for entries on fiction to present non-propositional know\-ledge as being one of its hallmarks. For this aim, a special concept is developed that is designed to function as the study's formal object: "`Erle\textbf{s}nis"' (which in German is a pun that combines "`Erlebnis"' -- adventure experience -- and "`lesen"' -- reading). It is defined as non-propositional know\-ledge that has been acquired in an individual reading process. Writing about one's own \textit{Erlesnis} in new ways is being tried out in essays on \textit{Traveling on One Leg} (1989) by Herta Mueller, \textit{Save the Reaper} (1998) by Alice Munro, \textit{Alfred and Emily} (2008) by Doris Lessing and \textit{rein GOLD} (2013) by Elfriede Jelinek respectively. Ideally, an \textit{Erlesnis} is based on a text's literariness (see Ulrike Draesner's re-reading of \textit{Felix Krull}, Thomas Mann's last novel). On de.wikipedia.org an experiment is conducted to find out what community members think about the idea of including, in entries on fiction, sections specifically designed to report about what people felt like when reading a certain text. Finally, a draft typology of \textit{Erlesnis}-writing is suggested. This contribution is the first of its kind internationally to deal with Wi\-ki\-pe\-dia from the point of view of transfer of know\-ledge on li\-te\-ra\-tur\-e to the public. For the theory of this field of research some new aspects are offered for debate.


\newpage
\thispagestyle{empty}

¸\newline
\newline
\newline
\newline
\newline
\newline
\newline
\newline

\textbf{Abstract}

Ausgehend von der Pr\"amisse, dass Wi\-ki\-pe\-dia-Eintr\"age durch ihr vorz\"ugliches Ran\-king bei Google gro{\ss}en Einfluss auf die Wahrnehmung von li\-te\-ra\-rischen Texten haben, ist die Li\-te\-ra\-tur\-ver\-mitt\-lungsfor\-schung gefordert, sich mit web\"of\-fent\-lichen Inhalten sowie gemeinn\"utzigen Aspekten auseinanderzusetzen. Aus einer Binnenperspektive des multidisziplin\"aren Aushandlungsraums Wi\-ki\-pe\-dia, der vorwiegend pro\-po\-si\-ti\-o\-na\-les Wissen pr\"asentiert, wird argumentiert, dass in Eintr\"agen zu li\-te\-ra\-rischen Werken als deren erlebtem Charakteristikum al\-ler\-dings nicht-pro\-po\-si\-ti\-o\-na\-les Wissen aus Li\-te\-ra\-tur\-lekt\"ure essentiell w\"are. Das zu diesem Zweck ent\-wickelte Formalobjekt "`Erle\textbf{s}nis"' be\-zeichnet, was beim Lesen dem eigenen Empfinden nach an nicht-pro\-po\-si\-ti\-o\-na\-lem Wissen erworben wurde. \mbox{Vier} eigene Interpretationen erproben neue Schreibweisen zu \textit{Erlesnissen}: \textit{Reisende auf einem Bein} (1989) von Herta M\"uller, \textit{Save the Reaper} (1998) von Alice Munro, \textit{Alfred and Emily} (2008) von Doris Lessing und \textit{rein GOLD} (2013) von Elfriede Jelinek. Im Idealfall wird \"uber ein \textit{Erlesnis} aufgrund der Literarizit\"at eines Textes berichtet (Ulrike Draesner zu ihrer Re-Lekt\"ure von \textit{Felix Krull}). Mit einem Experiment in der deutschspra\-chi\-gen Wi\-ki\-pe\-dia-Community zu expliziten "`Leseerlebnis"'-Abschnitten in Li\-te\-ra\-tur\-eintr\"agen werden Daten zur aktuellen Akzeptanz-Tendenz zutagegef\"ordert. Der Entwurf einer Typologie von Schreibweisen \"uber \textit{Erlesnisse} inklusive Rezensionsnotizen der Plattform \textit{per\-len\-tau\-cher.de} rundet die Arbeit ab. Es handelt sich international um die erste Arbeit, die sich aus einer Li\-te\-ra\-tur\-ver\-mitt\-lungsperspektive mit Wi\-ki\-pe\-dia befasst, und f\"ur die Theorie der Li\-te\-ra\-tur\-ver\-mitt\-lung werden Ankn\"upfungspunkte aufgezeigt.


\newpage
\pagenumbering{roman}

\tableofcontents

\onehalfspacing

\newpage
\clearpage
\pagenumbering{arabic}

\section{Zum Einstieg}
\label{sec:1}

Im M\"arz 1995 wurde das erste Wiki f\"ur die Web\"of\-fent\-lichkeit verf\"ugbar gemacht. 1987 war f\"ur die Entwicklung von Wiki-Software das Konzept von Entwurfsmustern aus der Architektur aufgenommen worden, wo sie Ende der 1970er Jahre dazu dien\-ten, Bewohner\textsuperscript{\tiny *} in den Entwurfsprozess k\"unftiger Geb\"aude einzubeziehen. Dem Enzy\-klop\"adieprojekt Wi\-ki\-pe\-dia, im Januar 2001 gestartet, liegt ebenfalls eine Wiki-Software zugrunde.

Ausgangspunkt der vorliegenden Arbeit war die Beobachtung, dass zwei wichtige kulturelle Institutionen einander weitgehend ignorieren: Li\-te\-ra\-tur\- und Wi\-ki\-pe\-dia. Im Blickwinkel der gew\"ahlten Fragestellung sind beide im \"ubertragenen Sinn als Bewirtschaftungsweisen eines je spezifischen Geb\"audes anzusehen. Deren jeweilige Entwurfsmuster weisen nicht nur vom Alter her Unterschiede auf, und darin k\"onnte ein Grund liegen, warum man zwar benachbart lebt, die Fensterseiten der Geb\"aude aber nicht einander zugewandt sind -- und auch ansonsten l\"auft man sich kaum \"uber den Weg.\footnote{Viele andere Blickwinkel auf Wi\-ki\-pe\-dia w\"aren m\"oglich und f\"ur For\-schung interessant. Da Erfah\-rungswerte des eigenen Arbeitens auf einer Wiki-Plattform in li\-te\-ra\-tur\-wissenschaftlichen Kreisen vermutlich derzeit nicht vorausgesetzt werden k\"onnen, wurde zum Zwecke argumentativer Koh\"arenz ein gewisses Ma{\ss} an Einf\"uh\-rung zu Wi\-ki\-pe\-dia f\"ur notwendig erachtet, insbesondere mit der Binnenschau auf \textit{Objektebene I} sowie in Abschnitt \textit{5.2 Wissen \"uber Li\-te\-ra\-tur bei Wi\-ki\-pe\-dia}. Die Plattform wird \mbox{hier} aus dem speziellen Blickwinkel meiner Fragestellung vorgestellt. Da es in der Wikipedistik wenig For\-schung zur deutschspra\-chi\-gen Wi\-ki\-pe\-dia gibt und keine zu den sogenannten Li\-te\-ra\-tur\-artikeln, wird Aussagen einzelner \mbox{User} relativ viel Raum gew\"ahrt, weil ich ihnen Belegcharakter zuspreche. Ich verwende "`\mbox{User}"' anstelle der deutschspra\-chi\-gen Be\-zeichnung "`Be\-nut\-zer"', da sich "`\mbox{User}"' genderneutraler lesen l\"asst.}

Allgemein lie{\ss}e sich die Ausgangslage etwa so formulieren: Den Besonderheiten der jeweils anderen Seite wird nicht gen\"ugend Bedeutung f\"ur die eigenen Anliegen bei\-ge\-mes\-sen. Die Bedingungen, die in dieser Arbeit untersucht werden, h\"angen mit Handlungsweisen von Akteuren\textsuperscript{\tiny *} auf beiden Seiten zusammen. Ich arbeite mit gegenseitiger Kontextualisierung und versuche in meinem Research Design den Besonderheiten beider Seiten Rechnung zu tragen. Ein weiterer Aspekt der Gestaltung einer For\-schungsarbeit der vorliegenden Art ist, dass essayistisch gehaltene einf\"uhrende Informationen zum Kontext bestimmter Bedingungen bei Wi\-ki\-pe\-dia einen verh\"altnism\"a{\ss}ig gro{\ss}en Raum einnehmen, da Kenntnisse bez\"uglich der funktionalen Grundlagen der mediawiki-Software sowie der \"ublichen Arbeits\-weisen bei Wi\-ki\-pe\-dia als nicht voraussetzbar eingesch\"atzt wurden.
 
Jede der beiden Institutionen verwendet auch in ihrer Arbeits\-weise bestimmte Entwurfsmuster -- das \mbox{hier} verwendete Arbeits\-konzept zur Auslotung m\"og\-li\-cher \"Uber\-g\"an\-ge zwischen Li\-te\-ra\-tur\- und Wi\-ki\-pe\-dia. Im Falle eines Wiki kann jede einzelne Seite als ein Entwurfsmuster angesehen werden, an dem fortlaufend mit neuen Versionen weitergearbeitet wird, und im Fall von Li\-te\-ra\-tur sehe ich li\-te\-ra\-rische Texte als die Entwurfsmuster an, mit denen Prozesse in Gang gebracht beziehungsweise am Laufen gehalten werden. Die in der Institution Li\-te\-ra\-tur g\"angigen Entwurfsmuster, li\-te\-ra\-rische Texte, werden durch Leseprozesse in Emp\-fin\-dun\-gen verwandelt, auf deren Basis anschlie{\ss}end etwas entsteht. Eine der zu er\"orternden Fragen ist, wie jene Emp\-fin\-dun\-gen -- nachdem sie in sprachliche Form gebracht wurden -- als Belege f\"ur Aussagen in Wi\-ki\-pe\-dia-Artikeln auffindbar sind, denn ohne dass sie den Status einer belegten Aussage haben, werden sie im Kontext eines Wi\-ki\-pe\-dia-Eintrags nicht als relevant angesehen. 

Sowohl f\"ur Li\-te\-ra\-tur als auch f\"ur Wi\-ki\-pe\-dia gilt aus dem Blickwinkel der vorliegenden Arbeit: Wer sich wie beteiligt, gestaltet auch bis\-her schon das jeweilige Geb\"aude und dessen Bewirtschaftungsweise mit und leistet dabei eventuell Vorarbeiten f\"ur die n\"achsten Entw\"urfe. Sowohl f\"ur die Institution Li\-te\-ra\-tur als auch f\"ur das Projekt Wi\-ki\-pe\-dia kann der Entwurfsprozess in seinen Grundz\"ugen als abgeschlossen gelten. Zwar unter verschiedenen Gesichtspunkten, aber dennoch in \"ahnlicher Weise, kann das Sprechen und Schreiben \"uber Li\-te\-ra\-tur ebenso wie das Arbeiten bei und mit Wi\-ki\-pe\-dia als ein offenes Collaborative angesehen werden -- wie Marko Demantowsky und Christoph Pallaske eine zeittypische Nutzung am Bei\-spiel ge\-schicht\-li\-chen Wis\-sens beschreiben:

\singlespacing
\begin{quote}
\href{http://www.degruyter.com/view/books/9783486858662/9783486858662-001/9783486858662-001.xml}{"`Besonders die Lebenswelt junger Menschen, denen als \flq digital Natives\frq \\bereits eine eigene Generationenbe\-zeichnung zugeschrieben wird, hat sich in den letzten wenigen Jahren rasant und radikal ver\"andert. Altersty\-pische Kommunikationen finden via Instant-Messanger [sic] oder in so\-zi\-a\-len Netzwerken statt, Informationen werden offenen Collaboratives (Wi\-ki\-pe\-dia) oder Web-Angeboten entnommen, die nicht den herk\"omm\-li\-chen Reputationsregeln unterliegen (Weblogs etc.). An ge\-sell\-schaft\-li\-chen Wand\-lun\-gen und Ereignissen nimmt man spontan, \"of\-fent\-lich und in Echtzeit teil (Twitter, Online-Threads etc.). Das Ende der Gutenberg-Galaxis scheint in der Generation der heute Heranwachsenden bereits eine Tatsache zu sein."'} (\cite{DemantowskyPallaske2015}:VII)
\end{quote}
\onehalfspacing

Auf Basis der Ergebnisse der vorliegenden Arbeit werden Interessierte beider Seiten (Li\-te\-ra\-tur und Wi\-ki\-pe\-dia) eingeladen, sich am weiteren Entwurfsprozess zu beteiligen. Zu Wi\-ki\-pe\-dia sei dabei vorausgeschickt, dass die Bewirtschaftungsweise stark abh\"angig ist von den Absichten des Managements (Wikimedia Foundation) und deren Auswirkungen auf die Bereitschaft der Einzelnen in der Community, ihre Freizeit mit Wi\-ki\-pe\-dia zu verbringen, von den Arbeiten der Sysops\textsuperscript{\tiny *}, von der Anzahl und den Nutzungsgewohnheiten der Akteure\textsuperscript{\tiny *} und nicht zu\-letzt davon, wer sich bereitfindet, Wissen zusammenzutragen und beizusteuern. Zu beobachten ist, dass sich die Zusammensetzung der adhoc-Teams, die die Community letzt\-lich ausmachen, st\"andig ver\"andert, Tag f\"ur Tag, je nachdem, wer sich womit und woran beteiligt, und, um im Bild zu bleiben: wer sich "`zuhause"' f\"uhlt, wer als Gast, und wer eventuell in beiden Rollen, je nachdem. Nicht unerheblich ist, wer wie kommt, bleibt oder geht, und ob es andere mitbekommen und dar\"uber diskutieren.

\textit{Nicht-pro\-po\-si\-ti\-o\-na\-les Wissen aus Li\-te\-ra\-tur\-lekt\"ure und Bedingungen seiner Darstellbarkeit in Eintr\"agen bei Wi\-ki\-pe\-dia} arbeitet mit einer gegenseitigen Kontextualisierung von Wi\-ki\-pe\-dia und Schreibweisen \"uber Li\-te\-ra\-tur, unter anderem anhand von eigenen Interpretationen zu \mbox{vier} li\-te\-ra\-rischen Texten:
\begin{quote} 
\textit{Reisende auf einem Bein} (1989) von Herta M\"uller\\
\textit{Save the Reaper} (1998) von Alice Munro\\
\textit{Alfred and Emily} (2008) von Doris Lessing\\
\textit{rein GOLD} (2013) von Elfriede Jelinek
\end{quote}

Zu jedem der \mbox{vier} Werken be\-steht inzwischen ein Eintrag in der deutschspra\-chi\-gen Version von Wi\-ki\-pe\-dia, gr\"o{\ss}tenteils von mir verfasst. Gemeinsam ist den ausgew\"ahlten Werken und der Plattform, dass sie we\-sent\-lich textbasiert funktionieren und in der Gegenwart der letzten drei{\ss}ig Jahre verortet sind. Mit dem spezi\-fi\-schen Blickwinkel meiner Fragestellung arbeite ich sowohl in Bezug auf den Rahmen der weit \"uberwiegend nicht-Li\-te\-ra\-tur-bezogenen Plattform Wi\-ki\-pe\-dia, die einen enzy\-klop\"adischen Anspruch vertritt, als auch in Bezug auf den Rahmen Li\-te\-ra\-tur, und zwar zun\"achst in den \mbox{vier} Interpretationen von Werken aus den Jahren zwi\-schen 1989 und 2013. Da Bedingungen der Darstellbarkeit nicht-pro\-po\-si\-ti\-o\-na\-len Wissens auf beiden Seiten auszumachen sind, kann nur dann schl\"ussig argumentiert werden, wenn das Schrei\-ben \"uber li\-te\-ra\-rische Texte mit Wi\-ki\-pe\-dia kontextualisiert wird und umgekehrt das Schrei\-ben bei Wi\-ki\-pe\-dia mit Li\-te\-ra\-tur\textsuperscript{\~.\~.}. Der Begriff Li\-te\-ra\-tur\textsuperscript{\~.\~.} (mit \textsuperscript{\~.\~.}) be\-zeichnet im Rahmen dieser Arbeit das Entstehen von Emp\-fin\-dun\-gen im Lese- und/oder Zuh\"orkontakt mit li\-te\-ra\-rischen Texten. Dabei soll nicht von Belang sein, zu welchem Zeitpunkt oder aus welchem Anlass jemand dieser Emp\-fin\-dun\-gen gewahr wird. Vielmehr geht es im Kontext der vorliegenden Argumentation allein darum, dass ihr Entstehen einer bestimmten Situation zugeschrieben wird, in der ein li\-te\-ra\-rischer Text eine gewisse Rolle spielte, dessen Titel eventuell konkret benannt wird -- und vor allem: wie diese Emp\-fin\-dun\-gen in Worte gefasst werden. Mit Li\-te\-ra\-tur\-\textsuperscript{\~.\~.} ist in dieser Arbeit also keine Institution gemeint, sondern ein Geschehen, ein Ereignis, und zwar betrachtet als ein Prozess, bei dem eine un\-be\-stimm\-bare Menge an Emp\-fin\-dun\-gen entsteht. "`Lekt\"ure"' wiederum be\-zeichnet wie \"ublich die Handlung des Lesens. Der im Titel verwendete Terminus "`Li\-te\-ra\-tur\-lekt\"ure"' ist herk\"ommlich zu verstehen und meint Li\-te\-ra\-tur im Sinne von "`li\-te\-ra\-rische Werke"', hingegen ist der Begriff "`Li\-te\-ra\-tur\textsuperscript{\~.\~.}\-lekt\"ure"' folgenderma{\ss}en zu lesen: Das Nachsinnen \"uber den Prozess, dem ein Leser\textsuperscript{\tiny *} das Entstehen bestimmter Emp\-fin\-dun\-gen zuschreibt. Erst durch ein Nachsinnen \"uber die eigenen Emp\-fin\-dun\-gen aus Anlass von Lekt\"ure ergeben sich Voraussetzungen daf\"ur, dass neben Behauptungswissen in Leseberichten auch nicht-pro\-po\-si\-ti\-o\-na\-les Wissen in Worte gefasst w\"urde -- eine Art von Wissen, die vor allem in einem multidisziplin\"aren Kontext wie dem der Platt\-form Wi\-ki\-pe\-dia besonders hervorzuheben ist, weil bis\-her trotz des Anspruchs "`all human know\-ledge"' zu (re)pr\"asentieren kaum Augenmerk darauf gelegt zu werden scheint. \"Ahnlich wie das Filme-Schauen erm\"oglicht das Lesen von Li\-te\-ra\-tur den Erwerb nicht-pro\-po\-si\-ti\-o\-na\-len Wissens. Dass es sich aus diesem Grund lohnen kann, auch dem Lesen Zeit zu widmen, sollte demzufolge Inhalt von Li\-te\-ra\-tur\-ver\-mitt\-lungsinitiativen sein. 

Nun kann man von Wikpedia halten, was man will: Da die Eintr\"age bei Google vorzugsweise gerankt werden und in der Ergebnisliste ein Wi\-ki\-pe\-dia-Artikel gelegentlich erstplatziert vor Verkaufs- oder Rezensionsseiten zu finden ist (vgl. \textit{Reisende auf einem Bein}), erfordert allein dieses Faktum eine Auseinandersetzung mit den bei Wi\-ki\-pe\-dia frei im Web verf\"ugbaren Inhalten, denn sie sind einflussreich. Aus meiner Sicht ist f\"ur die Theorie und Praxis der Li\-te\-ra\-tur\-ver\-mitt\-lung inbesondere relevant, unter welchen Bedingungen nicht-pro\-po\-si\-ti\-o\-na\-les Wissen aus Li\-te\-ra\-tur\-lekt\"ure in Wi\-ki\-pe\-dia-Eintr\"agen darstellbar ist. In dieser Arbeit beobachte, gestalte, beschreibe, ana\-lysiere und diskutiere ich das multidisziplin\"ar angelegte Projekt Wi\-ki\-pe\-dia aus dem Blickwinkel der Fragestellung, welche Aspekte bez\"uglich li\-te\-ra\-rischer Werke als relevant genug und als spezifisch anzusehen sind, um in einem Eintrag dargestellt zu werden. Nicht-pro\-po\-si\-ti\-o\-na\-les Wissen aus Li\-te\-ra\-tur\-lekt\"ure ist auf Basis der \mbox{hier} vorliegenden Ergebnisse dazuzuz\"ahlen. Ich sehe demzufolge Aussagen \"uber das im eigenen Leseprozess erworbene nicht-pro\-po\-si\-ti\-o\-na\-le Wissen f\"ur jene Eintr\"age als spezifisch an, die bei Wi\-ki\-pe\-dia der Kategorie "`Li\-te\-ra\-risches Werk"' zugeordnet sind. Der Erwerb nicht-pro\-po\-si\-ti\-o\-na\-len Wissens und die formulierbaren Ergebnisse eines solchen Prozesses sollten als Teil der Rezeptionsgeschichte meines Erachtens in keinem Eintrag zu einem li\-te\-ra\-rischen Text fehlen -- sofern es dazu Belege gibt.

An dieser Stelle kommt neben dem li\-te\-ra\-rischen Text und dem Wi\-ki\-pe\-dia-Artikel ein drittes Entwurfsmuster ins Spiel, dessen Konzept ich f\"ur meine Argumentation in der vorliegenden Arbeit erfunden habe: "`Erlesnis"' (die Be\-zeichnung \mbox{selbst} verdanke ich Charlotte "`Charly"' Rudolf). \textit{Erlesnis} ist definiert als: was bei Li\-te\-ra\-tur\-lekt\"ure dem eigenen Empfinden nach an nicht-pro\-po\-si\-ti\-o\-na\-lem Wissen erworben wurde. Wird ein \textit{Erlesnis} in pu\-bli\-zierten Leseberichten beschrieben, kann es als Br\"ucke zwischen dem Wi\-ki\-pe\-dia-Artikel und dem li\-te\-ra\-rischen Text fungieren, denn der Lesebericht ist als Beleg in einem Werkartikel (Artikel \"uber ein Werk, zum Beispiel ein li\-te\-ra\-ri\-sches) verwendbar geworden. Unter recht pragmatischen Gesichtspunkten sind also \textit{Erlesnis}-Autoren\textsuperscript{\tiny *} gefragt und ihre \"Au{\ss}erungen sollten zitierbar sein. In diesem Blickwinkel nehmen bestimmte Textstellen in einem Essay von Ulrike Draesner (2013) einen besonderen Platz ein. Um f\"ur meine \"Uberlegungen Anhaltspunkte bez\"uglich der Akzeptanzneigung innerhalb der deutschspra\-chi\-gen Wi\-ki\-pe\-dia-Community zu erhalten, habe ich mir ein Experiment ausgedacht, bei dem Beschreibungen von \textit{Erlesnissen} nicht lediglich im bereits g\"angigen Abschnitt "`Rezeption"' miterw\"ahnt werden, sondern in einer neuen Art von Abschnitt, dessen Titel "`Leseerlebnis"' lautet, so dass sie als dezidiert nennenswert he\-rausgestellt werden k\"onnen -- zusammen mit anderen Arten von Leseerlebnissen.

Ich teile die Auffassung von Anne Katrin Lo\-renz be\-z\"ug\-lich der be\-son\-de\-ren Rol\-le von Li\-te\-ra\-tur an der Schnitt\-stel\-le zw\-ischen Pri\-vat\-heit und \"Of\-fent\-lich\-keit. Lorenz hat die kommunikativen Bedingungen der freim\"utigen Rede (\textit{parrhesia}) anhand li\-te\-ra\-rischer Beispiele untersucht (\cite{Lorenz2012}:360 und passim). W\"ahrend Lorenz neben der Ana\-lyse von "`parrhesiastischen"' Szenen in li\-te\-ra\-rischen Werken auch die spezifisch li\-te\-ra\-rische Ver\-mitt\-lung von \textit{parrhesia} auf der textuellen Metaebene in den Blick genommen hat, richtet sich die Fragestellung der vorliegenden Arbeit auf nicht-pro\-po\-si\-ti\-o\-na\-les Wissen durch das Entstehen von Emp\-fin\-dun\-gen bei der Lekt\"ure, und auf die Ver\-mitt\-lung dieser Art von Wissen in einer nicht \"asthetisch stilisierten Sprache in der Web\"of\-fent\-lichkeit. Es gilt nun, das ausgesprochene Selbstgef\"uhl des \textit{Erlesnisses} f\"ur schriftliche Kommunikation im web\"of\-fent\-lichen Raum in Worte zu fassen.

Schrei\-ben \"uber Li\-te\-ra\-tur\textsuperscript{\~.\~.} als ein Weg, "`sich von den Autoren inspirieren zu lassen, statt zu glauben, sie kategorisieren zu m\"ussen"' (\cite{Wertheimer2005}:76), wird im Folgenden erstens mittels eigener Interpretationen von \mbox{vier} Texten der Gegenwartsli\-te\-ra\-tur\- erprobt, zweitens in Augenschein genommen anhand von Leseberichten an\-de\-rer: in Essayb\"anden der deutschspra\-chi\-gen Autoren\textsuperscript{\tiny *} Ulrike Draesner, Hans Ulrich Gumbrecht und Ina Hartwig, drittens in anonym pu\-bli\-zierten Rezensionsnotizen bei \textit{per\-len\-tau\-cher.de} sowie viertens in web\"of\-fent\-liche Aushandlungen eingebracht in Form zusammenfassenden Dar\-stel\-lens der Aussagen an\-de\-rer im Rahmen des Entwurfsmusters Werkartikel bei Wi\-ki\-pe\-dia. Auszuhandeln ist die enzy\-klop\"adische Le\-gi\-ti\-mi\-t\"at bestimmter Aussagen Einzelner, die, obwohl subjektiv, in enzy\-klop\"adischem Stil im Rahmen der Wi\-ki\-pe\-dia-Grunds\"atze "`neutral"' dargestellt werden sollen. Die vorliegende Arbeit baut also auf der Arbeit von Lorenz auf, indem aus einem li\-te\-ra\-tur\-pragmatischen Blickwinkel der Raum f\"ur \textit{Erlesnisse} als einer neuen Art von Entwurfsmuster erkundet wird.

Die Li\-te\-ra\-tur\-ver\-mitt\-lungsfor\-schung sieht Anhaltspunkte daf\"ur, dass Wertungsroutinen in der nachb\"urgerlichen Wissens- und Erlebnisge\-sell\-schaft unter ver\"anderten, medial gepr\"agten Marktbedingungen zustandekommen. Der Erlebnischarakter von Lekt\"ure ger\"at zunehmend in den Blick. Als eine weitere Voraussetzung daf\"ur, dass neben Behauptungswissen in Leseberichten auch nicht-pro\-po\-si\-ti\-o\-na\-les Wissen in Worte gefasst w\"urde, wird in dieser Arbeit angesehen, dass indirekte Aus\-drucks\-weisen "`parrhesiastisch"' in direktere gewandelt worden sind, um nachb\"urgerlichen Verh\"altnissen entsprechend Rechnung zu tragen, im sprachlichen Ausdruck ebenso wie in einer leichteren Zug\"ang\-lichkeit zu dieser Art von Schil\-derungen -- wie sie in Leseforen bereits zu finden sind. Ein Zusammenhang zwischen Li\-te\-ra\-tur\-ver\-mitt\-lung und einer Plattform wie Wi\-ki\-pe\-dia be\-steht darin, dass die Ziele beider eng mit medialer Aufmerksamkeits\"okonomie (Georg Franck) verbunden sind. Besonders deutlich wird dies, wenn die Zugriffszahlen f\"ur einzelne Eintr\"age in die H\"ohe schnellen, sobald es in einem anderen Rahmen ein Ereignis gab, durch das Aufmerksamkeit f\"ur Autor\textsuperscript{\tiny *} oder Werk generiert wurde, etwa durch Auszeichnungen. Dass im Folgenden anhand einiger Werke von Nobelpreistr\"agern\textsuperscript{\tiny *} sowie von Lydie Salyvare (Prix Goncourt 2014) gearbeitet wird, gr\"undet in der \"Uberlegung, dass aufgrund der Prominenz vonseiten der Wi\-ki\-pe\-dia-Community mit der gr\"o{\ss}ten Bereitschaft zu rechnen ist, \"uber ein durchschnittliches Ma{\ss} hinaus au{\ss}er f\"ur biografische Daten von zu Ruhm gelangten Pers\"onlichkeiten auch an Spezifika einzelner Werke Interesse zu zeigen -- und im Zuge dessen eventuell sogar an deren Literarizit\"at, dem Kern dessen, was an Experten\textsuperscript{\tiny *}wissen \"uber Li\-te\-ra\-tur in die Aufmerksamkeit der \"Of\-fent\-lichkeit gebracht werden kann. \textit{Nicht-pro\-po\-si\-ti\-o\-na\-les Wissen aus Li\-te\-ra\-tur\-lekt\"ure und Bedingungen seiner Darstellbarkeit in Wi\-ki\-pe\-dia-Eintr\"agen zu li\-te\-ra\-rischen Werken} schl\"agt eine neue Gestaltung von Werkeintr\"agen bei Wi\-ki\-pe\-dia vor. \mbox{Selbst} wenn ich in dieser Arbeit weniger theo\-retisch als vielmehr experimentell verfahre, so liegt mir in Sachen Li\-te\-ra\-tur\-ver\-mitt\-lung \"uberwiegend daran, auf Basis meiner Ergebnisse neue Aspekte f\"ur eine Theorie der Li\-te\-ra\-tur\-ver\-mitt\-lung zusammenzutragen, um m\"ogliche neue Schlussfolgerungen anzubieten, mit der die Debatte von Grundlagen voranzutreiben w\"are.

\textit{Objektebene I} bietet eine Einf\"uh\-rung zu Wi\-ki\-pe\-dia-Spezifika f\"ur die vorliegende Fragestellung, unter anderem, indem ich Wi\-ki\-pe\-dia mit ausgew\"ahlten Aspekten der \textit{Encyclop\'{e}die ou Dictionnaire raisonn\'{e} des sciences, des arts et des m\'{e}tiers} vergleiche, wobei markante \"Ahnlichkeiten und Unterschiede aufgezeigt werden. Auf \textit{Metaebene I} er\"ortere ich Konzepte meiner Argumentation, stelle zweitens anhand von Aussagen, die mir auf beiden Seiten als ty\-pisch erscheinen, das Feld dar, auf das sich mein Erkenntnisinteresse richtet und f\"ur das ich die Ergebnisse der Arbeit fruchtbar machen m\"ochte. Drittens f\"uhre ich anhand eigener Interpretationen von \mbox{vier} zeitgen\"ossischen li\-te\-ra\-rischen Texten vor, wie \textit{Erlesnisse} in schriftlicher Form Teil eines Beitrags sein k\"onnten, der in pu\-bli\-zierter Form f\"ur einen Wi\-ki\-pe\-dia-Eintrag als belegf\"ahig gelten k\"onnte. \textit{Objektebene II} dient dazu, mein Experiment in der deutschspra\-chi\-gen Wi\-ki\-pe\-dia-Community im Detail vorzustellen, zu erl\"autern und auszuwerten, um die unausgesch\"opften Potenziale aufzuzeigen. Im abschlie{\ss}enden Kapitel des Hauptteils, \textit{Metaebene II}, stelle ich die Bedeutung von \textit{parrhesia} im Schrei\-ben \"uber Li\-te\-ra\-tur\textsuperscript{\~.\~.} dar und er\"ortere Schreibweisen, mit denen \"uber \textit{Erlesnisse} berichtet wird, in typologischer Hinsicht. Darauf aufbauend formuliere ich Erkenntnisse f\"ur die Theorie der Li\-te\-ra\-tur\-ver\-mitt\-lung und stelle sie zur Debatte. Auf den beiden Objektebenen befasse ich mich also aus literaturwissenschaftlichem Blickwinkel mit Wikipedia, hingegen auf den beiden Metaebenen mit Literatur sowie mit Literaturwissenschaft, Literaturkritik und Literaturvermittlungsforschung.

\textit{Nicht-pro\-po\-si\-ti\-o\-na\-les Wissen aus Li\-te\-ra\-tur\-lekt\"ure und Bedingungen seiner Darstellbarkeit in Wi\-ki\-pe\-dia-Eintr\"agen zu li\-te\-ra\-rischen Werken} erforscht \textit{Erlesnisse} als Entwurfsmuster f\"ur die Darstellung nicht-pro\-po\-si\-ti\-o\-na\-len Wissens -- bezogen auf das Verh\"altnis zwischen Wi\-ki\-pe\-dia als einem kollektiv zu bewirtschaftenden Geb\"aude, das mit Wiki-Software betrieben wird, und Li\-te\-ra\-tur\textsuperscript{\~.\~.} als einem individuellen prozessualen Ereignis, bei dem eine unbestimmbare Menge an Emp\-fin\-dun\-gen entsteht -- mit einer Kombination aus drei theo\-retischen Perspektiven: Nicht-pro\-po\-si\-ti\-o\-na\-les Wissen, \textit{parrhesia} und Li\-te\-ra\-tur\-ver\-mitt\-lung.

\pagebreak

\section{Research Design}
\label{sec:2}

\subsection{Erkenntnisziele}
\label{subsec:2.1}

Die Fragestellung der vorliegenden Arbeit zielt erstens darauf he\-rauszufinden, unter welchen Bedingungen sich im multidisziplin\"aren Aushandlungsraum Wi\-ki\-pe\-dia nicht-pro\-po\-si\-ti\-o\-na\-les Wissen aus Li\-te\-ra\-tur\-lekt\"ure dar\-stel\-len l\"asst, also auf Basis von \textit{Erlesnissen} Dritter. Dass zu diesem Zweck auf der Seite der Li\-te\-ra\-tur ge\-eig\-nete \"Au\-{\ss}e\-run\-gen pu\-bli\-ziert worden sein m\"ussen, liegt auf der Hand. Im Zuge der Fragestellung wird aber damit begonnen, an einer Typologie von \textit{Erlesnissen} zu arbeiten, um kooperationsbereiten Li\-te\-ra\-tur\textsuperscript{\~.\~.}Experten\textsuperscript{\tiny *} erste Erkenntnisse an die Hand zu geben. 

He\-rausfinden m\"ochte ich zweitens, ob ich den von mir neu gepr\"agten Begriff \textit{Erlesnis} f\"ur ge\-eig\-net halte, um ihn in der Li\-te\-ra\-tur\-ver\-mitt\-lungsfor\-schung in weitergehende Debatten einzubringen. Drittens soll erwiesen werden, dass die gew\"ahlte Methode ge\-eig\-net ist, um die gestellten Fragen zu l\"osen.

\subsection{Pr\"amissen}
\label{subsec:2.2}

Ausgangspunkt meines For\-schungsinteresses war, dass ich die von mir beobachtete Ignoranz auf beiden Seiten des Verh\"altnisses von Li\-te\-ra\-tur und Wi\-ki\-pe\-dia als ein interessantes Ph\"anomen von gro{\ss}er ge\-sell\-schaftlicher Relevanz erachtete, denn f\"ur das kulturelle Leben sind die Impulse beider nicht mehr wegzudenken. Meine Motivation war daher, eine n\"ahere Verbindung zu suchen, die pragmatisch ausgestaltet werden k\"onnte.

Zweitens hatte ich die Einsch\"atzung, dass es sich bei Wi\-ki\-pe\-dia um eine Plattform handelt, die daf\"ur ge\-eig\-net sein w\"urde, die \"of\-fent\-liche Wahrnehmung li\-te\-ra\-rischer Kunst in ihrer Vielf\"altigkeit und Deutungsoffenheit zu f\"ordern, vor allem auf Seiten von Lesern\textsuperscript{\tiny *}, die mit einer Websuche erste Informationen \"uber ein be\-stimmtes li\-te\-ra\-risches Werk zu finden hoffen. Auf Basis meiner Beobachtungen seit April 2005, als ich \mbox{selbst} begann, bei Wi\-ki\-pe\-dia mitzuschrei\-ben, vertrete ich die Auffassung, dass Wi\-ki\-pe\-dia zu selten und zu wenig fachlich fundiert daf\"ur genutzt wird, in einem multidisziplin\"aren Kontext wie diesem erstens die Funktionsweise der Li\-te\-ra\-tur\-wissenschaftlichen Be\-weis\-f\"uh\-rung mittels Arbeit am Text in Eintr\"agen zu li\-te\-ra\-rischen Werken anhand von Textstellen aufzuzeigen, zweitens meh\-rere interpretatorische Stimmen zu demselben Werk vorzustellen und drittens Berichte von Lesern\textsuperscript{\tiny *} \"uber ihr Erleben w\"ahrend der Lekt\"ure zusammenzufassen. Dabei k\"onnten gerade auf dieser Plattform leicht verst\"andliche Einsch\"atzungen zu einem Werk zur Geltung gebracht werden, um weniger bewanderte Leser\textsuperscript{\tiny *} dar\"uber zu informieren, warum andere Li\-te\-ra\-tur lesen und warum gerade dieses Werk. Erfahrene Leser\textsuperscript{\tiny *}, die bereits wissen, warum sie lesen, sind es gewohnt, sich an Stimmen der Li\-te\-ra\-tur\-kritik oder der Li\-te\-ra\-tur\-wissenschaft zu orientieren, denen allgemein von der gebildet lesenden \"Of\-fent\-lichkeit aufgrund des Publikationsorgans oder der Namhaftigkeit schon vorab viel Bedeutung bei\-ge\-mes\-sen wird. Vermutbar ist, dass auch diese Gruppe der Interessierten bei Wi\-ki\-pe\-dia erste und weiterf\"uhrende Informationen \"uber ein be\-stimmtes Werk sucht, und ich denke, sie sollten ebenfalls gut bedient werden -- so die \"Uberlegungen im Hintergrund meines Zugangs zum Thema der vorliegenden Arbeit.

Eine weitere Pr\"amisse ist, dass f\"ur ein kritisches Nachdenken zum \"of\-fent\-lich vor\-herr\-schenden Machtdiskurs (bei Wi\-ki\-pe\-dia derjenige, dass (allein) pro\-po\-si\-ti\-o\-na\-les Wissen pr\"asentierbar sei) in der Privatsph\"are des Individuums ein k\"unstlerischer Freiraum bestehen muss und Li\-te\-ra\-tur\-lekt\"ure als ein Gestaltungsmittel f\"ur diesen Freiraum anzusehen ist (\cite{Lorenz2012}:369). Mit einer li\-te\-ra\-tur\-pragmatischen Ausrichtung gehe ich davon aus, das im Beispielkontext Wi\-ki\-pe\-dia ein solcher Freiraum trotz des \href{https://de.wikipedia.org/wiki/Wikipedia:Neutraler_Standpunkt}{"`NPOV"'}-Grund\-satzes (web)\"of\-fent\-lich vermittelbar ist, auch wenn die Forderung, einen "`neutraler Standpunkt"' einzunehmen, als eine gewisse H\"urde f\"ur die Ver\-mitt\-lung von nicht-pro\-po\-si\-ti\-o\-na\-lem Wissen aus Li\-te\-ra\-tur\-lekt\"ure wahrgenommen werden kann.

Die vierte Pr\"amisse betrifft den multidisziplin\"aren Aushandlungsraum: Zwar be\-an\-sprucht die Wi\-ki\-me\-dia Foun\-da\-tion seit 2007, in Zukunft \href{http://wikimediafoundation.org/w/index.php?title=Vision\&oldid=21860}{"`The sum of all know\-ledge"'} bzw. \href{https://en.Wikipedia.org/w/index.php?title=Wikipedia:Prime_objective&oldid=497516953}{"`The sum of all human know\-ledge"'} als "`Prime objective"' von Wi\-ki\-pe\-di\-a zu bieten -- wobei als Quelle f\"ur diese Formulierung ein Interview mit Co-Gr\"under Jimmy Wales von Juli 2004 in \textit{Slashdot} angegeben wird --, um welches Wissen es dabei gehen soll, ist al\-ler\-dings nicht n\"aher definiert worden. Der einzige mir bis\-her bekannt gewordene Einwand zu diesem Punkt wurde 2010 von Finn \r{A}rup Nielsen erhoben: Daten seien auch Wissen, f\"anden bei Wi\-ki\-pe\-dia aber keine angemessene Darstellung (\cite{Nielsen2010}). Die vorliegende Arbeit setzt an dem Punkt an, dass "`all human know\-ledge"' nicht ausschlie{\ss}lich im Sinne von "`gro{\ss}e Menge"' gelesen, sondern das unbestimmte Nummeral "`all"' im We\-sent\-lichen qualitativ interpretiert wird: "`alle Arten des Wissens"'. Folgende \"Uberlegung liegt dem zugrunde, gespeist aus li\-te\-ra\-tur\-philosophischer Sicht: Wenn die M\"oglichkeit, eine besondere Art von Wissen zu erwerben, ein Spezifikum von Li\-te\-ra\-tur\-lekt\"ure ist, m\"ussen in Wi\-ki\-pe\-dia-Eintr\"agen zu li\-te\-ra\-rischen Werken \"uber Beispiele des Effekts von nicht-pro\-po\-si\-ti\-o\-na\-lem Wissen einige Informationen zu finden sein, sofern Li\-te\-ra\-tur\-ver\-mitt\-lung eine allgemeine \"Of\-fent\-lichkeit davon \"uberzeugen soll, dass sich Lesen von Li\-te\-ra\-tur lohnt. Die vorliegende Arbeit zeigt auf, in welchem Ma{\ss}e Wi\-ki\-pe\-dia als eine Plattform angesehen werden kann, auf der sich in einem multidisziplin\"aren Kontext und in allgemeinverst\"andlicher Sprache neben anderen Li\-te\-ra\-tur\-spezifischen Aspekten vor allem dieser besondere Effekt von Li\-te\-ra\-tur\textsuperscript{\~.\~.}lekt\"ure dar\-stel\-len l\"asst. Wenn also in Wi\-ki\-pe\-dia-Eintr\"agen zu li\-te\-ra\-rischen Werken Angaben \"uber den Effekt der Lekt\"ure zu finden w\"aren, w\"urde erstens nicht nur deutlich, dass Li\-te\-ra\-tur mehr als nur etwas Objekthaftes ist (\cite{Attridge2004}). Je gr\"o{\ss}er also die Variabilit\"at und Bandbreite der wiedergegebenen Eindr\"ucke zu einem li\-te\-ra\-rischen Text, desto eindr\"ucklicher kann die Art und Tiefe der Wirkm\"achtigkeit von Li\-te\-ra\-tur in einem enzy\-klop\"adischen Eintrag der Marke Wi\-ki\-pe\-dia zum Ausdruck gebracht werden, so mein Ausgangspunkt. In Bezug auf die Darstellbarkeit nicht-pro\-po\-si\-ti\-o\-na\-len Wissens aus Li\-te\-ra\-tur\-lekt\"ure wird argumentiert, dass es al\-ler\-dings auch Bedingungen gibt, die nicht allein auf Seiten der Wi\-ki\-pe\-dia-Community erf\"ullt sein m\"ussen, damit die Ver\-mitt\-lung der Bedeutsamkeit dieser Besonderheit von Li\-te\-ra\-tur\-lekt\"ure anhand von Beispielen gelingen kann.

Ich gehe ferner davon aus, dass mittels guter Darstellungen bei Wi\-ki\-pe\-dia das Lesen von Li\-te\-ra\-tur f\"ur den Blickwinkel einer allgemein interessierten \"Of\-fent\-lichkeit plausibilisiert werden kann. Indem vielf\"altige und einander widersprechende Interpretationen einzelnder Werke pr\"asentiert werden, kann indirekt auch die Bedeutsamkeit der ge\-sell\-schaftlichen Rolle von Li\-te\-ra\-tur gesteigert werden. Gute Darstellungen w\"urden indirekt auch das ge\-sell\-schaftliche Demokratisierungspotential aufzeigen, das in offener -- und vor allem web\"of\-fent\-licher -- Debatte \"uber Li\-te\-ra\-tur zur Geltung gebracht werden kann, erst recht in einem Enzy\-klop\"adieartikel, denn diesem Format wird allgemein ein hoher Geltungsanspruch zugestanden. Ich fragte mich demzufolge, wie am besten zu argumentieren w\"are, dass es sich lohnt, Strategien zu entwickeln, um erwerbsorientierte Akteure\textsuperscript{\tiny *} der Li\-te\-ra\-tur\-ver\-mitt\-lung daf\"ur zu gewinnen, Ressourcen in ein web\"of\-fent\-lich zug\"ang\-liches Projekt zu investieren, mit dem Ziel des Verstehbar- und Zug\"ang\-lichmachens li\-te\-ra\-rischer Qualit\"aten, so dass andere ahnen, was sie an eigenen Lekt\"ureerlebnissen verpassen k\"onnten. Daf\"ur w\"urde nicht zu\-letzt das Lesen von Li\-te\-ra\-tur als eine anregende Besch\"aftigung erscheinen m\"ussen. Dies aufzuzeigen w\"are insbesondere anhand von nicht-pro\-po\-si\-ti\-o\-na\-lem Wissen m\"oglich, dessen Erwerb in (schriftlichen) \"Au{\ss}erungen der Lekt\"ure eines bestimmten Werks zugeschrieben worden ist. Ein namhafter \textit{Erlesnis}-Beleg im Wiki\-pedia-Ein\-trag "`Bekenntnisse des Hoch\-stap\-lers Felix Krull"' w\"are zum Beispiel Ulrike Draesners Essay "`Gedanken zum Altern anl\"asslich der Wiederlekt\"ure von Thomas Manns letztem Roman"' (\cite{Draesner2013}). 

Ich argumentiere in diesem Zusammenhang, dass, auch wenn \mbox{(Nicht-)}\-Lese\-ge\-wohn\-hei\-ten durch das Pr\"asentieren an\-de\-rer Inhalte im Web vermutlich kaum ver\"andert werden k\"onnen, es bei Wi\-ki\-pe\-dia Gestaltungsm\"oglichkeiten gibt, deren Effekt f\"ur das Ansehen von Lekt\"ure positiv ist, sei es, weil ein gr\"o{\ss}eres Ma{\ss} an Anerkennung f\"ur eine T\"atigkeit wie Li\-te\-ra\-tur\-lekt\"ure hervorgerufen wird oder sei es, um anzuzeigen, wie einem besonderen Wissen, das durch Lekt\"ure entstehen kann, durch Leute, die bereit sind, sich \"uber ihre Lekt\"ureerfah\-rung \"of\-fent\-lich zu \"au{\ss}ern, Wertsch\"atzung entgegengebracht wird. Eine These, die ich \mbox{hier} als Pr\"amisse einsetze, zu der ich aber lieber eine Studie lesen w\"urde, lautet: Durch gute Wi\-ki\-pe\-dia-Eintr\"age kann zumindest ein Interesse an dem gesteigert werden, was andere \"uber Li\-te\-ra\-tur beziehungsweise sogar \"uber Li\-te\-ra\-tur\textsuperscript{\~.\~.} \"of\-fent\-lich zu sagen bereit sind, \mbox{selbst} wenn die Anzahl der Menschen nicht zunehmen wird, die nicht in jedem Fall dem Rezipieren von Filmen den Vorzug geben, sondern habituell \mbox{selbst} Li\-te\-ra\-tur lesen oder sogar Li\-te\-ra\-tur\textsuperscript{\~.\~.}lekt\"ure betreiben. 

Als die jeweiligen Kernbereiche von Li\-te\-ra\-tur\- und Wi\-ki\-pe\-dia sehe ich an, dass Akteure\textsuperscript{\tiny *} im Bereich Li\-te\-ra\-tur vor allem anstreben, dass li\-te\-ra\-rische Texte gelesen werden und Gegenstand kultureller Debatten sind, und Akteure\textsuperscript{\tiny *} im Bereich Wi\-ki\-pe\-dia vor allem, dass "`The Sum of All Know\-ledge"', die Summe allen Wissens, in allgemein verst\"andlicher Form dargestellt und frei im Web verf\"ugbar wird. Der Titel der Arbeit fokussiert auf Bedingungen der Darstellbarkeit nicht-pro\-po\-si\-ti\-o\-na\-len Wissens aus Li\-te\-ra\-tur\-lekt\"ure in Eintr\"agen bei Wi\-ki\-pe\-dia, wodurch es so scheinen k\"onnte, als ob das eingangs beschriebene Ph\"anomen damit erkl\"arbar geworden w\"are. Die Pro\-blem\-stel\-lung und Eingrenzung des Gegenstands der Arbeit wurde aber vorgenommen, um an einem bestimmten Punkt in der Gemengelage die Chance f\"ur ein ver\"andertes Verh\"altnis zu erkunden: Aus Sicht der Institution Li\-te\-ra\-tur im Rahmen von "`Li\-te\-ra\-tur\-ver\-mitt\-lung"' -- bzw. in der \mbox{hier} argumentierten notwendigen Konzeption als Li\-te\-ra\-tur\textsuperscript{\~.\~.}ver\-mitt\-lung -- und aus Sicht der Marke Wi\-ki\-pe\-dia mit besseren Inhalten f\"ur den vermuteten Informationsbedarf derjeinigen, die sich freuen, bei einer Websuche zu sehen, dass es zu diesem Text schon einen eigenen Eintrag bei Wi\-ki\-pe\-dia gibt.

\subsection{Eingrenzung des Themas}
\label{subsec:2.3}

Techniken und Dynamiken des Sich-Beteiligens bei Wi\-ki\-pe\-dia sind ein eigenes Thema, dem in dieser Arbeit nicht nachgegangen wird. Dazu ist bereits vielsei\-tig ge\-forscht worden, wenn auch zu selten in Bezug auf die Kooperations- und Streitkulturen in der deutschspra\-chi\-gen Wi\-ki\-pe\-dia-Version. Grundlagen f\"ur das notwendige Minimum an Erl\"auterungen zum Verst\"andnis dessen, was anschlie{\ss}end argumentiert werden soll, sch\"opfe ich aus meiner Kenntnis einiger Studien zu Wi\-ki\-pe\-dia, aus Dis\-kus\-sio\-nen auf der For\-schungsmailing\-liste und auf diversen Projektdis\-kus\-si\-onsseiten der Wikimedia Foundation sowie aus eigenen Beobachtungen vor allem in der Phase seit 2012, der Zeit, in dem dieses Projekt seinen Anfang nahm. Aber ich \mbox{selbst} beforsche diese Beteiligungsprozesse nicht als solche, sondern beschreibe und kommentiere sie dort, wo sie f\"ur ein Verst\"andnis des unmittelbaren Kontexts relevant sind, in dem das "`Experiment"' durchgef\"uhrt wird.

Prinzipiell stimme ich mit Ruth Page \"uberein, wenn es mit Bezug zu Wi\-ki\-pe\-dia in der Frage "`how contemporary counter narratives might emerge and function"' hei{\ss}t:

\singlespacing
\begin{quote}
"`The mechanisms by which counter and dominant narratives negotiate their position through structure and tellership will be dependent on context, and requires an ana\-lytical approach that can take account of the process of narrative production and interaction, not just a textual ana\-lysis of the narrative as a decontextualized product."' (\cite{Page2014}:74-75)
\end{quote}
\onehalfspacing

Innerhalb des Kontextes der Plattform Wi\-ki\-pe\-dia ana\-lysiere ich den\-noch \"uber\-wie\-gend Texte (und nicht \"uberwiegend Handlungen), denn mein Erkenntnisinteresse sind Schreibweisen (\"uber Li\-te\-ra\-tur im Rahmen von Li\-te\-ra\-tur\-ver\-mitt\-lung) und nicht Handlungsweisen (etwa zur Durchsetzung einer m\"oglichst breiten Akzeptanz f\"ur das Pr\"a-sentieren auch nicht-pro\-po\-si\-ti\-o\-na\-len Wissens in Werkeintr\"agen). Auch wenn diese eventuell ma{\ss}geblich zu Akzeptanz beitragen k\"onnen, sind sie in der vorliegenden Arbeit nur insofern Thema, als im Rahmen meines Experiments die Klaviatur der verschiedenen Wi\-ki\-pe\-dia-Seiten teils virtuos bedient werden musste -- durch alle Beteiligten. Zu den Handlungsoptionen, die in diesem Kontext spezifisch sind, z\"ahlt ein Gew\"ahren-Lassen (Siehe hierzu auch die Abschnitte \textit{~\ref{subsubsec:6.1.4} "`Pas pleurer"' (Salvayre 2014)} und \textit{~\ref{subsubsec:6.1.11} "`Gr\"aser der Nacht"' (Modiano 2012)}).

Abgegrenzt ist meine Fragestellung auch von der Thematik "`Li\-te\-ra\-tur im Netz"'. Zwar steht im Fokus der Arbeit als Publikationsort ebenfalls das Web, aber nicht Texte, die einen \"asthetisch stilisierten Anspruch haben und \"uberwiegend mit fiktionalen Mitteln arbeiten. Im \mbox{hier} gew\"ahlten spezifischen Blickwinkel werden li\-te\-ra\-rische Texte lediglich als Anlass f\"ur das informierende Schrei\-ben "`\"uber sie"' betrachtet. Genauer gesagt ist in textueller Hinsicht allein dasjenige von Interesse f\"ur meine Untersuchung, was Per\-so\-nen, die schrei\-ben, \"uber ihren nicht-pro\-po\-si\-ti\-o\-na\-len Wissenserwerb durch das Lesen berichten, \"uber neue Erkenntnisse, die sie dem Lek\-t\"u\-re\-pro\-zess zuschrei\-ben.

Diese Arbeit hat ebensowenig mit Lesefor\-schung zu tun -- auch nicht mit Leseprozessen oder der Entstehung von Lesemotivationen derjenigen, die \"uber ein \textit{Erlesnis} schrei\-ben. In kultursoziologischer Perspektive hat Christian Dawidowski 2009 he\-rausgefunden, dass Verstehen als Lekt\"ureerlebnis eine Rolle spielt und in manchen F\"allen Fiktion an sich ausreichend Material bietet -- wie er in einem von 20 narrativen Interviews he\-rausgearbeitet hat: 

\singlespacing
\begin{quote}
"`Die Art und Weise, in der [die Interviewte] Anteil an der fiktionalen Welt nimmt, beinhaltet jedoch nicht -- wie im Fall der anderen Informanten -- die Minimierung des Skopus: Sie versucht nicht, die Strukturen und Inhalte der fiktionalen Welt auf die eigene Lebenswelt herunterzubrechen und ihr zu applizieren, sondern sie l\"asst diese Welt in ihrer Eigenheit bestehen und bewegt sich auf sie zu, um sie zu erkunden und \flq auf Zeit\frq\,\,zu bewohnen [...] sie flexibilisert und mobilisert die eigene Person, nicht die textliche Gegenwelt."' (\cite{Dawidowski2009}:264)
\end{quote} 
\onehalfspacing

Zuvor hatte Werner Graf in seiner Untersuchung \textit{Der Sinn des Lesens. Modi der li\-te\-ra\-rischen Rezeptionskompetenz} 2004 he\-rausgestellt, dass sich gerade f\"ur Lesemodi, die auf Erkenntnis und auf \"asthetischen Genuss hin orientiert sind, wenige empirische Belege finden lassen: 

\singlespacing
\begin{quote}
"`Vom interesse- und theoriegeleiteten Konzeptlesen \"offnen sich \"Uber\-gangs\-punk\-te zu zwei weiteren Lesemodi, f\"ur die sich empirisch wenige, aber eindrucksvolle Belege finden. Was die Li\-te\-ra\-tur\-wissenschaft fast als einzige Form des Lesens w\"urdigt, liegt in repr\"asentativ quantitativen Erhebungen kaum \"uber der Nachweisgrenze. Aber es treten Lesende auf, die beim Lesen z.B. die Sch\"onheit der Sprache erfreut, die Texte \"asthetisch wahrnehmen. Neben \"asthetischen Erfah\-rungen werden diskursive Erkenntnism\"oglichkeiten beschrieben, die sich auf unterschiedliche Textsorten beziehen. Der Reiz geht von Neuem aus, dem Versprechen, mit der Li\-te\-ra\-tur Erfah\-rungen machen zu k\"onnen -- auch irritierende. Das Lesen als Erkenntnismodus setzt spezifische textana\-lytische Lesekompetenzen voraus."' (\cite{Graf2004}:22)
\end{quote} 
\onehalfspacing

Dies d\"urfte sich dank derjenigen, die etwa in Leseforen eingehend pers\"onlich berichten, ins Positive gewandelt haben. In welchem Lesemodus oder welcher Bewegungsrichtung zwischen Leser\textsuperscript{\tiny *} und Text jemand zu dem gelangt, wor\"uber der Text verfasst wird, bleibt in meiner Er\"orterung au{\ss}en vor. Ich gehe davon aus, dass Leser*, die \"uber ein \textit{Erlesnis} schrei\-ben, etwas erlebt haben, und wie sie es bis zu ihrem Erlebnis gebracht haben, ist entweder Teil des Berichts oder nicht. Dass sich in einem Lesemodus, bei dem ein li\-te\-ra\-risches Werk als Sprachkunst wahrgenommen wird, \textit{Erlesnisse} ergeben k\"onnen wie ich sie f\"ur Li\-te\-ra\-tur\-lekt\"ure als spezifisch ansehe (und daher als besonders relevant f\"ur einen enzy\-klop\"adischen Eintrag in einem multidisziplin\"aren Kontext), sei dabei nicht in Abrede gestellt, sondern vielmehr erhofft.

Wie Geschriebenes zustandekommt ist Gegenstand vieler Untersuchungen im Feld der Schreibfor\-schung. Auch von diesem umfassenden Aspekt verwandter For\-schungs\-inte\-res\-sen, bei denen unter anderem mit psychologischen Methoden gearbeitet wird, ist die Fragestellung der vorliegenden Arbeit abgegrenzt. Eine vergleichende Ana\-lyse der jeweiligen Produktionslogiken des Schrei\-bens als gemeinsamer Basis von Wissen(schaft) und Li\-te\-ra\-tur (vgl. \cite{Kammer2013}) w\"are als Hintergrund f\"ur meine Arbeit sehr interessant -- vor allem wenn das Schrei\-ben von enyzklop\"adischen Eintr\"agen im web\"of\-fent\-lichen Raum Teil einer solchen Studie w\"are --, ich unternehme sie aber \mbox{hier} nicht \mbox{selbst}.

Ich befasse mich auch nicht damit, ob es bestimmte Voraussetzungen gibt oder Arten von Bereitschaft auf Leser\textsuperscript{\tiny *}seite, die grundlegend sind, um \textit{Erlesnisse} in schriftlicher Form dar\-stel\-len zu k\"onnen -- wie sie etwa von Terry Eagleton formuliert wurden:

\singlespacing
\begin{quote}
"`There is no literary work, as opposed to certain material objects known as books, without the \flq actualisations\frq \,\,of a reader, but this activity is not self-determining. Though by no means prescribed by the structures of the texts themselves, it is nonetheless cued, guided, and constrained by them. (This, one might note, is the key difference between Eco's approach and the bold-faced philosophical idealism of Stanley Fish.) In decoding the work, the reader brings to bear on it a certain general competence; but this rule-governed set of capacities is realised in a unique and distinctive way through the actual performance of the reading, to the point where competence and performance become hard to distinguish."' (\cite{Eagleton2012}:192)
\end{quote}
\onehalfspacing

Wodurch auch immer Leser\textsuperscript{\tiny *} in die Lage versetzt werden, li\-te\-ra\-rische Texte mit Genuss zu rezipieren -- f\"ur mein Erkenntnisinteresse ist vor allem interessant, wie dar\"uber geschrieben wird, und am besten \"uber das, was bei Li\-te\-ra\-tur\-lekt\"ure dem eigenen Empfinden nach an nicht-pro\-po\-si\-ti\-o\-na\-lem Wissen erworben wurde, also wie vermutlich Li\-te\-ra\-tur\textsuperscript{\~.\~.}lekt\"ure betrieben wurde. Die Erfah\-rung zeigt, dass sich neue Genres leichter entwickeln, wenn diejenigen, die in diesen Formaten Texte produzieren wollen, schon wissen, f\"ur welche Zusammenh\"ange ihre Ergebnisse willkommen sind. Mein Interesse ist, \textit{Erlesnisse} als neue Inhalte f\"ur Leseberichte hervorgebracht zu sehen, denn ohne dass sie als Belege vorhanden sind, ist es mir nicht m\"oglich, he\-rauszufinden, unter welchen Bedingungen und auf welche Weise bei Wi\-ki\-pe\-dia \"uber \textit{Erlesnisse} auf einzelne li\-te\-ra\-rische Texte bezogen informiert werden kann.

Naheliegenderweise m\"ussen f\"ur meinen Zweck Leseberichte geschrieben worden sein, denn ein Schrei\-ben \"uber Li\-te\-ra\-tur mittels Aussagen an\-de\-rer in Wi\-ki\-pe\-dia-Artikeln ist davon abh\"angig, welche Belege herangezogen werden k\"onnen -- von \"Au{\ss}erungen hinreichend namhafter Per\-so\-nen an publizistisch oder wissenschaftlich hinreichend namhaften Stellen. Daher erschien es mir auch notwendig, die "`Au{\ss}enseite"' genauer in Augenschein zu nehmen und Beispiele ver\"of\-fent\-lichter Texte zu er\"ortern.

Nicht hingegen befasse ich mich mit der Frage nach den Ursachen daf\"ur, dass \textit{Erlesnisse} in namhaft ver\"of\-fent\-lichter Form allem Anschein nach kaum zu finden sind. Pierre Bourdieu thematisiert den zeitgen\"ossischen wissenschafts\-soziologischen Kontext mit seinen Wertsch\"a\-tzungs\-be\-dingungen bzw. -- unter Bezugnahme auf Bateson -- die evokative Macht des gew\"ahlten sprachlichen Stils f\"ur die Sozialfor\-schung bzw. historische Sozialfor\-schung\footnote{"`[L]e \flq s\'{e}rieux\frq, dans la science commune aillieurs, est une vertue typiquement sociale, et ce n'est pas par hazard qu'on l'accorde par priorit\'{e} \`{a} ceux qui, dans leur style de vie que comme leur style de travaux, donnent les garanties de pr\'{e}visibilit\'{e} et de calculabilit\'{e} caract\'{e}ristiques des gens \flq responsables\frq, pos\'{e}, rang\'{e}s."'\\
"`Quant aux sociologues, ils trahissent souvent leur pr\'{e}tension \`{a} l'h\'{e}gemonie (inscrite d\`{e}s l'origine dans la classification comtiste des sciences) en empruntant alternativement ou simultan\'{e}ment aux rh\'{e}toriques les plus puissantes dans les deux champs par rapport auxquels ils se sont oblig\'{e}s de se situer, celle de la math\'{e}matique, souvent utilis\'{e}e comme signe ext\'{e}rieur de scientificit\'{e}, ou celle der la philosophie, souvent r\'{e}duite a les effets de lexique."'\\
\textsuperscript{35} "`Ceci ne signifie pas que la rechercher proprement \flq litt\'{e}raire\frq \,\,ne puisse trouver une signification scientifique. Ainsi, comme le remarquai Bateson \`{a} propos de l'ethnologue, la puissance \'{e}vocatrice du style constitue une des formes ind\'{e}passables de l'accomplissement scientifique lorsqu'il s'agit d'objectiver les traits pertinents d'une configuration sociale et de livrer par l\`{a} les principes de l'appr\'{e}hension syst\'{e}matique d'une n\'{e}cessit\'{e} historique: quand l'historien du Moyen Age \'{e}voque, par l'\'{e}fficacit\'{e} propre du langage, l'isolement et la d\'{e}solation de ces paysans qui, repli\'{e}s sur les ilots de terroir d\'{e}frich\'{e}s, sont livr\'{e}s \`{a} toutes les terreurs, il vise d'abord \`{a} reproduire pour le lecteur, dans et par des mots capables de produire un effet de r\'{e}alit\'{e}, le renouvellement de la vision qu'il a d\^{u} op\'{e}rer, contre les concepts-\'{e}crans et les automatismes de pens\'{e}e, pour acc\'{e}der \`{a} une compr\'{e}hension juste des \'{e}tranget\'{e}s de la culture carolignienne"' (\cite{Bourdieu1984}:46/Fn. 35).} und -- bezogen auf das Feld der Li\-te\-ra\-tur\-for\-schung -- formuliert Peter Lamarque zudem die Vermutung einer Sensibilit\"atsab\-nah\-me durch Verlauf und Dauer interpretatorischer Arbeit als einen weiteren m\"oglichen Grund daf\"ur, warum in der Li\-te\-ra\-tur\-for\-schung kaum \"Au{\ss}erungen \"uber eigene Gef\"uhls\-re\-ak\-tio\-nen zu lesen sind.\footnote{"`[B]roadly speaking, academic literary critics tend to give little attention to (psycholo\-gically real) emotional responses to li\-te\-ra\-ture. No doubt there are a number of reasons for this: partly that such critics are wary of responses that might seem subjective, variable and unmeasurable; partly, and perhaps more deeply, because if you study a work in detail, with repreated re-readings, perhaps doing fine-grained ana\-lyses of particular passages or exploring social, political and ethical forces at work, any strong emotional responses associated with a first-time reading are likely to have worn off"' (\cite{Lamarque2014}:196).} Ich belasse es an diesem Punkt dabei, diese Aussagen als Fu{\ss}noten anzuf\"uhren, denn ich bringe sie im Rahmen dieser Arbeit nicht weiter zum Einsatz. Als Hypothesen f\"ur eine speziellere Studie k\"onnten sie jedoch interessant sein.

Ebenso richtet sich mein Erkenntnisinteresse nicht auf die Frage, ob Texte Wissen enthalten k\"onnen oder nur Per\-so\-nen Wissen haben. Es geht mir also nicht um die Kl\"arung erkenntnistheo\-retischer Fragen, sondern ich lege zugrunde, was Per\-so\-nen an Erlebnissen bzw. \textit{Erlesnissen} in Worte fassen und \"uberlasse es der eigenen Einsch\"atzung der Verfasser\textsuperscript{\tiny *}, welches die Quelle des nicht-pro\-po\-si\-ti\-o\-na\-len Wissens ist, \"uber das sie schrei\-ben. Al\-ler\-dings gehe ich davon aus, dass im Lese- und/oder Zuh\"orkontakt mit li\-te\-ra\-rischen Texten Emp\-fin\-dun\-gen entstehen und ich lege meinem Konzept \textit{Erlesnis} die Annahme zugrunde, dass w\"ahrend der Li\-te\-ra\-tur\-lekt\"ure nicht-pro\-po\-si\-ti\-o\-na\-les Wissen erworben werden kann. Ob die Lekt\"ure in diesem Fall der einzige Faktor ist, interessiert mich nur in dem Ma{\ss}e wie die \textit{Erlesnis}-Autoren* es schil\-dern: Geht aus ihrem Text zum Beispiel hervor, dass sie der Auffassung sind, dass der Text, den sie gelesen haben, Eigenschaften hat, die zu ihrem \textit{Erlesnis} gef\"uhrt haben, kann dar\"uber in einem Wi\-ki\-pe\-dia-Eintrag berichtet werden, ohne dass ich entscheiden muss, ob ich das Zuschrei\-ben von Eigenschaften angemessen finde oder nicht. Funktionszuschreibungen wie zum Beispiel "`Li\-te\-ra\-tur veranschaulicht Wissen"' (\cite{Koppe2011}:6) oder "`Li\-te\-ra\-tur ist ein wirksames Mittel gegen Fundamentalismen"' (\cite{Wertheimer2013}) \"uberlasse ich der jeweiligen Einsch\"atzung der \textit{Erlesnis}-Autoren\textsuperscript{\tiny *}.

Ob die \textit{Erlesnisse} aufgrund der (nicht-)fiktionalen oder der fiktionalen Anteile li\-te\-ra\-rischer Texte entstehen, soll ebenfalls nicht in dieser Arbeit erforscht werden. Benjamin Gittel h\"alt es f\"ur fragw\"urdig, wenn bestimmte Erlebnisse als Wissen ausge\-zeichnet werden oder wenn f\"ur Erlebnisse bzw. Emotionen angenommen wird, dass mit ihnen etwas gerechtfertigt werden k\"onnte, zum Beispiel, dass Li\-te\-ra\-tur Wissen vermittelt (\cite{Gittel2013}:424). Dem halte ich in meiner Argumentation entgegen, dass in der Zuschreibung des Empfindens zu einem bestimmten Leseprozess vor allem dann ein Li\-te\-ra\-tur-bezogenes Wissen entsteht, wenn dar\"uber geschrieben wird. In pragmatischer Hinsicht l\"asst sich dies erweisen, indem Leseberichte in Wi\-ki\-pe\-dia-Eintr\"agen re\-ferenziert werden k\"onnen und als Wissen insofern Akzeptanz finden, als \"Anderungen an Eintr\"agen, mit denen Informationen zu dieser Art von Erlebniswissen aus Li\-te\-ra\-tur\-lekt\"ure erg\"anzt wurde, nicht r\"uckg\"angig gemacht werden -- was einer Infragestellung gleichk\"ame -- sondern stehenbleiben. Sie\-he hier\-zu auch \textit{~\ref{subsec:6.2} Experimentieren mit Leseberichten. Konzeption, Hergang, Auswertung}. 

Nicht zu\-letzt soll der Abgrenzungspunkt genannt werden, dass einem Schrei\-ben \"uber Li\-te\-ra\-tur\textsuperscript{\~.\~.} und der Produktion an\-de\-rer Schreibweisen vor allem dann gro{\ss}e Bedeutung bei\-ge\-mes\-sen werden k\"onnte, wenn sich bis\-herige Inhalte und konventionelle Diskursformen zur Erreichung bestimmter Ziele der Li\-te\-ra\-tur\-ver\-mitt\-lung als ungen\"ugend erweisen. Al\-ler\-dings ist es nicht Anliegen dieser Arbeit, zu ermitteln, ob bei bestimmten Ma{\ss}nahmen der erwartete Erfolg ausbleibt und daher andere bevorzugt eingesetzt werden sollten. Mir liegt in Sachen Li\-te\-ra\-tur\-ver\-mitt\-lung im Rahmen dieser Arbeit vielmehr daran, neue Aspekte f\"ur eine Theorie der Li\-te\-ra\-tur\-ver\-mitt\-lung zusammenzutragen, um die Debatte der Grundlagen zu bef\"ordern, auch wenn ich auf dem Weg dorthin weniger theo\-retisch als experimentell verfahre.

\subsection{Leitfragen}
\label{subsec:2.4}

¸* Welche Aspekte sind in einem multidisziplin\"aren web\"of\-fent\-lichen Kontext wie Wi\-ki\-pe\-dia f\"ur Li\-te\-ra\-tur so spezifisch, dass deren detaillierte Darstellung als enzy\-klop\"adisch relevant gelten kann?

¸* Auf welche Weise ist es m\"oglich, in einem Nachschlagewerk eine Wissensart zur Geltung zu bringen, deren Erwerb nur durch eigene Lekt\"ure entsteht und die anzuregen zu den spezifischen St\"arken von Sprachkunstwerken gez\"ahlt wird? Kann damit indirekt erwiesen werden, dass "`Wissen, wie sich etwas anf\"uhlt"' als glei\-cherma\-{\ss}en wertvoll gilt wie "`Wissen, dass etwas der Fall ist"'?

¸* Unter welchen Bedingungen gilt die Darstellung nicht-pro\-po\-si\-ti\-o\-na\-len Wissens aus Li\-te\-ra\-tur\-lekt\"ure bei Wi\-ki\-pe\-dia als relevant genug? 

¸* Kann eine Mehrheit der an diesen Fragen interessierten Akteure\textsuperscript{\tiny *} der deutschspra\-chi\-gen Wi\-ki\-pe\-dia-Community derzeit als offen genug eingesch\"atzt werden, um im Be\-reich Li\-te\-ra\-tur ein gewisses Abweichen von den Grundprinzipien zu tolerieren?

¸* Welche Schreibweisen f\"ur \textit{Erlesnisse} lassen sich differenzieren? 

¸* Welche Aspekte von Li\-te\-ra\-tur\-ver\-mitt\-lung sind in der For\-schungsli\-te\-ra\-tur bis\-her nicht ausreichend bedacht worden?

\subsection{Theoretischer Rahmen}
\label{subsec:2.5}

In den folgenden Bereichen stehen bereits Kenntnisse zur Verf\"ugung, aus denen f\"ur die vorliegende Arbeit ein begrifflicher Rahmen geschaffen wurde: Theorie des Erwerbs nicht-pro\-po\-si\-ti\-o\-na\-len Wissens aus Li\-te\-ra\-tur\-lekt\"ure, Theorie der Li\-te\-ra\-tur\-ver\-mitt\-lung und Theorie der \textit{parrhesia} bez\"uglich Li\-te\-ra\-tur. An neuem Wissen erarbeite ich erstens, dass es auch im Schrei\-ben \"uber Li\-te\-ra\-tur einen Bereich gibt, der ohne \textit{parrhesia} nicht formulierbar zu sein scheint, und zweitens, dass im Nachdenken \"uber Grundlagen der Li\-te\-ra\-tur\-ver\-mitt\-lung auch Aspekte, die f\"ur gemeinn\"utzige Arbeit konstitutiv sind, in den Blick genommen werden m\"ussen, damit effektive Wege gegangen werden k\"onnen, um Informationen, die Anhaltspunkte f\"ur eigene Einsch\"atzungen bez\"uglich des ge\-sell\-schaftlichen Werts des Lesens in der Web\"of\-fent\-lichkeit vermitteln, dort zu platzieren, wo sie leicht zug\"ang\-lich und weithin sichtbar sind. 

\"Ahnlich wie in kulturwissenschaftlich orientierten Ans\"atzen der Li\-te\-ra\-tur\-wis\-sen\-schaft wird in dieser Arbeit eine Beziehung zwischen li\-te\-ra\-rischen und nicht-li\-te\-ra\-rischen Texten sowie \"ubergreifenden kulturellen Ph\"anomenen plausibilisiert, indem \"Ahnlichkeiten der untersuchten Felder beziehungsweise Ph\"anomene nachgewiesen werden (\cite{KoppeWinko2013}:249) Al\-ler\-dings geht es nicht darum, eine kausale Analogie herzustellen und es wird auch kein Analogieschluss verwendet, denn es soll nicht bewiesen werden, dass neben den festgestellten \"Ahnlichkeiten weitere \"Ahnlichkeiten existieren (\cite{Bartha2013}) -- au{\ss}er dass beide Ph\"anomene textbasiert in Erscheinung treten und aus derselben Zeit stammen. Analogiebildung kommt also lediglich zum Einsatz, um bestimmte Ph\"anomene in zwei Systemen von Objekten hervorzuheben, die im Rahmen eines Vergleichs als einander so \"ahnlich angesehen werden, dass sie f\"ur die vorliegende Fragestellung interessant sind. Die Kontextualisierung erweist sich als ge\-eig\-net, da in den zu interpretierenden Werken ebenfalls Verfahren dieser Art thematisiert werden und weil deren li\-te\-ra\-risch gestaltete Spezifik mit diesem Vergleich besser verdeutlicht werden kann als ohne eine solche kontrastive Beleuchtung im \mbox{hier} gew\"ahlten Kontext. Damit wird al\-ler\-dings keine Hypothese vertreten, die besagen w\"urde, dass die gew\"ahlten Werke in eine bestimmte historische Situation "`eingebunden"' w\"aren (wie eine Studie etwa im Rahmen von \textit{New Historicism} gearbeitet w\"urde), wohl aber wird Wi\-ki\-pe\-dia als Kontext gew\"ahlt, um neue und interessante Aspekte einzelner li\-te\-ra\-rischer Texte der Gegenwart erkennbar zu machen. Dar\"uber hinaus wird der interpretierte Text als Kontext an\-de\-rer Texte gesehen, woraus folgt, dass im gleichen Zuge die Plattform Wi\-ki\-pe\-dia durch die \mbox{vier} Interpretationen kontextualisiert werden kann. Al\-ler\-dings geht es weder in der einen noch in der anderen Richtung darum, direkte Einwirkungen nachzuweisen, sondern es werden Aspekte der Verfahrensgestaltung einander gegen\"ubergestellt mit der Absicht, die je spezifischen Eigenheiten zu verdeutlichen.

\subsection{Methodenwahl}
\label{subsec:2.6}

Dariusz Jemielniak hat j\"ungst bereits die Methode der partizipativen qualitativen Beobachtung f\"ur Wikipedistik zur Anwendung gebracht. Indem ich seinen Ansatz f\"ur meine nicht-ethnografische Arbeit \"ubernehme, unterst\"utze ich seine Argumentation, dass Wikipedistik auch mit meiner Fragestellung nur von innerhalb des Systems betrieben werden kann, da erstens die als Voraussetzung notwendigen Kenntnisse anders nicht erworben werden k\"onnen -- \"ahnlich wie es bei Wissenschaftsfor\-schung der Fall ist -- und zweitens auf keinem anderen Wege das Vertrauen derjenigen zu erwerben ist, die sich in ihrer Freizeit an dem Projekt beteiligen (\cite{Jemielniak2014}:193-194).

Methodisch soll \mbox{hier} aber beiden Seiten Rechnung getragen werden. Daher kommen sowohl Interpretationen (Li\-te\-ra\-tur) als auch ein Experiment (Wi\-ki\-pe\-dia) zum Einsatz. In der gegenseitigen Bespiegelung von Li\-te\-ra\-tur und Wi\-ki\-pe\-dia wird doppelgleisig verfahren. Zum einen interpretiere ich \mbox{vier} sprachk\"unstlerische Werke aus Sicht von Wi\-ki\-pe\-dia-Konzepten und stelle -- inspiriert durch \textit{parrhesia} -- neue Lesarten vor, die \textit{Erlesnisse} enthalten. Ich suche f\"ur \textit{Erlesnisse} eine Form, die pu\-bli\-zierbar w\"are. 

Zum anderen unternehme ich in der deutschspra\-chi\-gen Wi\-ki\-pe\-dia-Community ein Experiment mit dem Ziel, das aktuelle Ma{\ss} an Akzeptanz f\"ur neue Abschnitte mit dem Titel "`Leseerlebnis"' zu erkunden. Meint man in einem For\-schungsprozess zu der Erkenntnis gelangt zu sein, dass die Fragestellung so neu ist, dass sie nur mit einem innovativen Vorgehen zu bearbeiten w\"are, ist nicht von anderen zu erwarten, dass sie die notwendigen Fakten schaffen (zum Beispiel Texte produzieren oder editorische Handlungen t\"atigen). Vielmehr liegt es nahe, dem zu beobachtenden System durch eigene Beteiligung Impulse zu geben, infolge derer die der Fragestellung impliziten Hypothesen auf eben dieser Systemebene ausprobiert werden k\"onnen, also: ein Experiment zu wagen -- im metaphorischen Sinne, wie in den Geisteswissenschaften \"ublich. Mit dem Experiment wird versucht, Aussagen m\"oglichst vieler an\-de\-rer Wi\-ki\-pe\-dia-User hervorzurufen (Daten zu kreieren), auf deren Basis die Bedingungen der Darstellbarkeit nicht-pro\-po\-si\-ti\-o\-na\-len Wissens in Eintr\"agen zu li\-te\-ra\-rischen Werken bei Wi\-ki\-pe\-dia untersuchbar gemacht werden sollen. Das Experiment umfasst meine Mitwirkung bei dem Vorschlagen, der Dis\-kus\-si\-on und der Auswahl von Teasern f\"ur die Hauptseitenrubrik "`Schon gewusst?"' sowie das Konzipieren von Leseerlebnis-Abschnitten, die Reaktionen aus der Community hervorrufen sollen. Ich platziere diese neuen Abschnitte in einigen Eintr\"agen zu Werken der Gegenwartsli\-te\-ra\-tur und schaue, was passiert. Zu diesem Zweck muss eine gr\"o{\ss}ere Aufmerksamkeit erzeugen als sie sonst Li\-te\-ra\-tur\-artikeln geschenkt wird. Wird ein gewisser Einspruch ge\"au{\ss}ert, beginne ich eine Dis\-kus\-si\-on, mehr oder weniger explizit, al\-ler\-dings ohne theo\-retische Begrifflichkeiten wie "`nicht-pro\-po\-si\-ti\-o\-na\-les Wissen"' zu verwenden, denn da sich meiner bis\-herigen Erfah\-rung nach grunds\"atzlichere Debatten nicht gewinnbringend f\"uhren lassen, weil die beharrenden Kr\"afte sich auf einer solchen Ebene mit ihrer verwaltungsorientierten Energie zu stark positionieren, bleibe ich mit meiner Rolle ebenso wie mit meiner Redeweise w\"ahrend des Experiments auf der all\-tags\-prak\-ti\-schen Ebene des Artikelverfassens -- nicht ohne nach Ende einer gr\"o{\ss}eren Debatte am \mbox{24. Januar} 2015 (als \mbox{User}:C.Kolt\-zen\-burg) doch einen Versuch unternommen zu haben, den theo\-retischeren Aspekt verschiedener Wissensarten wenigstens anzudeuten, worauf seitens dieses Users oder seitens an\-de\-rer, die dessen\textsuperscript{\tiny *} Be\-nut\-zerdis\-kus\-si\-onsseite lesen, auf dieser Ebene nicht weiter eingegangen wurde: 

\singlespacing
\begin{quote}
\href{https://de.wikipedia.org/w/index.php?title=Benutzer_Diskussion:Magiers&diff=prev&oldid=138104304}{"`Hallo Magiers, sag mal bitte, wie hast du die Gr\"aser-Rezension von Aufenanger verstanden: Beschreibt er seinen Wissenszuwachs durch die Lekt\"ure? Ich habe n\"amlich Leute gefunden, die meinen, dass bei Filmen und Li\-te\-ra\-tur vor allem ein \flq Wissen wie\frq\,\,erworben wird, nicht ein \flq Wissen was\frq. Interessant! Und dazu w\"usste ich gern die Meinung an\-de\-rer (bei Grillenwaage). Jetzt kam mir Aufenangers Beschreibung in den Sinn, was meinst du dazu? Gru{\ss}, --C.Kolt\-zen\-burg (Dis\-kus\-si\-on) 09:54, 24. Jan. 2015 (CET)"'} 
\end{quote}
\onehalfspacing

Dass ich ein Experiment durchf\"uhre, lege ich nicht offen, weil sonst der Austausch nicht als Teil der \"ublichen Routine ablaufen w\"urde. Wie in Jemielniak 2014 dargelegt, nutze auch ich die Usernamen in der Form wie sie seitens der Beteiligten web\"of\-fent\-lich verwendet werden, denn alle \mbox{User} wissen ohnehin, dass sie web\"of\-fent\-lich schrei\-ben und daher mit ihren Aussagen zitierbar sind.

Als entscheidend hat es sich erwiesen, im Fluss des Geschehens innerhalb der Wikipedia-Community die gr\"o{\ss}tm\"ogliche Flexibilit\"at zu bewahren und als g\"unstig einge\-sch\"atzte Gelegenheiten beim Schopfe zu packen, um zu sehen, ob sich daraus etwas entwickeln lassen w\"urde. Hierzu sei ein Beispiel genannt: Als das beste Ana\-lysebeispiel im Kontext dieser Arbeit sollte sich ein Eintrag he\-rausstellen, dessen Lemma ich zun\"achst lediglich als Unterabschnitt im Eintrag zur Autorin\textsuperscript{\tiny *} verlinkt hatte: "`La Vie commune"' von Lydie Salvayre, der Prix Goncourt-Preistr\"agerin\textsuperscript{\tiny *} von 2014 -- und zwar gerade rechtzeitig zum Tag der Pr\"asentation des Artikels "`Lydie Salvayre"' in der Hauptseitenrubrik "`Schon gewusst?"' --, dessen Eintragung in die Hauptseitenvorlage ich am Tag zuvor gesehen hatte. Auch im Artikel zur Autorin\textsuperscript{\tiny *} hatte ich bereits ein als solches benanntes Leseerlebnis-Beispiel platziert, dazu gab es am Pr\"asentationstag al\-ler\-dings keine Reaktion, sondern lediglich zum verlinkten Werkartikel zu einem nicht aktuell preisgekr\"onten Werk derselben Autorin\textsuperscript{\tiny *}, das von 1991 ist und 2007 zu\-letzt neu aufgelegt worden war. Der Eintrag "`Lydie Salvayre"' war etwa einen Monat zuvor von einem anderen Autor\textsuperscript{\tiny *} angelegt worden, mit einem gut ausbauf\"ahigen biografischen Ger\"ust. Ein Zusammenarbeits\-modell dieser Art war schon einige Wochen zuvor, Mitte September 2014, einmal erfolgreich, als derselbe andere Autor\textsuperscript{\tiny *} den biografischen Eintrag "`Yuri Herrera"' anlegte und ich mich veranlasst sah, schnellstens li\-te\-ra\-tur\-spezifische Informationen zu recherchieren und zu erg\"anzen, weil es kurzzeitig so aussah, als w\"urde dieser Artikelvorschlag von den gerade aktiven Usern des "`Schon gewusst?"'-adhoc Teams als nicht gen\"ugend ausgearbeitet abgelehnt. Im Fall von "`Lydie Salvayre"' einige Wochen sp\"ater, als ich das Konzept der Leseerlebnis-Abschnitte meinerseits schon f\"ur so weit gediehen hielt, dass die Zeit reif war, es zu erproben, hatte ich den Eintrag des anderen Autors\textsuperscript{\tiny *} f\"ur eine Pr\"asentation in der Hauptseitenrubrik "`Schon gewusst?"' vorgeschlagen, um au{\ss}er in Artikeln zu Werken des seit dem 8. Oktober aktuellen Nobelpreistr\"agers\textsuperscript{\tiny *} noch weitere Testballons zu haben, bei denen auch innerhalb der Wi\-ki\-pe\-dia-Community mit einem gewissen Ma{\ss} an aktueller Aufmerksamkeitsbereitschaft zu rechnen sein w\"urde.

Zum Dritten entwickele ich auf \textit{Metaebene II} ansatzweise eine Typologie des Schrei\-bens \"uber \textit{Erlesnisse}, f\"ur die ich neben Schreibweisen in Formaten wie Rezensionen und li\-te\-ra\-tur\-wissenschaftlichen Essays auch Rezensionsnotizen bei \textit{Per\-len\-tau\-cher.de} sowie eigene Formulierungen in Wi\-ki\-pe\-dia-Eintr\"agen in Betracht ziehe (siehe \textit{~\ref{subsec:7.1} \textit{Parrhesia} im Schrei\-ben \"uber Li\-te\-ra\-tur\textsuperscript{\~.\~.}: Ans\"atze einer Typologie}). Ich erprobe damit eine neue Art des beteiligten Beobachtens, insofern ich \mbox{selbst} erstelltes Material in eine typologische Ana\-lyse mit einbeziehe. Diese Entscheidung hat fachliche Gr\"unde: Nachdem ich zu dem Eindruck gelangt war, dass \textit{Erlesnisse} in Leseberichten an\-de\-rer nicht in einer Explizitheit formuliert wurden wie sie aus Wi\-ki\-pe\-dia-Perspektive gedacht am besten ge\-eig\-net w\"aren, habe ich in eigenen Interpretationen erprobt, \textit{Erlesnisse} im Rahmen pu\-bli\-zierbarer Essays zu formulieren. Im typologisch arbei\-tenden Abschnitt auf \textit{Metaebene II} beziehe ich die einige Monate zuvor formulierten Passagen aus eigenen Interpretationen quasi als Werkstattbericht in die Ana\-lyse mit ein, in einem vierten Schritt, also nachdem ich Vergleichsmaterial besprochen habe, das bereits pu\-bli\-ziert worden ist. 

Die vorliegende Studie hat nicht nur in der Hinsicht Pilot\textsuperscript{\tiny *}charakter, dass \"uber Li\-te\-ra\-tur\-eintr\"age bei Wi\-ki\-pe\-dia noch keine For\-schung pu\-bli\-ziert worden ist, sondern auch methodisch, indem ich ein be\-stimmtes Formalobjekt (\textit{Erlesnis}) in einer Verfah\-rens\-kombination erforsche, die jeweils einer der beiden Seiten eines Verh\"altnisses zweier kultureller Institutionen zueinander, das mir vorwiegend von gegenseitiger Ignoranz gepr\"agt zu sein scheint, gem\"a{\ss} ist.

\subsection{Konzepte}
\label{subsec:2.7}

Es wird mit den folgenden Konzepten gearbeitet: \textit{Erlesnis}, definiert als dasjenige, was bei Li\-te\-ra\-tur\-lekt\"ure dem eigenen Empfinden nach an nicht-pro\-po\-si\-ti\-o\-na\-lem Wissen erworben wurde, Li\-te\-ra\-tur\textsuperscript{\~.\~.}, womit das Entstehen von Emp\-fin\-dun\-gen im Lese- und/oder Zuh\"orkontakt mit li\-te\-ra\-rischen Texten gemeint ist, und drittens, demzufolge, mit Li\-te\-ra\-tur\textsuperscript{\~.\~.}lekt\"ure, was das Nachsinnen \"uber einen Prozess be\-zeichnet, dem ein Leser\textsuperscript{\tiny *} das Entstehen bestimmter Emp\-fin\-dun\-gen zuschreibt. Den \"Ubergang von einem Lek\-t\"u\-re\-pro\-zess zum eigenen Schrei\-ben \"uber \textit{Erlesnisse} erachte ich als flie{\ss}end, die Grenzen zwischen Wissensarten ebenso wie zwischen Schreibweisen. Auf dem Weg zu diesen Konzepten habe ich in eigenen Interpretationen zu \mbox{vier} li\-te\-ra\-rischen Werken Schreibweisen f\"ur Beschreibungen eigener \textit{Erlesnisse} erprobt. Mit den Be\-zeichnungen "`Leseerlebnis"' und \textit{Erlesnis} ist jeweils die in Sprache gefasste Form gemeint, \"ahnlich wie es im Gebrauch des Begriffs "`Lekt\"ure"' \"ublich ist, wenn es im Feuilleton hei{\ss}t, jemand pr\"asentiere Lekt\"uren (Beispiel: "`Zugleich aber bietet er zarte, mitf\"uhlende, fast liebevolle Lekt\"uren an"' (\cite{Martus2011}:50) Es k\"onnen auch Wendungen vorkommen wie "`Schreibweisen \"uber \textit{Erlesnisse}"'. Damit ist dann gemeint: "`wie in \textit{Erlesnissen} (bzw. genauer gesagt: in \textit{Erlesnis}-Berichten) \"uber \textit{Erlesnisse} geschrieben wird."' 
Ist mit "`Leseerlebnis"' die neue Art von Abschnitten gemeint, die es in manchen Wi\-ki\-pe\-dia-Eintr\"agen gibt, nutze ich die Be\-zeichnung "`Leseerlebnis-Abschnitte"'.

Konzept und Begriff \textit{Erlesnis} nutze ich als mein Formalobjekt (\cite{Meyenetal2011}:55-59). Ich habe es ent\-wickelt, indem ich in vorhandenem Material nach einem gr\"o{\ss}eren Zusammenhang gesucht habe, der \"uber das konkret fassbare Materialobjekt hinauswies. Mit \textit{Erlesnis} kann ich nunmehr auf ein dahinterliegendes Pro\-blem in Form einer Hypothese aufmerksam machen, n\"amlich, dass aufseiten Li\-te\-ra\-tur\-bewanderter im deutschspra\-chi\-gen Raum eine gewisse Scheu be\-steht, sich \"uber eigene \textit{Erlesnisse} schriftlich zu \"au{\ss}ern. Mit der vorliegenden Arbeit k\"onnen ansatzweise die nachteiligen Folgen dieser Zur\"uckhaltung f\"ur die gemeinn\"utzige Li\-te\-ra\-tur\-ver\-mitt\-lung aufgezeigt werden.

Wie bei der Erfindung des Wiki \"ubernehme auch ich aus der Architektur zus\"atzlich das Konzept "`Entwurfsmuster"'. Im objektorierten Programmieren wurde mithilfe von Entwurfsmustern die Wieder\-ver\-wert\-bar\-keit von Soft\-ware\-tech\-no\-lo\-gien ver\-bes\-sert (\cite{Stigler2009}:203-212). Das Konzept findet im Rahmen meiner Arbeit da Anwendung, wo ich eine aus meiner Sicht li\-te\-ra\-tur\-vermittelnde Abfolge zu denken versuche: von Entwurfsmuster 1 (pu\-bli\-zierte li\-te\-ra\-rische Texte) zu Handlung 1 (Lesen) zu Entwurfsmuster 2 (\textit{Erlesnis}) zu Handlung 2 (Leseberichte pu\-bli\-zieren) zu Entwurfsmuster 3 (Wi\-ki\-pe\-dia-Eintr\"age) und Handlung 3 (\textit{Erlesnisse} in "`Leseerlebnis"'-Abschnitten zusammenfassend dar\-stel\-len), in schematischer Ansicht, als Flowchart zu lesen: 

¸* Entwurfsmuster 1 = pu\-bli\-zierte literarische Texte\\
¸* Handlung 1 = Lesen\\
¸* Entwurfsmuster 2 = \textit{Erlesnis}\\ 
¸* Handlung 2 = als Leseberichte pu\-bli\-zieren\\
¸* Entwurfsmuster 3 = Wi\-ki\-pe\-dia-Eintr\"age\\
¸* Handlung 3 = \textit{Erlesnisse} in "`Leseerlebnis"'-Abschnitten zusammenfassend dar\-stel\-len; Zitate m\"oglich\\

In diesem Modell wird keine Aussage dar\"uber getroffen, welche Akteure\textsuperscript{\tiny *} sich an den jeweiligen Schritten beteiligen. Letzt\-lich k\"onnte ein Autor\textsuperscript{\tiny *} pu\-bli\-zierter li\-te\-ra\-rischer Texte von Entwurfsmuster 1 bis zu Handlung 3 alle Schritte vornehmen -- was selten der Fall sein wird, aber nicht undenkbar ist. We\-sent\-lich h\"aufiger wird zum Beispiel mein Fall sein, in dem ab Handlung 1 (Lesen) zwei Schritte \"ubersprungen werden und erst bei Entwurfsmuster 3 (Wi\-ki\-pe\-dia-Eintr\"age) der Prozess vom li\-te\-ra\-rischen Text zum Sichtbarmachen eines \textit{Erlesnisses} in einem Leseerlebnis-Abschnitt wieder aufgenommen wird. 

F\"ur die Li\-te\-ra\-tur-seitigen Bedingungen der Darstellbarkeit nicht-pro\-po\-si\-ti\-o\-na\-len Wissens ist in dieser Hinsicht die Abfolge zwischen Handlung 1 (Lesen), Entwurfsmuster 2 (\textit{Erlesnis}) und Handlung 2 (als Leseberichte pu\-bli\-zieren) entscheidend. Gibt es \mbox{hier} nicht gen\"ugend zitierbare Quellen, kommt der Prozess ins Stocken. Li\-te\-ra\-tur\textsuperscript{\~.\~.}lekt\"ure allein wird also f\"ur gemeinn\"utzige Li\-te\-ra\-tur\-ver\-mitt\-lung wie sie in dieser Arbeit in den Blick ger\"uckt wird nicht ausreichen. Werden Leseberichte, die \textit{Erlesnisse} enthalten, namhaft pu\-bli\-ziert, gelangen aus meiner Sicht sehr wertvolle Aussagen in den \"of\-fent\-lichen Raum. Manchen ist dies eventuell nur mit einem "`parrhesiatischen"' Schritt m\"oglich, denn je nach Empfinden in Bezug auf das \"Au{\ss}ern von \textit{Erlesnissen} jenseits von Freund\textsuperscript{\tiny *}eskreisen kann der Schritt, \textit{Erlesnisse} unter eigenem Namen Fremden gegen\"uber zu berichten, als eine Preisgabe mit gro{\ss}en "`Freim\"utigkeitsanteilen"' empfunden werden, so meine Vermutung, wenn ich von mir \mbox{selbst} ausgehe und insofern ich die Kapitel\"uberschrift "`Vom Heldenmut des Lesers"' zum \textit{Erlesnis}-Beitrag von Ulrike Draesner richtig interpretiere (\cite{Draesner2013}).

Festzuhalten sind im We\-sent\-lichen die Konzepte zu den Termini "`Entwurfsmuster"' und \textit{Erlesnis}, alle weiteren Konzepte und deren Begrifflichkeiten werden sich im Laufe meiner weiteren Argumentation leicht aus dem jeweiligen direkten Zusammenhang erkl\"aren.

\subsection{Aufbau des Hauptteils}
\label{subsec:2.8}

Der Hauptteil der vorliegenden Arbeit gestaltet sich mithilfe zweimaligen Wechselns zwischen Objektebene und Metaebene. Diese Anregung verdanke ich einer sprachwissenschaftstheo\-retischen Publikation von Brigitte Schlieben-Lange zu den Arbeiten einer Gruppe von Ideenwissenschaftlern\textsuperscript{\tiny *} mit erkenntnistheo\-retischen und se\-mio\-tischen Schwerpunkten, die in der napoleonischen Zeit in Frankreich ein p\"ada\-go\-gi\-sches Programm der Breitenaufkl\"arung zur Vorbeugung gegen eine neue Schre\-ckens\-herr\-schaft ins Werk setzten, mit dem Titel \textit{\flq Id\'{e}ologie\frq. Zur Rolle von Kategorisierungen im Wissenschaftsproze{\ss}} (2000). Brigitte Schlieben-Lange erl\"autert die Vorgehensweise in der Vorbemerkung wie folgt [in Klammern habe ich ihre Bezeichnungen der Ebenen nochmals erg\"anzt]: "`1. [\textit{Objektebene I}] In einem ersten Schritt will ich das For\-schungsprogramm der \textit{id\'{e}ologie} skizzieren. Einigen von Ihnen wird es bekannt sein; trotzdem ist eine solche pointierte Skizze f\"ur mein Vorhaben unerl\"asslich. 2. [\textit{Metaebene I}] In einem zweiten Schritt will ich mich dann dem Klassifikator \textit{id\'{e}ologie} zuwenden, seiner Setzung und seiner Geschichte. 3. [\textit{Objektebene II}] Dann kehre ich, drittens, zu den \textit{Id\'{e}ologistes} zur\"uck, um nun auf deren unausgesch\"opfte Potenziale zu sprechen zu kommen. 4. [\textit{Metaebene II}] Und schlie{\ss}lich komme ich wieder auf das theo\-retische Problem der Kategorisierung zur\"uck, sozusagen also in einer spiralf\"ormigen Bewegung in zwei Durchg\"angen von der Objektebene zur Metaebene"' (\cite{Schlieben-Lange2000}:1-2). Wo Schlieben-Lange sich auf der Metaebene anhand des Klassifikators \textit{id\'{e}ologie} und seines historischen Umfeldes mit dem theo\-retischen Problem von Kategorisierungen im Wissenschaftsprozess befasst, richte ich mein Augenmerk anhand des neuen Konzepts \textit{Erlesnis} am Beispiel von Gegenwartsli\-te\-ra\-tur auf das Problem der Vermittelbarkeit pers\"onlicher Erkenntnisse und erforsche, wie ein \textit{Erlesnis} in einem anderen Programm der Breitenaufkl\"arung im web\"of\-fent\-lichen Umfeld der Gegenwart (Wi\-ki\-pe\-dia) Wirkung entfalten k\"onnte. 

Auf den beiden Objektebenen befasse ich mich aus literaturwissenschaftlichem Blickwinkel mit Wikipedia, auf den beiden Metaebenen mit Literatur sowie mit Litera\-tur\-wis\-sen\-schaft, Literaturkritik und Literaturvermittlungsforschung. Mithilfe dieses Aufbaus l\"asst sich zum einen die gegenseitige Bespiegelung von Literatur und Wiki\-pe\-dia auch strukturell leichter darstellen, zweitens kann deutlich werden, dass sich meine vier Interpretationen auf \textit{Metaebene I} nicht direkt auf mein Experiment bei Wikipedia beziehen (\textit{Objektebene II}), sondern vielmehr auf meinen Vorschlag auf \textit{Metaebene II}, dass meine Erkenntnisse ebenso das literaturkritische Schreiben be\-rei\-chern k\"onn\-ten wie sie eventuell f\"ur eine Debatte zu den Grundlagen der Li\-te\-ra\-turvermittlungsforschung neue Impulse geben. 

\newpage

\section{Forschungsstand}
\label{sec:3}

\subsection{Nicht-pro\-po\-si\-ti\-o\-na\-les Wissen aus Literaturlekt\"ure}
\label{subsec:3.1}

Dass durch Li\-te\-ra\-tur\-lekt\"ure insbesondere nicht-pro\-po\-si\-ti\-o\-na\-les Wissen zu erlangen ist, wurde in j\"ungster Zeit verst\"arkt he\-rausgearbeitet. Li\-te\-ra\-rische Darstellungen generieren ein Wissen, das sich zum begrifflichen Wissen komplement\"ar verh\"alt und emergent ist (\cite{Klinkert2011}). In li\-te\-ra\-rischer Vergegenw\"artigung erm\"oglicht das "`Wissen, wie es w\"are"' eine Vertrautmachung mit bestimmten Erfah\-rungen und vervielf\"altigt unsere Perspektiven auf die Welt (\cite{Ferran2014}). Dass es sich bei Li\-te\-ra\-tur um sprachliche Kunstwerke handelt, ist auch in dieser Hinsicht bedeutsam (\cite{Huemer2014}), und dass mit li\-te\-ra\-rischen Darstellungen eine Aufforderung an den Leser\textsuperscript{\tiny *} als Interpreten\textsuperscript{\tiny *} einhergeht, f\"ur sie eigent\"umlich (\cite{Vesper2014}). Li\-te\-ra\-rische Werke k\"onnen sowohl kognitive als auch nicht-kognitive Funktionen verschiedenster Art erf\"ullen (\cite{Reicher2014}), sind auf keinen \flq Wissensbereich\frq \,\,festgelegt und k\"onnen "`zur Quelle von Wissen \"uber alles M\"ogliche werden"' (\cite{Koppe2011}:7). Ob es sich um fiktionale oder nicht-fiktionale Texte handelt, um sprachliche oder nicht-sprachliche Artefakte, das Erkennen von nicht-pro\-po\-si\-ti\-o\-na\-lem Wissen ist methodologisch glei\-cherma\-{\ss}en pro\-blematisch (\cite{DannebergSpoerhase2011}). Nicht-pro\-po\-si\-ti\-o\-na\-le Wissensinhalte k\"onnen dennoch teilweise in eine pro\-po\-si\-ti\-o\-na\-le Form \flq \"ubersetzbar\frq \,\,sein, sie k\"onnen aber in ihrer Eigenart nur auf dem Wege indirekter Ver\-mitt\-lung mittels \"asthetischer Gestaltungselemente gezeigt werden (\cite{Albrecht2011}). Briefe beispielsweise sind \href{https://www.uni-marburg.de/fb03/philosophie/institut/mitarbeiter/vendrell/form-inhalt_vendrell-wille.pdf}{auf drei Ebenen lesbar}: neben denen des Gegenstands und des Themas, auf denen die Inhalte explizit oder implizit sein k\"onnen, drittens auch auf einer Reflexionsebene, auf der Ton und Stimmung transportiert werden und die anregt zu Vergleichen und allgemeineren Schlussfolgerungen (\cite{FerranWille2012}). Auch in Li\-te\-ra\-tur ist inzwischen eine F\"ulle an Wissensformen und wissenschaftshistorischen Ph\"anomenen entdeckt worden und in der For\-schung wird gelegentlich von einem Erlebnis gesprochen, das Autoren\textsuperscript{\tiny *} und Leser\textsuperscript{\tiny *} verbindet (\cite{Erhart2014}:152;173). Dies l\"asst sich von Stil und Argumentationsweise her in einigen Beispielen li\-te\-ra\-tur\-wissenschaftlicher Darstellungen wiederfinden (siehe \textit{Abschnitt~\ref{subsec:7.1} \textit{Parrhesia} im Schrei\-ben \"uber Li\-te\-ra\-tur\textsuperscript{\~.\~.}: Ans\"atze zu einer Typologie}). 
In anderen F\"allen wird allein der Text als Gegen\"uber konzipiert, bei Anne Katrin Lorenz etwa in einem Szenario, aus dem sich Haltungen der \textit{parrhesia} ergeben k\"onnen: \href{https://publikationen.uni-tuebingen.de/xmlui/handle/10900/47049}{"`Im intimen Moment des Leseprozesses wird der Text zur Reflexionsfl\"ache und zum Experimentierfeld, die den Leser auf sich \mbox{selbst} zur\"uckwerfen und seine Position zu den vor\-herr\-schen\-den Verh\"altnissen neu justieren k\"onnen."'} (\cite{Lorenz2012}:190) Zu nicht-pro\-po\-si\-ti\-o\-na\-lem Wissen werden in erkenntnistheo\-retisch fundierten Debatten in folgenden Teilfragen unterschiedliche Annahmen diskutiert: wie es entsteht, inwiefern bestimmte Voraussetzungen notwendig sind, um nicht-pro\-po\-si\-ti\-o\-na\-les Wissen bei der Lekt\"ure li\-te\-ra\-rischer Texte erwerben zu k\"onnen, und ob es erforschbar ist. Benjamin Gittel referiert aus der Erkenntnistheorie drei Arten nicht-pro\-po\-si\-ti\-o\-na\-len Wissens, die sich dadurch aus\-zeichnen, dass dieses Wissen nicht in Form von beweisbaren Aussagen seinen Wert erh\"alt: "`\textit{know\-ledge by acquaintance} (etwas oder jemanden kennen), \textit{knowing how} und \textit{knowing what it is like}"'. Ebenso wie die Intuition, aus Li\-te\-ra\-tur gelernt zu haben, scheine es den anderen drei Arten eigen zu sein, so Gittel, dass das erworbene Wissen als schwer for\-mulierbar empfunden wird. Als weitere Beispiele gibt er an: "`Man kann etwa die Musik von Chopin kennen, wissen, wie man Tango tanzt oder wie es sich anf\"uhlt, nach langer Zeit wieder verliebt zu sein, ohne dass man in der Lage sein muss, dieses Wissen propositional auszudr\"ucken"' (\cite{Gittel2013}:281-82). Gittel h\"alt die Aussage, dass ein Leser\textsuperscript{\tiny *} nur infolge eigener lebensweltlicher Erlebnisse bei der Lekt\"ure ein qualitativ neues "`knowing what it is like"' empfinden kann, f\"ur eine schwer \"uberpr\"ufbare Hypothese, zu der seines Wissens noch nicht empirisch gearbeitet worden ist. Ebensowenig werde bis\-her ein Szenario erforscht, bei dem davon ausgegangen wird, dass beim Lesen "`ein Wissen \"uber die Auswirkungen des Kontigenzbewusstseins in Bezug auf die eigenen \"Uberzeugungen erworben wird"', womit die Annahme gest\"utzt werden k\"onnte, dass Li\-te\-ra\-tur exklusives Wissen biete (\cite{Gittel2013}:294;424).

Das jeweilige Ma{\ss} an Plausibilit\"at oder Begr\"undbarkeit bestimmter Zuschreibungen soll im Rahmen dieser Arbeit al\-ler\-dings nicht weiter er\"ortert werden, da es vom gew\"ahlten Ansatz her lediglich um dasjenige geht, was als Effekt von Lek\-t\"u\-re\-pro\-zessen geschil\-dert wird. Die jeweilige Einsch\"atzung der Autoren\textsuperscript{\tiny *} selbst soll f\"ur die Erkenntnisziele dieser Arbeit eine ausreichende Grundlage f\"ur die Plausibilit\"at einer Schil\-de\-rung sein. Inwiefern und in welcher Form anderen ein Eindruck davon vermittelt werden kann, muss an dieser Stelle offenbleiben. Ebenso, ob, und falls ja, mit welchen Mitteln es erforschbar w\"are. F\"ur die pragmatische Ausrichtung der vorliegenden Arbeit wird davon ausgegangen, dass nicht-pro\-po\-si\-ti\-o\-na\-les Wissen aus Li\-te\-ra\-tur\-lekt\"ure mindestens in Berichten \"uber Erlebnisse aufgrund von oder aus Anlass von Li\-te\-ra\-tur\-lek\-t\"u\-re (genauer gesagt: in \textit{Erlesnissen}) vermittelbar ist.

\subsection{Theorie der Literaturvermittlung}
\label{subsec:3.2}

In der deutschspra\-chi\-gen Li\-te\-ra\-tur\-ver\-mitt\-lungsfor\-schung ist in letzter Zeit zu ge\-eig\-neten Orten und Methoden gearbeitet worden und der Erlebnischarakter von Lekt\"ure kommt zunehmend auch f\"ur eine Theo\-retisierung in den Blick. Im Band \textit{(Ver)F\"uh\-rungen. R\"aume der Li\-te\-ra\-tur\-ver\-mitt\-lung} (2012) wird auf Basis der Pr\"a\-mis\-se einer li\-te\-ra\-rischen Attraktion vorgegangen, "`die (Ver)F\"uh\-rung im Sinne einer r\"aum\-li\-chen wie qualitativen Ann\"aherung, einer Hinwendung und eines Sich-Einlassens, erst er\-m\"oglicht"' (\cite{Disoskietal2012}:9). Es wird mit einem engeren und einem weiteren Begriff von Li\-te\-ra\-tur\-ver\-mitt\-lung gearbeitet, wobei der engere Bereiche umfasst, die sich auf Erwerbst\"atigkei\-ten richten (Berufsfelder), und im weiter gefassten Begriff auch Li\-te\-ra\-tur\-ver\-mitt\-lung durch "`nicht-professionelle Leser"' in Betracht gezogen wird (\cite{NeuhausRuf2010}:9).

Es wird "`eine auf die Steigerung von Erlebnisqualit\"aten abonnierte Ge\-sell\-schaft"' beobachtet (\cite{Beileinetal2012}:11) und dass "`das Lekt\"ure\-verhalten in der \flq Erlebnisge\-sell\-schaft\frq"' darauf ausgerichtet ist, ein "`Bed\"urfnis nach pers\"onlichen Leseerlebnissen durch das kollektive Erleben stimuliert zu erhalten"' (\cite{Beileinetal2012}:13), was dieser Einsch\"atzung zufolge "`die verbreitete Nutzung von Kulturformen mit hohem \flq Erlebnisgehalt\frq"' (\cite{Beileinetal2012}:12) zur Folge hat. Die Erlebniskomponente ist f\"ur die Theorie der Li\-te\-ra\-tur\-ver\-mitt\-lung auch aus wissenschaftssoziologischen Gr\"unden als relevant zu erachten. Im Ausblick seiner Musil-Studie hat sich auch Benjamin Gittel mit Fragen der Li\-te\-ra\-tur\-ver\-mitt\-lung befasst, wo er im Kapitel "`Aktualit\"atsbezug"' ausgehend von einem hochschuldidaktischen Beispiel die von ihm wahrgenommene Diskrepanz auf der Ebene von Hintergrundfaktoren folgenderma{\ss}en formuliert: 

\singlespacing
\begin{quote}
"`Die lebendige Erkenntnis, die der Student aus seiner Heimlekt\"ure mitbringt, ist in Seminardis\-kus\-si\-onen nicht zu verteidigen und \flq ver\-schwin\-det\frq, insofern sie nicht thematisiert wird. Dieser \mbox{hier} nicht empirisch belegte, jedoch grunds\"atzlich praxeologisch untersuchbare Zusammenhang hat folgende kurios anmutende Konsequenz: Einerseits ist die Phi\-lo\-lo\-gie bzw. Li\-te\-ra\-tur\-wissenschaft selbst, als li\-te\-ra\-rische Sozialisationsinstanz we\-sent\-liches Hemmnis f\"ur einen auf den Erwerb lebendiger Erkenntnis zielenden Umgang mit Li\-te\-ra\-tur. An\-de\-rerseits negiert sie im Kampf um sym\-bo\-lisches Kapital ihren Funktionswandel von einem sinnerzeugenden Kulturtr\"ager zu einer spezialisierten Wissenschaft auf einer Ebene zweiter Ordnung, indem sie ihrem Gegenstand genau die ko\-gni\-ti\-ven Qua\-lit\"aten zuerkennt, die unter den Bedingungen moderner Wissenschaft zunehmend weniger sichtbar werden."' (\cite{Gittel2013}:439)
\end{quote}
\onehalfspacing

Die Sichtbarmachung der Erlebnisqualit\"aten von Li\-te\-ra\-tur\textsuperscript{\~.\~.} in schriftlicher Form wird durch die Fragestellung der vorliegenden Arbeit zu einer Kernfrage. Nicht in der \"Au{\ss}erung selbst, sondern weit \"uberwiegend im Au{\ss}enraum hat das Ereignishafte von Li\-te\-ra\-tur in der Ver\-mitt\-lung seinen Ort bei Britta Hochkirchen (\cite{Hochkirchen2015}). Jedoch an an\-de\-rer Stelle in der For\-schung habe ich zwei Stimmen gefunden, die mir in diesem Zusammenhang erw\"ahnenswert erscheinen, weil sie Hinweise auf Aspekte geben, die unter dem Blickwinkel von innerem Erleben und dem Sich-\"Au{\ss}ern \"uber Effekte des Kontakts mit Li\-te\-ra\-tur in Betracht gezogen werden k\"onnen. Susanne Hochreiter merkt in einem Beitrag \"uber das Spielen mit Texten in der Improvisation an, dass in der Theatertheorie kaum ber\"ucksichtigt wird, was \textit{embodiment} "`f\"ur den/die SchauspielerIn \mbox{selbst}"' bedeutet, wenn in neueren Verfahren das Verh\"altnis von Darsteller\textsuperscript{\tiny *} und Rolle umgekehrt, Cross-Casting erprobt oder die Unzul\"ang\-lichkeit und Verletzlichkeit des K\"orpers sichtbar wird, und falls doch, dann als philosophische oder \"asthetische Frage oder in psychologischen oder p\"adagogischen Untersuchungen, jedoch nicht im Rahmen der Li\-te\-ra\-tur\-ver\-mitt\-lungsfor\-schung. F\"ur alle Lern- und Lebensprozesse sei es wichtig, leibliche Erfah\-rung als besondere Qualit\"at zu erkennen, durch die "`in Texten, in anderen Per\-so\-nen, im eigenen Selbst"' wichtige Prozesse stattfinden: "`Neues erfahren, Fremdes kennenlernen, Ungewohntes ausprobieren, \"Uberraschendes entdecken, Transformationen erleben"' (\cite{Hochreiter2012}). G\"unther Stocker berichtet von einer All\-tagsbeobachtung, wenn er in einer Fu{\ss}note schreibt: "`Lesende im \"of\-fent\-lichen Raum machen mit den sichtbaren Buchumschl\"agen auch Werbung f\"ur das gelesene Buch und verf\"uhren andere ebenfalls zur Lekt\"ure"' (\cite{Stocker2012}:Fn. 11). F\"ur die Li\-te\-ra\-tur\-ver\-mitt\-lungsfor\-schung w\"are es meines Erachtens relevant, Aspekte dieser Interaktion bei der Er\-\"or\-te\-rung theo\-retischer Grundlagen zu ber\"ucksichtigen, ebenso wie die von Daniela Strigl nebenbei formulierten Aspekte der "`egalit\"are[n] Schmutzkonkurrenz"', die "`der professsionellen, der gedruckten Li\-te\-ra\-tur\-kritik durch das Rezensionswesen im Internet [...] erwachsen"' ist (\cite{Strigl2012}) -- und dass Strigl Li\-te\-ra\-tur\-kritik nicht als "`Lekt\"urepartnerver\-mitt\-lung"' ansieht, denn sie stehe schlie{\ss}lich historisch im Erbe der Aufkl\"arung und im Idealfall gehe es bei Li\-te\-ra\-tur\-kritik um Erkenntnis.

Gittels Einsch\"atzung zufolge wird in Musils Konzeption lebendiger Erkenntnis die Erlebniskomponente insofern notwendig, als diese Art von Wissen "`sich von dem als wissenschaftlich ausge\-zeichneten durch seine begrenzte rationale Verhandelbarkeit"' unterscheidet (\cite{Gittel2013}:440). Dass die Erlebniskomponente nicht zu\-letzt aus diesem Grund einiges an Gewicht erhalten sollte, gilt aus meiner Sicht auch f\"ur die Darstellbarkeit nicht-pro\-po\-si\-ti\-o\-na\-len Wissens in Wi\-ki\-pe\-dia-Eintr\"agen. Anzumerken ist, dass Gittel zu Fragen der Li\-te\-ra\-tur\-ver\-mitt\-lung zwar \"Uberlegungen anstellt, von denen er meint, dass sie im Falle ihrer Stichhaltigkeit einen starken Anreiz b\"oten "`das Verh\"altnis von akademischer Li\-te\-ra\-tur\-wissenschaft einerseits und schulischer Li\-te\-ra\-tur\-sozialisation an\-de\-rerseits grundlegend zu \"uberdenken"' (\cite{Gittel2013}:439), auff\"allig ist aber, dass er die Plattform Wi\-ki\-pe\-dia f\"ur diesen Zusammenhang allem Anschein nach als nicht erw\"ahnenswert erachtet (siehe auch Abschnitt \textit{~\ref{subsec:5.3} Aussagen zu Wi\-ki\-pe\-dia in li\-te\-ra\-tur\-wissenschaftlichen Beitr\"agen}), obwohl sie f\"ur Millionen von Sch\"ulern\textsuperscript{\tiny *}, Lehrern\textsuperscript{\tiny *}, unterst\"utzenden Eltern\textsuperscript{\tiny *} ebenso wir f\"ur angehende wie praktizierende Li\-te\-ra\-tur\-wissenschaftler\textsuperscript{\tiny *} ein viel ge\-nutz\-tes aktuelles Nachschlagewerk ist -- auch \"uber li\-te\-ra\-risches Wissen. Man darf annehmen, dass die Plattform aufgrund ihrer leichten Erreichbarkeit und ihrer Offenheit f\"ur Perspektiven unterschiedlicher ge\-sell\-schaftlicher Gruppen schon seit Jahren eine implizite Br\"uckenfunktion zwischen den von Gittel genannten Li\-te\-ra\-tur\-ver\-mitt\-lungs\-instan\-zen Universit\"at und Schule einnimmt.

Sigrid Me{\ss}ner stellt fest, dass Li\-te\-ra\-tur im Netz neue Begriffsbildungen erfordert, dass durch hypertextuelle Verfahren bestehende Begriffe Transformationen ausgesetzt sind, dass in Autoren*foren neue Publikationspraxen entstehen und dass das Schrei\-ben im Medium Web "`in mehrfacher Hinsicht einem Wandel unterworfen ist"' (\cite{Messner2012}). F\"ur meine Fragestellung von Belang ist \mbox{hier}, dass auch das Schrei\-ben \"uber Li\-te\-ra\-tur sich ver\"andert, wenn es -- wie bei Wi\-ki\-pe\-dia -- in einem Raum entsteht, der durch die systemische Potenzialit\"at laufender Aushandlungen gekennzeichnet ist, weil es das we\-sent\-liche Charakteristikum der zugrundeliegenden Schreibumgebung ist, einem Wiki. Ob es bereits Studien dar\"uber gibt, wie in Leseforen oder Blogs \"uber Effekte der eigenen Li\-te\-ra\-tur\-lekt\"ure bzw. Li\-te\-ra\-tur\textsuperscript{\~.\~.}lekt\"ure geschrieben wird, ist mir nicht bekannt. Schon 2003 wies Christiane Heibach auf etwas hin, was aus der Sicht mei\-ner Fragestellung f\"ur eine Theo\-retisierung der Li\-te\-ra\-tur\-ver\-mitt\-lung relevant ist, weil eine weitere Art der Transformation des Li\-te\-ra\-tur\-umfeldes benannt wird:

\singlespacing
\begin{quote}
"`Stellt man die \textit{Vernetzung} in den Mittelpunkt, dann geht es um die Flexibilisierung der Rollenverteilungen, um die Funktionen von Autor und Rezipient sowie m\"ogliche neue Formen von Autorschaft und Kreati\-vi\-t\"at. Die Strukturen kollektiver Kreativit\"at, die im elektronischen Raum beobachtet werden k\"onnen, bilden dann die Hintergrundfolie, von der aus traditionelle li\-te\-ra\-rische Ph\"anomene betrachtet werden -- nat\"urlich wiederum unter Ber\"ucksichtigung der spezifischen Medialit\"aten: Li\-te\-ra\-tur ist dann ein Vernetzungsph\"anomen, in dem Sprachspiele, Konversationskunst, Briefli\-te\-ra\-tur, Gespr\"achsprotokolle etc. in Beziehung zu den Werken gesetzt werden, mit denen sie in Zusammenhang stehen. Der Einsatz leiblicher Medien ist dabei ebenso zu ber\"ucksichtigen wie die Transformation von face-to-face-Situationen in die Schriftform."' (\cite{Heibach2003}:268-269)
\end{quote}
\onehalfspacing

Die textuellen Aspekte des Bloggens hat Matti Trau{\ss}neck als \href{http://www.literaturkritik.de/public/rezension.php?rez_id=15119}{"`Grauzone"'} zwischen Li\-te\-ra\-tur\-wissenschaft und den Unterhaltungs\-me\-di\-en be\-schrie\-ben (\cite{Traussneck2010}), Michael Huter \mbox{nennt} die Grenzen zwischen der wissenschaftlichen und der praktischen Li\-te\-ra\-tur\-ver\-mitt\-lung in textlicher Hinsicht flie{\ss}end und macht eine grobe Aufstellung unter textlinguistischen ebenso wie schreibpraktischen Gesichtspunkten (\cite{Huter2010}:53) -- siehe \textit{Tabelle 1}.

\label{tab:1}
\textbf{Tabelle 1:\,Textsorten der wissenschaftlichen Li\-te\-ra\-tur\-ver\-mitt\-lung\,(Huter)}\\\newline
\begin{tabular}{p{2,5cm}|p{3,5cm}p{3,5cm}p{3,5cm}}
& Forschung & Lehre & Popularisierung \\\hline
&&& \\
Funktion & Wissensproduktion & Wissenstransfer & Wissenstransfer \\
&&& \\
Form & wissenschaftlich & didaktisch & popul\"ar \\
&&& \\
Inhalt & Neues & Gesichertes & Interessantes \\
&&& \\
Zielgruppen & Fachleute & Studierende & Laien \\
&&& \\\hline
&&& \\
Beispiele & Monographie, \newline wiss. Artikel & Lehrbuch, \newline Hypertexte \newline & Sachbuch, \newline pop. Artikel \\
\end{tabular}

Ich nutze \textit{Tabelle 1} als Grundlage f\"ur meine Darstellung in Abschnitt \textit{~\ref{subsec:7.2} Update f\"ur die Theorie der Li\-te\-ra\-tur\-ver\-mitt\-lung}, wo ich sie in Hinblick auf das \textit{Erlesnis} weiter ausarbeite (\textit{Tabelle 2: Das Entwurfsmuster \textit{Erlesnis} in seinem Umfeld (Kolt\-zen\-burg)}, um aufzuzeigen, an welchen Punkten der Li\-te\-ra\-tur\-ver\-mitt\-lungstheorie ich Entwicklungpotenzial sehe in Bezug auf Effekte der Lekt\"ure als \"Au{\ss}erungen von Lesern\textsuperscript{\tiny *} in einem Umfeld, in dem au{\ss}erli\-te\-ra\-rische Diskurse dominant sind. 

Im Beispielkontext Wi\-ki\-pe\-dia, in dem laufend pragmatische L\"osungen erforderlich sind und in offenen Beteiligungsformen diskutiert werden, habe ich zu diesem Zweck ein Experiment durchgef\"uhrt (siehe \textit{~\ref{subsec:6.2} Experimentieren mit Leseberichten. Konzeption, Hergang, Auswertung}). Die Erkundung weiterer neuer (oder neu ins Blickfeld genom\-me\-ner) li\-te\-ra\-tur\-vermittelnder Ak\-ti\-vi\-t\"aten oder Effekte scheint mir mit anderen For\-schungsmethoden ebenso vielversprechend zu sein und \mbox{hier} sehe ich einigen Spielraum f\"ur weitere For\-schung.

\subsection{Wikipedistik zu Eintr\"agen in der Kategorie "`Literarisches Werk"'}
\label{subsec:3.3}

Zu Li\-te\-ra\-tur\-artikeln bei Wi\-ki\-pe\-dia gibt es bis\-her keine wissenschaftlichen Studien, zumindest konnte ich keine Publikationen mit dieser Thematik zutage f\"ordern, we\-der aufgrund eigener Recherchen noch in einer anschlie{\ss}enden \"Uberpr\"ufung durch Anfragen in Wi\-ki\-pe\-dia-Kontexten. Auf der For\-schungsmailing\-liste der Wikimedia Foundation gab es keinerlei Antwort zu meiner Anfrage und auf Aus\-kunftssei\-ten einiger gro{\ss}er Sprachversionen der Wi\-ki\-pe\-dia (en, de, fr, it, pl) zwar einige spekulative Antworten und immerhin konkrete Hinweise, aber keine Ideen und Informationen, die ich nutzen konnte. Auf der Seite des "`Re\-fe\-rence Desk Humanities"' der eng\-lischspra\-chi\-gen Version auf zwei verschiedene meiner Anfragen hin in beiden F\"allen genannt -- und in einem davon zitiert -- wurde jedoch das Abstract eines kommunikationswissenschaftlichen Artikels, in dem der Standpunkt vertreten wird, dass Fiktionales enzy\-klop\"adisch nicht relevant sei und daher bei Wi\-ki\-pe\-dia nichts zu suchen habe. Und dies, obwohl zuvor im selben Abstract argumentiert worden war, Wi\-ki\-pe\-dia habe die Verh\"altnisse auf dem Buchmarkt abzubilden. (Von den Verkaufszahlen her macht "`Fiction"' einen gro{\ss}en Anteil aus.)

Zu Li\-te\-ra\-tur-Eintr\"agen bei Wi\-ki\-pe\-dia scheint es diesen Ergebnissen zufolge bis\-her keine dezidierte For\-schung zu geben, was mir durch die Aussage von Dariusz Jemielniak best\"atigt wurde, dem Autor der bei Stanford University Press verlegten j\"ungsten Studie zu Wi\-ki\-pe\-dia, sowie durch \mbox{User}:Edith Wahr (\href{https://de.wikipedia.org/w/index.php?title=Benutzer_Diskussion:Edith_Wahr\&diff=138421804\&oldid=138421510}{"`Untersuchungen da\-r\"u\-ber, wie Wi\-ki\-pe\-dia-Artikel zur Li\-te\-ra\-tur ausfallen, hab ich noch keine gesehen."'}, 2. Februar 2015) und durch Mesgari et al. 2014, wo zumindest aufgrund der Daten f\"ur die ersten zehn Jahre von Wikipedia (bis einschlie{\ss}lich Juni 2011) keine Studien \"uber Li\-te\-ra\-tur\-artikel genannt werden (\cite{Mesgarietal2014}).

Die Bibliographien \textit{MLA} und \textit{BDSL} verzeichneten am 1. M\"arz 2015 lediglich drei Beitr\"age zu Li\-te\-ra\-tur\-eintr\"agen, genauer gesagt zu Literaten\textsuperscript{\tiny *} (Brecht, Feuchtwanger, Shakespeare), wobei ich keinem dieser Artikel eine in engerem Sinne wissenschaftliche Bedeutung beimesse.

Zur deutschspra\-chi\-gen Wi\-ki\-pe\-dia sind zwei Studien bekannt geworden, die sich auf einzelne Eintr\"age beziehen. In einer fallbasierten Studie zum Handlungsmuster "`Verfassen eines Wikieintrags"' im Rahmen des kooperativen Schrei\-bens bei Wi\-ki\-pe\-dia gelangen Christian Kohl und Thomas Metten 2006 zu der Einsch\"atzung, dass ein Umsetzen der NPOV-Vorgabe als \textsc{Mehrheitsakzep\-tables Dar\-stel\-len} beschrieben werden kann und sie f\"uhren eine \"Au{\ss}erung von \mbox{User}:Markus Mueller auf der Dis\-kus\-si\-onsseite zum Eintrag "`Philosophie"' am 3. August 2005 als Beleg an: 

\singlespacing
\begin{quote}
\href{https://de.wikipedia.org/w/index.php?title=Diskussion:Philosophie\&diff=next\&oldid=7928221}{"`Besonderen Wert habe ich auf eine m\"oglichst neutrale und von den meisten philosophischen \flq Richtungen\frq\, akzeptable Darstellung gelegt."'} (\cite{UserMarkusMueller2005} zitiert in (\cite{KohlMetten2006}:188)
\end{quote}
\onehalfspacing

Der Aspekt \textsc{Mehrheitsakzeptables Dar\-stel\-len} umfasst nach Kohl und Metten zwei Handlungsfelder: \textsc{Integrativ schrei\-ben} (auf der Artikelseite) und \textsc{Kontroverse aus\-handeln} (auf der Dis\-kus\-si\-onsseite zum Artikel). Diese lassen sich nach ihrer Ansicht aufgrund einer Wi\-ki\-pe\-dia-ty\-pischen Funktion parallellaufend konzipieren, denn w\"ahrend auf der Dis\-kus\-si\-onsseite eines Artikels Kontroversen ausgetragen werden, kann auf der zugeh\"origen Artikelseite etwas ge\"andert werden und umgekehrt. Dem Handlungsfeld \textsc{Integrativ schrei\-ben} lassen sich ihrer Ana\-lyse zufolge die Teilhandlungen \textsc{Aus\-sa\-gen re\-la\-ti\-vie\-ren}, \textsc{Aus\-sa\-gen dif\-fe\-ren\-zie\-ren} sowie \textsc{Sach\-ver\-hal\-te ex\-pli\-zie\-ren} zuordnen und dem Handlungsfeld \textsc{Kon\-tro\-ver\-se aus\-handeln} verschiedene Teilhandlungen wie \textsc{Kri\-tisch kom\-men\-tie\-ren}, \textsc{\"An\-de\-run\-gen vor\-schla\-gen}, \textsc{Po\-si\-ti\-o\-nen ar\-gu\-men\-ta\-tiv st\"ut\-zen} sowie \textsc{Ko\-ope\-ra\-tiv handeln}. Bei der Frage, ob die Texthandlungen in einer bestimmten Ent\-ste\-hungs\-phase des Eintrags "`Philosophie"' auch f\"ur andere Beitr\"age nachweisbar sind und wie sie sich auf die Artikelqualit\"at auswirken, wird bei Kohl und Metten die Vorgabe eines NPOV nicht in Zweifel gezogen. Ihr Ziel ist es lediglich, "`m\"ogliche Vorteile des kooperativen Schrei\-bens und der damit verbundenen kontroversen Prozesse zu erkennen"' (\cite{KohlMetten2006}:190). 

Mit dem deutschspra\-chi\-gen Eintrag zu den Angriffen auf die Twin Towers in New York am 11. September 2001 hat sich Ren\'{e} K\"onig befasst, inklusive der Aushandlungen auf der Artikeldis\-kus\-si\-onsseite, und festgestellt, dass sich langfristig jene Positionen durchsetzen, die dem Mainstream am ehesten entsprechen und die die bestehenden Hierarchien st\"arken. Er gibt zu bedenken: "`Of course, having a neutral, meaning actually \textit{no} point of view is simply impossible"' (\cite{Konig2013}:163) und referenziert (\cite{Reagle2012}:54), dessen Arbeit sich (leider nur f\"ur die eng\-lischspra\-chi\-ge Wi\-ki\-pe\-dia) vor allem mit den Aushandlungen zur NPOV-Policy befasst, also mit der Wi\-ki\-pe\-di\-a-Seite "`WP:NPOV"' und der Dis\-kus\-si\-onsseite dazu. Immerhin mit einem bestimmten Lemma in zwei Sprachversionen (en, pl) hat sich Dariusz Jemielniak im Rahmen seiner ethnografischen Studie befasst, in der er unter anderem die Rolle von Wi\-ki\-pe\-dia-B\"urokratie beforscht, und zwar anhand der Konflikte um "`Gdansk"'/ "`Danzig"' (\cite{Jemielniak2014}).

Aus weiteren allgemeiner angelegten Wi\-ki\-pe\-dia-Studien zitiere ich, sofern in den Abschnitten zur Wi\-ki\-pe\-dia-Einf\"uh\-rung die jeweiligen Aspekte zur Sprache kommen.

\subsection{\textit{Parrhesia} und Gegenentw\"urfe zur Ordnung des Sagbaren}
\label{subsec:3.4}

Anne Katrin Lorenz hat im Anschluss an eine begriffsgeschichtliche Darlegung meh\-rerer Bedeutungsstr\"ange die kommunikativen Bedingungen der \textit{parrhesia} anhand li\-te\-ra\-rischer Beispiele untersucht und neben der Ana\-lyse von "`parrhesiastischen"' Szenen in li\-te\-ra\-rischen Werken auch die spezifisch li\-te\-ra\-rische Ver\-mitt\-lung von \textit{parrhesia} auf der textuellen Metaebene in den Blick genommen. Lorenz konstatiert eine besondere Rolle der Li\-te\-ra\-tur an der Schnittstelle zwischen Privatheit und \"Of\-fent\-lichkeit sowie die zentrale Rolle der \textit{parrhesia} in diesem Feld. Li\-te\-ra\-tur sei ein k\"unstlerisches Experimentierfeld f\"ur ge\-sell\-schaftlich bedeutsame Vorg\"ange. Die Aufgabe bestehe darin, die Art und Weise zu bestimmen, in der \textit{parrhesia} dynamisch bleibt und ein je vor\-herr\-schen\-des diskursives System unterl\"auft (\cite{Lorenz2012}:100, Fn. 331 u. a.). Lorenz arbeitet he\-raus, dass durch \textit{parrhesia} weniger ein bestimmter Inhalt in Frage gestellt wird, als vielmehr ein Mechanismus, der in seiner Tonart statische Machtverh\"altnisse offenbart und jede dynamische Entwicklung der Kommunikation verbietet. F\"ur die Praxis der \textit{parrhesia} ist eine dynamische Struktur charakteristisch, die sich einer abschlie{\ss}enden Form der Wahrheit widersetzt. 

\singlespacing
\begin{quote}
\href{https://publikationen.uni-tuebingen.de/xmlui/handle/10900/47049}{"`[D]er in \textit{parrhesia} angelegte Widerspruch zwischen der unvermittelten \"Au{\ss}erung einer \flq ge\-f\"uhlten\frq \,\,subjektiven Wahrheit und dem rhetorischen Moment einer \"of\-fent\-lich formulierten Kritik ist im k\"unstlerisch ausgeformten Text und in der Intimit\"at seines individuellen Nachvollzugs durch den Leser aufgehoben."'} (\cite{Lorenz2012}:191)\\
\href{https://publikationen.uni-tuebingen.de/xmlui/handle/10900/47049}{"`Eine literarische Form der \textit{parrhesia} erm\"oglicht die zweifache \flq Grenz\-arbeit\frq, die sich im Subjekt zum einen als die innere Grenzerfah\-rung seiner Selbst vollzieht und zum anderen die Schnittstelle zwischen diesem privaten Diskursmodus der freundschaftlichen Korrektur und der \"of\-fent\-lichen \"Au{\ss}erung seines kritischen Selbstge\-f\"uhls bildet."'}\\ (\cite{Lorenz2012}:369)
\end{quote}
\onehalfspacing

Benjamin Gittel befasst sich in einer Musil-Studie mit Wissenskonzeptionen und kommt in sei\-nem Ka\-pi\-tel "`Le\-ben\-di\-ge Er\-kennt\-nis als Gegen\-be\-we\-gung"' zu fol\-gen\-dem Schluss: "`Heutzutage sind intuitionistische Wissenskonzeptionen sowohl in aka\-de\-mi\-schen De\-bat\-ten als auch im All\-tag marginalisiert"' (\cite{Gittel2013}:430). Sofern Gittel mit seiner Auffassung Recht hat, kann gerade die von ihm diagnostizierte Marginalisiertheit den Einsatz "`parrhesiastischer"' Verfahren f\"ur die Produktion an\-de\-rer Schreibweisen vor allem im Rahmen einer Plattform wie Wi\-ki\-pe\-dia sinnvoll erscheinen lassen.

Nicholas Gilewicz und Fran\c{c}ois Allard-Huver stellen fest, dass anonymes oder pseu\-donymes Schrei\-ben (speech) im Netz sowohl demokratische als auch propagandistische Ziele bef\"ordern kann. Sie verwenden \textit{parrhesia} im Sinne von "`wahrhaftige Rede"' (truth-telling) und sind der Frage nachgegangen, wie im web\"of\-fent\-lichen Raum eine Glaub\-w\"urdigkeit wieder hergestellt werden kann angesichts der verbreiteten Verwendung von Astroturfing (unlautere Nutzung der Glaub\-w\"urdigkeit von Graswurzel-Bewegungen durch Lobbies, etwa wenn in einer bestimmten Sache Risikofaktoren aufgebauscht werden sollen, diese Anliegen aber von keiner tats\"achlichen Bewegung zur Sprache gebracht werden). Sie vertreten die Ansicht, dass, wer frei ist, wahrhaftig zu sprechen, sich auch im Netz entsprechend \"au{\ss}ern sollte, da die \"Of\-fent\-lichkeit einer Transparenz in politischen Diskursen bed\"urfe, an denen pers\"onliche Rede we\-sent\-lichen Anteil habe (\cite{GilewiczAllard-Huver2013}).

Andreas Hetzel untersucht Geltungsanspr\"uche von Gegenentw\"urfen zur jeweiligen hegemonialen Ordnung des Sagbaren unter anderem anhand von Jacques Ranci\`{e}re, in dessen Denken Wortergreifung -- auf politische Repr\"asentationsregime bezogen -- "`f\"ur die Erfindung einer Sprache von Anteilslosen [steht], die sich Geh\"or verschaffen, ohne sich den Regeln eines etablierten Diskurses zu unterwerfen"' (\cite{Hetzel2012}:244). Redeereignisse, die unm\"ogliche Forderungen artikulieren, k\"onnen dabei subversiv und unvernehmbar sein und sich dennoch Geh\"or verschaffen. Zwischenrufe k\"onnen eingespielte Diskurse unterbrechen und die Situation transformieren, der sie entspringen. Hetzels Ausf\"uh\-rungen sind insofern erkenntnisreich, als die offene Verhandelbarkeit der Ordnung des Sagbaren bei Wi\-ki\-pe\-dia auf Schreibweisen zu beziehen w\"are, mit denen das Be\-rich\-ten \"uber nicht-pro\-po\-si\-ti\-o\-na\-les Wissen aus Li\-te\-ra\-tur\-lekt\"ure m\"oglich w\"urde. Allein schon Leseerlebnisse in Eintr\"agen als solche zu deklarieren, erschien mir als Gegenentwurf, und erst recht, wenn tats\"achlich von \textit{Erlesnissen} berichtet w\"urde, die als subjektives Wissen nicht vereinbar w\"aren mit enzy\-klop\"adischen Vorstellungen im Kontext von Wi\-ki\-pe\-dia -- wie im folgenden Kapitel bez\"uglich der dort weitgehend anerkannten \href{https://de.wikipedia.org/wiki/Wikipedia:Neutraler_Standpunkt}{Neutralit\"atsdoktrin} darzulegen ist.

\newpage

\section{Objektebene I/ Binnenschau: Wikipedia und Diderots \flq Encyclop\'{e}die\frq}
\label{sec:4}

Wi\-ki\-pe\-dia tr\"agt den Untertitel "`Die freie Enzy\-klop\"adie"'. Damit stellt die \textit{Wikimedia Foundation} Marke und Produkt Wi\-ki\-pe\-dia in eine publizistische Tradition, deren Begrifflichkeit im Westen als gefestigt gilt, seit die \textit{Encyclop\'{e}die ou Dictionnaire raisonn\'{e} des sciences, des arts et des m\'{e}tiers} (35 B\"ande, 1751-1780), inhaltlich ma{\ss}geblich von Denis Diderot, Jean d'Alembert und Louis de Jaucourt betrieben, auf den Markt gebracht wurde (siehe Wi\-ki\-pe\-dia-Eintrag \href{https://de.wikipedia.org/w/index.php?title=Geschichte_und_Entwicklung_der_Enzyklop\%C3\%A4die\&oldid=134462044}{"`Geschichte und Entwicklung der Enzyklop\"adie"'}).

\subsection{\"Ahnlichkeiten}
\label{subsec:4.1}
Obwohl der Beginn dieser beiden Enzyklop\"adieprojekte 250 Jahre auseinanderliegt, sind sie sich in manchen Aspekten sehr \"ahnlich: Beide Projekte wurden von Gesch\"aftsleuten in Gang gebracht, die sich f\"ur die inhaltliche und redaktionelle Arbeit andere gesucht haben. Als Vorlage fungierte in beiden F\"allen eine Idee aus dem eng\-lischspra\-chi\-gen Raum. Bei der \textit{Encyclop\'{e}die} war es die \textit{Cyclopedia: or, An Universal Dictionary of Arts and Sciences}, erschienen in London ab 1728, mit f\"unf Ausgaben binnen 18 Jahren, das Werk des Journalisten Ephraim Chambers (1680-1740) (\cite{Blom2004}:xxiv-xxv), im Falle der deutschspra\-chi\-gen Wi\-ki\-pe\-dia die bereits von vielen verfasste eng\-lischspra\-chi\-ge Version desselben Unternehmens. 
F\"ur beide Enzyklop\"adien ist festgestellt worden, dass die Qualit\"at der Artikel schwankt. Von Diderot wird dieser Aspekt selbstironisch und in eleganter Prosa zum Besten gegeben (\cite{Blom2004}:153), hingegen bei Wi\-ki\-pe\-dia in profaner Weise von Vielen beklagt, so dass sich als eines der Zauberworte \textit{"`\href{https://de.wikipedia.org/wiki/Wikipedia:QS}{QS}"'} (Qualit\"atssicherung) als Ma{\ss}nahmenpaket etabliert hat, auf dessen Basis vor allem neu angelegte Artikel automatisiert gepr\"uft werden. Sie werden auf als solche definierte Fehlfunktionen hin durchsucht und ein sogenannter Baustein wird im Artikel platziert, in dem zu lesen ist, was am Artikel aus einer bestimmten Perspektive gesehen konkret ver\"andert werden muss, bis der QS-Baustein wieder entfernt werden kann. Die aktuellen F\"alle werden pro Tag auf einer bestimmten Seite gelistet. Wer meint, dass die \"Uberarbeitung des Artikel ausreichend ist, entfernt den Baustein und setzt auf der QS-Seite eine bestimmte Markierung, so dass jemand anderes noch Widerpruch einlegen kann. Gilt das Pro\-blem als gel\"ost, wird der Austausch dar\"uber ins \href{https://de.wikipedia.org/wiki/Wikipedia:Qualit\%C3\%A4tssicherung/Archiv#Alte_QS-Seiten}{Archiv der QS-Seite} verschoben. Dass die Artikelqualit\"at unter\-schied\-lich sein kann, ist ein erwartbares Charakteristikum gemeinschaftlicher Pu\-blikationen mit Beitr\"agen unterschiedlicher Autoren\textsuperscript{\tiny *}. Nicht darin also unterscheiden sich diese beiden enzyklop\"adischen Projekte, sondern lediglich in der Frage, wie mit dieser Beobachtung im weiteren Verlauf umgegangen wird. Beide Projekte nutzen zudem Querverweise (Print), respektive Verlinkungen (Web). Bei der \textit{Encyclop\'{e}die} wird \mbox{hier} der spielerische Aspekt betont, indem Autoren* die M\"oglichkeit erhalten, mit diesem Mittel ihrem Einfallsreichtum Ausdruck zu verleihen. An\-de\-rerseits meinte Diderot, dass das Werk auf diese Weise \"uber die Zeit an innerer Kraft und heimlicher Einheitlichkeit gewinne (\cite{Blom2004}:155). Auf der Meta-Ebene in der deutschspra\-chi\-gen Version von Wi\-ki\-pe\-dia hingegen sind mir noch keine Aussagen bez\"uglich tiefsinniger Motive beim Setzen von Wikilinks aufgefallen. Es sind noch weitere redaktionelle Aspekte \"uber die Jahrhunderte \"ahnlich geblieben. Editorische Leitlinien inhaltlicher Art entstanden bei der \textit{Encyclop\'{e}die} durch Debatten in den einflussreichen Salons von Paris (\cite{Blom2004}:145). Einflussreiche Salons gibt es auch in der deutschspra\-chi\-gen Wi\-ki\-pe\-dia, manche sind selbsternannt -- und w\"aren gern einflussreicher --, andere werden schonmal als \href{https://de.wikipedia.org/w/index.php?title=Benutzer:Simplicius\%2FDiderot-Club_II\&diff=next\&oldid=136141510}{"`Sumpfsalon"'} betitelt. Der \textit{Diderot-Club II} ist auf einer Unterseite von \mbox{User}:Simplicius angesiedelt, die \textit{Grillenwaage} ist eine Dis\-kus\-si\-onsseite des gleichnamigen Userkontos, dessen Gemeinschaftsaccount von \mbox{vier} Usern zu genau diesem Zweck angelegt worden ist. Auch die Debatten, die w\"ahrend der redaktionellen Arbei\-ten zur Erstellung des Wi\-ki\-pe\-dia-internen Periodikums "`Kurier"' laufen, haben zeitweise Club-Atmosph\"are. Manche der Salons sind offiziell administrativer Natur und setzen sich aus gew\"ahlten Usern zusammen. \mbox{Hier} meine ich den Club der sogenannten "`Administratoren"'. \"Uber deren Geschlecht sei hierbei nichts ausgesagt, so lautete vor Kurzem erneut die Auffassung der Mehrheit derjenigen \mbox{User} bei der deutschspra\-chi\-gen Wi\-ki\-pe\-dia, die sich dazu \"au{\ss}ern. In der deutschspra\-chi\-gen Version \"ubernehmen derzeit \href{https://de.wikipedia.org/wiki/Wikipedia:Liste_der_Administratoren}{etwa 250 \mbox{User}} diese Funktion. Zu den \href{https://de.wikipedia.org/wiki/Wikipedia:Administratoren\#Administrator-Funktionen}{Aufgaben} von Administratoren\textsuperscript{\tiny *} geh\"ort es: Sei\-ten zu sch\"utzen, Konten zu sperren, L\"oschdis\-kus\-si\-onen zu entscheiden und nicht mehr ben\"otigte Sei\-ten zu l\"oschen. \mbox{User}:Neon bringt die Admin-Salonbildungs-Methoden bei Wi\-ki\-pe\-dia mit Genderaspekten in Verbindung:

\singlespacing
\begin{quote}
\href{https://commons.wikimedia.org/w/index.php?title=File:Allesueberwikipedia.pdf\&page=42}{"`Ein weiteres Pro\-blem ist der Umgang der Wi\-ki\-pe\-dia-Administratoren mit den gew\"ohnlichen Usern. Die Administratoren sind sich einig: \flq Wi\-ki\-pe\-dia ist keine Demokratie\frq. Bei ihrer Sanktionierung von normalen Usern entscheiden sie in der Regel nicht nach rechtsstaatlichen Grunds\"atzen, sondern nach einem engen utilitaristischen Kosten-Nutzen-Kal\-k\"ul f\"ur Wi\-ki\-pe\-dia. Wenn sie davon ausgehen, einen unerw\"unschten St\"orer vor sich zu haben, der zudem mit ihrer Meinung nicht \"ubereinstimmt, sperren sie h\"aufig beim geringsten Anlass. Altgediente \flq Qualit\"atsauto\-ren\frq \,\,k\"on\-nen sich dagegen eine Menge he\-rausnehmen und werden bestenfalls mit Watteb\"ausch\-chen beworfen. Diese Willk\"ur f\"uhrt aber zu einem ver\-gif\-te\-ten Arbeitsklima und schreckt viele Menschen von der Mitarbeit bei Wi\-ki\-pe\-dia ab.\\ Die meisten Admins werden gew\"ahlt, weil sie in einem Fachgebiet Bedeutendes geleistet haben. Das bedeutet aber noch lange nicht, dass sie besonders dazu ge\-eig\-net sind, Konflikte zwischen den Usern zu l\"osen. Im Allgemeinen sind sie es nicht. H\"aufig versch\"arfen sie durch ihr Agieren diese Konflikte eher noch zus\"atzlich.\\ Nicht zu\-letzt dieses h\"aufig unbefriedigende Arbeitsklima bewirkt, dass Frauen in der Wi\-ki\-pe\-dia sehr unterrepr\"asentiert sind."'}\\ (\cite{BenutzerNeon022011}:42)
\end{quote}
\onehalfspacing

Eine weitere Gemeinsamkeit zwischen der \textit{Encyclop\'{e}die} und Wi\-ki\-pe\-dia scheint also zu sein, dass sich Beteiligte m\"annlichen Geschlechts so gut wie "`unter sich"' f\"uhlen. Bei der \textit{Encyclop\'{e}die} vor 250 Jahren wurden zumindest keine Frauen als Artikelverantwortliche benannt -- obwohl die Autor\textsuperscript{\tiny *}schaft von mindestens einer Gelehrten bekannt geworden ist (\cite{Blom2004}:143). Heute l\"asst die \textit{Wikimedia Foundation} gelegentlich Umfragen starten, aus denen sie glaubt zu wissen, wie viel Prozent der User m\"annlich sind. Nach allem, was ich \"uber die Jahre auf Wikimedia-Mailing\-listen und -Websei\-ten dazu in Erfah\-rung gebracht habe, sehe ich die Erhebungsmethoden, die bis\-her gew\"ahlt worden sind, als f\"ur diesen Kontext in der Summe hochgradig unprofessionell an, weil psychologische Faktoren in der Regel nicht ausreichend bedacht zu werden scheinen. In seiner Studie vom 2014 fasst Dariusz Jemielniak in dieser Hinsicht relevante Aspekte folgenderma{\ss}en zusammen: 

\singlespacing
\begin{quote}
"`According to \textit{Wi\-ki\-pe\-dia Editors Study}, published in 2011, 91 percent of all Wikipedia editors are male [...]. This figure may not be accurate, since it is based on a voluntary online survey advertised \mbox{to  31,699} registered users and resulting in 5,073 complete and valid responses [...] it is possible that male editors are more likely to respond than female editors. Similarly, a \mbox{study of  self-declarations} of gender showing only 16 percent are female editors (Lam et al. 2011) may be distorted, since more females may choose not to reveal their gender in a community perceived as male dominated."' (\cite{Jemielniak2014}:14-15)
\end{quote} 
\onehalfspacing

In einem meiner Dis\-kus\-si\-onsbeitr\"age auf der wiki-research-l Mailing\-liste der Wikimedia Foundation mit Betreff "`a cautious note on gender stats"' (14. Februar 2015) zitiere ich diese Passage aus Jemielniak und gebe zu bedenken: \href{https://lists.wikimedia.org/pipermail/wiki-research-l/2015-February/004151.html}{"`additionally, asserting status and flaunting seniority (also described by Jemielniak at the end of the paragraph previous to the one quoted above) is gene\-rally perceived to be a commonly employed trick to resist any changes; and, last but not least, one might argue that the group perceived as \flq in power\frq\,\,might feel to find strongly unbalanced outcomes most rewarding, and hence might tend to publish them as widely as possible and not least quote from them persistently, too..."'} Ich fragte abschlie{\ss}end, ob es dazu Widerspruch sei\-tens mitlesender Statistik-Experten\textsuperscript{\tiny *} gibt. Aaron Shaw schrieb daraufhin: \href{https://lists.wikimedia.org/pipermail/wiki-research-l/2015-February/004172.html}{"`My paper with Mako\footnote{Gemeint ist: Mako Hill und Aaron Shaw A (2013), \href{http://journals.plos.org/plosone/article?id=10.1371/journal.pone.0065782\#pone-0065782-t002}{"`The Wi\-ki\-pe\-dia Gender Gap Revisited: Characterizing Survey Response Bias with Propensity Score Estimation"'}, \textit{PLoS ONE} 8(6): e65782, Open Access.} has a lot of detail about why the 2008 editor survey (and all subsequent editor surveys, to my know\-ledge) has some profound limitations. The identification and estimation of the effects of particular causes and mechanisms that drive the gender gap (and related participation gaps) presents an even tougher challenge for researchers and is an area of active inquiry"'} (\cite{Shaw2015}).

In wieder anderen Salons haben User aus meiner Sicht eine lediglich sym\-bo\-lische Funktion, etwa in der Gesch\"aftsf\"uh\-rung der Wikimedia Foundation Deutschland e.V. (WMDE). Dar\"uber hinaus gibt es Verflechtungen der Art, dass Leute, die offiziell Vereinsmitglieder sind, dennoch Admin-Aufgaben \"ubernehmen, mit denen die Arbei\-ten derjenigen, die sich ohne Bezahlung beteiligen, reguliert werden. Dabei erwirtschaften die unbezahlbaren User letzt\-lich die Spendeneinnnahmen durch das Renommee der Inhalte -- und damit das Einkommen derjenigen, die im Auftrag oder als Angestellte bei WMDE t\"atig sind. Einige User haben dieses wirtschaftliche Verfahren unl\"angst kommentiert: 

\singlespacing
\begin{quote}
\href{https://de.wikipedia.org/w/index.php?title=Benutzer_Diskussion:Grillenwaage\&diff=135299468\&oldid=135298878}{"`Zumindest die WMF ist m.E. l\"angst soweit, dass unsere Textspenden gar nicht mehr interessieren. Die Marke "`Wi\-ki\-pe\-dia"' hat heute ihren Wert ganz ohne unsere Arbeit. \mbox{Selbst} das Einstellen aller Wartungsarbei\-ten br\"auchte lange, bis es die Marke besch\"adigt. Viel mehr als f\"ur die Substanz interessiert sich die WMF f\"ur Zahlen (Zugriffszahlen, egal wie erzeugt, und Artikelwachstum, lieber per Bot als durch aufs\"assige Autoren -- deswegen hei{\ss}t das neue Lieblingskind ja auch Wikidata) und nat\"urlich f\"ur die j\"ahrlichen Spendeneinnahmen. Das ist der einzige Punkt, wo man sie wirklich treffen k\"onnte: nicht keine Texte mehr spenden, sondern den Geldspendern klar machen, wof\"ur sie spenden und vor allem wof\"ur nicht (n\"amlich f\"ur die Artikel). Gru{\ss}"'} (\cite{UserMagiers2014a})
\end{quote} 
\onehalfspacing

oder 

\singlespacing
\begin{quote}
\href{https://de.wikipedia.org/w/index.php?title=Benutzer:Simplicius/Diderot-Club_II\&diff=prev\&oldid=135727987}{"`Solange Administratoren \& Funktion\"are des die Freiwilligkeit der Wi\-ki\-pe\-dianer auspl\"undernden Vereins Wikimedia ihre erhaltenen Spendengelder nicht offenlegen m\"ussen, bleiben viele Fragen zu deren Vorgehen in der Vergangenheit offen."'} (\cite{UserReinerStoppok2014}) 
\\\href{https://de.wikipedia.org/w/index.php?title=Benutzer:Simplicius/Diderot-Club_II\&diff=prev\&oldid=135727987}{"`PS: Um \flq Artikelarbeit\frq \,\,geht es \mbox{hier} jedenfalls schon lange nicht mehr."'}
\end{quote} 
\onehalfspacing 

oder 

\singlespacing
\begin{quote}
\href{https://de.wikipedia.org/w/index.php?title=Benutzer:Simplicius/Diderot-Club_II\&curid=4336672\&diff=135838088\&oldid=135786639}{"`Als Ehemaliger ist man bei gelegentlichen Stippvisiten auf den einschl\"agigen Sei\-ten erschrocken, was man sich \mbox{hier} an \flq Kollegenschaft\frq \,\,in seiner Freizeit angetan hat. Vom Umstand, dass \mbox{hier} mit freiwilliger unentgeltlicher Arbeit eine Funktion\"arsgruppe durchgef\"uttert wird, mal abgesehen."'} (\cite{UserJezza2014})
\end{quote} 
\onehalfspacing 

aber auch: 

\singlespacing
\begin{quote}
\href{https://de.wikipedia.org/w/index.php?title=Benutzer_Diskussion:Grillenwaage\&diff=135281359\&oldid=135281196}{"`Du solltest dir al\-ler\-dings bewusst sein, dass du deine Texte \mbox{hier} nicht der Foundation spendest, sondern der Allgemeinheit -- du stellst sie ja unter eine freie Lizenz und alle k\"onnen sie \"uberall nutzen. Du spendest sie via Wi\-ki\-pe\-dia und damit \"uber ein von der WMF betriebenes Projekt, aber die gesamte Enzyklop\"adie ist freies Wissen, das nicht zwingend an die WMF gebunden ist."'} (\cite{UserGestumblindi2014})
\end{quote} 
\onehalfspacing 

Jedoch sind meiner Einsch\"atzung nach in der deutschspra\-chi\-gen Wi\-ki\-pe\-dia diejenigen Salons am einflussreichsten, die nicht offiziell als solche zu erkennen sind. Ich verstehe sie als ein Gegenst\"uck zu dem, was im \mbox{Bereich} des wissenschaftlichen Pu\-bli\-zierens "`Zitierkartell"' genannt wird. Eine "`Salonbildung"' dieser Art gab es sicherlich auch im 18. Jahrhundert in denjenigen Kreisen, aus denen die Ideen f\"ur Lemmata und Inhalte der \textit{Encyclop\'{e}die} stammten.

Ich komme zu den Unterschieden, anhand derer sich nicht zu\-letzt die Variabilit\"at des Genres Enzyklop\"adie aufzeigen l\"asst. 

\subsection{Unterschiede}
\label{subsec:4.2}
W\"ahrend Mitte des 18. Jahrhunderts in Frankreich unter den Autoren ber\"uhmte Denker\textsuperscript{\tiny *} wie Voltaire, Rousseau und Montesquieu zu finden sind (\cite{Lih2009}:15) -- wobei Montesquieu \href{http://www.nzz.ch/nachrichten/kultur/literatur_und_kunst/ordnung-und-unordnung-des-wissens-1.4941730}{"`letzt\-lich gar nichts zur \textit{Encyclop\'{e}die} beisteuerte"'} (\cite{Hirschi2010}) -- kommt bei Wi\-ki\-pe\-dia eine sogenannte Schwarmintelligenz zum Zuge: 

\singlespacing
\begin{quote}
\href{http://de.wikipedia.org/w/index.php?title=Benutzer_Diskussion:Simplicius\&diff=prev\&oldid=136386468}{"`Es darf, wenn man sich sozialwissenschaftliche Studien insbesondere zur Wi\-ki\-pe\-dia ansieht, wohl als best\"atigt angesehen werden, dass die Schwarmarbeit Vorz\"uge beim Sammeln des Materials, aber Schw\"achen bei der Produktion wirklicher Qualit\"at hat. Nur: h\"atte man dem Schwarm gesagt, du darfst dein Material nur in dieses oder jenes Fach des enzyklop\"adischen Baustellenlagers einr\"aumen, um die (evtl.!) Verwertung k\"ummert sich eine Redaktion, w\"are der Schwarm wohl nur als Schw\"arm\-chen gelandet..."'} (\cite{User93.184.136.22014})
\end{quote} 
\onehalfspacing 

Die bekanntesten zeitgen\"ossischen Denker\textsuperscript{\tiny *} unserer Tage haben meines Wissens bis\-her nicht verlauten lassen, ob sie sich \mbox{selbst} zur Schwarmintelligenz in diesem Sinne z\"ahlen. 

\singlespacing
\begin{quote}
\href{https://de.wikipedia.org/w/index.php?title=Benutzer_Diskussion:Simplicius\&diff=next\&oldid=136226937}{"`Man k\"onnte sagen, bei Diderot arbeiteten die kritischen Geister mit. In der Wi\-ki\-pe\-dia wohl eher viele Kleingeister?"'} (\cite{UserSimplicius2014a}) 
\end{quote} 
\onehalfspacing

Die \"Of\-fent\-lichkeit wei{\ss} nichts davon, ob "`an der Wi\-ki\-pe\-dia"' auch 10 Jahre nach ihrem Start tats\"achlich noch Experten\textsuperscript{\tiny *} mitschrei\-ben, die in wissenschaftlichen Kreisen anerkannt w\"aren, unabh\"angig davon, ob sie mit ihrem Klarnamen dabei sind oder sich f\"ur ihre Beitr\"age einen anderen Usernamen ausgedacht haben. 

Einer meiner Eindr\"ucke verdient in diesem Zusammenhang eine Anmerkung: Bis vor einigen Jahren hat zumindest ein Geisteswissenschaftler\textsuperscript{\tiny *}, der unter anderem einen Mathematik-Hintergrund hat und in der Li\-te\-ra\-tur\-wissenschaft t\"atig ist, Artikel beigesteuert als \mbox{User}:Bunia und vor allem den Eintrag "`Fiktion"' bereichert. In seiner mengenm\"a{\ss}ig aktivsten Zeit bei Wi\-ki\-pe\-dia war von Remigius Bunia in einem Presseartikel bez\"uglich Li\-te\-ra\-tur folgender Standpunkt zu lesen: \href{http://litwiss.bunia.de/folders/litwiss/bunia_esra.pdf}{"`Li\-te\-ra\-tur ist ein besonders wichtiger ge\-sell\-schaftlicher Faktor, wenn sie nicht blo{\ss} unterhaltsam in \flq fremde Welten\frq \,\,entf\"uhrt, sondern Tatsachen in der realen Welt schafft"'} (\cite{Bunia2007}:39). Ich habe mich vergewissert, dass in dieser Aussage das "`nicht blo{\ss} unterhaltsam"' wirklich keine Anf\"uh\-rungszeichen hat und denke, dass \mbox{hier} eine Haltung gegen\"uber Li\-te\-ra\-tur zum Ausdruck kommt, die nach meinem Eindruck in manchen Phasen den "`Wi\-ki\-pe\-dia-Mainstream"' perfekt wiederzugeben scheint.

Bei Wi\-ki\-pe\-dia sind Beitr\"age in essayistischem Stil nicht willkommen, denn unter anderem enth\"alt dieser Stil Aussageformen, die als nicht enzyklop\"adisch abgelehnt werden.

\singlespacing
\begin{quote}
\href{https://de.wikipedia.org/w/index.php?title=Benutzer_Diskussion:Edith_Wahr\&diff=138430132\&oldid=138428015}{"`@C.Kolt\-zen\-burg: Was mir zum Thema \flq NPOV\frq gerade einf\"allt: St\"o\-ber mal ein bisschen in der Versionsgeschichte des Artikels Hans Henny Jahnn. Da hat Ulrich Greiner mal einen langen und -- zugegeben nicht un\-be\-dingt \flq enzyklop\"adischen\frq \,\,-- Abschnitt \"uber das Werk eingef\"ugt. Auf der Dis\-kus\-si\-onsseite wurde dann gleich mit L\"oschung gedroht, in\-zwi\-schen ist der Abschnitt halbwegs \flq wikifiziert\frq \,\,worden. Es ist aber ein recht ty\-pi\-scher Umgang mit Li\-te\-ra\-tur \mbox{hier}. Alles, was \"uber die neutralen und trocken beschriebenen Lebensfakten hinausgeht, wird mit Argwohn betrachtet. Lieber \"uberhaupt keinen Werkabschnitt als irgendwie einen Hauch von Meinung, Interpretation und eigenem Stil. Gru{\ss}"'} (\cite{UserMagiers2015c})
\end{quote}
\onehalfspacing

\singlespacing
\begin{quote}
\href{https://de.wikipedia.org/w/index.php?title=Benutzer_Diskussion:Simplicius&diff=next&oldid=136395255}{"`Wir haben sehr mangelhafte \"Ubereink\"unfte \"uber Wissen, Wissenschaft\-lich\-keit, Nachvollziehbarkeit und Zitierbarkeit. Es fehlen oftmals die re\-dak\-tio\-nel\-len F\"ahigkei\-ten, gute Gliederungen zu finden sowie Zusammenh\"ange zu formulieren und darzustellen. Da finden dann auch ideologische Grabenk\"am\-pfe statt. Also mindestens drei Haken."'}\\ (\cite{UserSimplicius2014b})
\end{quote}
\onehalfspacing 

Dass bei Wi\-ki\-pe\-dia Autoren\textsuperscript{\tiny *} von Diderots Format aktiv w\"aren, ist mir noch nicht aufgefallen -- zumindest in der aktuellen Phase scheint das nicht der Fall zu sein. In seinem \textit{Jacques le Fataliste} setzt Diderot inversive Techniken f\"ur die textuelle Doppelstruktur seiner selbst\-referentiellen Texte ein und f\"uhrt auf diese Weise eine Metaisierung der Texte herbei. Er \href{https://publikationen.uni-tuebingen.de/xmlui/handle/10900/47049}{"`setzt dem bestehenden Machtdiskurs der Ge\-sell\-schaft auf diese Weise eine Wahrheit entgegen, die oszilliert zwischen ihrer Form und den Voraussetzungen dieser Form und deshalb niemals ganz im realen Machtspiel aufgehen kann"'} (\cite{Lorenz2012}:366). Bei der Sprachverwendung in Wi\-ki\-pe\-dia-Eintr\"agen hingegen hat man den Eindruck, dass sich nur wenige mit Sprache an sich auskennen:

\singlespacing
\begin{quote}
\href{https://de.wikipedia.org/w/index.php?title=Benutzer_Diskussion:Epipactis\&diff=133460313\&oldid=133455572}{"`Wie auch immer die Sache ausgeht -- solche Dis\-kus\-si\-onen m\"ussen schon mal ausgetragen werden, denke ich. Denn, vielleicht ist es dir schon ebenso aufgefallen wie mir: in Wi\-ki\-pe\-dia herrscht ein eklatanter Mangel an Mit\-ar\-bei\-tern, die sich wirklich fundiert mit Sprache auskennen, oder sich wenigstens von Fall zu Fall etwas eingehender damit zu befassen gewillt sind. (Wobei ich mich bestenfalls zu den Letzteren rechnen darf.)"'} (\cite{UserEpipactis2014})
\end{quote}
\onehalfspacing 

Vergegenw\"artigt man sich den Unterschied zwischen Diderot und denjenigen, die in gro{\ss}er Zahl als Nichtliteraten\textsuperscript{\tiny *} bei Wi\-ki\-pe\-dia aktiv sind, dann ist es eventuell nicht verwunderlich, wenn sich zu li\-te\-ra\-rischen Werken viele Eintr\"age finden, die nach Schema X verfasst worden sind und leblos wirken, keinen Esprit haben -- \mbox{selbst} wenn Artikelqualit\"at immer mit Einzelnen zusammenh\"angt und daher allgemeinere Aussagen nicht in jedem Fall aussagekr\"aftig sind. Am bei Wi\-ki\-pe\-dia geforderten enzyklop\"adischen Stil allein kann es nicht liegen, wenn gleichzeitig vie\-ler\-lei Tipps zu finden sind, wie ein Artikel "`gut lesbar"' geschrieben und gestaltet werden kann. Wenn es in der \textit{Encyclop\'{e}die} bereits Eintr\"age zu einzelnen li\-te\-ra\-rischen Werken gegeben h\"atte (und auch Li\-te\-ra\-tur\-wissenschaften schon), w\"are sicher der Anteil der Lemmata, die sich mit ihnen als Sprachkunstwerken befassen -- und nicht lediglich als Geschichten mit irgendeiner Handlung -- we\-sent\-lich h\"oher gewesen als in der Wi\-ki\-pe\-dia. Es l\"asst sich \mbox{hier} argumentieren, "`dass gerade das Verstehen der sprachlichen Dimension li\-te\-ra\-rischer Texte ein we\-sent\-licher Bestandteil des Bildungsverst\"andnisses des Hochkulturenschemas ist"' (\cite{Dawidowski2009}:179) und dieses Verst\"andnis von li\-te\-ra\-rischen Texten in Eintr\"agen bei Wi\-ki\-pe\-dia nicht durchgehend zu finden ist: Selten wird dargelegt, was einen li\-te\-ra\-rischen Text bemerkenswert oder lesenswert macht, \"ahnlich selten gibt es Informationen dazu, auf welche Weise er verschieden interpretiert worden ist -- dies betrifft zumindest die meisten der neuen Artikel seit Mitte 2013. Wenn es darum ginge, Wi\-ki\-pe\-dia (oder den akademischen Ableger "`Wikiversity"') als Arbeitsplatt\-form bei der Ver\-mitt\-lung li\-te\-ra\-tur\-wissenschaftlicher Fachkenntnisse einzusetzen, w\"urde ich argumentieren, dass Wi\-ki\-pe\-dia, gerade weil \mbox{hier} we\-sent\-lich textbasiert gearbeitet wird, eine ideale Umgebung darstellt, um die Bedeutsamkeit der Arbeit am Text (nicht zu\-letzt als Spezifikum li\-te\-ra\-tur\-wissenschaftlicher Be\-weis\-f\"uh\-rung) allgemeinverst\"andlich darzulegen. In Eintr\"agen zu li\-te\-ra\-rischen Werken kann direkt mit Textbeispielen gearbeitet werden, anhand derer sogar verschiedene In\-ter\-pre\-ta\-tions\-me\-tho\-den und Zugangsweisen und ihre Einsichten aufgezeigt werden k\"onnen. Als enzyklop\"adisch wertvoll w\"urde es mir erscheinen, he\-rauszustellen, durch welche Interpretation beziehungsweise unter welchem Blickwinkel sich die Rezeption eines Textes im Laufe der Zeit wie ver\"andert hat. Als Belege w\"aren zumindest Publikationen wie \textit{Kafkas "`Urteil"' und die Li\-te\-ra\-tur\-theorie: Zehn Modellana\-lysen} (2002) ge\-eig\-net, zu denen kontrastierende Zusammenfassungen im Artikeltext we\-sent\-liche Einsichten vermitteln k\"onnen. Aber auch einzelne Textbeispiele werden als akzeptabel angesehen, wie ich durch einen Test erweisen konnte, sowohl im \href{https://de.wikipedia.org/w/index.php?title=Reisende_auf_einem_Bein&oldid=137914805#Stil}{Abschnitt "`Stil"'} des Eintrags "`Reisende auf einem Bein"' als auch im Eintrag "`Herta M\"uller"' im \href{https://de.wikipedia.org/w/index.php?title=Herta_M\%C3\%BCller&oldid=137918366\#Sprache_und_Poetologie_Herta_M.C3.BCllers}{Abschnitt "`Werk und W\"urdigungen"'/"`Sprache und Poetologie Herta M\"ullers"'/"`Sprache als Werkzeug, \flq umgekehrte Ingenieurskunst\frq"'}. Web\"of\-fent\-liches li\-te\-ra\-tur\-wissenschaftliches Arbei\-ten kann auf diesem Wege Nachweise \"uber seinen Wert f\"ur allgemeine ge\-sell\-schaftliche Kontexte erbringen.

An\-de\-rerseits erfordert bei Wi\-ki\-pe\-dia die Transparenz ebenso wie die m\"ogliche Hartn\"ackigkeit der Beteiligten bei Aushandlungen einiges an Chuzpe und Durchhal\-te\-ver\-m\"ogen. Jeden Monat werden allein in der deutschspra\-chi\-gen Wi\-ki\-pe\-dia Ak\-ti\-vi\-t\"aten von etwa 20.000 verschiedenen User-Accounts gez\"ahlt -- \"Anderungen, die von Rechner-IPs aus, also ohne registrierten Wi\-ki\-pe\-dia-Account get\"atigt werden, nicht eingerechnet. Angesichts der schieren Menge an m\"oglichen Gegenreden ist es also fair, diejenigen Wi\-ki\-pe\-diaautoren\textsuperscript{\tiny *} hervorzuheben, die nicht nur Artikel verfassen oder verbessern helfen, sondern die Aushandlungen dazu aktiv beglei\-ten. Dies sind Anforderungen, denen sich Autoren\textsuperscript{\tiny *} der \textit{Encyclop\'{e}die} vor 250 Jahren trotz aller Begeisterung f\"ur Demokratie in Bezug auf jedes Detail ihrer Beitr\"age zur \textit{Encyclop\'{e}die} nicht stellen mussten -- al\-ler\-dings drohen Wi\-ki\-pe\-dianern\textsuperscript{\tiny *} f\"ur ihre \textit{parrhesia} auch keine Gef\"angnisstrafen. 

Im Folgenden soll kurz ein Streitgespr\"achbeispiel vorgestellt werden, das aus meiner Sicht unter denjenigen, die ich mitbekommen habe, zu den interessantesten der letzten Zeit z\"ahlt, weil \mbox{hier} durch \mbox{User}:Goesseln f\"ur einen biografischen Artikel nahezu investigativ gearbeitet worden war und die Debatte auf \href{https://de.wikipedia.org/wiki/Wikipedia_Diskussion:Hauptseite/Schon_gewusst/Diskussionsarchiv/2014/Dezember#Eigenvorschlag:_Hans_Haffenrichter_.284._Dezember.29_.28erl..29}{zwei} \href{https://de.wikipedia.org/wiki/Wikipedia_Diskussion:Hauptseite/Archiv/2015/Januar#Kein_Beleg}{verschiedenen} Sei\-ten ausgetragen wurde, mit unterschiedlichen Beteiligten -- wenn sich auch manche von ihnen auf beiden dieser Sei\-ten \"au{\ss}ern. Neben der Thematik (biografische Fakten zu einem K\"unstler\textsuperscript{\tiny *} der NS-Zeit sind auch Jahrzehnten noch unklar -- und wie diese Tatsache enzyklop\"adisch darstellbar w\"are) ist auch der Verlauf interessant, weil er als f\"ur Wi\-ki\-pe\-dia charakteristisch angesehen werden kann, auch wenn viele der Debatten weit l\"anger sind und weniger konstruktiv gef\"uhrt werden als in diesem Fall. Anlass war, dass ein neuer Artikel f\"ur die entsprechende Rubrik ("`Schon gewusst?"') auf der Hauptseite vorgeschlagen wurde und es darum ging, einen passenden Teaser zu finden, und zwar, nachdem der Vorschlag bereits f\"ur den \"ubern\"achsten Tag eingetragen worden war. Es bestand also ein gewisser Zeitdruck, um die Sache zu kl\"aren. Zwischen dem Artikelvorschlag am 27. Dezember 2014 und der ersten kritischen Wortmeldung 29. Dezember 2014 vergehen knapp zwei Tage, eine erl\"auternde Antwort des Artikelinitiators\textsuperscript{\tiny *} -- und in diesem Fall derselbe \mbox{User} wie der Vorschlagende -- folgt erst am 19. Januar 2015. Drei Tage sp\"ater meldet sich ein an\-de\-rer \mbox{User} etwas emotionaler und merkt an, dass der Teaser trotz des ersten Einwandes f\"ur den \"ubern\"achsten Tag in der Hauptsei\-tenvorlage platziert worden sei. Zwei Tage sp\"ater, am Morgen des ers\-ten Pr\"asentationstags antwortet \mbox{User}:C.Kolt\-zen\-burg auf diesen Einwand, dass die Erl\"auterung des Initiators\textsuperscript{\tiny *} eingeleuchtet habe. Ferner kommentierte \mbox{User}:C.Kolt\-zen\-burg das Verwirrende am Inhalt des Artikels noch auf eine Weise, mit der dem Kritiker\textsuperscript{\tiny *} indirekt Wertsch\"atzung daf\"ur zukam, dass seine Fragen zurecht zus\"atzlich auf etwas Interessantes hinweisen (dass es Fragen seien, die an die Geschichtsschreibung gestellt werden m\"ussten), und nutzte dennoch die Gelegenheit, um die zuvor eingenommene Position zu best\"atigen, dass es (auch deshalb) ein brillianter Teaser und ein guter Artikel sei. Der Kritiker\textsuperscript{\tiny *} wundert sich, wie verschieden man das Pro\-blem sehen kann, etwas belustigt, be\-steht auf einer Kl\"arung zweier Punkte im Detail und zitiert Regeln. Der beanstandete Teaser ist w\"ahrenddessen bereits auf der Hauptseite zu sehen. Etwa eine halbe Stunde sp\"ater meldet sich ein weiterer \mbox{User} und sagt, es sei regelgem\"a{\ss} verfahren worden, und erg\"anzt moderierend, dass "`wir"' Protest aushalten (es darf angenommen werden, dass die Beteiligten wissen, dass dieser \mbox{User} auch die Rolle eines deWP-Administrators innehat). Auf erneute Nachfrage bez\"uglich des schon an\-ge\-spro\-chenen Details antwortet derselbe User, dass der Artikel den Sachverhalt aus\-rei\-chend gut darstelle und der Teaser so stimme. Etwa eine Stunde sp\"ater meldet sich der Artikelinitiator\textsuperscript{\tiny *} wieder und legt die Gr\"unde f\"ur die gew\"ahlte Darstellungsweise dar, mit dem Fazit: \href{https://de.wikipedia.org/w/index.php?title=Wikipedia_Diskussion:Hauptseite/Schon_gewusst&diff=prev&oldid=138107562}{"`Also: deutsche Geschichte bleibt spannend, denke ich mal, und ist nicht immer in ein paar d\"urren Teaser-Worten abschlie{\ss}end zu behandeln"'} (\cite{UserGoesseln2015a}). \mbox{User}:C.Kolt\-zen\-burg schreibt ein "`+1"' unter diesen Beitrag, um Zustimmung zu sig\-na\-li\-sieren. Am Nachmittag des ersten Pr\"asentationstages wird auf der Dis\-kus\-si\-onsseite der Hauptseite unter dem Titel "`Kein Beleg"' ein formaler Einwand ge\"au{\ss}ert (denn Aussagen im Teaser m\"ussen im Artikel belegt sein). Ein Admin weist auf die Dis\-kus\-si\-onsseite hin. Der Artikelinitiator\textsuperscript{\tiny *} meldet sich auch auf dieser Seite und erl\"autert am selben Tag abends in an\-de\-rer Form und eingehender als in der ersten Debatte, wie es zu diesem Teaser kommt und schlie{\ss}t mit: \href{https://de.wikipedia.org/w/index.php?title=Wikipedia_Diskussion:Hauptseite&diff=prev&oldid=138120673}{"`Vielleicht hast Du ja einen besseren Vorschlag, wie man so etwas formulieren soll, wenn es 66 Jahre braucht, bis so langsam etwas mehr Licht ans Geschehen kommt"'} (\cite{UserGoesseln2015b}). Der Vorredner\textsuperscript{\tiny *} spricht zwei Punkte an, von denen der erste ein "`m. E."' enth\"alt, formuliert ein Fazit in Form einer Bitte, betont die konstruktive Absicht des Einwands, denn die Aussage sei "`\"au{\ss}erst seltsam formuliert"' und das habe ihn\textsuperscript{\tiny *} verwundert, und formuliert eine Zusammenfassung der Aspekte des Artikels (die im Teaser genannt sind), was al\-ler\-dings f\"ur einen Teaser zu lang ist. Etwa eine Stunde sp\"ater meldet sich der vorige Kritiker\textsuperscript{\tiny *} auf der ersten Seite wieder und be\-steht auf einer konkreten Angabe zu einem zuvor an\-ge\-spro\-chenen Detail. Der Artikelinitiator\textsuperscript{\tiny *} geht zwei Stunden sp\"ater erneut darauf ein, erl\"autert weitere Umst\"ande und spekuliert etwas ("`meine TF"'), um mit der Aussage zu schlie{\ss}en, dieser K\"unstler habe eben einen Bauhaus-Nimbus. Am Morgen des zwei\-ten Pr\"asentationstages sig\-na\-li\-siert der Kritiker\textsuperscript{\tiny *} von der ersten Seite "`100\% Zustimmung"' mit den Ausf\"uh\-rungen des Vorredners\textsuperscript{\tiny *} und beschwert sich \"uber die Leute, die im Team von "`Schon gewusst?"' aktiv sind (zu denen in einer Kultur mit offenen Gruppenstrukturen der Kritiker\textsuperscript{\tiny *} \mbox{selbst} auch z\"ahlt, sich aber wohl nicht z\"ahlen mag). Der Vorredner\textsuperscript{\tiny *} merkt etwas sp\"ottisch an, dass er\textsuperscript{\tiny *} zwar gesucht, aber nicht gefunden hat, auf was sich der Kritiker\textsuperscript{\tiny *} bezieht, \"au{\ss}ert sich bedauernd \"uber die Arbeitsweise ("`Unwissenschaftlich und journalistisch unsauber"') und hinterfragt abschlie{\ss}end eine "`man"'-Formulierung des Kritikers\textsuperscript{\tiny *} mit \href{https://de.wikipedia.org/w/index.php?title=Wikipedia_Diskussion:Hauptseite&type=revision&diff=138151607&oldid=138138024}{"`Und wer bitte ist eigentlich \flq man\frq \,\,bei Wi\-ki\-pe\-dia?"'} (\cite{User188.106.99.552015}). Am Ende des zwei\-ten Tages schlie{\ss}t ein Admin die Debatte mit zwei formalen Bemerkungen. Der Teaser wurde nicht ver\"andert. F\"ur meine Fragestellung ist an dieser Debatte inhaltlich interessant, dass an zwei Stellen auf Wissenschaftliches Bezug genommen wird und zu zwei weiteren Punkten betont wird, dass Aussagen eindeutig sein m\"ussen. Darauf wird in einem der folgenden Abschnitte noch einzugehen sein, denn

\singlespacing
\begin{quote}
"`a vital element in all literature is indistinctness, and this empowers the reader. The reader, that is, not only can but must come to some kind of accomodation with the indistinctness in order to take meaning from the text. For that, the imagination must operate."' (\cite{Carey2005}:214)
\end{quote}
\onehalfspacing

Bei "`Diderots Projekt"' gab es von Anfang an die Redaktionsrichtlinie, dass sich Aussagen in verschiedenen Artikeln widersprechen k\"onnen (\cite{Blom2004}:145). Bei Wi\-ki\-pe\-dia hingegen gibt es einen QS-Baustein, mit dem bem\"angelt werden kann, dass Aussagen in diesem Artikel mit denen in einem anderen nicht \"ubereinstimmen.

\singlespacing
\begin{quote}
\href{https://de.wikipedia.org/w/index.php?title=Wikipedia_Diskussion:Redaktion_Musik\&diff=137580712\&oldid=137579206}{"`Es gibt in WP keine zentrale Redaktion, die festlegt welche Artikel am wichtigsten sind und zuerst geschrieben werden m\"ussten, denn WP beruht auf Freiwilligenarbeit und die Autoren schrei\-ben dementsprechen [sic] wor\"uber sie wollen, solange sie bestimmte Mindestkriterien einhalten. Dementsprechend legen die in WP existierenden Redaktionen nur gewisse Rahmenbedingungen vor, innerhalb deren die Autoren selber entscheiden was sie machen. Ein Thema wird da eben immer erst dann bearbeitet, wenn ein Autor auftaucht der dar\"uber schrei\-ben will. In diesem Sinne ist WP weitgehend hierarchielos und funktioniert \flq bottom up\frq \,\,im Gegensatz zum redaktionellen \flq top down\frq \,\,in konventionellen Enzyklop\"adien. Das ergibt sich letzt\-lich mhr [sic] oder weder [sic] zwangsl\"aufig aus der offenen hierarchiearmen Struktur die auf freiwilliger Mitarbeit beruht."'} (\cite{UserKmhkmh2015})
\end{quote}
\onehalfspacing 

Andere Aspekte \"ahnlicher Faktoren sieht \mbox{User}:Magiers, wenn es hei{\ss}t: 

\singlespacing
\begin{quote}
\href{https://de.wikipedia.org/w/index.php?title=Benutzer:Magiers/Wer_hat_Angst_vorm_Hauptautor\%3F&oldid=136460924#Platzhirsch_vs._Platzhirsch}{"`Denn die flache Hierarchie im Projekt existiert nur auf dem Papier: in Wahrheit wird die Wi\-ki\-pe\-dia von einer vielschichtigen Hierarchie von Administratoren, Sichtern, Ein\-gangs\-kon\-trol\-leu\-ren, Qua\-li\-t\"ats\-si\-che\-rern, von Re\-dak\-ti\-o\-nen, Pro\-jek\-ten, Stamm\-ti\-schen und \sout{Seilschaften} [sic] Be\-kannt\-schaf\-ten regiert."'} (\cite{UserMagiers2014b1})
\end{quote}
\onehalfspacing

und:

\singlespacing
\begin{quote}
\href{https://de.wikipedia.org/w/index.php?title=Benutzer:Magiers/Wer_hat_Angst_vorm_Hauptautor\%3F&oldid=136460924#Fazit}{"`Die Wi\-ki\-pe\-dia ist ein Projekt von einzelnen, unterschiedlichen Autoren, die auf ihre ganz eigene Art Artikel schrei\-ben und f\"ur das Ergebnis auch Verantwortung tragen. Kollaborative Zusammenarbeit, das vielbeschwo\-rene Wiki-Prinzip, kann auf Dauer nur auf einer Grundlage funktionieren: Respekt f\"ur die Arbeit des anderen."'} (\cite{UserMagiers2014b2})
\end{quote}
\onehalfspacing

In welchem Stil und auf welchem Wege Leser\textsuperscript{\tiny *} zu mehr Wissen gelangen, hat Diderot oftmals die Verfasser\textsuperscript{\tiny *} \mbox{selbst} entscheiden lassen (\cite{Blom2004}:154). Auch bei Wi\-ki\-pe\-dia k\"onnen \mbox{User} beim Verfassen von Artikels ihren Stil entfalten, solange dieser auch von anderen f\"ur "`enzyklop\"adisch"' genug gehalten wird, aber Widerspr\"uchlichkei\-ten zwischen Artikeln werden h\"ochstwahrscheinlich als Fehler angesehen, die es "`zu beheben"' gilt -- sofern sie jemandem auffallen, der sich entsprechend dazu \"au{\ss}ert oder gleich \mbox{selbst} Hand anlegt.

Zwar wird ein be\-stimmtes Charakteristikum von Wikis bereits verstanden: dass es auch in dieser Arbeitsumgebung bei enzyklop\"adischen Eintr\"agen Wandlungen von "`Auflage"' zu "`Auflage"' gibt bzw. wie traditionell bei Debatten, die in Zeitschriften ein Forum erhalten, von Ausgabe zu Ausgabe -- was in etwa den Dis\-kus\-si\-onssei\-ten bei Wi\-ki\-pe\-dia entspricht. Vor allem ein Begriff scheint dazu avanciert zu sein, das Wesen des Arbei\-tens in einem Wiki zu repr\"asentieren: Edit War, eine Be\-zeichnung, die eine k\"ampferische Motivation hervorhebt. "`Wohlwollend betrachtet stellen Edit Wars letzt\-lich nichts anderes als falsch kanalisierte Versuche der Aushandlung dar,"' fasste Daniela Pscheida es 2010 zusammen (\cite{Pscheida2010}:385). Durch Wi\-ki\-pe\-dia wurde zwar die Existenz und Funktionsweise von Wikis weithin bekannt, doch dass sich Teile eines Wiki im Gegensatz zu nicht-kollaborativ angelegten Websites versionsgenau zitieren l\"asst, hat auch im 15. Jahr von Wi\-ki\-pe\-dia noch nicht alle erreicht. Im Navigationsrahmen jedes Artikels wird inzwischen bei "`Werkzeuge"' sogar "`Seite zitieren"' als Funktion angeboten. Drei Formate stehen zur Auswahl: eine einfache Variante "`zum Kopieren"', eine "`bibliografische"', aus der auf die Versionsgeschichte der Seite als Bestandteil der Angabe verlinkt wird und eine im BibTeX-Format, mit einer Erl\"auterung. (Verlinkungen auf die jeweilige Versionsgeschichte eines Artikels innerhalb einer bibliografischen Angabe soll das Nennen von Autoren\textsuperscript{\tiny *}namen er\"ubrigen -- was nicht zu\-letzt in recht\-li\-cher Hinsicht kritisch zu sehen ist, denn das Urheberrecht liegt nach wie vor bei den Einzelnen und nicht bei der Wikimedia Foundation oder ihren Chapters, auch wenn das Layout der Sei\-ten dies suggerieren soll, weil die Autoren\textsuperscript{\tiny *}namen nicht zu sehen sind. In dieser Hinsicht ist die Anzeigevariante des Projekts \href{http://wikibu.ch}{\textit{wikibu.ch}} interessant, das Wi\-ki\-pe\-dia-Daten importiert und auf andere Weise darstellt. Unter anderem wird \mbox{hier} angezeigt, wer in letzter Zeit ma{\ss}geblich am Artikel gearbeitet hat und auch weitere Autoren\textsuperscript{\tiny *} werden genannt.) 

Die Versionsgeschichte jeder Seite dient dazu, mittels Protokollierung die jeweiligen Ver\"anderungen Schritt f\"ur Schritt nachvollziehbar zu machen. Im Gegensatz zu Aktualisierungen, die an Auflagendruck gebunden sind, entstehen "`Auflagen"' in einem Wiki lediglich in h\"oherer Aktualisierungsfrequenz -- wenn das in Aussicht gestellte Ergebnis auch nicht ganz so brilliant ist wie es in dieser Selbst\-preisung im Eintrag "`Wikipedia:About"' klingt: \href{https://en.wikipedia.org/w/index.php?title=Wikipedia:About\&oldid=642064642\#Wikipedia_vs_paper_encyclopedias}{"`A paper encyclopedia stays the same until the next edition, whereas editors can update Wi\-ki\-pe\-dia at any instant, around the clock, to help ensure that articles stay abreast of the most recent events and scholarship"'} -- und jede \"Anderung wird transparent nachweisbar angezeigt. Bester wissenschaftlicher Stil also, k\"onnte man sagen. 

Allgemeiner gesehen ist f\"ur redaktionelle Prozesse in adhoc-Teams eine Ver\-fah\-rens\-wei\-se "`erst machen, dann reden"' oftmals am effizientesten. Je nachdem, wieviele \mbox{User} sich in der letzten Zeit f\"ur diesen Artikel interessiert haben (oder aktuell auf einen Streit\-punkt aufmerksam geworden), ist f\"ur den Informationswert des Artikels eventuell mehr erreicht, wenn man eine neue Version herstellt, die besser erscheint, und erst anschlie{\ss}end schaut, ob es Widerspruch bez\"uglich dieser \"Anderung gibt. Wie zutr\"aglich dieses Verfahren f\"ur welchen Themenbereich und zu welchem Zeitpunkt ist, h\"angt vom Umstrittenheitsfaktor ab. J\"ungst wurde eine sanftere Haltung in Worte gefasst von \mbox{User}:Sasso H\"u{\ss}elmann, wenn es hei{\ss}t: 

\singlespacing
\begin{quote}
\href{https://de.wikipedia.org/w/index.php?title=Diskussion:Stimmungen_lesen._\%C3\%9Cber_eine_verdeckte_Wirklichkeit_der_Literatur\&diff=next\&oldid=139801823}{"`Ich versuche immer erst zu verstehen was sich ein Autor beim Aufbau seines Artikels gedacht hat -- bevor ich einfach nach Gutd\"unken ohne R\"ucksprache massiv in den Textaufbau eingreife oder am Aufbau des Textes rumschraube. L\"asst sich zwar alles immer wieder zur\"ucksetzen -- aber Missverst\"andnisse und VMs sind dann (trotz guter Absichten) leider oftmals vorprogrammiert..."'} (\cite{UserSalmanSasso2015})
\end{quote}
\onehalfspacing

W\"ahrend es in der \textit{Encyclop\'{e}die} vom Prinzip her keine Eintr\"age zu einzelnen Per\-so\-nen gab -- zumindest nicht bis zu dem relativ sp\"aten Zeitpunkt, ab dem Louis de Jaucourt, der naturwissenschaftlich spezialisierte Au{\ss}enseiter und dritter Hauptautor, einige verfasste -- machen biografische Artikel bei Wi\-ki\-pe\-dia einen sehr gro{\ss}en Anteil aus.
Dass die damalige Entscheidung angesichts limitierter personeller Ressourcen nachzuvollziehen ist, l\"asst sich f\"ur die deutschspra\-chi\-ge Wi\-ki\-pe\-dia am Beispiel des Eintrags "`Herta M\"uller"' verdeutlichen. \mbox{Hier} haben sich seit der ersten Version am 1. Dezember 2003 missg\"unstige Absichten im Artikel niedergeschlagen.
Die Projektionsfl\"achen, als die vor allem Informationen \"uber lebende Per\-so\-nen h\"aufig genutzt werden, wurden in der \textit{Encyclop\'{e}die} konterkariert, indem in den Eintr\"agen zu "`Duc"' und "`Roi"' an erster Stelle V\"ogel dieses Namens besprochen wurden, erst danach folgt eine Abhandlung zu Mitgliedern der Aristokratie dieses Ranges -- im Allgemeinen (\cite{Blom2004}:154). Ironie war zudem ein wichtiges Mittel, um Wissen zu vermitteln, mit dem Zensurma{\ss}nahmen unterlaufen werden sollten (\cite{Blom2004}:155). Mitte des 18. Jahrhunderts fand Zensur offiziell statt und traf die \textit{Encyclop\'{e}die} und ihre Autor\textsuperscript{\tiny *}en. Die \textit{Encyclop\'{e}die} wurde von Intellektuellen ins Leben gerufen, die wussten, dass es keine neutralen Standpunkte geben kann. Diderot war es wichtig, dass Leute dazu angeregt werden, an demjenigen zu zweifeln, was ihnen als Wissen geboten wird (\cite{Blom2004}:154). Bei Wi\-ki\-pe\-dia Anfang des 21. Jahrhunderts hingegen findet Zensur offiziell nicht statt und es ist von einem Wissensbegriff ausgegangen worden, der sich von der Ansicht herleitet, dass es so etwas wie neutral formuliertes "`gesichertes Wissen"' geben kann. Daf\"ur zu sorgen, dass nur eine bestimmte Sicht der Dinge als g\"ultig angesehen wird und dass diese widerspruchsfrei zu sein scheint, erledigt sich mittels der Doktrin des "`neutralen Standpunkts"' -- aber nur solange, wie gen\"ugend Leute in ihrer Freizeit dazu bereit sind, diese spezifische Sicht der Dinge von Fall zu Fall durchzusetzen. \mbox{User}:Neon02 formuliert seine Sicht dieser Dynamik folgenderma{\ss}en: 

\singlespacing
\begin{quote}
\href{https://commons.wikimedia.org/w/index.php?title=File:Allesueberwikipedia.pdf\&page=41}{"`[D]iejenigen Be\-nut\-zer, deren Meinung mit dem gerade existierenden Mainstream \"ubereinstimmt, [k\"onnen] behaupten, sie vertr\"aten die reine Wahrheit, w\"ahrend alle anderen \flq \href{https://de.wikipedia.org/wiki/Wikipedia:Neutraler_Standpunkt}{POV}-Pusher\frq \,\,oder \flq Men on a Mission\frq \,\,seien, die von Wi\-ki\-pe\-dia ferngehalten werden m\"ussen. Nur derjenige, der aus der Position der ge\-sell\-schaftlich dominierenden Ideologie bzw. des dominierenden Wissens spricht, also der Doxa im Sinne von Bourdieu, kann diesen Vorwurf \"au{\ss}ern. Das bedeutet, dass sich in den Artikeln langfristig diejenigen Positionen durchsetzen werden, die in der Ge\-sell\-schaft gerade dominant sind. Al\-ler\-dings noch nicht einmal in der Gesamtge\-sell\-schaft, sondern in der Gruppe der Wi\-ki\-pe\-dia-Autoren, also vor allem der jungen, gut gebildeten, m\"annlichen Naturwissenschaftler."'} (\cite{BenutzerNeon022011}:41)
\end{quote}
\onehalfspacing

Benutzer:Neon02 f\"uhrt diese Dynamik und ihre Mechanismen darauf zur\"uck, dass zu dem Zeitpunkt, als die Grundprinzipien der Wi\-ki\-pe\-dia formuliert wurden, Jimmy Wales sich \href{https://commons.wikimedia.org/w/index.php?title=File:Allesueberwikipedia.pdf\&page=41}{"`an einem eng positivistischen Wahrheitsbegriff orientiert"'} habe, \href{https://commons.wikimedia.org/w/index.php?title=File:Allesueberwikipedia.pdf\&page=41}{"`der in den Naturwissenschaften noch akzeptabel sein mag, aber in den Sozialwissenschaften, wo soziale Interessen For\-schungsprogramme und Erkenntnisse beeinflussen, gro{\ss}e Pro\-ble\-me bereitet. Nicht zuf\"allig drehen sich die gro{\ss}en Methodendebatten in der Soziologie um Fragen der m\"oglichen oder unm\"oglichen Werturteilsfreiheit"'} (\cite{BenutzerNeon022011}:41). In seinem Beitrag "`Brecht on Wiki"' von 2012 wiederum bescheinigt Mautpreller der Wi\-ki\-pe\-dia seit ihrer Gr\"undung eine "`Schlagseite zum \flq Faktischen\frq"' (\cite{Mautpreller2012}:35). Positionen, von denen aus Fakten als neutral gelten, scheinen bei Wi\-ki\-pe\-dia also in einem Ma{\ss}e zu dominieren, dass aus einer anderen Warte das Gegenteil von Ausgegeglichenheit oder Gleichgewicht zu beobachten ist: sogar eine "`Schlagseite"'.

\singlespacing
\begin{quote}
"`W\"ahrend der klassische Enzyklop\"adie-Betrieb eine Expertokratie ist, hat Wi\-ki\-pe\-dia seine Wurzeln eher im Anarchismus und in der Basisdemokratie.\\ Neben dieser Ideologie der Offenheit spielt die Idee des neutralen Standpunkts eine ganz entscheidende Rolle. Nur,[sic] wenn wir jedem Be\-nut\-zer das Ge\-f\"uhl geben k\"onnen, dass seine Sicht der Dinge nach fairen Grunds\"atzen behandelt wird, k\"onnen wir verhindern, dass sich Menschen mit den unterschiedlichen religi\"osen und politischen Ansichten die K\"opfe einschlagen."' (Erik M\"oller), zitiert in \cite{SchliekerLehmann2007}:259)
\end{quote} 
\onehalfspacing 

F\"ur meine Fragestellung interessant ist vor allem der Unterschied, dass vor 250 Jahren bei der \textit{Encyclop\'{e}die ou Dictionnaire raisonn\'{e} des sciences, des arts et des m\'{e}tiers} ein herrschendes Dogma infrage gestellt werden sollte (\cite{Lih2009}:15) bzw. in er\-kennt\-nis\-the\-o\-\-re\-ti\-scher Hinsicht eine grundlegend neue Architektur von Wissensgebieten und ihrem Bezug zueinander erarbeitet und dargestellt wird (\cite{Darnton1972}). Bei Wi\-ki\-pe\-dia hingegen wird tendenziell die jeweils herrschende Perspektive ge\-st\"arkt, wenn der \href{https://de.wikipedia.org/wiki/Wikipedia:Neutraler_Standpunkt}{NPOV}-Doktrin Folge geleistet wird. Wie die Wikipedistik sich mit dem Ph\"anomen "`NPOV"' und dessen Folgen befasst hat, soll im folgenden Abschnitt dargestellt werden.

Wales soll 2006 bez\"uglich der von seinem Mitgr\"under Larry Sanger eingef\"uhrten (\cite{Lih2009}:6-7) NPOV-Policy gesagt haben: \href{http://reason.com/archives/2006/08/15/the-neutrality-of-this-article}{"`One of the great things about NPOV is that it is a term of art and a community fills it with meaning over time."'} (\cite{Mangu-Ward2006}) Hier\-zu zwei Ein\-sch\"at\-zungen, wie sich dieses Prinzip in der Praxis zeigt:

\singlespacing
\begin{quote}
\href{https://de.wikipedia.org/w/index.php?title=Wikipedia_Diskussion:Wiki-Dialoge\%2FQualit\%C3\%A4t\%2FWissenschaftliches_Schreiben\&diff=131040372\&oldid=131030900}{"`Ich habe gesehen wie einzelne Autoren miteinander gerungen haben einen Artikel NPOV zu bekommen. Als das erreicht war, tauchte eine neue Person auf und rief: POV! Also, was eben noch NPOV war, ist jetzt POV. Das hei{\ss}t, dass ein Artikel nicht per se neutral oder nicht neutral ist, sondern das ist eine Zuschreibung von au{\ss}en."'} (\cite{UserGoldzahn2014})
\end{quote}
\onehalfspacing

\singlespacing
\begin{quote}
\href{https://de.wikipedia.org/w/index.php?title=Wikipedia_Diskussion:Kurier\&diff=prev\&oldid=137785144}{"`Nebenbei glaube ich, dass der eigentliche Grund f\"ur die hemmungslose Verteufelung von \flq Feminismus\frq \,\,und \flq Frauengruppen\frq \,\,etc. ist, dass der Feminismus eine Provokation f\"ur die Wi\-ki\-pe\-dia darstellt. Er h\"alt n\"amlich fest, dass es die eine ungeteilte rationale Wirklichkeit gar nicht gibt. Ich bin kein Feminist und habe Zweifel daran, dass der feministische Blick wirklich so viel leisten kann wie behauptet. Aber die Provokation f\"ur das abstrakte, geschlechtslose rationale Subjekt, das den Konstruktionen des NPOV zugrunde liegt, ist heilsam. Blo{\ss} wird sie eben von manchen nicht vertragen."'} (\cite{UserMautpreller2015a})
\end{quote}
\onehalfspacing

Gelegentlich werden neue Einsch\"atzungen formuliert, wie mit Belegen umzugehen ist. Meist geschieht dies aus Anlass von konkreten F\"allen, in denen sich Leute nicht einig sind. Eine solche Formulierung von Januar 2015 ist f\"ur meine Fragestellung unmittelbar relevant, daher sei sie \mbox{hier} zitiert -- ohne absch\"atzen zu k\"onnen, wie stark sie von welchen einflussreichen weiteren Usern der deutschspra\-chi\-gen Community l\"angerfristig mitgetragen w\"urde: 

\singlespacing
\begin{quote}
\href{https://de.wikipedia.org/w/index.php?title=Benutzer_Diskussion:Grillenwaage\&diff=137961506\&oldid=137943802}{"`Nehmen wir als Beispiel das Schlo{\ss} oder ein paar andere Sachen von Kafka: V\"ollig unm\"oglich, diese alptraumhafte Atmosph\"are einer offenbar naturgesetzlichen Vergeblichkeit mit einer reinen Inhaltszusammenfassung r\"uberzubringen, und auch die akademischen Ausdeutungen verm\"o\-gen das nicht zu leisten. Wie sollten sie auch? Denn es sind ja nicht die gelehrten Deutungen und Schlussfolgerungen, sondern eben diese ge\-f\"uhlte Atmosph\"are, die den Reiz und das Wesen des Werks ausmachen.\\
Was aber, wenn es keine entsprechenden Quellen gibt und man sich diese sozusagen \flq atmosph\"arische\frq \,\,Charakterisierung \mbox{selbst} aus den Fingern saugen mu{\ss}? Ich finde, da{\ss} man auf diesem speziellen Gebiet die Ma{\ss}gaben von \href{https://de.wikipedia.org/wiki/Wikipedia:Keine_Theoriefindung}{TF} und \href{https://de.wikipedia.org/wiki/Wikipedia:Neutraler_Standpunkt}{POV} zumindest au{\ss}ergew\"ohnlich kulant handhaben sollte."'} (\cite{UserEpipactis2015b})
\end{quote}
\onehalfspacing

Eine "`normale"' Webseite und eine Seite in einem Wiki m\"ogen auf den ersten Blick kaum zu unterscheiden sein. Eine Wikiseite jedoch bietet durch automatische Versionierung pro get\"atigter \"Anderung eine entwicklungsgeschichtliche Transparenz, die bemerkenswert ist. In Wikis zu lesen und zu arbei\-ten bedeutet, sich \"uber jeweils neue Inhalte mittels Versionsgeschichten zu informieren. Ein Wi\-ki\-pe\-dia-Eintrag etwa gibt immer auch die Montiertheit seiner Inhalte preis -- f\"ur ein Projekt, das allgemein auf das Sammeln und Pr\"asentieren von "`Wissen"' angelegt ist, in epistemologischer Hinsicht hochinteressant, weil deutlich wird, dass die Inhalte eines Wi\-ki\-pe\-dia-Eintrags durch Montage zustandegekommen sind. Wer also die \"Anderungen an einem Wi\-ki\-pe\-dia-Eintrag laufend mitbekommen m\"ochte, liest dessen Versionsgeschichte. Auch aktuelle Entwicklungen auf Dis\-kus\-si\-onssei\-ten werden von erfahreneren Wiki-Usern \"uberwiegend via Versionsgeschichte gelesen. F\"ur wen dies m\"oglich ist, h\"angt auch von Navigationskenntnissen ab, also davon, ob ich verstehe, wohin ich klicken muss, um zu diesen Protokollen zu kommen, und wenn das geklappt hat, w\"urde es als N\"achstes darum gehen, he\-rausgefunden zu haben, wie die einzelnen \"Anderungen, die im Protokoll gelistet sind, miteinander verglichen werden k\"onnen. 

Vielleicht l\"asst sich der gr\"o{\ss}te Unterschied zwischen den beiden enzyklop\"adischen Projekten an einem Punkt ausmachen, den ich zwischen Kultursoziologie und Pu\-bli\-zistiktheorie verorten w\"urde: War unter den vielen Autoren\textsuperscript{\tiny *} der \textit{Encyclop\'{e}die} vor 250 Jah\-ren Louis de Jaucourt mit einem naturwissenschaftlichen Hintergrund in einer Au{\ss}enseiter\textsuperscript{\tiny *}position und trug als neue Art von Eintr\"agen biografische Artikel zum Gesamten bei, sind im Projekt Wi\-ki\-pe\-dia Eintr\"age zu Per\-so\-nen v\"ollig im Normbe\-reich, aber das Dar\-stel\-len nicht-pro\-po\-si\-ti\-o\-na\-len Wissens f\"uhlt sich wie eine subversive T\"atigkeit an -- und dies betrifft vor allem Kunst in allen ihren Formen und Feldern. Nun war diese \textit{Encyclop\'{e}die} nicht die einzige ihrer Zeit und vergliche man Wi\-ki\-pe\-dia mit anderen enzyklop\"adischen Projekten der Zeit der Aufkl\"arung w\"urde vermutlich ein anderes Bild entstehen.

Schauen wir voraus: Wird der \textit{Encyclop\'{e}die} unter anderem die Wirkung bescheinigt, f\"ur die nachfolgende Generation die Franz\"osische Revolution vorbereitet zu haben, so l\"asst sich f\"ur die m\"ogliche Wirkung von Wi\-ki\-pe\-dia-Inhalten aus einer Studie zitieren: 

\singlespacing
\begin{quote}
\href{http://www.degruyter.com/view/books/9783486858662/9783486858662-002/9783486858662-002.xml}{"`Sch\"uler/innen [nutzen] das Internet als Nachschlagewerk [...] und [be\-vor\-zu\-gen] dabei Textstellen [...], die frei von historischen Deu\-tungs\-mus\-tern sind, also nach ihrer Ansicht \flq neutrales Wissen\frq\,\,enthalten. Dieses Wissen nehmen sie als Allgemeingut wahr, das lediglich gesammelt und in eine gut rezipierbare Form gebracht werden muss. Dazu werden die Textstellen kopiert und nach zwei Mustern zusammengef\"ugt: 1. meh\-rere unterschied\-liche, meist unbenannte Textstellen werden mit kurzen eigenen Ein\-sch\"u\-ben wie in einem Mosaik zusammengef\"ugt. 2. Unterschiedliche Elemente wie eine Zeitleiste, ein Wi\-ki\-pe\-dia-Artikel, Bilder werden an\-ein\-an\-der\-ge\-h\"angt und als Basisinformation zu einem m\"und\-li\-chen Vortrag genutzt."'} (\cite{Alavi2015}:9)
\end{quote}
\onehalfspacing
 
\pagebreak

\section{Metaebene I}
\label{sec:5}

In diesem Kapitel er\"ortere ich erstens Konzepte, die ich f\"ur meine Argumentation nutze (\textit{~\ref{subsec:5.1}}). Ich stelle zwei\-tens anhand von Aussagen, die mir auf beiden Sei\-ten als ty\-pisch erscheinen, das Feld dar, auf das sich mein Erkenntnisinteresse richtet und f\"ur das ich die Ergebnisse der Arbeit fruchtbar machen m\"ochte (\textit{~\ref{subsec:5.2}} und \textit{~\ref{subsec:5.3}}). Drittens produziere ich \textit{Erlesnisse} in schriftlicher Form im Rahmen eigener Interpretationen von \mbox{vier} zeitgen\"ossischen li\-te\-ra\-rischen Texten (\textit{~\ref{subsec:5.4}}). Eine Zusammenfassung beschlie{\ss}t das Kapitel.

\subsection{Literatur\textsuperscript{\~.\~.}lekt\"ure, Leseerlebnis, \textit{Erlesnis}, Lesebericht}
\label{subsec:5.1}

Im Rahmen dieser Arbeit werden einige Li\-te\-ra\-tur\-bezogene Begriffe wie gewohnt verwendet, einige werden neu definiert und infolgedessen in einer Abwandlung genutzt, und \textit{Erlesnis} wird neu eingef\"uhrt. Wie gewohnt zu verstehen sind "`Li\-te\-ra\-tur"' und "`lesen"' ebenso wie "`Leseprozess"'. In der deutschspra\-chi\-gen For\-schungs\-li\-te\-ra\-tur werden "`Li\-te\-ra\-tur\-lekt\"ure"', "`Lekt\"ureerfah\-rung"', "`Lekt\"ureerlebnis"' und "`Leseerlebnis"' verschiedentlich verwendet. Begrifflich definiert habe ich sie in neueren Studien bis\-her nicht finden k\"onnen.

Der Begriff Li\-te\-ra\-tur\textsuperscript{\~.\~.} (mit \textsuperscript{\~.\~.}) be\-zeichnet im Rahmen dieser Arbeit das Entstehen von Emp\-fin\-dun\-gen im Lese- und/oder Zuh\"orkontakt mit li\-te\-ra\-rischen Texten. Dabei soll nicht von Belang sein, zu welchem Zeitpunkt oder aus welchem Anlass jemand dieser Emp\-fin\-dun\-gen gewahr wird. Vielmehr geht es im Kontext der vorliegenden Argumentation allein darum, dass ihr Entstehen einer bestimmten Situation zugeschrieben wird, in der ein li\-te\-ra\-rischer Text eine gewisse Rolle spielte -- und wie diese Empfin\-dungen in Worte gefasst werden. W\"ahrend manche Akteure\textsuperscript{\tiny *} der Institution Li\-te\-ra\-tur eher den Begriff "`li\-te\-ra\-risches Werk"' prozessual definieren (\cite{Attridge2004}), wird \mbox{hier} als Br\"uckenfunktion zwischen dem Denkstil, der bei Wi\-ki\-pe\-dia vor\-herrscht, und li\-te\-ra\-tur\-wissenschaftlich gepr\"agten Denkstilen der Begriff "`Li\-te\-ra\-tur"' in etwas Pro\-zessu\-ales umgewandelt, zu Li\-te\-ra\-tur\textsuperscript{\~.\~.} (denn das Denkkollektiv (\cite{Moller2007}, \cite{WernerZittel2011}:16-24) der deutschspra\-chi\-gen Wi\-ki\-pe\-dia-Community macht auf mich ei\-nen we\-ni\-ger flexiblen Eindruck als die anzunehmende Leser\textsuperscript{\tiny *}schaft der vorliegenden Arbeit, die zum einen gewohnt ist, mit Diskursvarianten bewusst umzugehen und zum anderen mei\-ner Annahme nach f\"ur Gedankenexperimente offen ist). Mit Li\-te\-ra\-tur\textsuperscript{\~.\~.} ist in dieser Arbeit also keine Institution gemeint, sondern ein Geschehen, ein prozessuales Ereignis, bei dem eine unbestimmbare Menge an Emp\-fin\-dun\-gen entsteht.

"`Lekt\"ure"' wiederum be\-zeichnet wie \"ublich die Handlung des Lesens. Der im Titel verwendete Terminus "`Li\-te\-ra\-tur\-lekt\"ure"' ist herk\"ommlich zu verstehen und meint Li\-te\-ra\-tur im Sinne von "`li\-te\-ra\-rische Werke"', hingegen ist der Begriff "`Li\-te\-ra\-tur\textsuperscript{\~.\~.}lek\-t\"u\-re"' folgenderma{\ss}en zu lesen: Das Nachsinnen \"uber einen Prozess, dem ein Leser\textsuperscript{\tiny *} das Entstehen bestimmter Emp\-fin\-dun\-gen zuschreibt. Erst durch ein Nachsinnen \"uber die eigenen Emp\-fin\-dun\-gen aus Anlass von Lekt\"ure ergeben sich Voraussetzungen daf\"ur, dass neben Behauptungswissen in Leseberichten auch nicht-pro\-po\-si\-ti\-o\-na\-les Wissen in Worte gefasst w\"urde -- eine Art von Wissen, die vor allem in einem multidisziplin\"aren Kontext wie dem der Plattform Wi\-ki\-pe\-dia besonders hervorzuheben ist. In der \mbox{hier} vorgestellten Argumentation geh\"ort diese \"Uberlegung zu den Kernpunkten, weil bis\-her kaum Augenmerk darauf gelegt zu werden scheint. Aussagen \"uber das im eigenen Leseprozess erworbene nicht-pro\-po\-si\-ti\-o\-na\-le Wissen sind im Vergleich zu den meisten anderen Lemmata Spezifika f\"ur Eintr\"age der Kategorie "`Li\-te\-ra\-risches Werk"'. Au{\ss}ergew\"ohnlich wissenswert wird es erst durch Li\-te\-ra\-tur\textsuperscript{\~.\~.}lekt\"ure, so dass Informationen dar\"uber bei keinem Lemma dieser Kategorie fehlen sollten, sofern es dazu Belege gibt.

Bei \textit{Erlesnis} handelt es sich um eine Wortsch\"opfung, die dasjenige be\-zeichnet, was bei Li\-te\-ra\-tur\-lekt\"ure dem eigenen Empfinden nach an nicht-pro\-po\-si\-ti\-o\-na\-lem Wissen erworben wurde. Dem Konzept und dem Begriff \textit{Erlesnis} kommt in meiner Argumentation also besonderes Gewicht zu. \textit{Erlesnis} ist mein Formalobjekt, das ich ent\-wickelt habe, indem ich in vorhandenem Material nach einem gr\"o{\ss}eren Zusammenhang gesucht habe, der \"uber das konkret fassbare Materialobjekt hinausweist. \textit{Erlesnis} macht auf ein dahinterliegendes Pro\-blem aufmerksam, das ich in Form einer Hypothese vorbringen m\"ochte, n\"amlich, dass aufsei\-ten Li\-te\-ra\-tur\-bewanderter im deutschspra\-chi\-gen Raum eine gewisse Scheu be\-steht, sich \"uber eigene \textit{Erlesnisse} schriftlich zu \"au{\ss}ern. Mit der vorliegenden Arbeit werden ansatzweise die nachteiligen Folgen dieser Zur\"uckhaltung f\"ur gemeinn\"utzige Li\-te\-ra\-tur\-ver\-mitt\-lung aufgezeigt. 

Zwar gehe ich davon aus, dass im Lese- und/oder Zuh\"orkontakt mit li\-te\-ra\-rischen Texten Emp\-fin\-dun\-gen entstehen und ich lege meinem Konzept \textit{Erlesnis} die Annahme zugrunde, dass w\"ahrend der Li\-te\-ra\-tur\-lekt\"ure nicht-pro\-po\-si\-ti\-o\-na\-les Wissen erworben werden kann. Ob al\-ler\-dings nicht-pro\-po\-si\-ti\-o\-na\-les Wissen im Augenblick der Li\-te\-ra\-tur\-lekt\"ure oder erst im Zuge des Nachsinnens dar\"uber (also bei Li\-te\-ra\-tur\textsuperscript{\~.\~.}lekt\"ure) entsteht, ist \mbox{hier} nicht erheblich, denn es kommt f\"ur die vorliegende Fragestellung allein darauf an, was andere \"uber ihre \textit{Erlesnisse} schrei\-ben (und nicht, was sie dar\"uber denken, wie, warum oder wann es entstanden ist). 

Weitgehend in einer anderen Pr\"agung eingef\"uhrt wird der Begriff "`Lesebericht"'. Er wird \mbox{hier} als ein Oberbegriff verwendet f\"ur ein in Worten aufge\-zeichnetes Produkt, das inhaltlich von "`Lekt\"ureerfah\-rung"' handelt. Die Be\-zeichnung schlie{\ss}t verschiedene Formate ein: einen wissenschaftlichen Beitrag, eine Rezension, in der die erste Person Singular meist vermieden wird, verknappte Darstellungen jeweiliger Besprechungen aus \"uberregionalen Tageszeitungen (als Format bei \textit{per\-len\-tau\-cher.de} bekannt, wo sie als "`Rezensionsnotizen"' pro Werk zusammengestellt werden) oder einen Leseforen-Beitrag, der oft in einem pers\"onlichen und bewusst subjektiven Stil in Ich-Form formuliert wird (wie traditionell eine briefliche Mitteilung, heute im Stil einer Mail, aber im Web f\"ur unbekannte Adressaten\textsuperscript{\tiny *} ver\"of\-fent\-licht). Auch "`Leseerlebnis"' im \mbox{hier} definierten Sinne ist ein Lesebericht-Format: Als dezidierter Abschnitt kann es modular in einen Wi\-ki\-pe\-dia-Eintrag eingepasst werden. Objekte des \mbox{hier} definierten Formats "`Lesebericht"' sind in Worte gefasst und sollen m\"oglichst im Web mit direkter Verlinkung referenzierbar sein (Text, Audio oder Film). 

In Kapitel \textit{5.4 Interpretationen} zeigt sich das \textit{Erlesnis} in Aktion.

\pagebreak

\subsection{Wissen \"uber Literatur bei Wikipedia}
\label{subsec:5.2}

Das Fundament des bei Wi\-ki\-pe\-dia bereits Vorhandenen, auf dem die Fragestellung der Arbeit ent\-wickelt wurde, l\"asst sich aus verschiedenen Gr\"unden kaum genauer bestimmen. Weil noch keine anderen For\-schungsergebnisse zum \mbox{Bereich} Li\-te\-ra\-tur bei Wi\-ki\-pe\-dia vorliegen, sollen \mbox{hier} tentativ erstens einige Quellen f\"ur Antworten aufgezeigt, zwei\-tens an einem Punkt eine aktuelle Auswertung vorgenommen und drittens M\"oglichkeitsr\"aume er\"offnet werden. Nach einf\"uhrenden eigenen \"Uberlegungen sowie Einsch\"atzungen an\-de\-rer -- zum Beispiel zu vermuteten Interessen von Wi\-ki\-pe\-dia-Lesern\textsuperscript{\tiny *}, die durch eine Websuche bei einem Wi\-ki\-pe\-dia-Eintrag landen -- argumentiere ich im zwei\-ten Teil dieses Ab\-schnitts, dass sich in Debatten um die Eignung bestimmter \href{https://de.wikipedia.org/wiki/Teaser}{Teaser} f\"ur die Wi\-ki\-pe\-dia-Hauptseite eine Tendenz ausmachen l\"asst, die f\"ur eine erste Einsch\"atzung bez\"uglich der Frage relevant ist, welche Bedingungen der Darstellbarkeit nicht-pro\-po\-si\-ti\-o\-na\-len Wissens in Eintr\"agen zu li\-te\-ra\-rischen Werken als aktuell geltend angenommen werden k\"onnen. Dabei geht es um die Art der Aussagen in den Teasern, also die Pr\"asentation von Li\-te\-ra\-tur\-artikeln in einem Satz, und genauer: in welchem Ma{\ss} bereits \textit{Erlesnisse} formuliert werden. Das Teaser-Format kann in gewissem Sinne als Negativ-Steigerung folgender Beobachtung angesehen werden: "`Innerhalb kurzer Zeit werden von den Kritikern li\-te\-ra\-rische Werke gesichtet, die von Schriftstellern in jahrelanger Arbeit geschrieben wurden; kommentiert werden sie auf kleinstem Platz, der kaum ausreicht, um nur ann\"ahernd die Lekt\"ureerfah\-rung zu erz\"ahlen, die eine Li\-te\-ra\-tur\-kritik immer auch ist"' (\cite{Schwens-Harrant2008}:147-148). Diese Tendenz wird anhand von Beispielen aus der Zeit zwischen Oktober 2013 und Januar 2015 erhoben und f\"ur das Anliegen der vorliegenden Arbeit ana\-lysiert. Im dritten Teil er\"ortere ich, wie bei Wi\-ki\-pe\-dia Wissen \"uber Li\-te\-ra\-tur dargestellt werden k\"onnte und warum Informationen zu einem Sprachkunstwerk etwas anderes sein sollten als etwa Informationen zu einem Protein.

Erste allgemeine Anhaltspunkte w\"urden die Inhalte der Eintr\"age geben. Am inte\-ressantesten w\"aren Antworten auf die Frage, aus welcher -- vermeintlich neutralen -- li\-te\-ra\-tur\-theo\-retischen Perspektive Wissen \"uber Li\-te\-ra\-tur dargestellt wird und ob zum Beispiel eine bestimmte Richtung \"uberwiegt. Mein Eindruck ist, dass li\-te\-ra\-rische Texte in vielen Werkartikeln pr\"asentiert werden als seien sie lediglich Geschichten mit Handlungen und nicht Sprachkunstwerke. Von Ende 2014 bzw. Anfang 2015 lauten \"Au{\ss}erungen auf Wi\-ki\-pe\-dia-Dis\-kus\-si\-onssei\-ten etwa so: 

\singlespacing
\begin{quote}
\href{https://de.wikipedia.org/w/index.php?title=Benutzer_Diskussion:Grillenwaage\&diff=137653987\&oldid=137652648}{"`Ich unterscheide drei Artikelarten: 1.) Die nur Katalogk\"artchen und Klappentext abschrei\-ben; 2.) Die von Autoren, die das li\-te\-ra\-rische Werk gelesen und sich um Sekund\"arli\-te\-ra\-tur\- bem\"uht, verstanden und eingebaut haben; 3.) Die, wo alle ihr geballtes Fachwissen zur Li\-te\-ra\-tur und den be\-han\-del\-ten \,li\-te\-ra\-rischen \,und \,ausserli\-te\-ra\-ri\-schen The\-men ein\-ge\-bracht ha\-ben."'} (\cite{UserEmeritus2015})
\end{quote}
\onehalfspacing

oder: 

\singlespacing
\begin{quote}
\href{https://de.wikipedia.org/w/index.php?title=Wikipedia_Diskussion:Hauptseite\&diff=prev\&oldid=136449008}{"`Meist ist ja die WP heillos \"uberfordert, wenn es um Belletristik geht, wie oft genug auch bei Werken der bildenden K\"unste. Da liest man dann auch \"uber li\-te\-ra\-rische Klassiker Nacherz\"ahlungen wie von einem Grundsch\"uler. Dazu oft genug lange Listen von Per\-so\-nen."'}\\ (\cite{UserHummelhum2014})
\end{quote}
\onehalfspacing

oder:

\singlespacing
\begin{quote}
\href{https://de.wikipedia.org/w/index.php?title=Benutzer_Diskussion:Edith_Wahr\&diff=138421804\&oldid=138421510}{"`Untersuchungen dar\"uber, wie Wi\-ki\-pe\-dia-Artikel zur Li\-te\-ra\-tur ausfallen, hab ich noch keine gesehen. W\"urden wohl ohnehin nicht sonderlich schmeichelhaft ausfallen, wenn man den allgemeinen Zustand der Artikel \mbox{hier} so betrachtet..."'} (\cite{UserEdithWahr2015a})
\end{quote}
\onehalfspacing

und: 

\singlespacing
\begin{quote}
\href{https://de.wikipedia.org/w/index.php?title=Benutzer_Diskussion:Edith_Wahr\&diff=prev\&oldid=138424233}{"`Wie die Li\-te\-ra\-tur\-artikel in der Wi\-ki\-pe\-dia von au{\ss}erhalb gesehen werden: amateurhaft, unvollst\"andig und in der Sprache eines Kleing\"art\-ner\-mit\-tei\-lungs\-bl\"att\-chens. Ich hoffe, bei dem speziellen Artikel hat es sich ein wenig zum Guten ge\"andert, aber ganz Allgemein kann man das sicher immer noch unterschrei\-ben."'} (\cite{UserMagiers2015b})
\end{quote}
\onehalfspacing

Vermutlich beziehen sich diese Aussagen nur auf bekannte Werke ab dem 19. Jahrhundert, denn \"altere Werke, erst recht die der Medi\"avistik, sind meinem Eindruck nach \"uberdurchschnittlich gut dargestellt. Enormes an Wissen \"uber Li\-te\-ra\-tur kann sich in der Versionsgeschichte von Eintr\"agen befinden, quasi im Archiv eines Artikels, das dessen Entstehungsweise dokumentiert.
 
Zum Dritten w\"are relevant, was auf den dazugeh\"origen Dis\-kus\-si\-onssei\-ten bereits debattiert worden ist, auch \mbox{hier} inklusive der Versionsgeschichte und des Archivs. Nach meiner Erfah\-rung zeigt sich 
vier\-tens Wissen \"uber Li\-te\-ra\-tur vor allem, wenn kenntnisreiche Leute ihre Einsch\"atzungen artikelunabh\"angig auf interdisziplin\"aren Wi\-ki\-pe\-dia-Projektdis\-kus\-si\-onssei\-ten zur Spra\-che brin\-gen.
F\"unf\-tens gibt es ein "`Portal Li\-te\-ra\-tur"', das sich mit dieser Frage ana\-lysieren lie{\ss}e. Sechstens k\"onn\-ten die sogenannten L\"oschdis\-kus\-si\-onen als Material f\"ur Ana\-lysen sehr interessante Einblicke gew\"ahren, denn \mbox{hier} werden Vorbehalte gegen (die jeweils aktuellen Inhalte bestimmter) Artikel besonders pr\"agnant zum Ausdruck gebracht: Jemand \"au{\ss}ert, das Lemma dieses Artikels sei nicht relevant oder die Relevanz des Lemmas sei nicht gut genug dargestellt. Andere meinen erwartbarerweise, dass der Artikel nicht gel\"oscht werden soll und machen daf\"ur wiederum Gr\"unde geltend, die aus ihrer Sicht am stichhaltigsten sind. Eine solche L\"oschdis\-kus\-si\-on, die besonders eingehend gef\"uhrt wurde, richtete sich gegen Eintr\"age auch zu weniger bekannten Kurzgeschichten von Wolfgang Borchert. Die Debatte wurde 2007 schlie{\ss}lich von einigen Usern, die sich im Li\-te\-ra\-tur\-bereich auskannten, zum Anlass genommen, Richtlinien f\"ur Artikel zu li\-te\-ra\-rischen Werken abzufassen (F\"ur diese Informationen meinen Dank an \cite{UserMagiers2015a}). Auf die "`Causa Borchert"' wird gelegentlich wei\-terhin Bezug genommen und den \href{http://de.wikipedia.org/wiki/Wikipedia:Richtlinien_Literarische_Werke}{"`Richtlinien Li\-te\-ra\-rische Werke"'} seither einiges an Gewicht zugestanden, vor allem in Abgrenzung gegen\"uber Anspr\"uchen "`fachfremder"' Anliegen, aber auch in Relevanz- und Qualit\"atsfragen. In der Frage "`Wissen \"uber Li\-te\-ra\-tur bei Wi\-ki\-pe\-dia"' w\"urde mich -- siebtens -- in diskurspragmatischer Hinsicht (Foucault, \cite{Hetzel2012}:237) das Reservoir des gelegentlichen Austauschs auf Be\-nut\-zerdis\-kus\-si\-onsei\-ten am meisten interessieren, also was von all dem Wissen, das dort zur Sprache kommt, in welcher Weise in Artikeltexten (nicht) wiederzufinden ist: welche Prozesse von (Dis-)Artikulation \mbox{hier} beschreibbar w\"aren (\cite{Michaelis2011}:2).

Eine weitere Antwort zu "`Wissen \"uber Li\-te\-ra\-tur"' k\"onnte alle Lemmata zu Fachbegriffen der Li\-te\-ra\-tur\-wissenschaften einbeziehen, ebenso wie bestimmte Teile bio\-grafischer Eintr\"age zu Autoren\textsuperscript{\tiny *} li\-te\-ra\-rischer Werke, denn Informationen zu einzelnen Werken werden oftmals im Artikel zur Person in K\"urze formuliert, unabh\"angig davon, ob es zu dem Werk ein eigenes Lemma gibt oder nicht. Die Frage, welches Wissen bei Wi\-ki\-pe\-dia \"uber Li\-te\-ra\-tur\- vorhanden ist (im Sinne von nachlesbar), m\"usste auch eine Antwort zu Artikeln beinhalten, in denen es zum Beispiel einen Abschnitt "`In der Li\-te\-ra\-tur"' oder "`In der Belletristik"' gibt, was bei Eintr\"agen \"uber Orte der Fall sein k\"onnte, Wissen \"uber Li\-te\-ra\-tur also anders kon\-tex\-tu\-a\-li\-siert wird. Ebenso in biografischen Eintr\"agen (Per\-so\-nenartikeln), denn auch \mbox{hier} k\"onnte der eine oder andere intertextuelle Bezug als relevant genug erachtet worden sein, etwa \href{https://de.wikipedia.org/w/index.php?title=Harold_Pinter\&oldid=138481028\#In_der_Literatur}{im Eintrag "`Harold Pinter"' die Erw\"ahnung, dass Pinter in "`Save the Reaper"' von Alice Munro einen fiktionalisierten Cameo-Auftritt hat}. Und es gibt Eintr\"age \"uber einzelne Li\-te\-ra\-tur\-preise, wo eventuell von Jahr zu Jahr die Angaben zu den Namen bzw. Werken der Preistr\"ager\textsuperscript{\tiny *} aktualisiert werden. Oder in einem Abschnitt wie "`Pers\"onlichkei\-ten"' (oder "`S\"ohne und T\"ochter der Stadt"') in Ortsartikeln findet sich Wissen \"uber Li\-te\-ra\-tur, in einem wei\-ten Sinne: \"uber die Autoren\textsuperscript{\tiny *} li\-te\-ra\-rischer Werke, vor allem \href{https://de.wikipedia.org/w/index.php?title=Ni\%C8\%9Bchidorf\&oldid=131299136\#Pers.C3.B6nlichkeiten}{dann}, wenn der verlebten Zeit am Geburtsort gro{\ss}es Gewicht f\"ur das k\"unstlerische Schaffen bei\-ge\-mes\-sen wird. Nicht zu\-letzt kann "`bei Wi\-ki\-pe\-dia"' bedeuten, "`mit welchem Wissen potenziell zu rechnen ist"', zum Beispiel in Antworten bei einer Anfrage zu bestimmten Aspekten von Li\-te\-ra\-tur auf der allgemeinen Wi\-ki\-pe\-diaseite \href{https://de.Wikipedia.org/w/index.php?title=Wikipedia:Auskunft\&oldid=137900868}{"`Aus\-kunft"': "`\textsc{Willkommen}. Du konntest eine Information in Wi\-ki\-pe\-dia trotz Benutzung der Suchfunktion der Wi\-ki\-pe\-dia, einer Suchmaschine und des Archivs dieser Seite (Suchfeld unten) nicht finden? Auf dieser Seite beantworten Wi\-ki\-pe\-dianer allgemeine Wissensfragen."'}. Auf dieser Wikipediaseite werden zwar selten Fragen "`zu Li\-te\-ra\-tur"' gestellt, aber die potenziellen Antworten k\"onnten immerhin als ein we\-sent\-liches "`Wissen \"uber Li\-te\-ra\-tur bei Wi\-ki\-pe\-dia"' angesehen werden.

Trotz gro{\ss}er Zeitn\"ahe kann die Frage nach Wissen \"uber Li\-te\-ra\-tur in einem stark frequentierten Wiki allein aufgrund des Publikationsformats nur tentativ sein, denn was gestern noch nicht m\"oglich war, kann heute schon Akzeptanz finden. Oder ein Artikel, den es gestern noch nicht "`gab"', ist eventuell heute fr\"uh aus dem Be\-nut\-zernamensraum, wo er \"uber eine gewisse Zeit hin im "`Off"' verfasst worden ist, in den Artikelnamensraum geschoben worden -- und wurde erst daraufhin mittels Websuche auffindbar. \mbox{Selbst} das junge Projekt Wi\-ki\-pe\-dia hat seit 2001 be\-reits einige Phasen durchlebt und User, die schon meh\-rere Jahre stetig mitarbei\-ten oder eher von einer Halbau{\ss}enperspektive die Entwicklungen beobachten, t\"atigen gelegentlich \"Au{\ss}erungen, die darauf schlie{\ss}en lassen, dass in der deutschspra\-chi\-gen Wi\-ki\-pe\-dia-Community schon soetwas wie "`fr\"uhere Zei\-ten"' wahrgenommen werden, dieses Projekt also schon "`Geschichte"' hat. Wie lange die "`gegenw\"artige Zeit"' also in einem solchen flie{\ss}enden Projektgeschehen bei einem be\-stimmten Thema gegenw\"artig bleiben wird, kann ebenfalls von Fall zu Fall verschieden sein. In diesem Sinne kann es vorteilhaft sein, wenn im Li\-te\-ra\-tur\-bereich nicht besonders viele \"Anderungen get\"atigt werden: Au{\ss}er dass in der allgemeinen Wahrnehmung -- gelegentlichen Aussagen Einzelner nach zu urteilen, die schon lange bei Wi\-ki\-pe\-dia als angemeldete \mbox{User} mitarbei\-ten und fr\"uhe Phasen des Projekts mit der aktuellen Situation aus eigenem Erleben vergleichen -- bei Artikeln zu "`Li\-te\-ra\-tur\-themen"', wie sie auch manchmal genannt werden, nicht viel ge\"andert wird, also "`nicht so viel los"' ist: 

\singlespacing
\begin{quote}
\href{https://de.wikipedia.org/w/index.php?title=Benutzer_Diskussion:Grillenwaage&diff=prev&oldid=137132932}{"`Das mit dem unterschiedlich langen Geduldfaden [sic] glaube ich auch. Es gibt Menschen, die sind gelassener oder eloquenter und k\"onnen vielleicht besser mit einem (oft auch satirisch) trollenden Verhalten umgehen als andere, denen schnell der Geduldsfaden rei{\ss}t oder die sich schlicht hilflos f\"uhlen. Es hat auch was mit den Artikel [sic] zu tun [sic] die man schreibt: In der Li\-te\-ra\-tur\-ecke l\"asst sich's besser kuscheln als in Artikeln aus der WiPo-Ecke, solchen um den Ukraine-Konflikt oder aktuell PEGIDA. [...] Gru{\ss}"'} \cite{UserMagiers2014c}
\end{quote}
\onehalfspacing

Es sind auch kaum Beispiele f\"ur anhaltende Streitereien \"uber Inhalte von Artikeln zu li\-te\-ra\-rischen Werken auszumachen. Dar\"uber, als wie interessant die Informationen aus Leser\textsuperscript{\tiny *}sicht empfunden werden, sagen diese Beobachtungen allein keineswegs etwas aus. 

Infolge geringer Debattendichte wirkt Wissen zu Li\-te\-ra\-tur in Artikeln mit Informationen \"uber li\-te\-ra\-rische Werke (also auch in bio\-gra\-fi\-schen Artikeln zu Autoren\textsuperscript{\tiny *}) im Vergleich zu Artikeln in manchen anderen Themengebieten einerseits vom Schreibstil her monologischer (da vielfach ein einzelner \mbox{User} gro{\ss}en Anteil am Entstehungsprozess des Artikels hat), an\-de\-rerseits k\"onnten Eintr\"age in diesem thematischen Feld der Enzyklop\"adie aus diesem Grund schon zum jetzigen Zeitpunkt weniger auf Neutralit\"at hin ausgerichtet sein. \mbox{Hier} zeigt sich vermutlich ein Effekt dessen, dass sich eine Mehrheit in der Community mit li\-te\-ra\-rischen Werken nicht auskennt. Ob nur deswegen keine gr\"o{\ss}eren Kontroversen entstehen, ist nicht so leicht einzusch\"atzen. Diese finden eher bei Artikelgruppen statt, die in der deutschspra\-chi\-gen Wi\-ki\-pe\-dia-Community bei einer gr\"o{\ss}eren Anzahl von Usern auf Interesse sto{\ss}en und deren Thema interdisziplin\"ar debattierbar ist. Im besten Fall weisen Artikel zu li\-te\-ra\-rischen Werken eine Bandbreite an Positionen auf, wenn die Ergebnisse verschiedener Interpretationsrichtungen nebeneinander und glei\-cherma\-{\ss}en ausf\"uhrlich besprochen werden.

Eine Frage, die sich auf Quantitives richtet, k\"onnte zun\"achst damit beantwortet werden, dass die archivierte Hauptseite des 1. Februar 2015 als Gesamtzahl der Artikel \href{https://de.wikipedia.org/wiki/Wikipedia:Hauptseite/Archiv/1._Februar_2015}{1.807.687} angibt (1. Februar 2015). Wieviele davon keine Eintr\"age im engeren Sinne sind, sondern lediglich etwa Begriffskl\"arungssei\-ten, ist mir nicht bekannt, k\"onnte aber erfragt werden. Es sind also f\"ur die deutschspra\-chi\-ge Version von Wi\-ki\-pe\-dia ann\"ahernd 2 Millionen Eintr\"age verzeichnet. An Artikeln, denen die \href{https://de.wikipedia.org/wiki/Kategorie:Literarisches_Werk}{Kategorie "`Li\-te\-ra\-risches Werk"'} zugeordnet worden ist, wurde zu demselben Datum die Anzahl 9114 verzeichnet (1. Februar 2015). Diese Angabe kann ich \mbox{hier} lediglich mit einer f\"ur Wi\-ki\-pe\-dia\-verh\"altnisse relativ unspezifischen Quellenangabe machen, denn auf der Seite wird laufend fortgez\"ahlt und ich habe keine Quelle gefunden, bei der zu dieser Seite t\"aglich eine Seite archi\-viert w\"urde so wie es seit Anfang 2011 f\"ur die Hauptseite der Fall ist. 

Es ist also eine gro{\ss}e Menge an Informationen erwartbar, auch wenn Artikel, die der "`Kategorie:Li\-te\-ra\-risches Werk"' nur -- aber immerhin -- sch\"atzungsweise 0,5 Prozent aller Eintr\"age ausmachen. Warum bez\"uglich der Anzahl an Eintr\"agen zu li\-te\-ra\-rischen Werken in der deutschspra\-chi\-gen Wi\-ki\-pe\-dia keine genauere Angabe m\"oglich ist, kann meh\-rere Gr\"unde haben, die teils als f\"ur Wi\-ki\-pe\-dia ty\-pisch anzusehen sind. Als Charakteristikum von Denkweisen in Enzyklop\"adien inklusive Wi\-ki\-pe\-dia gilt: \href{http://pierrelevyblog.files.wordpress.com/2013/05/stevejankowski_thesis_v18.pdf}{"`[E]ncyclopaedism values know\-ledge that is consistently reduced to essential and defining qualities"'} (\cite{Jankowski2013}:123). Bez\"uglich Wi\-ki\-pe\-dia verweist er auf die Seite "`Wi\-ki\-pe\-dia:Categorization"': 

\singlespacing
\begin{quote}
\href{https://en.wikipedia.org/w/index.php?title=Wikipedia:Categorization\&oldid=644735441}{"`The central goal of the category system is to provide navigational links to all Wi\-ki\-pe\-dia pages in a hierarchy of \textit{categories} which readers, knowing essential -- \textit{defining} -- characteristics of a topic, can browse and quickly find sets of pages on topics that are defined by those characteristics."'} ("`Wikipedia:Categorization"')
\end{quote}
\onehalfspacing

Konkreter bedeutet dies f\"ur die auf Multidisziplinarit\"at ausgelegte Artikel-Da\-ten\-bank von Wikipedia:

\singlespacing
\begin{quote}
\href{https://en.wikipedia.org/w/index.php?title=Wikipedia:Categorization\&oldid=644735441}{"`A central concept used in categorising articles is that of the [[defining]] characteristics of a subject of the article. A defining characteristic is one that [[reliable sources]] \textit{commonly} and \textit{consistently} define ({\footnotesize in prose, as opposed to a tabular or list form}) the subject as having -- such as nationality or notable profession (in the case of people), type of location or region (in the case of places), etc."'} ("`Wikipedia:Categorization"')
\end{quote}
\onehalfspacing

Die genannten Beispielthemen geben einen Eindruck davon, was in der eng\-lischspra\-chi\-gen Version als ein ty\-pisches Thema f\"ur einen Wi\-ki\-pe\-dia-Eintrag angesehen wird -- dies gilt auch f\"ur die \href{https://de.wikipedia.org/wiki/Wikipedia:Kategorien}{deutschspra\-chi\-ge Version} -- : Per\-so\-nen und Orte. Und ob ein Eintrag zu einem li\-te\-ra\-rischen Werk als ein solcher erkannt worden ist, das hei{\ss}t: die Zuordnung zu der Kategorie "`Li\-te\-ra\-risches Werk"' als dessen "`essential -- \textit{defining} -- characteristics of a topic"' angesehen wurde, l\"asst sich nicht zuverl\"assig sagen, denn daran arbei\-ten \mbox{User} vermutlich gr\"o{\ss}tenteils in ihrer Freizeit und \mbox{selbst} wenn nicht wenige von ihnen bei Wi\-ki\-pe\-dia ihrer Leidenschaft der Gewissenhaftigkeit viel Raum geben, kann ihnen leicht ein Artikel entgangen sein, etwa w\"ahrend allgemeiner Ferienzei\-ten. Zahlenm\"a{\ss}ig k\"onnen sich in einer solchen Phase des Jahres die Proportionen zwischen denjenigen stark ver\"andern, die neue Artikel verfassen, und denjenigen, die neue Eintr\"age routinem\"a{\ss}ig nach systematischen Gesichtspunkten durchsehen, so dass sich ein ungewohnter \"Uberhang ergeben kann und in dieser Zeit mehr Einzelhei\-ten \"ubersehen werden als sonst. Au{\ss}erdem ist davon auszugehen, dass es vermutlich keine li\-te\-ra\-tur\-wissenschaftlich ausgebildeten Leute sind, die einen Eintrag der Kategorie "`Li\-te\-ra\-risches Werk"' zuordnen, weil es ein Buch zu sein scheint, aber eventuell beispielsweise nicht eindeutig ein Sachbuch. Dann wurde "`li\-te\-ra\-risches Werk"' viel\-leicht als die bessere Zuordnung angesehen, denn in der deutschspra\-chi\-gen Wi\-ki\-pe\-dia gibt es keine Kategorie wie etwa in der eng\-lischspra\-chi\-gen Version: "`Book"'. \mbox{Selbst} wenn f\"ur das "`Portal Li\-te\-ra\-tur"' in der deutschspra\-chi\-gen Wi\-ki\-pe\-dia der Begriff "`Li\-te\-ra\-tur"' eventuell genauer definiert wurde, bedeutet dies also nicht, dass die Kategorisierung in allen F\"allen zutreffend ist -- denn Wi\-ki\-pe\-dia ist keine wissenschaftliche Bibliothek mit Personal, das im Katalogisieren nach professionell g\"angigen Schlagwortthesauri ausgebildet worden ist und sich st\"andig in bezahlter Arbeitszeit darin fachspezifisch fortbildet -- \mbox{selbst} wenn anzunehmen ist, dass sich einige Bibliothekare\textsuperscript{\tiny *} an Wi\-ki\-pe\-dia beteiligen. Die am 1. Februar 2015 mit 9114 verzeichnete Anzahl der Eintr\"age zu li\-te\-ra\-rischen Werken ist mindestens aus diesen Gr\"unden als ein N\"aherungswert anzusehen. Wollte jemand sich tats\"achlich wissenschaftlich mit allen Lemmata der deutschspra\-chi\-gen Wi\-ki\-pe\-dia befassen, die ein li\-te\-ra\-risches Werk zum Inhalt haben (oder eine Gruppe von Werken, etwa die Text-Bild-Collagen von Herta M\"uller), m\"usste bei der Fehlersuche eine Standardabweichung definiert werden, um pr\"aziser angeben zu k\"onnen, wie gro{\ss} vermutbarerweise die Anzahl an Artikeln ist, die man nicht als solche hat identifizieren k\"onnen. An Zuwachs bei Eintr\"agen zu li\-te\-ra\-rischen Werken in der deutschspra\-chi\-gen Wi\-ki\-pe\-dia lassen sich f\"ur den Zeitraum zwischen dem 30. Juli 2013 und dem 13. November 2014 im Schnitt rund 1,3 pro Tag \href{https://de.wikipedia.org/w/index.php?title=Wikipedia:Projektdiskussion/Mehr_Artikel_zu_literarischen_Werken&oldid=135954188}{angeben}, es waren in diesem Zeitraum also 621 neue Eintr\"age.

Gemessen an den vermuteten Interessen der Leser\textsuperscript{\tiny *} w\"are das "`Wissen bei Wi\-ki\-pe\-dia \"uber Li\-te\-ra\-tur"' nochmals anders einzusch\"atzen. Aus diesem Blickwinkel kommen Aspekte der Li\-te\-ra\-tur\-ver\-mitt\-lung in den Blick, und speziell M\"oglichkei\-ten von Li\-te\-ra\-tur\textsuperscript{\~.\~.}\-ver\-mitt\-lung. Bez\"uglich der Darstellbarkeit nicht-pro\-po\-si\-ti\-o\-na\-len Wissens sind neue Fragen zu stellen, deren Antworten weit \"uber Wi\-ki\-pe\-dia hinaus von Bedeutung sein k\"onnen. J\"ungst sind in einer Wi\-ki\-pe\-dia-Debatte Fragen formuliert worden, von denen an\-ge\-nom\-men wird, dass Leser\textsuperscript{\tiny *} sie bei einer Websuche im Kopf haben, wenn sie eventuell bei Wi\-ki\-pe\-dia landen. \mbox{User}:Epipactis formulierte am 19. Januar 2015 folgendes, \href{https://de.wikipedia.org/w/index.php?title=Benutzer_Diskussion:Grillenwaage\&diff=137961506\&oldid=137943802}{"`(\"ubrigens v\"ollig spontan)"'}:

\singlespacing
\begin{quote}
\href{https://de.wikipedia.org/w/index.php?title=Benutzer_Diskussion:Grillenwaage\&diff=next\&oldid=137663350}{"`Was habe ich zu \textit{erwarten}, wenn ich mir das Werk ansehe? Sollte ich es mir \textit{g\"onnen}, wird es sich f\"ur mich \textit{lohnen}? Mu{\ss} ich es mir \textit{zumuten}, um \textit{mitreden} zu k\"onnen? Oder: Ich habe es gesehen, hat mich nicht sonderlich beeindruckt, vielleicht ist mir etwas aus Expertensicht \textit{Bemerkenswertes} entgangen, aber wenn ja -- was?"'} (\cite{UserEpipactis2015a})
\end{quote}
\onehalfspacing 

Ich halte diese Vermutungen \"uber in die Motivation von Lesern\textsuperscript{\tiny *}, sich -- selbst\-sozialisierend -- im Web nach Informationen umzusehen, auch au{\ss}erhalb von Wi\-ki\-pe\-dia f\"ur h\"ochst relevant und denke, es lie{\ss}en sich stichhaltige Begr\"undungen daf\"ur erarbei\-ten, warum Li\-te\-ra\-tur\-ver\-mitt\-lung auch in ihren herk\"ommlichen Schwerpunkten ein Augenmerk auf Fragen dieser Art legen sollte. Im Verlauf der Debatte kam der ein oder andere Gegenstandpunkt ebenfalls zur Sprache, was die Fragen von \mbox{User}:Epipactis f\"ur mein Erkenntnisinteresse umso relevanter macht. Meine Ausf\"uh\-rungen hierzu finden sich bei der Auswertung des Experiments auf \textit{Objektebene II} in Abschnitt \textit{~\ref{subsubsec:6.2.2} Debatte zu Leseerlebnis-Abschnitten}. Kommen wir zum zwei\-ten Teil dieses Abschnitts. 

Li\-te\-ra\-tur\-ver\-mitt\-lung kann auf einer Plattform wie Wi\-ki\-pe\-dia in gro{\ss}em Stil inszeniert werden. Die durch\-schnitt\-li\-chen Abrufzahlen der deutschspra\-chi\-gen Wi\-ki\-pe\-dia-Hauptseite werden mit nahezu 1 Million pro Tag \href{http://stats.grok.se/de/latest90/Wikipedia:Hauptseite}{angezeigt}. Ab einem bestimmten Zeitpunkt im Laufe meines For\-schungprojekts habe ich erwogen, die Hauptseite verst\"arkt f\"ur Artikel zu li\-te\-ra\-rischen Werken zu nutzen. Ab 2014 habe ich diese Option mit einer Steigerung zum Jahreswechsel hin genutzt. Zwar gab es nur einige wenige Reaktionen, aber diese haben f\"ur die Entstehung ausreichend interessanter Daten gereicht und in der Folge auch die Debatte bereichert, im Laufe derer \mbox{User}:Epipactis im Januar 2015 die oben genannten Fragen formulierte.
 
F\"ur neue Artikel kann eine gewisse Aufmerksamkeit bei nicht prim\"ar li\-te\-ra\-risch interessierten Mitautoren\textsuperscript{\tiny *} in der Community sowie bei der Web\"of\-fent\-lichkeit erzeugt werden, indem sie nach der Erstellung binnen 31 Tagen f\"ur die Hauptsei\-tenrubrik "`Schon gewusst?"' vorgeschlagen werden, mit einem Tea\-ser, in dem hinter dem Lemma des Artikels ein Link gesetzt ist. In den Debatten um die Eignung bestimmter Tea\-ser hat sich im Zeitraum zwischen Oktober 2013 und Januar 2015 eine Tendenz ausmachen lassen, die f\"ur eine erste Einsch\"atzung relevant ist, welche Bedingungen der Darstellbarkeit nicht-pro\-po\-si\-ti\-o\-na\-len Wissens in Eintr\"agen zu li\-te\-ra\-rischen Werken als aktuell geltend w\"urden angenommen werden k\"onnen. Bevor ich anhand von Teaser-Beispielen der letzten 16 Monate im Detail argumentiere, m\"ochte ich kurz darauf eingehen, wie sich die erzielte Aufmerksamkeit in Abrufzahlen darstellt, um eine ungef\"ahre Gr\"o{\ss}enordnung f\"ur mediale Effekte vor Augen zu haben: Wieviele Male wird ein bestimmtes St\"uck "`Li\-te\-ra\-tur\-ver\-mitt\-lung bei Wi\-ki\-pe\-dia"' mit einem Klick auf die Verlinkung genutzt? Ich nenne zun\"achst die Statistik f\"ur die \mbox{vier} Artikel, die Teil meiner Ausf\"uh\-rungen geworden sind, dann ein Vergleichsbeispiel, ebenfalls aus dem Li\-te\-ra\-tur\-bereich, und anschlie{\ss}end zum Vergleich die Gr\"o{\ss}enordnung der Abrufzahlen f\"ur \mbox{vier} biografische Artikel. Der Teaser zum Lemma "`Pour que tu ne te perdes pas dans le quartier"' wurde f\"ur den \href{https://de.wikipedia.org/wiki/Wikipedia:Hauptseite/Archiv/4._Dezember_2014}{4.} und den \href{https://de.wikipedia.org/wiki/Wikipedia:Hauptseite/Archiv/5._Dezember_2014}{5. Dezember 2014} auf der Hauptseite eingetragen, ohne Abbildung. F\"ur den ersten Pr\"asentationstag sind 9.740 Abrufe verzeichnet und f\"ur den zwei\-ten 3.805 Abrufe, macht zusammen 13.545 \href{http://stats.grok.se/de/latest90/Pour_que_tu_ne_te_perdes_pas_dans_le_quartier}{Abrufe} in zwei Tagen. Ein Teaser, unter anderem zum Romaninhalt von \textit{Pas pleurer}, "`[[Lydie Salvayre]] wurde f\"ur ihren Roman \"uber das Jahr 1936 im Spanischen B\"urgerkrieg mit dem Prix Goncourt 2014 ausge\-zeich\-net"', war mit Bild der Autorin\textsuperscript{\tiny *} und Verlinkung zum Eintrag "`Lydie Salvayre"' am 10. und 11. Dezember 2014 auf der Hauptseite zu sehen. Statistisch verzeichnet wurden 4.200 und 891, also 5.091 Abrufe in zwei Tagen (Diese Anzahl von rund 5.000 rief mir unwillk\"urlich in Erinnerung, dass vor 50 Jahren \textit{Le Figaro litt\'{e}raire} in der Berichterstattung \"uber das Event "`Que peut la litt\'{e}rature?"' am 9. Dezember 1964 in Paris schrieb: "`Dans la salle, trois mille \'{e}tudiants (et deux milles dehors) attendent la r\'{e}ponse"' (\cite{Louette2014}:36)). Als im Januar ein weiteres Werk derselben Autorin\textsuperscript{\tiny *} mit einem eigenen Lemma in dieser Rubrik von der Hauptseite verlinkt wurde, landeten beim biografischen Eintrag nochmals 284 + 123 = 407 \href{http://stats.grok.se/de/latest90/Lydie_Salvayre}{Abrufe}. Als der Eintrag "`La Vie commune"' am 13. und 14. Ja\-nu\-ar 2015 \mbox{selbst} an der Reihe war, wurde er der verwendeten Statistik nach 3.342 + 1.638 = 4.980 Male angeklickt. Als im Monat zuvor der Eintrag "`Lydie Salvayre"' zu sehen war, gelangten zu diesem Werkartikel 88 + 29 = 117 \href{http://stats.grok.se/de/latest90/La_Vie_commune}{Abrufe}. Eine am Artikel get\"atigte \href{https://de.wikipedia.org/w/index.php?title=La_Vie_commune&diff=next&oldid=136619097}{\"Anderung} wurde von mir zum Anlass genommen, eine \href{https://de.wikipedia.org/w/index.php?title=Diskussion:La_Vie_commune&oldid=138030860#Meinung_einzelner}{Dis\-kus\-si\-on zu beginnen}, die ich auf \textit{Objektebene II} in Abschnitt \textit{~\ref{subsubsec:6.2.1} Reaktionen auf einzelne Leseerlebnis-Abschnitte} mit Blick auf die gew\"ahlte Fragestellung ana\-lysiere. Bei der Hauptsei\-tenpr\"asentation des Eintrags "`Alfred and Emily"', mit Bild, aber ohne Nennung des Autor\textsuperscript{\tiny *}namens, kamen am 7. und 8. Ja\-nuar 2015 \href{http://stats.grok.se/de/latest90/Alfred_und_Emily}{Abrufe} in H\"ohe von 16.460 + 4.504 = 20.964 zusammen. Der Eintrag zur Autorin\textsuperscript{\tiny *}, "`Doris Lessing"', wurde an diesen beiden Tagen 2.173 + 578 Mal aufgerufen, also 2.751 Mal. Von den Vergleichsf\"allen mit Abbildung, deren Statistik ich mir angesehen habe, ist vor allem der Teaser bzw. das Artikelthema von "`The Lady of Shalott"' (Tennyson) bemerkenswert. Die Abbildung zum Teaser zeigte ein historisches Gem\"alde, das vom Sujet her als "`roman\-tisch"' wahrgenommen werden kann. Der Teaser lautete: "`[[\textit{The Lady of Shalott}]] trotzte ihrem Fluch und lie{\ss} daf\"ur ihr Leben."' Am ersten Tag wurden 35.646 \href{http://stats.grok.se/de/latest90/The_Lady_of_Shalott}{Abrufe} vermerkt, am zwei\-ten 11.072, zusammen also fast 50.000 (46.718), womit dieser Teaser zu den am h\"aufigsten aufgerufenen \"uberhaupt geh\"ort, was einem "`Li\-te\-ra\-tur\-thema"' selten gelingt. Auf die Teaser-Auswahlm\"oglichkei\-ten, die es zu diesem Vorschlag gegeben hatte, wird im Folgenden eingegangen -- im Vergleich zu der Auswahl bei anderen Vorschl\"agen. Im Mittelpunkt dieser Arbeit stehen Eintr\"age zu Werken, dennoch sei zur Abrundung noch auf Zugriffszahlen von \mbox{vier} biografischen Eintr\"agen hingewiesen. Im Zeitraum der 90 Tage vor einem willk\"urlich gew\"ahlten Datum, dem 6. Oktober 2014, belief sich die Menge an Abrufen f\"ur den Eintrag "`Alice Munro"' (Nobelpreis 2013) auf 10.848 (de, 126 pro Tag) bzw. 70.401 (en, 782 pro Tag), f\"ur "`Herta M\"uller"' (Nobelpreis 2009) auf 15.238 (de, 169 pro Tag), 14.781 (en, 164 pro Tag) und 2.637 (ro, 29 pro Tag), f\"ur den Eintrag "`Doris Lessing"' (Nobelpreis 2007, gestorben 2013) auf 9.263 (de, 103 pro Tag) bzw. 53.271 (en, 592 pro Tag) und f\"ur den Eintrag "`Elfriede Jelinek"' (Nobelpreis 2004) auf 14.697 (de, 163 pro Tag) bzw. 14.798 (en, 164 pro Tag). Es hat den Anschein, als ob biografische Eintr\"age mehr Interesse finden als Werkartikel. Da aber die Verweildauer nicht verzeichnet ist, kann letzt\-lich keine verl\"assliche Aussage gemacht werden. Auch Mehrfachbesuche, die auf ein vertieftes Interesse am Thema schlie{\ss}en lassen w\"urden, werden statistisch nicht erfasst. Ferner w\"aren die Abrufzahlen der pdf-Version eines Werkartikels interessant, denn \mbox{hier} w\"urden sich traditionelle Lesegewohnhei\-ten ausmachen lassen, die recht nahe an Li\-te\-ra\-tur\-lekt\"ure liegen.

In der Rubrik "`Schon gewusst?"' wurden im Laufe der letzten 16 Monate 60 li\-te\-ra\-rische Werke (inklusive essayistischer Schriften, Tagebuch, Autobiografie) sowie eine Sure, M\"archen und vereinzelt auch Sachb\"ucher mit Teasern vorgestellt, also rund 3\,\% der insgesamt 1.920 vorgestellten Artikel in diesem Zeitraum. \mbox{Selbst} wenn man einrechnet, dass etwa ein Drittel der Vorschl\"age von mir eingebracht wurde, man also bereinigt von einem Prozent weniger ausgehen sollte, so bel\"auft sich der Anteil von Artikeln \"uber li\-te\-ra\-rische Werke, zu denen Teaser auf der Hauptseite pr\"asentiert worden sind, mit 2\,\% statt 0,5\,\% dennoch auf das Vier\-fache des Anteils, den Artikel der "`Kategorie:Li\-te\-ra\-risches Werk"' am 1. Februar 2015 an der Gesamtartikelmenge hatten. Ob einer der \mbox{hier} ausschlaggebenden Faktoren als \href{https://de.wikipedia.org/w/index.php?title=Wikipedia_Diskussion:Kurier&diff=prev&oldid=138042996}{
"`Kulturtapeten"'}-Effekt beschrieben werden kann, m\"usste in Relation zur Anzahl der Teaser aus weiteren Themengebieten untersucht werden, die durch diese Rubrik in das Rampenlicht der Hauptseite gelangen.

An der Dis\-kus\-si\-on der Vorschl\"age f\"ur "`Schon gewusst?"' k\"onnen sich alle beteiligen -- wie \"ublich auch adhoc. Wer einen neuen Vorschlag einbringen will, legt auf der Dis\-kus\-si\-onsseite f\"ur die Rubrik einen neuen Abschnitt mit einer \"Uberschrift an, die aus drei Teilen be\-steht: dem Wort "`Vorschlag"', dem Lemma des Vorschlags und dem Datum, seit wann der Artikel neu ist, und macht gleich einen Teaser-Vorschlag. Meistens werden au{\ss}er dem ersten Teaser von anderen Usern alternative Ideen eingebracht, gelegentlich wird auch die Qualit\"at des Artikels diskutiert (anstatt auf der Artikeldis\-kus\-si\-onsseite). Ist der Artikel zu umfangreich, schon l\"anger als 31 Tage im Artikelnamensraum, oder hat eine Mehrheit derjenigen, die sich \"au{\ss}ern, Bedenken, diesen Artikel von der Hauptseite aus zu verlinken, gilt dieser Abschnitt als erledigt. \"Uber Ausnahmen wird verhandelt, was anschlie{\ss}end gelegentlich zu leichten \"Anderungen an den Spielregeln f\"uhrt, die wiederum ausgehandelt werden. 

In der Rubrik "`Schon gewusst?"' werden t\"aglich \mbox{vier} Artikel mit Teasern pr\"asentiert. Zwei davon wurden vom Vortag \"ubernommen, die anderen beiden durch zwei neue ersetzt. Die neuen Teaser werden an den ersten beiden Positionen platziert, die beiden Teaser vom Vortag auf Position 3 und 4 verschoben. Aus den 50-70 Vorschl\"agen der Dis\-kus\-si\-onsliste werden von irgendjemandem zwei neue Artikel pro Tag ausgesucht und in die Vorlage f\"ur die Hauptseite eingetragen. Bei der Formulierung von Teasern ebenso wie bei der darauffolgenden Auswahl spielt neben den f\"ur diese Rubrik ausgehandelten Regeln auch die Annahme dar\"uber eine Rolle, was Hauptsei\-tenleser\textsuperscript{\tiny *} interessiert. Ein stichhaltiger Einwand gegen Repr\"asentativit\"at w\"are, dass nur Teaser ausgew\"ahlt werden k\"onnen, die vorgeschlagen werden. Zwar trifft dies zu, aber mein Gegenargument ist, dass alle, die wollen, in der Dis\-kus\-si\-on andere Teaser vorschlagen k\"onnen. Teaser sollen erstens den jeweiligen Artikelinhalt wiedergeben, wenn auch nur in einem Punkt, zwei\-tens soll ein Teaser neugierig machen und drittens im Sinne der expliziten oder impliziten Spielregeln von "`Schon gewusst?"' als passend empfunden werden k\"onnen. Zu einer skeptischen Einsch\"atzung bez\"uglich der Aussagekraft von Teasern im Verh\"altnis zum vorgestellten Artikel gelangt \mbox{User}:Hans Castorp, wenn es am Ende in der Dis\-kus\-si\-on zu einem eigenen Li\-te\-ra\-tur\-artikelvorschlag hei{\ss}t: 

\singlespacing
\begin{quote}
\href{https://de.wikipedia.org/w/index.php?title=Wikipedia_Diskussion:Hauptseite/Schon_gewusst&diff=prev&oldid=130907996}{"`Dann bleibt es halt bei diesem Teaser, zumal der noch kurze Artikel ohnehin we\-sent\-licher ist, als diese nervigen, letzt\-lich nicht sonderlich re\-levanten Teaser- oder Bild-Fragen."'} (\cite{UserHansCastorp2014})
\end{quote}
\onehalfspacing

Ich komme zu meiner Frage: Welche Art von Aussagen in K\"urzestform wurden von den jeweiligen ad-hoc-Teams der "`Schon gewusst?"'-Dis\-kus\-si\-onsseite f\"ur Li\-te\-ra\-tur\-artikel als passend empfunden, um an zwei Tagen viel Aufmerksamkeit zu verdienen? Manche der ausgew\"ahlten Teaser beleuchte ich zusammen mit ihren Vorversionen, also den Alternativen. Mir geht es um die Frage, ob \mbox{hier} ein Schrei\-ben \"uber Li\-te\-ra\-tur\textsuperscript{\~.\~.} gelingt und nicht nur Aspekte der Handlung eines Werks zur Sprache kommen. Mich interessiert, ob es zumindest Andeutungen gibt, worin ein \textit{Erlesnis} bestehen k\"onnte. Dies nenne ich im Folgenden "`\textit{Erlesnis}-Chance"'.

Vorab dies: In den meisten der 61 F\"alle wird in Teasern keine Aussicht auf eine \textit{Erlesnis}-Chance angedeutet, so dass ich im Folgenden nur die etwas aussichts\-rei\-cheren Teaser einzeln bespreche. Daher weist die Liste in der Z\"ahlung L\"ucken auf. Mit einem Asterisk am Zeilenanfang sind diejenigen Vorschl\"age der chronologischen Reihung nach gelistet, aus denen pro Artikel h\"atte ausgew\"ahlt werden k\"onnen. Darunter ist der Text des Teasers, der auf der Hauptseite zu sehen war, in Anf\"uh\-rungs\-striche gesetzt. Die Fu{\ss}note am ausgew\"ahlten Teaser gibt das Datum des ersten von zwei Pr\"asentationstagen an. In manchen F\"allen wurde lediglich ein Teaservorschlag erarbeitet, wird \mbox{hier} aber mit kommentiert, denn bei Li\-te\-ra\-tur\-ver\-mitt\-lung im gro{\ss}en Stil z\"ahlt f\"ur das Erkenntnisinteresse in dieser Arbeit nicht allein die Entstehung dessen, was f\"ur eine Auswahl m\"oglich gewesen w\"are, sondern auch das Ergebnis, das auf der Hauptseite pr\"asentiert worden ist, \mbox{selbst} wenn es nur einen einzigen Teaservorschlag gab -- aus welchen Gr\"unden auch immer.

\singlespacing

[1] [[\textit{Sieben Jahre}]]
\singlespacing
\begin{quote}
¸* Peter Stamms Dreiecksgeschichte [[\textit{Sieben Jahre}]] erstreckt sich \"uber 18 Jahre.\\
¸* Peter Stamms Roman [[\textit{Sieben Jahre}]] nimmt die biblische Geschichte von den Ehen Jakobs anhand eines Architekten auf, der versucht sein Leben perfekt zu planen, aber in den gro{\ss}en Entw\"urfen scheitert.\\
¸* Peter Stamms alttestamtentarisch moti\-vier\-te Dreiecksgeschichte [[\textit{Sie\-ben Jahre}]] spielt im b\"urgerlichen Milieu eines Architekturb\"uros.\\
¸* In seiner Dreiecksgeschichte [[\textit{Sieben Jahre}]] l\"asst Peter Stamm seine Hauptfigur sowohl beruflich als Architekt scheitern als auch beim Streben nach einem perfekten Lebensplan.\\
¸* Peter Stamms Roman [[\textit{Sieben Jahre}]] verfolgt die Dreiecksbeziehung eines Architekten (mit biblischen Parallelen) \"uber einen Zeitraum von achtzehn Jahren.\\
¸* In Peter Stamms Roman [[\textit{Sieben Jahre}]] muss ein Architekt, der zwi\-schen zwei Frauen steht, Insolvenz anmelden.\\
¸* Peter Stamm stellt in seiner Dreiecksgeschichte [[\textit{Sieben Jahre}]] unterschiedliche Architekturstile und Beziehungsformen einander gegen\"uber.

F\"ur die Pr\"asentation \href{https://de.wikipedia.org/wiki/Wikipedia_Diskussion:Hauptseite/Schon_gewusst/Diskussionsarchiv/2013/Oktober#Eigenvorschlag:_Sieben_Jahre_.28Peter_Stamm.29_.282._Okt..29_.28erl..29}{ausgew\"ahlt} wurde:\\ 
"`In Peter Stamms Roman [[\textit{Sieben Jahre}]] muss ein Architekt, der zwi\-schen zwei Frauen steht, Insolvenz anmelden."' (\href{https://de.wikipedia.org/wiki/Wikipedia:Hauptseite/Archiv/17._Oktober_2013}{Hauptseite am 17.} und am \href{https://de.wikipedia.org/wiki/Wikipedia:Hauptseite/Archiv/18._Oktober_2013}{18. Oktober 2013})
\end{quote}
\onehalfspacing

Zu diesem Vorschlag wurden sieben Teaser erarbeitet. Dass in dem Werk nicht-pro\-po\-si\-ti\-o\-na\-les Wissen erworben werden kann, wird in zweien von ihnen angedeutet. In diesen beiden Teasern kann man ahnen, dass ein \textit{Erlesnis} darin bestehen k\"onnte erfahren zu haben, wie es sich anf\"uhlt, wenn jemand "`versucht sein Leben perfekt zu planen, aber in den gro{\ss}en Entw\"urfen scheitert"' (zweiter Teaser) oder "`sowohl be\-ruf\-lich als Architekt [zu] scheitern als auch beim Streben nach einem perfekten Le\-bens\-plan"' (vier\-ter Teaser). Aber es wird kein \textit{Erlesnis} explizit ausgedr\"uckt. Der siebte Teaser hebt auf Stil oder Bauweise des Werks ab und je nach Lekt\"ureerfahrenheit kann \mbox{hier} diese Aussage die Erwartung eines \textit{Erlesnisses} hervorrufen, \"ahnlich wie bei den drei Teasern, die einen biblischen Bezug des Werks erw\"ahnen. Al\-ler\-dings wurde keiner dieser sechs Teaser genommen, wohl aber derjenige, in dem sich als einzigem die Hauptaussage auf Wirtschaftliches bezieht. Mit diesem Teaser also kommt auf der Hauptseite an diesen zwei Tagen keine \textit{Erlesnis}-Chance zur Sprache.\\

[3] [[\textit{Der B\"ar kletterte \"uber den Berg}]]
\singlespacing
\begin{quote}
¸* Das Ende der Kurzgeschichte [[\textit{Der B\"ar kletterte \"uber den Berg}]] von Alice Munro wird h\"aufig kommentiert.\\
¸* In [[\textit{Der B\"ar kletterte \"uber den Berg}]] von Alice Munro wird Sex als Spielmarke zum Feilschen eingesetzt.\\
¸* Alice Munro l\"asst offen, wer der B\"ar in ihrer Short Story [[\textit{Der B\"ar kletterte \"uber den Berg}]] ist.\\
¸* In der Kurzgeschichte [[\textit{Der B\"ar kletterte \"uber den Berg}]] der Nobelpreistr\"agerin Alice Munro kommt weder ein B\"ar noch ein Berg vor.

F\"ur die Pr\"asentation \href{https://de.wikipedia.org/wiki/Wikipedia_Diskussion:Hauptseite/Schon_gewusst/Diskussionsarchiv/2013/Oktober#Vorschlag:_Der_B.C3.A4r_kletterte_.C3.BCber_den_Berg_.2820._Oktober_2013.29_.28erl..29}{ausgew\"ahlt} wurde:\\
"`Die Nobelpreistr\"agerin Alice Munro beschreibt in ihrer Kurzgeschichte [[\textit{Der B\"ar kletterte \"uber den Berg}]] weder einen B\"aren noch einen Berg."' (\href{https://de.wikipedia.org/wiki/Wikipedia:Hauptseite/Archiv/25._November_2013}{Hauptseite am 25.} und am \href{https://de.wikipedia.org/wiki/Wikipedia:Hauptseite/Archiv/26._November_2013}{26. November 2013})
\end{quote}
\onehalfspacing

Es wurden \mbox{vier} verschiedene Teaser vorgeschlagen und abgewandelt. Beim Ausw\"ah\-len kam die Be\-zeichnung "`beschrei\-ben"' als T\"atigkeit der Autorin\textsuperscript{\tiny *} neu hinzu, als jemand aus den Elementen der vorhandenen Vorschl\"age eine neue Version herstellte. Der Teaser verweist auf etwas R\"atselhaftes, also ist zu ahnen, dass ein \textit{Erlesnis} darin bestehen k\"onnte, ein R\"atsel gel\"ost zu haben oder es nicht gel\"ost zu haben oder keines von beidem. Eine \textit{Erlesnis}-Chance wird somit angedeutet.\\

[5] [[\textit{Im Rettungsboot}]]
\singlespacing
\begin{quote}
¸* Stephen Crane verarbeitete 1897 seine Erfah\-rungen im drei{\ss}ig\-st\"un\-di\-gen \"Uberlebenskampf als Schiffbr\"uchiger [[\textit{im Rettungsboot}]] auf offener See in einer Kurzgeschichte.\\
¸* Nach drei{\ss}igst\"un\-di\-gem \"Uberlebenskampf als Schiffbr\"uchiger [[\textit{in einem kleinen Boot}]] auf offener See verarbeitete der amerikanischer Schrifsteller [sic] Stephen Crane seine Erfah\-rungen in einer Kurzgeschichte.\\
¸* Als Schiffbr\"uchiger k\"ampfte Stephen Crane fast zwei Tage lang hilflos auf offener See in einem kleinen Boot verzeifelt [sic] um sein Leben und schrieb anschlie{\ss}end eine [[\textit{Kurzgeschichte}]].\\
¸* Als Schiffbr\"uchiger auf offener See k\"ampfte Stephen Crane fast zwei Tage lang hilflos und verzeifelt [sic] [[\textit{in einem kleinen Boot}]] um sein Leben. Danach verfasste er eine Kurzgeschichte.\\
¸* Stephen Crane geh\"orte zu den einzigen \mbox{vier} \"Uberlebenden nach einem Schiffbruch / dem Schiffbruch der Commodore. Seine [sic] drei\-{\ss}ig\-st\"un\-digen \"Uberlebenskampf auf offener See verarbeitete er in einer Kurz\-ge\-schich\-te.\\
¸* In der Erz\"ahlung [[\textit{Im Rettungsboot}]] verarbeitete der Korrespondent Stephen Crane seinen drei{\ss}igst\"undigen \"Uber\-lebens\-kampf als Schiff\-br\"u\-chiger vor Florida.\\
¸* In der Erz\"ahlung [[\textit{Im Rettungsboot}]] schil\-dert der Schriftsteller Stephen Crane seinen drei{\ss}igst\"undigen \"Uber\-lebens\-kampf als Schiff\-br\"uch\-iger vor Florida.

\href{https://de.wikipedia.org/wiki/Wikipedia_Diskussion:Hauptseite/Schon_gewusst/Diskussionsarchiv/2013/November#Eigenvorschlag:_Im_Rettungsboot_.2813._Oktober_2013.29_.28erl..29}{Ausgew\"ahlt} wurde:\\
"`In der Erz\"ahlung [[\textit{Im Rettungsboot}]] schil\-dert der Schriftsteller Stephen Crane seinen drei{\ss}igst\"undigen \"Uberlebenskampf als Schiffbr\"uchiger vor Florida."' (\href{https://de.wikipedia.org/wiki/Wikipedia:Hauptseite/Archiv/28._November_2013}{Hauptseite am 28.} und am \href{https://de.wikipedia.org/wiki/Wikipedia:Hauptseite/Archiv/29._November_2013}{29. November 2013})
\end{quote}
\onehalfspacing

Von sieben Teasern wurde der letzte genommen. Ein \textit{Erlesnis} war aufgrund der Dramatik des Gegenstands der Geschichte in allen Teasern zu ahnen: wie es sich anf\"uhlen w\"urde einen Schiffbruch zu \"uberleben. Im Gegensatz zu manchen der anderen Tea\-ser klingt der ausgew\"ahlte \"uberaus sachlich und es wird auf der Hauptseite keine \textit{Erlesnis}-Chance angedeutet. \\

[7] [[\textit{Der Mann, der die S\"unde erfand}]]
\singlespacing
\begin{quote}
¸* Sean O'Faolain schil\-dert in einer Erz\"ahlung, wie die S\"unde in Irland erfunden wurde.\\
¸* Sean O'Faolain verarbeitete seiner Erfah\-rungen im irischen Un\-ab\-h\"an\-gig­\-keits\-kampf in einer [[\textit{Kurzgeschichte}]].\\
¸* Trotz der Kritik an der Katholischen Kirche in seiner [[\textit{Kurzprosa}]] konvertierte Sean O'Faolain nach einer Italienreise zum r\"omisch-katholischen Glauben.\\
¸* Schwester Maria Magdalena in Sean O'Faolains [[\textit{Kurzgeschichte}]] s\"un\-digt genauso wie die \"ubrigen Geistlichen, hat aber keine moralischen Bedenken. / kein schlechtes Gewissen\\
¸* Die Ordensschwester Maria Magdalena s\"undigt ebenso wie die \"ubrigen Geistlichen in Sean O'Faolains [[\textit{Kurzgeschichte}]], jedoch ohne schlechtes Gewissen.\\
¸* Der Pfarrer in Sean O'Faolains [[\textit{Kurzgeschichte}]] tr\"agt teuflische Z\"uge. oder: ... diabolische Z\"uge\\
¸* In Sean O'Faolains Kurzgeschichte [[\textit{Der Mann, der die S\"unde erfand}]] s\"undigt Maria Magdalena ebenso wie die \"ubrigen Geistlichen, jedoch ohne schlechtes Gewissen.

\href{https://de.wikipedia.org/wiki/Wikipedia_Diskussion:Hauptseite/Schon_gewusst/Diskussionsarchiv/2013/November#Eigenvorschlag:_Der_Mann.2C_der_die_S.C3.BCnde_erfand_.283._November_2013.29_.28erl..29}{Ausgew\"ahlt} wurde:\\ 
"`In Sean O'Faolains Kurzgeschichte [[\textit{Der Mann, der die S\"unde erfand}]] s\"undigt Maria Magdalena ebenso wie die \"ubrigen Geistlichen, jedoch ohne schlechtes Gewissen."' (\href{https://de.wikipedia.org/wiki/Wikipedia:Hauptseite/Archiv/6._Dezember_2013}{Hauptseite am 6.} und am \href{https://de.wikipedia.org/wiki/Wikipedia:Hauptseite/Archiv/7._Dezember_2013}{7. Dezember 2013})
\end{quote}
\onehalfspacing

In dem von den sieben Vorschl\"agen\footnote{Unter einem anderen Blickwinkel w\"are das Produzieren vieler Teaservorschl\"age \mbox{selbst} als eine interessante Methode der Li\-te\-ra\-tur\-ver\-mitt\-lung anzusehen, weil es aufzeigt, auf wieviel verschiedene Arten allein in Kurzform \"uber ein Werk gesprochen werden kann. Die Dis\-kus\-si\-onsseite der Hauptsei\-tenrubrik "`Schon gewusst?"' wird rund 200 Mal pro Tag aufgerufen, al\-ler\-dings werden die meisten der \mbox{User} lediglich Versions\"anderungen lesen und an einem so ausf\"uhrlich mit Teasern bedachten Li\-te\-ra\-tur-Vorschlag nicht so h\"aufig zum Lesen verweilen wie es die Abrufzahlen anzudeuten scheinen; auch ist die Seite sehr umfangreich, das hei{\ss}t in diesem Fall: viele Abschnitte, auf die man einzeln klickt, wenn man ein paar Tage die \"Anderungen nicht per Versionsgeschichte verfolgt hat und nur bei den Vorschlagsthemen nachsehen will, die einen am meisten interessieren.} ausgew\"ahlten Teaser deutet sich an, dass ein \textit{Erlesnis} darin bestehen k\"onnte zu erfahren, wie es ist, ohne schlechtes Gewissen zu s\"undigen. Auch zwei der anderen Teaser h\"atten diesen Punkt an\-ge\-spro\-chen und von den \mbox{vier} weiteren h\"atten mindestens zwei ebenfalls ein potenzielles "`Wissen wie"' zum Ausdruck gebracht. Es wird eine \textit{Erlesnis}-Chance angedeutet.\\

[10] [[\textit{Das Krokodil}]]
\singlespacing
\begin{quote}
¸* Das wirtschaftliche Prinzip habe Vorrang, und was das bedeutet, stellt Dostojewski in seiner Satire [[\textit{Das Krokodil}]] dar.\\
¸* Dostojewskis Satire [[\textit{Das Krokodil}]] nimmt den entfessel\-ten Ka\-pi\-ta\-lis\-mus aufs Korn und blieb aus \"okomischen Gr\"unden unvollendet.\\
¸* Dostojewskis Satire [[\textit{Das Krokodil}]] nimmt den Ka\-pi\-ta\-lis\-mus aufs \mbox{Korn} und blieb aus \"okomischen Gr\"unden unvollendet.

\href{https://de.wikipedia.org/wiki/Wikipedia_Diskussion:Hauptseite/Schon_gewusst/Diskussionsarchiv/2013/Dezember#Vorschlag:_Das_Krokodil_.284._Dezember_2013.29__.28erl..29}{Ausgew\"ahlt} wurde:\\
"`Dostojewskis Satire [[\textit{Das Krokodil}]] nimmt den Kapitalis­mus aufs Korn und blieb aus \"okono­mischen Gr\"unden unvollendet."' (\href{https://de.wikipedia.org/wiki/Wikipedia:Hauptseite/Archiv/5._Januar_2014}{Hauptseite am 5.} und am \href{https://de.wikipedia.org/wiki/Wikipedia:Hauptseite/Archiv/6._Januar_2014}{6. Januar 2014})
\end{quote}
\onehalfspacing

Der gew\"ahlte, letzte Teaser enth\"alt wie die anderen beiden die Genrebe\-zeichnung Satire, was je nach Geschmack des Lesers\textsuperscript{\tiny *} \textit{Erlesnisse} in Mengen verspricht, bei denen man zum Beispiel mit Verfremdungseffekten und starkem Lachreiz fertigwerden muss. \textit{Erlesnis}-Chance vorhanden.\\

[11] [[\textit{Der Neugierige}]]
\singlespacing
\begin{quote}
¸* In der Erz\"ahlung [[\textit{Der Neugierige}]] von Hans Erich Nossack steht das "'Nichts"' sym\-bo\-lisch f\"ur das Unbekannte und Neue oder f\"ur das noch nicht Geschaffene.\\
¸* "'Woher kommt mir eigentlich diese Neugierde, die mich immer von neuem aus den T\"umpeln hochjagt?"' fragt sich der Erz\"ahler in [[\textit{Der Neugierige}]] von Hans Erich Nossack.\\
¸* Vollkommene Abgeschiedenheit sucht der Ich-Erz\"ahler von [[\textit{Der Neu\-gie\-ri\-ge}]] von Hans Erich Nossack, als er untr\"ostlich \"uber den Verlust eines Freundes ist.\\
¸* Hans Erich Nossacks Erz\"ahlung [[\textit{Der Neu\-gierige}]] wurde aufgrund seines universalen Anspruchs von der Kritik als "'neuer Mythos"' be\-zeichnet.\\
¸* Die Erz\"ahlung [[\textit{Der Neugierige}]] befasst sich mit der Einsamkeit in Grenzsituationen, der Suche nach dem eigenen Ich und dem Tod und greift damit die gro{\ss}en Themen Hans Erich Nossacks auf.\\
¸* Hans Erich Nossacks Erz\"ahlung [[\textit{Der Neugierige}]] f\"uhrt den Leser bis an die Grenze des Ertr\"aglichen und dar\"uber hinaus.\\
¸* Hans Erich Nossacks Erz\"ahlung [[\textit{Der Neugierige}]] \"uberschreitet die Grenze des Ertr\"aglichen.

\href{https://de.wikipedia.org/wiki/Wikipedia_Diskussion:Hauptseite/Schon_gewusst/Diskussionsarchiv/2013/Dezember#Vorschlag:_Der_Neugierige_.2811._Dezember_2013.29_.28erl..29}{Ausgew\"ahlt} wurde:\\
"`Hans Erich Nossacks Erz\"ahlung [[\textit{Der Neugierige}]] f\"uhrt den Leser bis an die Grenze des Ertr\"aglichen und dar\"uber hinaus."' (\href{https://de.wikipedia.org/wiki/Wikipedia:Hauptseite/Archiv/7._Januar_2014}{Hauptseite am 7.} und am \href{https://de.wikipedia.org/wiki/Wikipedia:Hauptseite/Archiv/8._Januar_2014}{8. Januar 2014})
\end{quote}
\onehalfspacing

F\"ur die Hauptsei\-tenpr\"asentation wurde aus sieben Teasern derjenige ausgew\"ahlt, der als einziger explizit von der Wirkung auf den Leser\textsuperscript{\tiny *}, spricht, das hei{\ss}t: ein Leser\textsuperscript{\tiny *} kommt als Akteur\textsuperscript{\tiny *} vor, von dem gesagt wird, dass er im Kontakt mit der Erz\"ahlung etwas empfindet. Impliziter spricht davon auch ein weiterer Teaser. Die Chance f\"ur ein \textit{Erlesnis} scheint sehr hoch zu sein: wie sich etwas anf\"uhlt, was mich im Lesekontakt mit diesem Text "`bis an die Grenze des Ertr\"aglichen und dar\"uber hinaus"' bringt.\\

[15] [[\textit{Save the Reaper}]]
\singlespacing
\begin{quote}
¸* Alice Munro hat von ihrer Geschichte [[\textit{Save the Reaper}]] im selben Jahr zwei sehr verschiedene Versionen pu\-bli\-ziert.\\
¸* In Alice Munros Geschichte [[\textit{Save the Reaper}]] merkt der siebenj\"ahrige Enkel Philip scheinbar, dass die Gro{\ss}mutter von ihren Erlebnissen nicht alles der Tochter erz\"ahlen will.

\href{https://de.wikipedia.org/wiki/Wikipedia_Diskussion:Hauptseite/Schon_gewusst/Diskussionsarchiv/2014/Januar#Eigenvorschlag:_Save_the_Reaper_.284._Januar_2014.29_.28erl..29}{Ausgew\"ahlt} wurde:\\
"`In Alice Munros Geschichte [[\textit{Save the Reaper}]] merkt der siebenj\"ahrige Enkel Philip scheinbar, dass die Gro{\ss}mutter der Tochter nicht alles von ihren Erlebnissen erz\"ahlen will."' (\href{https://de.wikipedia.org/wiki/Wikipedia:Hauptseite/Archiv/7._Februar_2014}{Hauptseite am 7.} und am \href{https://de.wikipedia.org/wiki/Wikipedia:Hauptseite/Archiv/8._Februar_2014}{8. Februar 2014})
\end{quote}
\onehalfspacing

Von zwei Teasern wurde derjenige ausgew\"ahlt, der eine \textit{Erlesnis}-Chance andeutet: Es k\"onnte zu erfahren sein, wie es sich anf\"uhlt, am Verschweigen von Geheimnissen teilzuhaben und dies in einer weiteren Perspektive gespiegelt zu erhalten.\\ 

[19] [[\textit{Goodbye, Columbus}]]
\singlespacing
\begin{quote}
¸* In Philip Roths Kurzroman [[\textit{Goodbye, Columbus}]] geht es um die Frage nach der G\"ultigkeit \"uberlieferter Normen und Werte.\\
¸* Die Infragestellung der G\"ultigkeit \"uberlieferter Normen und Werte ist ein Thema in Philip Roths Kurzroman [[\textit{Goodbye, Columbus}]].\\
¸* In Philip Roths Kurzroman [[\textit{Goodbye, Columbus}]] ging es 1959 nicht nur um den vor\-ehelichen Sex.

\href{https://de.wikipedia.org/wiki/Wikipedia_Diskussion:Hauptseite/Schon_gewusst/Diskussionsarchiv/2014/Februar#Vorschlag:_Goodbye.2C_Columbus_.288._Februar_2014.29_.28erl..29}{Ausgew\"ahlt} wurde:\\
"`In Philip Roths Kurzroman [[\textit{Goodbye, Columbus}]] ging es 1959 nicht nur um den vor\-ehelichen Sex."' (\href{https://de.wikipedia.org/wiki/Wikipedia:Hauptseite/Archiv/17._M\%C3\%A4rz_2014}{Hauptseite am 17.} und am \href{https://de.wikipedia.org/wiki/Wikipedia:Hauptseite/Archiv/18._M\%C3\%A4rz_2014}{18. M\"arz 2014})
\end{quote}
\onehalfspacing

In den beiden ersten Teasern h\"atte es die Andeutung einer \textit{Erlesnis}-Chance gegeben, im dritten, ausgew\"ahlten Teaser gibt es keine.\\

[20] [[\textit{Die hellen Tage}]]
\singlespacing
\begin{quote}
¸* Deutschland mutet im Bestseller [[\textit{Die hellen Tage}]] von Zsuzsa B\'{a}nk sonderbar und elegisch an.\\
¸* Deutschland mutet im Bestseller [[\textit{Die hellen Tage}]] von Zsuzsa B\'{a}nk sonderbar an.

\href{https://de.wikipedia.org/wiki/Wikipedia_Diskussion:Hauptseite/Schon_gewusst/Diskussionsarchiv/2014/M\%C3\%A4rz#Eigenvorschlag:_Die_hellen_Tage_.283._M.C3.A4rz_2014.29_.28erl..29}{Ausgew\"ahlt} wurde:\\
"`Deutschland mutet im Bestseller [[\textit{Die hellen Tage}]] von Zsuzsa B\'{a}nk sonderbar an."' (\href{https://de.wikipedia.org/wiki/Wikipedia:Hauptseite/Archiv/6._April_2014}{Hauptseite am 6.} und am \href{https://de.wikipedia.org/wiki/Wikipedia:Hauptseite/Archiv/7._April_2014}{7. April 2014})
\end{quote}
\onehalfspacing

Es wird an\-ge\-spro\-chen, welche Stimmung in einem Werk vorherrscht, was der Andeutung einer \textit{Erlesnis}-Chance gleichkommt.\\

[23] [[\textit{Annawadi oder Der Traum von einem anderen Leben}]]
\singlespacing
\begin{quote}
¸* Kritiker bescheinigen [[\textit{Annawadi oder Der Traum von einem anderen Leben}]] sicht\"andernde Qualit\"aten.\\
¸* Dem Sachbuch [[\textit{Annawadi oder Der Traum von einem anderen Leben}]] wird eine nachhaltige Wirkung auf den Leser bescheinigt.

\href{https://de.wikipedia.org/wiki/Wikipedia_Diskussion:Hauptseite/Schon_gewusst/Diskussionsarchiv/2014/M\%C3\%A4rz\#Eigenvorschlag:_Annawadi_oder_Der_Traum_von_einem_anderen_Leben__.2829._M.C3.A4rz_214.29_.28erl..29}{Ausgew\"ahlt} wurde:\\
"`Dem Sachbuch [[\textit{Annawadi oder Der Traum von einem anderen Leben}]] wird eine nachhaltige Wirkung auf den Leser bescheinigt."' (\href{https://de.wikipedia.org/wiki/Wikipedia:Hauptseite/Archiv/25._April_2014}{Hauptseite am 25.} und am \href{https://de.wikipedia.org/wiki/Wikipedia:Hauptseite/Archiv/26._April_2014}{26. April 2014})
\end{quote}
\onehalfspacing

In beiden Teasern wird eine \textit{Erlesnis}-Chance angedeutet, im ausgew\"ahlten Teaser kommt sogar der Leser\textsuperscript{\tiny *} als (Re)Akteur\textsuperscript{\tiny *} vor.\\

[25] [[\textit{Reisende auf einem Bein}]]
\singlespacing
\begin{quote}
¸* Herta M\"uller erz\"ahlt in [[\textit{Reisende auf einem Bein}]] sprunghaft und z\"ogerlich wie bei einer H\"upfbewegung.\\
¸* Herta M\"uller erz\"ahlt in [[\textit{Reisende auf einem Bein}]] von einer Aus\-l\"an\-de\-rin im Ausland.\\
¸* Herta M\"uller erz\"ahlt in [[\textit{Reisende auf einem Bein}]] poetisch-bizarr \"uber Heimatlosigkeit.

\href{https://de.wikipedia.org/wiki/Wikipedia_Diskussion:Hauptseite/Schon_gewusst/Diskussionsarchiv/2014/April#Eigenvorschlag:_Reisende_auf_einem_Bein_.287._April_2014.29_.28erl..29}{Ausgew\"ahlt} wurde:\\
"`Herta M\"uller erz\"ahlt in [[\textit{Reisende auf einem Bein}]] von einer Aus\-l\"an\-de\-rin im Ausland."' (\href{https://de.wikipedia.org/wiki/Wikipedia:Hauptseite/Archiv/10._Mai_2014}{Hauptseite am 10.} und am \href{https://de.wikipedia.org/wiki/Wikipedia:Hauptseite/Archiv/11._Mai_2014}{11. Mai 2014})
\end{quote}
\onehalfspacing

Von drei Teasern wurde derjenige ausgew\"ahlt, der keine Adjektive enth\"alt. Es wird ein Gegenstand des Werks benannt, von den anderen beiden Teasern aber nichts zur Erz\"ahlweise \"ubernommen, die die Andeutung einer \textit{Erlesnis}-Chance geboten h\"atten.\\ 

[47] [[\textit{Jenseits von Schuld und S\"uhne. Bew\"altigungsversuche eines \"Uberw\"altigten}]]
\singlespacing
\begin{quote}
¸* Radikale Selbstbefragung ist ein Kennzeichen der Essays in [[\textit{Jenseits von Schuld und S\"uhne. Bew\"altigungsversuche eines \"Uber\-w\"al\-tig\-ten}]] von Jean Am\'{e}ry.\\
¸* In [[\textit{Jenseits von Schuld und S\"uhne. Bew\"altigungsversuche eines \"Uber\-w\"al\-tig\-ten}]] befragt Jean Am\'{e}ry sich selbst.

\href{https://de.wikipedia.org/wiki/Wikipedia_Diskussion:Hauptseite/Schon_gewusst/Diskussionsarchiv/2014/Oktober#Vorschlag:_Jenseits_von_Schuld_und_S.C3.BChne_.2810._September.29_.28erl..29}{Ausgew\"ahlt} wurde:\\
"`In [[\textit{Jenseits von Schuld und S\"uhne. Bew\"altigungsversuche eines \"Uber\-w\"al\-tig\-ten}]] befragt Jean Am\'{e}ry sich selbst."' (\href{https://de.wikipedia.org/wiki/Wikipedia:Hauptseite/Archiv/16._Oktober_2014}{Hauptseite am 16.} und am \href{https://de.wikipedia.org/wiki/Wikipedia:Hauptseite/Archiv/17._Oktober_2014}{17. Oktober 2014})
\end{quote}
\onehalfspacing

Hier wurde der zweite Teaser ohne Dis\-kus\-si\-on genommen. In beiden Teasern geht es zwar um die Handlung, aber Selbstbefragung als Thema kann als Andeutung einer \textit{Erlesnis}-Chance angesehen werden.\\

[48] [[\textit{Das Sternbild des Ziegentur}]]

\href{https://de.wikipedia.org/wiki/Wikipedia_Diskussion:Hauptseite/Schon_gewusst/Diskussionsarchiv/2014/Oktober#Vorschlag:_Das_Sternbild_des_Ziegentur_.2814._Oktober.29_.28erl..29}{Ausgew\"ahlt} wurde:
\singlespacing
\begin{quote}
"`[[\textit{Das Sternbild des Ziegentur}]] l\"asst sich als Satire auf die Wirt\-schafts\-po\-li\-tik Chruschtschows deuten."' (\href{https://de.wikipedia.org/wiki/Wikipedia:Hauptseite/Archiv/2._November_2014}{Hauptseite 2.} und \href{https://de.wikipedia.org/wiki/Wikipedia:Hauptseite/Archiv/3._November_2014}{3. November 2014})
\end{quote}
\onehalfspacing

Es gibt die Andeutung einer \textit{Erlesnis}-Chance, erstens, weil im Teaser genannt wird, dass es sich um Satire handelt, zwei\-tens, weil Deutbarkeit zur Sprache kommt. \\

[60] [[\textit{La Vie commune}]]

\href{https://de.wikipedia.org/wiki/Wikipedia_Diskussion:Hauptseite/Schon_gewusst/Diskussionsarchiv/2015/Januar#Eigenvorschlag:_La_Vie_commune_.289._Dezember.29_.28erl..29}{Ausgew\"ahlt} wurde:
\singlespacing
\begin{quote}
"`In Lydie Salvayres Roman [[\textit{La Vie commune}]] liegt allt\"agliche Gewalt in gefl\"usterten Vertraulich­kei\-ten."' (\href{https://de.wikipedia.org/wiki/Wikipedia:Hauptseite/Archiv/13._Januar_2015}{Hauptseite am 13.} und am \href{https://de.wikipedia.org/wiki/Wikipedia:Hauptseite/Archiv/14._Januar_2015}{14. Januar 2015})
\end{quote}
\onehalfspacing

Der Teaser spricht nicht die Handlung des Romans, sondern eine Darstellungsweise an. Es wird eine \textit{Erlesnis}-Chance angedeutet, die als f\"ur Li\-te\-ra\-tur spezifisch an\-ge\-sehen werden kann.

Ich komme zur Auswertung von Teil 2 dieses Abschnitts. In 10 von 61 F\"allen wurde in Teasern, die auf der Hauptseite landeten, mehr oder weniger stark eine \textit{Erlesnis}-Chance angedeutet -- mit etwa 15 \% immerhin keine geringe Menge, aber es handelt sich um keine direkten \textit{Erlesnis}-Schil\-derungen, sondern nach meinem Empfinden nur um Andeutungen einer Chance. Zwar wissen erfahrene Leser\textsuperscript{\tiny{*}}, was man bei der Lekt\"ure erleben kann, aber mir ging es darum, anhand eines Kriteriums aufzuzeigen, in welchem Ma{\ss} und wie (un)deutlich in einem Teaser auf der Wi\-ki\-pe\-dia-Hauptseite der deutschspra\-chi\-gen Version Erlebnisse formuliert werden, die unter anderem f\"ur eine T\"atigkeit wie das Lesen von Li\-te\-ra\-tur spezifisch sind. In \mbox{vier} weiteren F\"allen wurde diese Chance verpasst, denn obwohl es entsprechende Teaser-Vorschl\"age ge\-ge\-ben hat, wurde ein an\-de\-rer Aspekt aus dem Artikel bevorzugt. Das hei{\ss}t aus meiner Sicht: \mbox{selbst} wenn es in einem Werkartikel schon Informationen gibt, aus denen \textit{Erlesnisse} zu schlie{\ss}en w\"aren, und diese tats\"achlich in Teaservorschl\"agen zur Sprache kommen, kann es sein, dass andere Aspekte bevorzugt werden -- aus welchen Gr\"unden auch immer. Eine der am deutlichsten formulierten \textit{Erlesnis}-Chancen bezieht sich: auf ein Sachbuch [23]. Ob es sich hierbei um einen Zufall oder eher um einen systemisch bedingten Treffer handelt, der f\"ur die Haltung von Usern, die sich an Arbei\-ten in diesem adhoc-Team beteiligen, erwartbar ist, k\"onnte in einer gr\"o{\ss}er angelegten Studie he\-rauszufinden sein -- wobei Schwankungen je nach Periode einzukalkulieren w\"aren, deren Unregelm\"a{\ss}igkeit ebenfalls als f\"ur Wi\-ki\-pe\-dia charakteristisch erscheinen w\"urde. Schon allein me\-tho\-disch w\"are es eine spannende Studie. Hinzunehmen k\"onnte man einen Vergleich der Sprachversionen -- wobei ich nicht wei{\ss}, wie hoch in anderen Wi\-ki\-pe\-dias der Anteil an Werkartikeln im \mbox{Bereich} Li\-te\-ra\-tur ist und auch nicht, ob und wie sie bei "`Schon gewusst?"' vorgestellt werden.

Im nun folgenden dritten Teil geht es um die Anforderung einer distanzierten Ausdrucksweise als Anzeichen f\"ur Neutralit\"at und was dies aus meiner Sicht f\"ur Eintr\"age zu einzelnen Sprachkunstwerken bedeutet. Wenn es sich bei Wissen, das Li\-te\-ra\-tur zu vermitteln in der Lage ist, in hohem Ma{\ss}e um nicht-pro\-po\-si\-ti\-o\-na\-les Wissen handelt, bei Wi\-ki\-pe\-dia aber pro\-po\-si\-ti\-o\-na\-le Formen von Wissen bereitgestellt werden sollen, wie m\"usste ein Eintrag zu einem li\-te\-ra\-rischen Werk aussehen? Genauso wie Eintr\"age mit Wissenswertem \"uber Bauwerke, Begriffe oder Per\-so\-nen? Ich meine: Nein, und stelle im Folgenden den Rahmen, den Wi\-ki\-pe\-dia f\"ur die Darstellung verschiedener Wissensarten bietet, detailierter dar, um anschlie{\ss}end meine Argumentation weiter auszuarbei\-ten.

Bei Wi\-ki\-pe\-dia werden auf ein- und derselben Web-Plattform mit vielen anderen auch Artikel \"uber li\-te\-ra\-rische Werke pu\-bli\-ziert, durch Beteiligte, die meist einander nicht kennen. Es sind kommunikativ konstituierte Gemeinschaften (\cite{Krotz2005}:261), die in adhoc-Teams arbei\-ten: Ein Artikel kommt zustande, indem diejenigen, die etwas zum Thema beitragen m\"ochten und wissen wie es geht, es tun, und zwar direkt im Web. Dies w\"are die Formulierung aus Sicht der Entstehung des jeweiligen Artikels auf das, was Wi\-ki\-pe\-dia "`ist"'. Aus Sicht von Usern, die mitschrei\-ben, k\"onnte es so beschrieben werden: Wenn mir in einem Artikel bei Wi\-ki\-pe\-dia etwas auff\"allt, wozu ich einen Verbesserungsvorschlag machen kann, dann tue ich dies unabh\"angig davon, wer schon an diesem Artikel gearbeitet hat. Es geht bei Wi\-ki\-pe\-dia darum, dass die je aktuelle Fassung eines Artikels m\"oglichst zuverl\"assige Aussagen in m\"oglichst neutralen Formulierungen aufweist. Daf\"ur muss in m\"oglichst hoher Frequenz und zu m\"oglichst vielen Themen das Wissen von Web-Nutzern\textsuperscript{\tiny *} und die unbezahlbare Be\-reit\-schaft, etwas zum Ganzen beizutragen, eingeworben werden. Der stete Wandel, das grundlegende Prinzip dieser Zusammenarbeit, muss dabei auf eine Weise ad\-mi\-ni\-stra\-tiv begleitet -- und personell wie finanziell gemanagt -- werden, dass laufende Aktualisierungen m\"oglich sind und auch get\"atigt werden.
Des weiteren sind einige formale und inhaltliche Spielregeln ausgehandelt worden. Auch sie werden st\"andig erg\"anzt und weiter ausgehandelt. Ihr Zweck ist, einen Rahmen zu bilden, innerhalb dessen beim Verfassen von Artikeln einerseits Standardisierungen erzielbar sind und an\-de\-rerseits gen\"ugend Spielraum bleibt, damit diejenigen, die etwas beitragen wollen, sich nicht so stark eingeschr\"ankt f\"uhlen, dass sie die Lust am Mitmachen verlieren. Nicht zu\-letzt muss zum Erhalt der Bekanntheit und Qualit\"at der Marke Wi\-ki\-pe\-dia eine gen\"ugend gro{\ss}e Anzahl an Per\-so\-nen die Artikel in Suchmaschinen gut genug positioniert auffinden k\"onnen und dar\"uber hinaus dem Projekt eine Wertsch\"atzung entgegenbringen -- die sich eventuell in Spenden ausdr\"uckt -- und mit m\"oglichst vielen anderen dar\"uber kommunizieren, auch wenn diese eventuell Wi\-ki\-pe\-dia "`nur zum Lesen"' nutzen. Denn von dieser Wertsch\"atzung zu erfahren moti\-viert wiederum diejenigen, die bei Wi\-ki\-pe\-dia "`auch schrei\-ben"', das hei{\ss}t dem Projekt Wi\-ki\-pe\-dia Wissen und Zeit durch ihre Beteiligung an Aushandlungen zugute kommen lassen, ohne mit dieser T\"atigkeit pers\"onlich ein Einkommen zu erwirtschaften.

Basiert eine Plattform auf Wiki-Software, bietet sie M\"oglichkei\-ten, die f\"ur den Austausch \"uber Li\-te\-ra\-tur sehr f\"orderlich sein k\"onnen, zum Beispiel \href{http://edocs.fu-berlin.de/docs/receive/FUDOCS_document_000000010457}{"`sich wechsel\-sei\-tig als Dialogpartner anzuerkennen und keinen Einwand gegen eigene Behauptungen, den Dritte vorbringen k\"onnten, von vornherein auszuschlie{\ss}en"'} (\cite{Schmidt2010}:28). Bekanntlich k\"onnen im Dialog verschiedene Perspektiven zur Sprache gebracht werden: Man tauscht sich aus und ist danach meist kl\"uger. Zumindest hat man vermutlich etwas dar\"uber erfahren, wie andere Leute eine bestimmte Sache sehen, und erh\"alt dadurch Gelegenheit, dar\"uber zu reflektieren, ob man die eigene Meinung ganz oder teilweise revidieren m\"ochte. Die wech\-sel\-sei\-tige Anerkennung als Dialogpartner\textsuperscript{\tiny *} ist auch in der Arbeit an Wi\-ki\-pe\-dia-Inhalten grundlegend, zum Beispiel bei Auseinandersetzungen \"uber eine der Grunds\"aulen des Selbstverst\"andnisses von Wi\-ki\-pe\-dia: ob es so etwas wie eine neutrale Perspektive gibt, einen "`neutral point of view"' (NPOV). Und falls ja, woran eine solche Perspektive zu erkennen w\"are. 

Konkret wird diese Frage immer dann, wenn man \mbox{selbst} einen neuen Artikel anlegen oder zu einem Artikel eine \"Anderung beitragen m\"ochte: Welche Formulierungen w\"ahle ich, damit mein Beitrag "`neutral"' klingt? Wessen Neutralit\"at ist gemeint und welcher Geltungsbereich ist relevant, wenn es sich um eine Enzyklop\"adie mit web-weitem Anspruch handelt?

Ich bin optimistisch bez\"uglich der Nutzung von Wiki-Software f\"ur Debatten, in denen es um dasjenige geht, was k\"unstlerische Werke ausmacht: das Vieldeutige, das wir im Lek\-t\"u\-re\-pro\-zess vor unseren Augen in Szene setzen und in \textit{Erlesnissen} wei\-ter bearbei\-ten. Indem wir uns dar\"uber austauschen wollen, feiern wir ihre Existenz gewisserma{\ss}en, die Werke ebenso wie unsere Lekt\"ureerfah\-rungen. 

Ina Hartwig formulierte 2012 ihre Empfindungen und Einsichten in einer besonderen Art von Lesebericht, dem Vorwort zu ihrem Essayband \textit{Das Geheimfach ist offen. \"Uber Li\-te\-ra\-tur}, folgenderma{\ss}en:

\singlespacing
\begin{quote}
"`Eher im Gegenteil verliert man sich doch im Lesen, vergisst sich, wird sich los f\"ur ein Weilchen. Und in diesem Sichverlieren, diesem Abwarten und gedehnten Nichtwissen, wird man reicher. Das gilt auch f\"ur die gezielt lesende Kritikerin, die sich innerlich frei macht, um das Gelesene aufzunehmen, zu begreifen, zu empfinden, zu ana\-lysieren, zu werten, und schlie{\ss}lich (oft nach erheblicher Anstrengung) als Essay oder Kritik, also als wiederum Geschriebenes, zur\"uckzugeben an ein hof\-fent\-lich interessiertes Publikum.\\ 
¸[...]\\
Das bedeutet, unter s\"akularen Bedingungen, Li\-te\-ra\-tur als Passion -- sich vergessen in und denken mit der Sprache des anderen. Am Ende dieser passionierten Bewegung will man aber doch zeigen, was man gefunden hat. Da tritt das Ego wieder hervor aus seiner Versenkung und ruft: \mbox{Hier} bin ich. Seht her, was ich mitbringe von meiner Reise."' (\cite{Hartwig2012}:16-17)
\end{quote}
\onehalfspacing

Je offener die Debatten gef\"uhrt werden k\"onnen, desto interessanter werden sie auch. Mit "`offen"' meine ich in diesem Zusammenhang, dass in einer solchen Debatte nichts vorab als relevanter gilt, sondern alles (her)vorkommen und (vor)gebracht werden kann. 

Um meine Fragen beantworten zu k\"onnen, habe ich mich entschieden, neue Werkartikel anzulegen, weil sich kritische Aufmerksamkeit am ehesten auf neue Artikel lenken l\"asst, so mein Eindruck aus den Beobachtungen der vorigen Jahre. 

Im Verlauf des kollektiven Verfassens wird f\"ur die jeweiligen Inhalte ein passendes Artikelformat erarbeitet. Die Inhalte der Artikel werden in Prosa dargestellt, Abbildungen dienen der Illustration des Textes. Damit entstehen die Inhalte \"uberwiegend in einem Modus, der demjenigen der Li\-te\-ra\-tur ebenso nah verwandt ist wie demjenigen des Schrei\-bens \"uber Li\-te\-ra\-tur: Text. Was von Autoren\textsuperscript{\tiny *} bei Wi\-ki\-pe\-dia an F\"ahigkei\-ten erwartet wird, ist dar\"uber hinaus weitgehend deckungsgleich mit dem, was 2009 in praxeologischer Hinsicht als "`spezifisches Gepr\"age"' li\-te\-ra\-tur\-wis\-sen\-schaft\-li\-cher Dis\-ziplinen diagnostiziert worden ist, n\"amlich: 

\singlespacing
\begin{quote}
"`Praxisformen des Textumgangs, der Begriffsbildung, der Themen\-fin\-dung, der Wissensordnung, der Validierung und Darstellung von Wis\-sens\-anspr\"uchen."' (\cite{MartusSpoerhase2009}:89)
\end{quote}
\onehalfspacing

Es gibt also eine handwerkliche N\"ahe zwischen Arbeitsroutinen in den Li\-te\-ra\-tur\-wissenschaften und bei Wi\-ki\-pe\-dia, da es um das Konzipieren, Ana\-lysieren, Beschrei\-ben (Kommentieren), Erarbei\-ten und Bearbei\-ten (Editieren) von \textbf{Texten} geht. W\"ahrend dies zun\"achst unabh\"angig vom Thema oder Wissensgebiet des jeweiligen Artikels ist, k\"onnten User mit li\-te\-ra\-tur\-wis\-sen\-schaft\-li\-chem Hintergrund auch aufgrund ihrer vermutbaren text\-kri\-ti\-schen Arbeitsweise bei Artikeln zu li\-te\-ra\-ri\-schen Werken we\-sent\-liche Beitr\"age leisten. Weitere und vor allem interdisziplin\"ar angelegte Ana\-lysen k\"onnten kl\"aren, ob li\-te\-ra\-tur\-wissenschaftliche Arbeitsroutinen bei Artikeln zu li\-te\-ra\-rischen Werken plausibler sind als diejenigen an\-de\-rer Disziplinen. Dass Li\-te\-ra\-tur weit mehr zu sein vermag als ein Objekt (zum Beispiel: Buch), ist auch Wi\-ki\-pe\-dia-Eintr\"agen zu entnehmen, denn die Materialit\"at der Buchausgabe steht in der Regel nicht im Mittelpunkt des Artikeltextes. Es finden sich so wenige Angaben, dass ein Eintrag in vielen F\"allen nicht einmal einen Link zu einer Abbildung des Co\-vers enth\"alt -- ebensowenig wie es eine Angabe zur Sei\-tenzahl (oder zum Kaufpreis) gibt.

Im weiteren Verlauf dieser Arbeit geht es um \textit{parrhesia} und die Produktion an\-de\-rer Schreibweisen. Mein Augenmerk richtet sich insbesondere auf das Sich-\"Au{\ss}ern zu Lekt\"ureerfah\-rungen und speziell darum, \"Au{\ss}erungen von \textit{Erlesnissen} zu wagen. Damit sie im Text eines Wi\-ki\-pe\-dia-Eintrags referiert werden k\"onnen, m\"ussen sie belegbar an an\-de\-rer Stelle zur Sprache gekommen sein. Auf welche Weise \"uber die eigene Lekt\"ure im Web berichtet werden kann, l\"asst sich anhand von Beitr\"agen in Leseforen zeigen, wo Leseberichte in pers\"onlicher Form nicht nur gestattet, sondern sogar erw\"unscht sind. In gr\"o{\ss}eren Leseforen wie \textit{bablio.com} oder \textit{goodreads.com}, in denen ich auf die Suche nach Aussagen gegangen bin, um im Zuge des Verfassens von Wi\-ki\-pe\-dia-Eintr\"agen relevante Aspekte bez\"uglich des subjektiv wahrgenommenen Effekts bei der Lekt\"ure li\-te\-ra\-rischer Werke zu finden, habe ich den Eindruck gewonnen, dass es in Leseforen m\"oglich ist und -- zumindest in eng\-lischer und franz\"osischer Sprache -- praktiziert wird, freim\"utig in gut lesbarem Stil die Freiheit zu nutzen, in pers\"onlicher Form aus dem eigenen Erleben zu berichten. Allem Anschein nach kommt es dabei auf die L\"ange einer Aussage nicht einmal an, sondern allein darauf, sich am Austausch \"uberhaupt zu beteiligen, und vielleicht mit dem Aspekt, der einem am wichtigsten erscheint -- und von dem man meint, dass die eigene Beschreibung f\"ur andere in der von mir gew\"ahlten Form interessant sein k\"onnte.

Wenn in Wi\-ki\-pe\-dia-Eintr\"agen zu li\-te\-ra\-rischen Werken Angaben \"uber den Effekt der Lekt\"ure zu finden w\"aren, w\"urde erstens nicht nur deutlich, dass Li\-te\-ra\-tur mehr als nur etwas Objekthaftes ist (\cite{Attridge2004}) und dass zwei\-tens beim Lesen eines Eintrags \mbox{selbst} M\"oglichkeitsr\"aume entstehen bzw. Leser\textsuperscript{\tiny *} unerwartet neue Anregungen erhalten (\cite{Traussneck2010}). Darin, scheint mir, liegt eine M\"oglichkeit, wie das durch ein Neutralit\"atsgebot entstandene System zu unterlaufen ist, denn, so meine Einsch\"atzung f\"ur das Anliegen, in einem dominant positivistischen Umfeld dennoch spezifisch Li\-te\-ra\-risches in seiner Eigenheit zur Geltung zu bringen: NPOV allein bietet \mbox{hier} keinen angemessenen Rahmen. Denn Li\-te\-ra\-tur ist Sprachkunst und es kann daher nicht erstrebenswert sein, sie in "`neutraler"' Art und Weise zu beschrei\-ben -- selbst wenn es ginge. Und warum sollte bei Wi\-ki\-pe\-dia li\-te\-ra\-rische Kunst weniger ihren Eigent\"umlichkei\-ten gem\"a{\ss} "`enzyklop\"adisch"' dargestellt werden als Bauwerke oder Proteine oder Transistoren oder Biografien von Pers\"onlichkei\-ten? In Artikeln zu li\-te\-ra\-rischen Werken m\"usste mit Verfahren, die von der \textit{parrhesia} her inspiriert sind, ein NPOV-Diktat unterwanderbar sein, um Wi\-ki\-pe\-dia f\"ur eine ansprechende Li\-te\-ra\-tur\-ver\-mitt\-lung zu nutzen. Wissen \"uber Li\-te\-ra\-tur k\"onnte auch f\"ur \"altere Werke in seiner Variabilit\"at nicht nur im R\"uckschaumodus dargestellt werden, sondern zus\"atzlich an\-hand ge\-gen\-w\"ar\-ti\-ger Reaktionen, zu denen es keine aktuellen Rezensionen mehr gibt. Dies w\"urde inbesondere f\"ur das Schrei\-ben in Wikis bedeuten, dass sich m\"oglichst viele \mbox{User} mit ihren je\-wei\-li\-gen Erfah\-rungen und Perspektiven am Zusammenstellen der Informationen zum jeweiligen Werk beteiligen, wozu auch die Auswahl der Belege geh\"ort, die als relevant gelten sollen. Denn nicht zu\-letzt aufgrund des Webs, unter anderem wegen Rezensionsplattformen und Leseforen, meinte Brigitte Schwens-Harrant 2008, "`hat die Li\-te\-ra\-tur\- heute mehr \"Of\-fent\-lichkeit denn je. Doch die Bedingungen des \"of\-fent\-lichen Gespr\"achs \"uber Li\-te\-ra\-tur haben sich ver\"andert, was auch auf die ra\-san\-te Entwicklung der Verlagsbranche und der Massenmedien zur\"uckzuf\"uhren ist: Zahlen z\"ahlen und dominieren dort wie da"' \cite{Schwens-Harrant2008}:9-10). Man k\"onnte einwenden, Wi\-ki\-pe\-dia sei von diesen Entwicklungen unabh\"angig, denn die Autoren\textsuperscript{\tiny *} arbei\-ten unentgeltlich, sind mit ihren Beitr\"agen zeitlich ungebunden und w\"ahlen Werke, zu denen sie Artikel anlegen, aufgrund pers\"onlicher Interessen aus. Zwar sind die genannten Aspek\-te zu\-tref\-fend und interessant\footnote{Die drei Faktoren: unentgeltlich t\"atige Autoren\textsuperscript{\tiny *}, Ausw\"ahlen von Werken allein aufgrund pers\"onlicher Interessen sowie Arbeit an Beitr\"agen bei freier Zeiteinteilung haben in der Li\-te\-ra\-tur\-ver\-mitt\-lungsfor\-schung bis\-her keine Beachtung gefunden, zumindest allem Anschein nach nicht unter Einbeziehung von Wi\-ki\-pe\-dia als Li\-te\-ra\-tur\-plattform, siehe dazu Abschnitt \textit{~\ref{subsec:7.2} Update f\"ur die Theorie der Li\-te\-ra\-tur\-ver\-mitt\-lung}}, al\-ler\-dings richten sich die Relevanzkriterien, die in der deutschspra\-chi\-gen Version von Wi\-ki\-pe\-dia aufgestellt wurden, dennoch nach Auf\-merk\-sam\-keits\-effekten, die Ergebnis marktf\"ormiger Gewichtungen sind: Anzahl der Rezensionen in \"uberregionalen Tageszeitungen, Li\-te\-ra\-tur\-preise, Verkaufszahlen etc. Der Einwand k\"onnte positiv gesehen f\"ur eine genauere Ausleuchtung genutzt werden, indem man der Frage nachgeht, ob die Relevanzkriterien um jeden Preis eingehalten werden m\"ussen und sich unter dieser Schwelle der Aufwand f\"ur die Arbeit an einem Artikel gar nicht lohnt: Jein, w\"are meine Einsch\"atzung dazu, denn erstens gibt es Tipps auf einer Seite, die mit "`Relevanzpr\"ufung"' betitelt ist, zwei\-tens werden neue Artikel gelegentlich sogar von den kritischsten Kritikern\textsuperscript{\tiny *} \"ubersehen (zum Beispiel wenn jene einige Tage nicht online sind), und drittens, falls doch eine L\"oschdis\-kus\-si\-on vom Zaun gebrochen wird (vorzugsweise gegen Eintr\"age zu Werken von Autoren\textsuperscript{\tiny *} mit weiblich klingenden Vornamen), k\"onnen Artikel in den Be\-nut\-zernamensraum genommen werden, bis die Relevanzkriterien erf\"ullt sind, oder der Artikel wird vom \href{http://meta.prepedia.org/wiki/Hauptseite}{Rettungswiki ''PrePedia''} adoptiert, oder beides. An Bedingungen des \"of\-fent\-lichen Gespr\"achs \"uber Li\-te\-ra\-tur, die sich speziell durch Wi\-ki\-pe\-dia ver\"andert haben, w\"are meiner Einsch\"atzung nach auf der Negativseite zu verbuchen, dass -- sehr knapp gesagt -- Aussagen, die Adjektive enthalten sowie pers\"onliche Einsch\"atzungen, die explizit als solche zur Sprache gebracht werden, bis\-her weniger Raum erhalten. Das betrifft insbesondere nicht-pro\-po\-si\-ti\-o\-na\-les Wissen aus Li\-te\-ra\-tur\-lekt\"ure, also \textit{Erlesnisse}. Um deren enzyklop\"adische Aussagekraft als Bereicherung f\"ur jeden Werkartikel geht es unter anderem in der vorliegenden Arbeit. Ob \textit{Erlesnisse} als eine ge\-eig\-nete Form des Widerstands zur Ausbalancierung gegen die bei Wi\-ki\-pe\-dia herrschende "`Schlagseite zum Faktischen"' auch f\"ur Artikel als akzeptabel angesehen werden, deren Werke sich seit einiger Zeit jenseits aktueller Medienaufmerksamkeit befinden, w\"are als N\"achstes bei literarhistorischen Namen wie Thomas Mann auszuprobieren, in dem Sinne, dass "`Klassiker nicht ist, was im Archiv verschwunden ist, sondern, was immer noch Aufmerksamkeit verdient"' (\cite{Franck2009}). Ulrike Draesners \textit{Erlesnis} zu \textit{Die Bekenntnisse des Hochstaplers Felix Krull} k\"onnte diesbez\"uglich ein guter Einstieg sein (\cite{Draesner2013}:339-353). Ich lese den Titel des Schlusskapitels als Zeichen der gegl\"uckten Umgehung einer (Dis-)Artikulation, eine Be\-griffs\-pr\"a\-gung von Beatrice Michaelis: "`Damit geht es weder um ein absolutes Schweigen noch um ein l\"uckenloses Reden, sondern um eine dialektische Verkn\"upfung von Verschweigen und Besprechen, die \"uber die Absicht, etwas nicht zu sagen, eine spe\-zi\-fi\-sche Form Form des Sprechens, die (Dis-)Artikulation, als Schweigeeffekt hervorbringt"' \cite{Michaelis2011}:2). Und weil \mbox{hier} und da Vorbehalte ge\"au{\ss}ert wurden -- wenn auch wenige, siehe Abschnitt \textit{~\ref{subsec:6.2} Experimentieren mit Leseberichten. Konzeption, Hergang, Auswertung} --, lohnt es sich, \textit{Erlesnisse} in enzyklop\"adischem Gewand als eine relativ unauff\"allige Widerstandsform aufmerksam im Auge zu behalten. 

Marianne Schuller \"uberlegte 2009, ob die von ihr diagnostizierte "`auf\-f\"al\-lige Un\-auf\-f\"al\-lig\-keit von \flq Widerstand\frq"' (gegen die aktuellen ge\-sell\-schaftlichen Krisen) ein "`Symptom f\"ur die Dehnbarkeit und Geschmeidigkeit des \flq Normalismus\frq "' sei, der als flexibles Integrationsverfahren die Grenzen von Normen flexibilisiere, und fragte: "`Ereignet sich Widerstand und Widerst\"andiges auf Wegen und in Me\-tho\-den, die \"uber die bekannten Formen hinausgehen, sie unter- oder \"uberschrei\-ten?"' (\cite{Schuller2009}:15). Meinerseits sind Zusammenfassungen von \textit{Erlesnissen} in Wi\-ki\-pe\-dia-Artikeln eine subversive T\"atigkeit in Hinblick auf die NPOV-Policy, denn infolge der Aushandlungen in einem solchen Rahmen wird ein Normalismus zur St\"utzung der jeweils he\-ge\-mo\-nia\-len Positionen bestimmend. \textit{Erlesnisse} in Li\-te\-ra\-tur\-artikeln zu pr\"asentieren erscheint mir ge\-eig\-net, um die Ordnung des Sagbaren (\cite{Hetzel2012}) bei Wi\-ki\-pe\-dia zu ver\"andern, im Sinne des Offenhaltens einer anderen Zukunftsf\"ahigkeit (\cite{Schuller2009}:16) in Diskursen \"uber Li\-te\-ra\-tur und Lesen, also Li\-te\-ra\-tur\textsuperscript{\~.\~.}lekt\"ure, transformiert in pers\"onliche Aussagen in Form von schriftlichen \textit{Erlesnissen}. Auf dem Weg dorthin halte ich als Fazit dieses Teilabschnitts fest: Je gr\"o{\ss}er die Variabilit\"at und Bandbreite der wiedergegebenen Eindr\"ucke in einem Werkartikel, desto eindr\"ucklicher kann die Art und Tiefe der Wirkm\"achtigkeit von Li\-te\-ra\-tur in einem enzyklop\"adischen Eintrag bei Wi\-ki\-pe\-dia zum Ausdruck gebracht und das Wissen \"uber Li\-te\-ra\-tur bei Wi\-ki\-pe\-dia bereichert werden. 

Auf Basis meiner Er\"orterungen im zwei\-ten Teil dieses Abschnitts lautet meine erste Einsch\"atzung dazu, welche Bedingungen der Darstellbarkeit nicht-pro\-po\-si\-ti\-o\-na\-len Wissens in Eintr\"agen zu li\-te\-ra\-rischen Werken als aktuell geltend angenommen werden k\"onnen, dass in letzter Zeit immerhin 15 \% der Teaser f\"ur Li\-te\-ra\-tur\-artikel im Ram\-pen\-licht der Hauptseite gestanden haben, die \textit{Erlesnis}-Chancen andeuteten, auch wenn in \mbox{vier} F\"allen einem aus meiner Sicht weit weniger Leseerlebnis-relevanten Teaser der Vorzug gegeben wurde. In der deutschspra\-chi\-gen Version wurde demnach zwi\-schen Oktober 2013 und Januar 2015 in basisdemokratischen Prozessen bei Wi\-ki\-pe\-dia nicht-pro\-po\-si\-ti\-o\-na\-les Wissen aus Li\-te\-ra\-tur\-lekt\"ure in Form einer \textit{Erlesnis}-Chance in gewissem Ma{\ss}e als auf der Hauptseite pr\"asentabel angesehen.

\pagebreak

\subsection{Aussagen zu Wikipedia in literaturwissenschaftlichen Beitr\"agen}
\label{subsec:5.3}

Zu den Bedingungen der Darstellbarkeit nicht-pro\-po\-si\-ti\-o\-na\-len Wissens in Li\-te\-ra\-tur\-eintr\"agen bei Wi\-ki\-pe\-dia geh\"ort die ge\"au{\ss}erte Haltung gegen\"uber der Plattform in Kreisen, die professionell mit Li\-te\-ra\-tur befasst sind. Allgemein lie{\ss}e sich im Blickwinkel meiner Fragestellung die Ausgangslage des Verh\"altnisses von Wi\-ki\-pe\-dia und Li\-te\-ra\-tur so formulieren, dass den Besonderhei\-ten der jeweils anderen Seite nicht gen\"ugend Bedeutung f\"ur die eigenen Anliegen bei\-ge\-mes\-sen wird. Lag im vorigen Abschnitt der Fokus auf Informationen, die sich \"uber Li\-te\-ra\-tur aufsei\-ten von Wi\-ki\-pe\-dia finden lassen, werden in diesem Unterkapitel Aussagen zu Wi\-ki\-pe\-dia in Li\-te\-ra\-tur\-wissenschaftlichen Beitr\"agen beispielhaft er\"ortert. Im Folgenden interpretiere ich auf \mbox{Deutsch} pu\-bli\-zierte Aussagen unter dem Gesichtspunkt, im Zuge welcher \"Uberlegungen Wi\-ki\-pe\-dia explizit erw\"ahnt wird und welche der an der Plattform und ihrer Nutzung wahrgenommenen Aspekte in einem bestimmten Zusammenhang hervorgehoben werden. Ich gehe davon aus, dass die ausgew\"ahlten Beteiligten als Autoren\textsuperscript{\tiny *} von ver\"of\-fent\-lichten Meinungen \"uber Li\-te\-ra\-tur wirken und auf diese Weise den aktuellen Wi\-ki\-pe\-dia-Diskurs zumindest innerhalb der Li\-te\-ra\-tur\-szene mit pr\"agen. Wenn diese Li\-te\-ra\-tur-nahen Stimmen auch nicht als repr\"asentativ angesehen werden k\"onnen, so denke ich, dass die Auswahl in etwa die Bandbreite der Jahre 2009 bis 2014 wiedergibt. 

Die ersten drei der Beispiele finden sich in einem deutschspra\-chi\-gen Essayband von 2013 mit dem Titel \textit{Zukunft der Li\-te\-ra\-tur} (\textit{Text + Kritik Sonderband 2013}) und stammen von Autoren\textsuperscript{\tiny *}, die am Ende des Bandes beruflich beschrieben werden als Promovend\textsuperscript{\tiny *}, Dozent\textsuperscript{\tiny *} f\"ur \mbox{Deutsch} als Fremdsprache, Verlagskooperator\textsuperscript{\tiny *} und \"Ubersetzer\textsuperscript{\tiny *} (geboren 1980), als Lyriker\textsuperscript{\tiny *} und Leiter\textsuperscript{\tiny *} eines K\"unstlerhauses (geboren 1988) und als Buchwissenschaftler\textsuperscript{\tiny *} (geboren 1972). Die weiteren drei Beispiele sind verschiedenen Publikationen der Jahre 2008-2014 entnommen.\newline
\newline

\textbf{Beispiel 1} ("`Schullesungen oder wo die Magie endet"'):
\singlespacing
\begin{quote}
"`Was noch harmlos und ertr\"aglich bei einem Sch\"uler mit Ahnungslosigkeit in der Gewandheit mit den Dingen der Welt abgetan werden kann, ist schmerzlich fehlerbehaftet und formlos bei den Lehrerinnen und Lehrern, die zum Teil \"ubernehmen f\"ur den Kollegen, der sich den Besuch des Autors hat einfallen lassen und der \flq heute leider verhindert\frq \,\,ist. So steht der Autor schutzlos in Formlosigkeit aufgel\"ost neben seinem unfreiwillig Einf\"uhrenden, der sich schmerzhafter Generalismen bedient -- \flq Ich wei{\ss} jetzt leider auch nicht, was der Autor XY so schreibt, aber nett, dass er gekommen ist\frq \,\,oder \flq Seid ruhig, jetzt erz\"ahlt euch der Autor XY eine Geschichte. Das machen Sie doch, oder, Herr XY?\frq \,\,-- und dann noch bei \mbox{selbst} verschuldeter Unkenntnis auf die verhasste \textbf{Wi\-ki\-pe\-dia}-Seite im Internet verweist, die \flq so einiges\frq \,\,verraten w\"urde \"uber den mysteri\"osen Gast, wie man gerade noch im Lehrerzimmer aufgeschnappt h\"atte. Oft werden die d\"ummsten Details bem\"uht, um den Autor nur ja wie Sterntaler aller Hemden, aller Talente, aller W\"urde beraubt stehen zu lassen. Auf das er dann \flq am besten sich \mbox{selbst}\frq \,\,vorstelle, [...]"' (\cite{Gomringer2013}:162) \textit{[Meine Hervorhebung]}
\end{quote}
\onehalfspacing

In einem essayistischen Beitrag zu Autoren\textsuperscript{\tiny *}lesungen an Schulen kommt zur Sprache, dass es bei Wi\-ki\-pe\-dia biografische Eintr\"age zu lebenden Per\-so\-nen gibt. Der ent\-spre\-chen\-de Eintrag zur Person wird als "`verhasst"' beschrieben. Es handelt sich um Kolportage, insofern die indirekt get\"atigte Aussage lautet, man sei Informationen zur eigenen Person bei Wi\-ki\-pe\-dia hilflos ausgesetzt -- was nicht den Tatsachen entspricht. Der Beitrag \mbox{selbst} verbreitet damit Ger\"uchte \"ahnlich denen, die laut Aussagen dieses Essays in polemischem Ton in "`\mbox{selbst} verschuldeter Unkenntnis"' durch einf\"uhrende Lehrer\textsuperscript{\tiny *} verbreitet werden, wenn sie andeuten, was sie sich von anderen \"uber den Inhalt des betreffenden Eintrags haben sagen lassen. Genutzt w\"urde Wi\-ki\-pe\-dia demnach, um an Informationen zu gelangen, \"uber die man von einem Autor, der vor einem steht, nichts erfahren w\"urde.

\textbf{Beispiel 2} ("`Die Zukunft der Literaturzeitschrift. Vom Tabloid zum Tablet? Ein Pl\"adoyer"'):
\singlespacing
\begin{quote}
"`Wichtig sind vor allem die ansprechende und informative Pr\"asentation der Zeitschrift auf ihrer Homepage sowie eine auff\"allige, verl\"assliche Verlinkung mit der Stelle, an der sie bezogen werden kann. Weiterhin ist die Einrichtung unkomplizierter Kauf- bzw. Download-Mechanismen f\"ur jede einzelne Ausgabe -- ob aktuell oder alt -- von gro{\ss}er Wichtigkeit. Dies vermindert die Wahrscheinlichkeit, potenzielle Bezieher durch zeit- und nervenaufreibende Kaufprozesse zu verlieren. Dagegen halte ich es f\"ur unn\"otig, dar\"uber zu spekulieren, ob sich die \flq \textbf{Wi\-ki\-pe\-dia}-Mentalit\"at\frq \,\, der nachwachsenden Generationen so auf ihre Haltung auswirken wird, dass sie erwarten, f\"ur online erscheinende Zeitschriften kein Geld ausgeben zu m\"ussen. Zum einen w\"are auch die gegenteilige Entwicklung denkbar: dass die Hemmschwelle, qualitativ hochwertige virtuelle Produkte zu kaufen, sinken wird, weil auch Filme und Musik vermehrt digital bezogen werden. Zum anderen kann man aber vor allem davon ausgehen, dass weiterin haupts\"achlich Universit\"aten Abonnenten von nun online ver\"of\-fent\-lichten Zeitschriften sein werden und Studenten somit ohnehin \"uber den Universit\"atsserver kostenlos auf sie zugreifen k\"onnen."' (\cite{Arnold2013}:176-177) \textit{[Meine Hervorhebung]}
\end{quote}
\onehalfspacing

In diesem Beispiel wird "`der nachwachsenden Generationen"' pauschalisierend eine Mentalit\"at zugeschrieben, die sich der Darstellung nach an (der Nutzung von) Wi\-ki\-pe\-dia festmachen l\"asst ("`Wi\-ki\-pe\-dia-Mentalit\"at"'). Die kostenfreie Zug\"ang\-lichkeit von Wi\-ki\-pe\-dia-Inhalten pr\"agt demnach die Erwartungshaltung einer ganzen Ge\-ne\-ra\-tion. Die Aussage des Essays k\"onnte an diesem Punkt indirekt so verstanden werden, dass Wi\-ki\-pe\-dia ein Publikationsorgan ist -- was zutrifft --, und dass dessen Gesch\"aftsmodell "`f\"ur online erscheinende Zeitschriften"' nachteilig ist, wenn jene Einnahmen durch Einzelper\-so\-nen erwarten, was im Essay aber relativiert wird durch die Vermutung, dass es zunehmend eine gr\"o{\ss}ere Bereitschaft geben wird, f\"ur "`qualitativ hochwertige virtuelle Produkte"' Geld auszugeben. R\"uckwirkend wird damit die Qualit\"at von Wi\-ki\-pe\-dia-Inhalten pauschalisierend geringgesch\"atzt, denn jene sind kostenfrei zug\"ang\-lich. In Zusammenhang mit \"Uberlegungen zur k\"unftigen Gestaltung finanzieller Grundlagen f\"ur Li\-te\-ra\-tur\-zeitschriften, die sich mit Printausgaben nicht mehr rentieren, ist die Perspektive nachvollziehbar, al\-ler\-dings richtet sich die Ablehnung \mbox{hier} verk\"urzt gegen frei zug\"ang\-liche Inhalte, wo sie sich gegen die \"Uber\-macht eines Modells richten sollte, mit dem durch Verbindungen zu globalen Kon\-zer\-nen wie Google unter Ausnutzung unbezahlter Arbeitsleistungen j\"ahrlich steigende Spendeneinnahmen erzielt werden.

\textbf{Beispiel 3} ("`Die Zukunft des Sachbuchs"'):
\singlespacing
\begin{quote}
"`Regelrechte Medienverb\"unde vom Sachbuch ausgehend\,fin\-den\,sich,\,wenn \"uber\-haupt, im \mbox{Bereich} der Kinder-und Jugendli\-te\-ra\-tur. [...] Das Sachbuch f\"ur Erwachsene ist dagegen oft nicht Ausgangspunkt oder auch nur zentraler Aufmerksamkeitskatalysator von Debatten, Themenkonjunkturen und Skandalen, sondern eher ein Epiph\"anomen -- al\-ler\-dings wei\-ter\-hin ein zentrales. Denn nicht nur Fernsehdokumentationen und medial pr\"asente Themen finden fr\"uher oder sp\"ater ins Buch, \mbox{selbst} \textbf{Wi\-ki\-pe\-dia} \mbox{nennt} nicht nur die von Nutzern zusammengestellten und im PDF-Format abrufbaren Artikelauswahlen \flq Buch\frq \,\,-- die \flq meist\-ge\-such\-ten Inhalte der freien Enzyklop\"adie\frq \,\,wurden 2008 bei Bertelsmann als gedrucktes Buch ver\"of\-fent\-licht."' (\cite{Oels2013}:66) \textit{[Meine Hervorhebung]}
\end{quote}
\onehalfspacing

In Zusammenhang mit der Vermarktbarkeit von Sachb\"uchern findet Wi\-ki\-pe\-dia als Quelle von Sachtexten Erw\"ahnung und wird als rechtfertigender Ma{\ss}stab daf\"ur genommen, wie wichtig ein Sachbuch in Buchform sei. Dass Wi\-ki\-pe\-dia-Inhalte in Printausgaben zur Verf\"ugung gestellt werden k\"onnen und Interesse finden, wird pa\-ral\-lel gesehen mit Aufmerksamkeitseffekten, die der Buchbranche durch "`Fern\-seh\-do\-kumentationen und medial pr\"asente Themen"' Vorteile verschaffen -- nur dass der spezifische Nutzen freier Inhalte von Wi\-ki\-pe\-dia ("`\mbox{selbst} Wikpedia"') entweder durch Kostenersparnis Privatper\-so\-nen zukommt oder genau einem bereits ein\-fluss\-rei\-chen Konzern. Die an Wi\-ki\-pe\-dia wahrgenommenen Aspekte sind in diesem Essay Informationen, die f\"ur Sachb\"ucher ge\-eig\-net sind. \"Uber deren Qualit\"at wird nichts ausgesagt, sondern nur, dass sogar die Inhalte von Wi\-ki\-pe\-dia in Printformat vermarktbar sind.

Netzerfahrene Li\-te\-ra\-tur\-interessierte suchen f\"ur den deutschspra\-chi\-gen Kontext even\-tuell direkt bei Per\-len\-tau\-cher.de, um Rezensionsnotizen zu Kritiken in \"uberregiona\-len deutschspra\-chi\-gen Tageszeitungen zu finden, sowie in weiteren bevorzugten Rezensionsblogs und Leseforen. Hierzu ist mir keine For\-schung bekannt, daher kann ich nur Vermutungen \"au{\ss}ern. Anzunehmen ist meines Erachtens, dass Li\-te\-ra\-tur\-interessierte, die nichts von dem verpassen wollen, was andere eventuell schon wissen, mit dem Titel des Werkes eine Websuche starten und in der Ergebnisliste auch Wi\-ki\-pe\-dia finden, nicht nur, falls es dort schon einen eigenen Eintrag zum gesuchten Werk gibt, sondern auch dann, wenn der Titel im m\"oglicherweise vorhandenen Eintrag zum Autor\textsuperscript{\tiny *} oder im Artikel zu einer Verfilmung genannt wird. Eine bekannte Pers\"onlichkeit in der Szene der Li\-te\-ra\-tur\-ver\-mitt\-lungsfor\-schung formuliert es hingegen so:

\textbf{Beispiel 4} (F\"ur welche Art von Information bei Wi\-ki\-pe\-dia nachgeschaut wird):

\singlespacing
\begin{quote}
"`Wer erste Informationen zu einer Autorin oder einem Autor sucht, schaut bei \textbf{Wi\-ki\-pe\-dia} nach, liest die Informationen auf der Verlagswebsite oder \flq googelt\frq \,\,die Homepage der betreffenden Person. Wer sich \"uber die li\-te\-ra\-tur\-kritische Aufnahme eines Buches informieren m\"ochte, nutzt den \flq Per\-len\-tau\-cher\frq."' (\cite{Neuhaus2010}:20) \textit{[Meine Hervorhebung]}
\end{quote}
\onehalfspacing

In diesem Beispiel wird Wi\-ki\-pe\-dia in der Einleitung zu einem wissenschaftlichen Band mit dem Titel \textit{Digitale Li\-te\-ra\-tur\-ver\-mitt\-lung} allein als Informationsquelle zu biografischen Daten genannt. Dabei geht es zwar dieser Einsch\"atzung nach lediglich um "`erste Informationen"', aber Wi\-ki\-pe\-dia wird zusammen mit "`Verlagswebsite"' und "`Homepage der betreffenden Person"' genannt. Ob implizit eine h\"ohere Wertsch\"at\-zung f\"ur die Inhalte gemeint ist, indem Wi\-ki\-pe\-dia von diesen Dreien zuerst genannt wird, bleibt offen. Deutlich davon abgegrenzt werden Informationen "`\"uber die li\-te\-ra\-tur\-kritische Aufnahme eines Buches"', f\"ur die aus der \mbox{hier} dargestellten Sicht allein eine Quelle in Frage kommt -- in vertrautem Szenejargon formuliert -- "`der Per\-len\-tau\-cher"' (etwa so wie andere "`die Wi\-ki\-pe\-dia"' sagen). Dass sich bei Wi\-ki\-pe\-dia teils ausf\"uhrliche Eintr\"age zu einzelnen Werken und ihrer Rezeptionsgeschichte finden, die weit umfassender ist als die anonymen Notizen bei per\-len\-tau\-cher.de, wird \mbox{hier} nicht erw\"ahnt.

\textbf{Beispiel 5} (\"Uberpr\"ufung des b\"urgerlichen Namens eines Pops\"angers):
\singlespacing
\begin{quote}
"`So zumindest der\,Eintrag auf \textbf{Wikipedia}. Siehe http://de.wikipedia.org
/wiki/Rex\_Gildo (25. 08. 2012). Best\"atigt wird dieser Eintrag von der Internetpr\"asenz des eurovision songcontest http://www.eurovision.de/teiln
ehmer/rexgildo113.html (25. 08. 2012). Eine Fan-Homepage f\"uhrt wiederum Ludwig Friedrich Hirtreiter als b\"urgerlichen Namen an. Siehe http://www.rex-gildo-fan.de/Startseite.htm (25. 08. 2012)"'\\ (\cite{Assmann2014}:433; Fn.289) \textit{[Meine Hervorhebung]}
\end{quote}
\onehalfspacing

Die vorige Einsch\"atzung wird in einem Punkt durch dieses Beispiel aus einer Studie \"uber die Thematisierung des Li\-te\-ra\-tur\-betriebs in fiktionalen Texten der Gegenwart best\"atigt. Wi\-ki\-pe\-dia findet \mbox{hier} in einer Fu{\ss}note Erw\"ahnung, aus der hervorgeht, dass die Plattform zur \"Uberpr\"ufung von Per\-so\-nendaten genutzt worden ist. Dabei wird Wi\-ki\-pe\-dia Vorrang vor einer Website gegeben, die sich auf berufliche Inte\-ressen der entsprechenden Person bezieht. Abgeglichen wurden diese Informationen mit einer Website, die von Fans des Pops\"angers betrieben wird und deren Informationen den beiden zuvor genannten widerspricht. Hervorgehoben wird von Wi\-ki\-pe\-dia allein der Aspekt der angenommenen relativ gr\"o{\ss}eren Zuverl\"assigkeit von Daten im Vergleich zu zwei weiteren genannten Quellen.

\textbf{Beispiel 6} (Belegstelle f\"ur einen Terminus zur Beschreibung der Art eines Unternehmens):
\singlespacing
\begin{quote}
"`... was sich im Netz unter Li\-te\-ra\-tur\-kritik im weitesten Sinne finden l\"asst. Das erste Beispiel bezieht sich auf Katharina Hackers Roman \textit{Die Habenichtse} und dessen Rezeption vor allem auf der Internetplattform des \flq Social-Commerce-Versandhaus\frq \,\,\textbf{(Wi\-ki\-pe\-dia)} \textit{Amazon}."'\\ (\cite{Wegmann2012}:181-182)
\end{quote}
\onehalfspacing

Auch in diesem Beispiel werden Informationen, die bei Wi\-ki\-pe\-dia zu finden waren, als Beleg angegeben. Zwar geht es in dem Beitrag um Li\-te\-ra\-tur\-kritik, Wi\-ki\-pe\-dia wird aber wegen etwas Anderem erw\"ahnt: einem Terminus f\"ur die Beschreibung der Unternehmensart eines Konzerns, auf dessen Website Rezensionen zu lesen sind, auf die in dem Beitrag unter anderem eingegangen wird. Wi\-ki\-pe\-dia wird zwar als Quelle angesehen (und in Klammern gesetzt genannt), aber ohne dass eine genauere Zitierung erfolgen w\"urde, obwohl aus Wi\-ki\-pe\-dia punktgenau zitiert werden kann. Umso \"uberraschender scheint mir dies, weil es sich um einen Beitrag zu einem Band handelt, der in einem Wissenschaftsverlag pu\-bli\-ziert worden ist. \"Uber die Gr\"unde der Unterlassung k\"onnte \mbox{hier} ohne Eigenaussagen nur spekuliert werden. Von an\-de\-rer Seite wird es im Wissenschaftsbereich sehr wohl als ein Charakteristikum angesehen, dass die Informationen bei Wi\-ki\-pe\-dia \mbox{selbst} immerhin versehen werden mit "`sources, further reading, footnotes showing where a specific piece of information came from (and allowing the researcher to backtrack the path of this \flq information\frq)"' (\cite{Hotz-Daviesetal2009}:x).

In diesen Beispielen deutschspra\-chi\-ger \"Au{\ss}erungen kommt zum Ausdruck, dass es weder eine Vertrautheit mit der Idee und Funktionsweise von Wi\-ki\-pe\-dia gibt noch Wertsch\"atzung f\"ur Li\-te\-ra\-tur\-eintr\"age bei Wi\-ki\-pe\-dia. Eine kommentierte \"Ubersicht zu Handb\"uchern und anderen li\-te\-ra\-tur\-wissenschaftlichen Publikationen, in denen ich Aussagen zu Wi\-ki\-pe\-dia erwartet h\"atte (und seien es negative), w\"urde den Rahmen dieser Arbeit sprengen. Best\"atigt wird durch diese Befunde meine Annahme, dass der vorliegenden Arbeit Pilot\textsuperscript{\tiny *}charakter zukommt. Welches die Gr\"unde sind f\"ur die in den Beispielen mehrheitlich zum Ausdruck kommende distanzierte Haltung gegen\"uber Wi\-ki\-pe\-dia sei\-tens Pers\"onlichkei\-ten, die in der Institution Li\-te\-ra\-tur erwerbsorientierte Ak\-ti\-vi\-t\"aten betreiben, w\"are eine eigene Studie wert.

\newpage

\subsection{\textit{Erlesnisse} in Aktion: Interpretationen}
\label{subsec:5.4}

Die Qualit\"at eines Textes k\"onnen normale Li\-te\-ra\-tur\-liebhaber\textsuperscript{\tiny *} und philologische Spezialisten\textsuperscript{\tiny *} glei\-cherma\-{\ss}en wenig in Worte fassen, meint Manfred Koch in sei\-ner Rezension von Hans Ulrich Gumbrechts Essayband \textit{Stimmungen lesen. \"Uber eine verdeckte Wirk\-lich\-keit der Li\-te\-ra\-tur} (2011). Er schreibt einlei\-tend, dass Texte eine "`auf uns eindringende Energie"' verstr\"omen und fragt: "`Was hei{\ss}t es nun, die Stimmung eines Textes und ihre Wirkung auf den Leser zu erkunden?"' (\cite{Koch2011}:66). Darauf antwortend gelangt er zu der Be\-zeichnung "`stimmungsvergegenw\"artigende Rede \"uber Li\-te\-ra\-tur"'. Konkret w\"urde es darum gehen, "`ein neues Gesp\"ur f\"ur Rhythmus und Klang, f\"ur den nerv\"osen oder melancholischen Grundtton eines Textes, ja f\"ur seine Farbe, seinen Geruch, seine Atmosph\"are zu entwickeln. Und dies darzustellen, ohne so\-gleich wieder zu fragen, was denn die tiefere Bedeutung sei."' Darzustellen w\"are dies auf einem Mittelweg, meint Koch, der zwischen staunenden Ausrufen und "`schau\-mi\-ge[m] Ergriffenheitsgefasel"' liegt -- was ein schwieriger Weg sei, weil er "`einige Ausdrucksnot"' bringe. Koch sieht es als "`die Leistung gro{\ss}er Li\-te\-ra\-tur"' an, dass sie nicht nur Handlung beschreibt, sondern etwas zu einem "`physisch sp\"urbaren Erlebnis werden"' l\"asst. Auch aus diesem Grund erscheine die Reduktion der "`sinnliche[n] Dimension eines Textes auf die Funktion einer blossen [sic] Illustration des geistigen Gehalts"' als eine "`alte Versuchung der Interpreten"', die damit abzuwehren sei, jenes neu ent\-wickelte Gesp\"ur darzustellen. Dass Gumbrecht dies in seinen "`exemplarischen Stimmungslekt\"uren"' noch nicht gelinge, sei darauf zur\"uckzuf\"uhren, dass die Essays zu kurz ausfallen. Mein Eindruck ist vielmehr, dass nicht Mangel an L\"ange der Grund ist, sondern fehlende \textit{parrhesia} -- au{\ss}er vielleicht im Essay \"uber die Freiheit in der Stimme von Janis Joplin in ihrem Song "`Me and Bobby McGee"'. Hierbei, so Stefan Hajduk in seiner Rezension desselben Bandes, handele es sich um eine generationsnostalgische Vergegenw\"artigung (\cite{Hajduk2012}), die Detlev Sch\"ottker in sei\-ner Rezension als "`eine von vielen bereicherndern Einsichten"' be\-zeichnet, die dieses Buch seinen Lesern zu vermitteln verm\"oge, und zwar der Aspekt, dass "`\mbox{selbst} der Meisterdenker unter den zeitgen\"ossischen Philologen die sentimentalische Stimmung nicht aus seinem Ge\-f\"uhlshaushalt verbannt hat"' (\cite{Schottker2011}).

"`Stimmungsvergegenw\"artigende Rede \"uber Li\-te\-ra\-tur"' scheint eine passende Umschreibung f\"ur jenen Bedarf zu sein, den \mbox{User}:Epipactis formuliert, wenn es im Rahmen einer Wi\-ki\-pe\-dia-Debatte von Januar 2015 hei{\ss}t: 

\singlespacing
\begin{quote}
\href{https://de.wikipedia.org/w/index.php?title=Benutzer_Diskussion:Grillenwaage\&diff=next\&oldid=137663350}{"`Was habe ich zu \textit{erwarten}, wenn ich mir das Werk ansehe? Sollte ich es mir \textit{g\"onnen}, wird es sich f\"ur mich \textit{lohnen}? Mu{\ss} ich es mir \textit{zumuten}, um \textit{mitreden} zu k\"onnen? Oder: Ich habe es gesehen, hat mich nicht sonderlich beeindruckt, vielleicht ist mir etwas aus Expertensicht \textit{Bemerkenswertes} entgangen, aber wenn ja -- was?"'} (\cite{UserEpipactis2015a})
\end{quote}
\onehalfspacing 

Ob auch ich Ausdrucksnot habe f\"ur die Beschreibung der Qualit\"at eines Textes, kann nun anhand der folgenden \mbox{vier} Interpretationen he\-rausgefunden werden. Sie beinhalten einige \textit{Erlesnisse}, indem ich sie als solche benenne und beschreibe. Das schriftliche Sich-\"Au{\ss}ern \"uber eigene Lekt\"ure ist ein Anlass, Grenzziehungen zu thematisieren sowie Grenzen zu \"uberschreiten, die durch Konventionen des Sprechens und Schrei\-bens als geltend behandelt werden -- je nach Kontext auch von mir weitgehend. Bei einer schriftlichen \"Au{\ss}erung \"uber eigene Lekt\"ure bin ich als Leserin\textsuperscript{\tiny *} gefordert, aus der Intimit\"at des individuellen Leseprozesses, der in einem machtreduzierten privaten Raum stattgefunden hat, he\-rauszutreten in Richtung des Machtgef\"alles der \"Of\-fent\-lichkeit (\cite{Lorenz2012}), wo das \"Au{\ss}ern meines Empfindens mit Risiko verbunden ist -- andernfalls, meiner Einsch\"atzung zufolge -- w\"urden genau dies Literaturbewanderte in gr\"o{\ss}erem Ma{\ss}e tun, macht doch allem Anschein nach das pers\"onliche Empfinden beim Lesen einen Gro{\ss}teil der Faszination von Literaturlekt\"ure aus.

Ausgew\"ahlt wurde je ein Text von Elfriede Jelinek (\textit{rein GOLD}, 2013, Wikipedia-Eintrag \href{https://de.wikipedia.org/wiki/Rein_Gold:_Ein_B\%C3\%BChnenessay}{"`Rein Gold: Ein B\"uhnenessay"'}), Doris Lessing (\textit{Alfred and Emily}, 2008, Wikipedia-Eintrag \href{https://de.wikipedia.org/wiki/Alfred_und_Emily}{"`Alfred und Emily"'}), Herta M\"uller (\textit{Reisende auf einem Bein}, 1989, Wikipedia-Eintrag \href{https://de.wikipedia.org/wiki/Reisende_auf_einem_Bein}{"`Reisende auf einem Bein"'}) und Alice Munro (\textit{Save the Reaper}, 1998, Wikipedia-Eintrag \href{https://de.wikipedia.org/wiki/Save_the_Reaper}{"`Save the Reaper"'}). Zu allen \mbox{vier} Werken sind seit August 2013 Wi\-ki\-pe\-dia-Eintr\"age entstanden, die \href{https://tools.wmflabs.org/wikihistory/wh.php?page_title=Rein_Gold:_Ein_B\%C3\%BChnenessay}{gr\"o{\ss}tenteils} \href{https://tools.wmflabs.org/wikihistory/wh.php?page_title=Alfred_und_Emily}{von} \href{https://tools.wmflabs.org/wikihistory/wh.php?page_title=Reisende_auf_einem_Bein}{mir} \href{https://tools.wmflabs.org/wikihistory/wh.php?page_title=Save_the_Reaper}{verfasst} wurden.

Zu \textit{Save the Reaper} verkn\"upfe ich in meiner Interpretation ein doppeltes \textit{Erlesnis} mit einer aktuellen Zusammenschau zu g\"angigen Versionierungsverfahren, um einen be\-stimmten Aspekt von Alice Munros Erz\"ahlkunst anhand von Textstellen interpretierend darzulegen.

In meinem Beitrag zu \textit{Alfred and Emily} interpretiere ich die inhaltliche Signifikanz des intertexuellen Bezugs zur \textit{London Encyclop{\ae}dia} 1983, und speziell zu dessen Eintrag "`Royal Free Hospital"', den Lessing 25 Jahre sp\"ater wortw\"ortlich an zentraler Stelle von \textit{Alfred and Emily} \"ubernimmt, womit mein \textit{Erlesnis} zusammenh\"angt.

Meine Interpretation von \textit{Reisende auf einem Bein} fu{\ss}t auf einem \textit{Erlesnis}, das mich um eine besondere Sorte von "`Wissen wie"' bereichert hat. Ich gehe einer bestimmten Spur nach, \"uber deren Bedeutsamkeit f\"ur das Thema des Textes ich in der mir bekannten For\-schungsli\-te\-ra\-tur zu \textit{Reisende auf einem Bein} bis\-her nirgends eine Erw\"ahnung gefunden habe.

\"Uber meine Lekt\"ure von \textit{rein GOLD} berichte ich anhand einer kulturgeschichtlichen Reise, in der Jelineks Variante von Gezanke mit Aushandlungen und Streit\-ge\-spr\"a\-chen an anderen Orten und zu anderen Zei\-ten verglichen wird. Mein \textit{Erlesnis} formuliere ich zu Beginn.

\newpage

\subsubsection{Fiktionalisierte Varianten von Versionierungsverfahren in Alice Munros \flq Save the Reaper\frq }
\label{subsubsec:5.4.1}

An Verfahren der Versionierung lassen sich verschiedenartige bestimmen und in Kurzgeschichten von Alice Munro wiederum fiktionalisierte Varianten bestimmter Verfahren. Im Folgenden werden zun\"achst \mbox{vier} Versionierungsverfahren besprochen, um anschlie{\ss}end Wi\-ki\-pe\-dia und Munro in einer Weise einander gegen\"uberzustellen, die f\"ur die Fragestellung der vorliegenden Arbeit relevant ist. 

Die Editionsphilologie hat mit Ergebnissen historischer Prozesse zu tun, in denen meh\-rere Per\-so\-nen an unterschiedlichen Orten und zu unterschiedlichen Zei\-ten glei\-che oder \"ahnliche Stoffe textlich bearbeitet haben. Auch Wi\-ki\-pe\-dia-Eintr\"age sind letzt\-lich Ergebnisse historischer Prozesse, in denen meh\-rere Per\-so\-nen von unterschiedlichen Orten aus zu unterschiedlichen Zei\-ten textlich arbei\-ten, al\-ler\-dings auf derselben Seite im Web und mit einer einzigen Versionsgeschichte, die protokollartig auf Basis eines time stamps gef\"uhrt wird.
In editionsphilologischen Darstellungsformen werden traditonell entweder verschiedene Fassungen zu einer einzigen Version verschmolzen oder Varianten nebeneinander sichtbar gemacht, so dass \"Anderungen als solche hervortreten -- oder es wird mit einer Kombination aus diesen beiden Verfahren editiert.
Dar\"uber, was eine Fassung ausmacht, be\-steht in der Edi\-tions\-philologie kein Konsens. Neben formalistischen Definitionen sind editionspraktisch orientierte oder auf den Autor\textsuperscript{\tiny *} bezogene zu finden (\cite{Kraftetal2001}:9-10,223-224) F\"ur den eng\-lischspra\-chi\-gen Raum fasste Allan Dooley 1992 die Lage folgenderma{\ss}en zusammen: 

\singlespacing
\begin{quote}
"`Scholars disagree to some extent about the minimum amount of difference required to define a variant, and to a great extent about the amount of difference necessary to make a variant significant or meaningful."' (\cite{Dooley1992}:160; zitiert in \cite{Ravelhofer2005}:158)
\end{quote}
\onehalfspacing

Bei einer editionspraktischen Definition wird das St\"uck Text erst durch den He\-raus\-geber\textsuperscript{\tiny *} (Editor\textsuperscript{\tiny *}) zu einer Fassung, die nicht selten erst im Zuge der He\-rausgabe einer Edition eines Werkes hergestellt wird, zum Beispiel in der Rekonstruktionsphilologie Karl Lachmanns Mitte des 19. Jahrhunderts (\cite{StarkeyWandhoff2008}:45) Dies trifft in an\-de\-rer Weise auch zu auf die traditionelle anglo-amerikanische Art des Edierens ("`copy-text editing"'): "`The text is a combination of readings from more than one document. The reading text has, by definition, never achieved material form before"', so Paul Eggert: "`The principal aim is to establish the wording and accidentals of the reading text"' (\cite{Eggert2005}:202).

Eine Edition ver\"andert ein Werk, im Laufe der Zeit bilden sich \"asthetisch-qualitative Unterschiede zwischen Fassungen he\-raus. Eine Edition stellt in dieser Hinsicht die grundlegende Kristallisationsform desjenigen geschichtlichen Prozesses dar, in dem sich die Bedeutung eines Kunstwerks mit der Relation zum historisch definierten Subjekt wandelt (\cite{Kraftetal2001}). Die formalistische Definition hebt darauf ab, dass sich eine Fassung durch mindestens eine Variante konstituiert und dass diese schon durch die \"Anderung nur eines Elements zustande kommt, weil dadurch neue Beziehungen entstehen, also ein neues System (\cite{Zeller1975}; zitiert in {Kraftetal2001}:224, Fn. 11). Bei "`Textfassungen"' handelt es sich in diesem Sinne um verschiedene Ausf\"uh\-rungen eines Werkes, die sich durch Textidentit\"at aufeinander beziehen und aufgrund von Textvarianz voneinander unterscheidbar sind (\cite{Scheibe1982}; zitiert in {Kraftetal2001}:224, Fn. 11). Weil Eigen\"ubersetzungen als Versionen angesehen werden k\"onnen, argumentierte Klaus Gerlach 1991, dass Textfassungen nicht allein aufgrund von Text\-identit\"at aufeinander beziehbar sind: Scheibes Definition von "`Fassung"' sei durch den Aspekt der Text\"aquivalenz zu erg\"anzen (\cite{VanHulle2005}). Auf den Autor\textsuperscript{\tiny *} bezogen sind De\-fi\-ni\-ti\-o\-nen, die betonen, dass "`Textfassungen"' im Kontext einer bestimmten Phase f\"ur den Autor\textsuperscript{\tiny *} das Werk dar\-stel\-len (\cite{Scheibe1982}; zitiert in {Kraftetal2001}:224, Fn. 11). Herbert Kraft, Diana Schilling und Gert Vonhoff meinen jedoch, dass sich die Orientierung am Autor\textsuperscript{\tiny *} nicht zur Definition einer textlichen Struktur eignet. Der Terminus "`Fassung"' akzentuiert den rezeptions\-orientierten Blick auf ein Werk, so R\"udiger Nutt-Kofoth. "`Fassung"' h\"ange insofern mit "`Variante"' zusammen, als gefragt werden m\"usse, "`ab welchem Variantenumfang oder ab wel\-cher Variantenintensit\"at statt von einer Fassung eines Werkes nun von einem neuen eigenst\"andigen Werk gesprochen werden muss."' Dar\"uber hinaus schl\"agt Nutt-Kofoth vor, in einer erg\"anzenden produktionsorientierten Perspektive statt von "`Varianten"' von "`\"Anderungen"' zu sprechen (\cite{Nutt-Kofoth2005}:137-138; mit Bezug zu \cite{ZellerSchildt1991} und \cite{Kanzog1991}). 1989 hatte es neue Ideen gegeben und eine "`New Philology"' kam auf.

\singlespacing
\begin{quote}
"`The immediate inspiration for this new philology came from Bernard Cerquiglini's polemical essay \textit{\'{E}loge de la variante} from 1989, which mark\-ed a clear turning point in the history of medieval textual studies by arguing that instability (variance) is a fundamental feature of chirographically transmitted li\-te\-ra\-ture: variation is what the medieval text is about."' (\cite{Driscoll2010}:Fn. 8)
\end{quote}
\onehalfspacing

"`Die "`New Philology"' ihrerseits hat zu einer differenzierteren Betrachtung varianter Fassungen von mittelalterlichen Texten gef\"uhrt. Sie hat aber auch das Interesse geweckt, die Variante als Objekt der Interpretation st\"arker zu ber\"ucksichtigen"', h\"alt Christa Jansohn fest (\cite{Jansohn2005}:2). Zum Ph\"anomen Variante gibt es verschiedene theo\-retische und methodische Konzepte. 
F\"ur die eng\-lischspra\-chi\-gen Begrifflichkei\-ten hat Christian Moraru eine Unterscheidung zwischen "`variant"' und "`version"' vorgeschlagen. Ein variant sei das Ergebnis der Frage nach dem WIE, eine version hingegen die Antwort auf die Frage WAS? Dazu z\"ahle auch ein abgeleiteter Status oder eine Abh\"angigkeit von einer vorherigen Narration "`which it both repeats and modifies"' (\cite{Moraru2005}). Diese Idee ist f\"ur meine Lekt\"ure von Munros Short Stories anregend geworden, ich habe mir dennoch angew\"ohnt, die Termini ohne n\"ahere begrifflich Unterscheidung zu nutzen, also verwende ich "`Variante"', "`Version"' und "'Fassung"' an verschiedenen Stellen in gleicher Bedeutung. 

Zunehmend betrifft das Nachvollziehen der Genese von Werken ab dem Beginn der Neuzeit -- und insbesondere mit der Etablierung des Autor\textsuperscript{\tiny *}begriffs -- auch die Werke Einzelner. Um einen solchen Fall, der zus\"atzlich fiktionalisierte Varianten von Versionierungsverfahren aufweist, geht es bei \textit{Save the Reaper} von 1998: Von dieser Kurzgeschichte hat Alice Munro im selben Jahr zwei verschiedene Fassungen ver\"of\-fent\-licht, eine davon am 22. Juni 1998 im \textit{New Yorker} und eine in ihrem neun\-ten Band mit Kurzgeschichten, \textit{The Love of a Good Woman}, der am 1. Oktober 1998 erschien. Bei diesem Beispiel lassen sich die Versionen nicht wie \"ublich nach Pu\-bli\-ka\-tions\-kontext unterscheiden (Zeitschrift und Buch), und auch bei Munro ist es unty\-pisch, so Ildik\'{o} de Papp Carrington 2002, dass nicht die Buchversion, sondern die Zeitschriftenversion sp\"ater f\"ur eine Anthologie ausgew\"ahlt wurde, f\"ur den Band \textit{The Best American Short Stories 1999}, he\-rausgegeben von Amy Tan: 

\singlespacing
\begin{quote}
\href{http://canlit.ca/site/getPDF/article/10625}{"`there are two significantly different versions of \flq Save the Reaper\frq. Munro often revises her stories after their initial New Yorker publication, sometimes retaining and sometimes rejecting the magazine's editorial revisions (Barber). The situation, here, however, seems somewhat atypical. First published in the New Yorker in June 1998, \flq Save the Reaper\frq\,\, was republished in the same form in The Best American Short Stories 1999. When it was published in a longer and differently structured form in \textit{The Love of a Good Woman}, Munro added an author's note: "`Stories included in the collection that were previously published in The New Yorker appeared there in a very different form"'} (\cite{dePappCarrington2002}:35-36)
\end{quote}
\onehalfspacing

Im Mittelpunkt des vorliegenden Beitrags stehen \"Uberlegungen bez\"uglich des Ver\-h\"alt\-nis\-ses zwischen Gegenstand und Thematik des Textes einerseits und der Existenz zweier pu\-bli\-zierter Fassungen an\-de\-rerseits, unter dem Gesichtspunkt der Gestaltungsweise von Versionierungsverfahren. Dabei sehe ich die beiden Versionen als gleichranging an und befasse mich in meiner Interpretation intensiver mit einer Stelle, die in beiden Versionen wortgleich ist. Bez\"uglich des zeitlichen Verh\"altnisses der beiden Versionen zueinander widersprechen sich die \"Uberlegungen, Informationen und Annahmen von de Papp Carrington 2002 (\cite{dePappCarrington2002}) und Duncan 2011 (\cite{Duncan2011}). Meine Abw\"agungen dazu finden sich auf der Dis\-kus\-si\-onsseite des deutschspra\-chi\-gen Wi\-ki\-pe\-dia-Eintrags im Abschnitt \href{https://de.wikipedia.org/w/index.php?title=Diskussion:Save_the_Reaper\&oldid=137099675#Buchversion_die_erste.2C_Zeitschriftenversion_ist_eine_gek.C3.BCrzte_Fassung_.28.3F.29}{"`Buchversion die erste, Zeitschriftenversion ist eine gek\"urzte Fassung (?)"'} (25. Dezember 2014).

Drittens gibt es auch heute ein Versionierungsverfahren, bei dem die neuere Version zwar als die bessere gilt, eine \"altere aber weiterhin im Umlauf bleibt, etwa in der Softwareproduktion und -distribution. Viertens ist ein Versionierungsverfahren bekannt, bei dem die jeweils neueste Version eine vorige \"uberschreibt, weil sie als ma{\ss}geblicher angesehen wird: Beim Revidieren von Vertr\"agen oder Testamenten ist in der Regel die j\"ungere Version die allein g\"ultige, bei Schreibsoftware wird mit dem Speichern in derselben Datei eine vorige Version ersetzt, die eventuell nicht wiederherstellbar ist. Eine Variante des dritten und vierten Verfahrens liegt zum Beispiel derjenigen Software zugrunde, die bei Wi\-ki\-pe\-dia verwendet wird, mediawiki: \mbox{Hier} ist es standardm\"a{\ss}ig die neueste Version, die angezeigt wird. Einerseits geht dem \"Of\-fent\-lich-Sichtbarwerden einer \"Anderung in manchen F\"allen aber ein Sichtungsprozess voraus, so dass es bereits eine neuere als die \"of\-fent\-lich angezeigte Version gibt, diese aber noch nicht f\"ur alle sichtbar freigeschaltet worden ist. An\-de\-rerseits bleiben \"altere Versionen bei Wi\-ki\-pe\-dia insofern von gewisser G\"ultigkeit, als sie mittels der protokollarischen Listung in der Versiongeschichte einer Seite prinzipiell einsehbar sind. In den \mbox{hier} genannten \mbox{vier} Versionierungsverfahren spielt ein Denken in vorher/nachher eine gewisse Rolle, \mbox{selbst} wenn die jeweilige Position auf einer Zeitachse relativ zu anderen Versionen nicht immer eindeutig festzulegen ist -- wie im Fall von \textit{Save the Reaper}. Bei Wi\-ki\-pe\-dia hat in der Versionsgeschichte einer Seite zwar der time stamp Priorit\"at vor anderen Faktoren, es wird also chro\-no\-lo\-gisch sortiert und nicht etwa alternativ alphabetisch nach Autor\textsuperscript{\tiny *}namen oder nach Beitragsgr\"o{\ss}e, aber die Versionsgeschichte einer Artikelseite wird mit derjenigen der dazugeh\"orenden Dis\-kus\-si\-onsseite nicht zusammengef\"uhrt. Ist beispielsweise auf einer Dis\-kus\-si\-onsseite ein Aspekt noch Teil einer laufenden Aushandlung, aber am Artikeltext wurde in\-zwi\-schen etwas ge\"andert, so scheint die Information im Artikeltext zu diesem Lemma \"alteren Datums zu sein als das, was auf der Dis\-kus\-si\-onsseite zu diesem Lemma debattiert wurde und und fr\"uher begann, aber erst sp\"ater in den Artikel \"ubernommen worden ist. Was als "`vorher"' in der Versionsgeschichte der Artikelseite verzeichnet ist, kann also \"alteren Datums sein als das, was laut time stamp "`sp\"ater"' war. Demzufolge kommt es bez\"uglich der Aussagekraft von time stamps bei Wi\-ki\-pe\-dia auf den Systemausschnitt an und die Datenlage ist keineswegs so leicht zu durchschauen wie es aufgrund des Layouts auf den ersten Blick zu sein scheint. 

Munro hat von \textit{Save the Reaper} im selben Jahr unter demselben Titel zwei Fassungen ver\"of\-fent\-licht -- wie von mindestens 27 anderen ihrer \href{https://en.wikipedia.org/w/index.php?title=List_of_short_stories_by_Alice_Munro&oldid=643370478}{mehr als 150 Short Stories}. 
M\"oglicherweise ist keine dieser beiden die erste oder die zweite Version -- falls etwa gleichzeitig an ihnen gearbeitet worden ist -- dies w\"are meiner Einsch\"atzung nach Munro durchaus zuzutrauen. Ob zwei Erz\"ahlungen denselben Inhalt haben, ist vom Frageinteresse abh\"angig (\cite{Lamarque2014}:10). Insofern hat mich aus einem "`Blickwinkel Wiki"' in diesem Fall das Versionsgeschichte-Lesen am meisten interessiert.

F\"ur einen Versionsvergleich ist interessant, was bez\"uglich des Verh\"altnisses von Philip und seiner Gro{\ss}mutter gesagt wird, nachdem Eve "`ihre Geschichte"' der Tochter und derem Ehemann in einer Fassung erz\"ahlt hat, die sie bewusst zensieren wollte:
 
\singlespacing
\begin{quote}
"`She would not mention the fragment of wall she had seen beyond the bushes. Why bother, when there were so many things she thought best not to mention? First, the game she had got Philip playing, overexciting him. And nearly everything about Harold and his companions. Everything, every single thing about the girl who had jumped into the car. [...]\\
She could say that the house smelled vile, and that the owner and his friends looked altogether boozy and disreputable, but not that Harold was naked and never that she herself was afraid. And never what she was afraid of."' (\cite{Munro1998B})\textit{[zitiert aus der Buchversion]}
\end{quote}
\onehalfspacing

Ob Eve diese Version erz\"ahlt hat, ist nicht zu erfahren. Gesagt wird zu Beginn des folgenden Ausschnitts nur, dass "`Eve's story"' zu einem bestimmten Zeitpunkt als "`was told"' gelten soll. In der Zeitschriftenversion befindet sich die folgende Passage im letzten Abschnitt, in der Buchversion geh\"ort sie zum vorletzten Abschnitt, weil direkt nach dieser Stelle ein neuer Abschnitt beginnt, dessen Anfang in beiden Fassungen gleichlautend ist, dessen Schluss aber nicht, und damit auch das Ende der Story nicht. Die Buchversion ist um einen Halbsatz l\"anger, in dem vermutet wird, dass in der Umgebung des hohl gewordenen Hauses der Mais eventuell aufgeh\"ort hat zu wachsen, auch wenn derzeit noch das ty\-pische Ger\"ausch zu h\"oren ist. 

Wie wird das Verh\"altnis von Philip und seiner Gro{\ss}mutter in den beiden Fassungen erz\"ahlt? Gibt es einen stillschweigenden Pakt bez\"uglich der von Eve pr\"asentierten Version der Geschichte -- welche auch immer es gewesen sein mag? In der Zeit\-schrif\-ten\-ver\-sion lautet die Passage so:

\singlespacing
\begin{quote}
"`Philip had added nothing to Eve's story and had not seemed to be concerned with the telling of it. [...]\\ But once it was told, [...], Philip did look up from the stooping and crawling he was doing around the adults' feet. He looked at Eve -- a flat look, a moment of conspirational blankness, a buried smile, that passed before there could be any demand for recognition of it. It seemed to mean that, however much or little he knew, he knew about the importance of keeping things to yourself. Eve got a jolt from that. She wished to deny that she was in any way responsible for it, but she couldn't disclaim it.\\
If the girl came looking for her [...]"' (\cite{Munro1998Z})
\end{quote}
\onehalfspacing

In der Buchversion bleibt offen, ob Eve Philips Blick sieht. Es wird lediglich von seinem Blick erz\"ahlt. Das Darauffolgende wird mit einer Frage eingeleitet und anders kommentiert als in der Zeitschriftenversion:

\singlespacing
\begin{quote}
"`Philip had added nothing to Eve's story and had not seemed to be concerned with the telling of it. [...]\\ But once it was told, [...], Philip did look up from his stooping and crawling work around the adults' feet. He looked at Eve. A flat look, a moment of conspirational blankness, a buried smile, that passed before there could be any need for recognition of it.\\
What did this mean? Only that he had begun the private work of storing and secreting, deciding on his own what should be preserved and how, and what these things were going to mean to him, in his unknown future.

If the girl came looking for her [...]"' (\cite{Munro1998B})
\end{quote}
\onehalfspacing

Judith Maclean Miller ist der Auffassung, dass in der Buchversion beschrieben wird, wie Eve das Vor\-enthalten, Nicht-Erz\"ahlen und F\"ur-sich-Behalten an Philip, und damit an die Enkelgeneration, weitergibt. Eve merke, dass Philip nichts sagt. Miller schlie{\ss}t daraus, dass "`What did this mean?"' Eves Kommentar ist (\cite{Miller2002}). Meine Einsch\"atzung ist, dass die Stelle in mehrfacher Hinsicht offen bleibt. Die Buchfassung bietet mehr Interpretationsspielraum, weil mit "`his unknown future"' eine weitere, gr\"o{\ss}ere Perspektive aufgemacht wird und zugleich von einem Anfang ("`he had begun"') die Rede ist, was nach meinem Empfinden auf die zuvor erz\"ahlte Situation zur\"uckstrahlt: \mbox{Hier} ist nichts klar.

\begin{figure}
\includegraphics[width=1.0\textwidth]{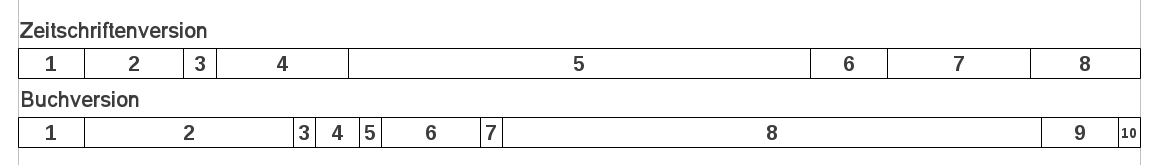}
\caption{\small{"`Save the Reaper"', Zeitschriftenversion 1998 (1255 Zeilen), Buchversion 1998 (34,5 Seiten), Vergleich von L\"ange und Position einzelner Abschnitte relativ zur Gesamtl\"ange}}
\label{fig:1}
\end{figure}

Eine Darstellung nach relativer Abschnittsl\"ange zeigt andere Aspekte des Ver\-sions\-un\-ter\-schieds auf (siehe Abbildung 1). In der Buchversion gibt es \mbox{vier} sehr kurze Abschnitte und die Varianz der Abschnittsl\"angen ist gr\"o{\ss}er als in der Zeitschriftenversion: Am kurzen Ende der Skala steht eine gr\"o{\ss}ere Anzahl an Abschnitten, die relativ viel k\"urzer sind als der einzelne lange Abschnitt am anderen Ende der Skala. Der l\"angste Abschnitt (mit der Szene in der Behausung) ist in der Buchversion prozentual noch l\"anger als in der Zeitschriftenversion. In der Zeitschriftenversion ist der 5. Abschnitt der l\"angste, in der Buchversion ist es der 8. Abschnitt. Bevor der lange Abschnitt beginnt, findet in der Buchversion h\"aufiger ein Szenenwechsel statt als in der Zeitschriftenversion. Nach dem l\"angsten Abschnitt gibt es in der Zeitschriftenversion 3 Abschnitte von eher mittlerer L\"ange, in der Buchversion nur noch zwei relativ kurze.

Die Buchversion geht erstens in R\"uckblenden we\-sent\-lich mehr ins Detail und zwei\-tens wirkt sie unruhiger, weil es \"ofter einen Szenenwechsel gibt und manche Abschnitte nicht mehr als eine halbe Seite lang sind. Als dann der lange Abschnitt beginnt, ist die Spannung in der Handlungsf\"uh\-rung und in den Dialogen in einem solchen Ma{\ss}e gestiegen, dass fehlende Abschnittswechsel nicht einmal mehr auf\-fal\-len, das hei{\ss}t: sie fehlen nicht.

Im l\"angsten Abschnitt l\"asst Munro die Protagonistin\textsuperscript{\tiny *} Eve mit ihren beiden Enkelkin\-dern in eine bedrohlich wirkende Situation geraten. In einem verwahrlosten Haus hat sie die Dreij\"ahrige auf dem Arm, den Siebenj\"ahrigen "`mute, pressed against her side"', w\"ahrend sie mit dem R\"ucken zur T\"ur in einem stickigen, stinkenden und vollgestopften Raum steht, wo einige alkoholisierte M\"anner an einem Tisch sitzend Karten spielen: 

\singlespacing
\begin{quote}
"`Her arm and shoulder ached from the weight of Daisy and from the tension which had got hold of her entire body. Yet she was thinking how she would describe this -- she'd say it was like finding yourself in the middle of a Pinter play. Or like all her nightmares of a stolid, silent, hostile audience."' (\cite{Munro1998B} und (\cite{Munro1998Z})
\end{quote}
\onehalfspacing

Wie werden die Mittel beschrieben, mit denen es der Protagonistin\textsuperscript{\tiny *} m\"oglich ist, wahrzunehmen, was sie gerade erlebt? Trotz k\"orperlichen Unwohlseins kann Eve ihre Reflexionsf\"ahigkeit aktivieren, indem sie auf Lekt\"ure- und B\"uhnenerfah\-rung zur\"uckgreift. Zum einen spiegelt sie sich die konkrete Lage in einer Art von bewusstem D\'{e}j\`{a}-vu, indem sie ihr aktuelles Empfinden vergleicht mit Erinnerungen an Ge\-f\"uhle in bereits erlebten, anderen Situationen. Zum zwei\-ten kann Eve diesen Switch \"uberhaupt erst herbeif\"uhren, weil es ihr gelingt, sich vorzustellen, wie sie die Situation beschrei\-ben w\"urde: "`how she would describe this"'. Sie fasst gedanklich in Worte, was sie gerade empfindet. Es bleibt offen, ob Eve sich vorstellt, w\"ahrend des Beschrei\-bens zu sich \mbox{selbst} zu sprechen oder zu jemand anderem (oder beides). Allein indem sie sich ihr Empfinden in Worte fasst, kann sie sich trotz der unangenehmen aktuellen Situation in eine Art Reflexionspose begeben. Und obwohl ihre Assoziationen keineswegs auf angenehme vergleichbare Situationen rekurrieren (denn sonst w\"urde sie diese schwerlich als vergleichbar assoziieren), erm\"oglicht sich Eve durch ein imagin\"ares reflektierendes He\-raustreten aus der spezifischen Situation, auf eine fiktionale Kraftreserve zuzugreifen. Sie hat allem Anschein nach die F\"ahigkeit, w\"ahrend einer gro{\ss}en Belastung Erinnerungen anzuzapfen und vor ihrem inneren Auge eine andere Version hinzuzuf\"ugen, um sich ihre situative emotionale Wirklichkeit zu vergegenw\"artigen. Bereits w\"ahrend sie sich also in der unangenehmen Situation befindet, stellt Eve sich das eigene Story-telling vor: wie sie das Ge\-f\"uhl beschrei\-ben w\"urde, in dieser Situation zu sein. Und sie ent\-wickelt zwei Varianten: eine verallge\-mei\-nernde Beschreibung, "`like finding yourself in the middle of a Pinter play"', wobei es sich um eine Erinnerung handelt, die ein \textit{Erlesnis} beinhaltet, und eine weitere aus subjektiver Sicht: "`like all her nightmares of a stolid, silent, hostile audience."' Als Verbindung zwischen diesen beiden nutzt Eve zu Beginn eines neuen Satzes die Konjunktion "`Or"', einen Versionierungsmarker, der den zwei\-ten Vergleich einleitet, am Ende eines Absatzes und mit einem grammatikalisch unvollst\"andigen Satz, als ob diese Assoziation allein nicht vollst\"andig w\"are oder Eve sich unschl\"ussig ist. Eventuell wurde sie nur in ihren Gedanken unterbrochen, denn jemand anderes spricht weiter.

Die erste assoziative \"Ahnlichkeitsversion erscheint merkw\"urdig lapidar: "`like finding yourself in the middle of a Pinter play"', denn wie es sich darin anf\"uhlt, kann ein Leser\textsuperscript{\tiny *} nur aus eigenem Erleben "`in the middle of a Pinter play"' wissen -- und es bleibt offen, ob gelesen oder in einer bestimmten Inszenierung oder beides. Mit "`like"' wird \mbox{hier} vorausgesetzt, dass andere sich vorstellen k\"onnen, was genau Eve in einem Pinter-Drama als \"ahnlich zu der Situation empfindet, in der sie und die beiden Enkelkinder sich in diesem Raum einer Gruppe Alkoholisierter gegen\"ubersehen. Die Formulierung, sich "`in the middle of"' wiederzufinden, stellt eine Verbindung her zwischen Eves aktueller Lage und derjenigen, die sie als eine ihrer Erfah\-rungen aus Lekt\"ure oder Theater erinnert. M\"oglicherwesie ist es ein "`in the middle of"' im Sinne eines Nicht-aufh\"oren-K\"onnens, weil man w\"ahrend einer Lekt\"ure trotz oder wegen einer bestimmten Atmosph\"are nicht unterbrechen will, sondern so lange "`mittendrin"' sein will, bis das St\"uck zuende gelesen ist. Als parallel gesetzt k\"onnte man in diesem Vergleich auch empfinden, dass Eve sich aus dieser Situation allem Anschein nach wegen H\"oflichkeitserwartungen nicht einfach befreit, \"ahnlich wie man einen Theatersaal m\"oglichst nicht vorzeitig verl\"asst, in jenem Zusammenhang aus R\"ucksicht gegen\"uber den anderen Zuschauern\textsuperscript{\tiny *}. Der Absatz hatte damit begonnen, dass Eve sagt: "`I'm really sorry"', und im darauffolgenden Absatz wird Eves Haltung der Intention nach best\"atigt: "`when she could not think of any further charming or apologetic thing to say"' -- wovon sie sich f\"ur die Bew\"altigung dieser Lage etwas zu versprechen scheint, obwohl oder weil die Szene sich f\"ur sie bedrohlich anf\"uhlt. Das lapidar Formulierte dieser ersten Assoziation l\"asst Vieles offen, w\"ahrend die zweite \"Ahnlichkeitsversion andere Anhaltspunkte gibt, weil Adjektive zum Einsatz kommen ("`stolid, silent, hostile"') sowie "`all her nightmares"'. Falls ein Leser\textsuperscript{\tiny *} an die erste Assoziation nicht ankn\"upfen kann, etwa weil die ty\-pische Atmosph\"are in Pinters Theaterst\"ucken kein Begriff ist, bietet diese zweite Assoziation eventuell einen leichteren Zugang. Man k\"onnte beim Lesen also die erste Assoziation als relativ nichtssagend \"uberspringen bis man sich eines Tages eventuell an diese Munro-Story erinnert f\"uhlt, wenn sich die Gelegenheit ergibt, Pinters Universum \mbox{selbst} zu sp\"uren zu bekommen. Man w\"urde dann quasi ein fiktionalisiertes \textit{Erlesnis} in R\"uckschau \mbox{selbst} erleben -- was f\"ur eine Versionierungsvariante! 

Vorerst gibt es in diesem Fall f\"ur das Pinter-Feeling nur Anhaltspunkte aus der Beschreibung dessen, was Eve in der Behausung wahrnimmt, sowie aus ihrer Reaktion einer Verspanntheit im ganzen K\"orper, von der berichtet wird, bevor ihre Assoziationen einsetzen. Eingeleitet wurde diese Szene mit der Schil\-derung einer Aushandlung zwischen Eve und Philip sowie seinem Versuch, Eve davon abzuhalten, aus dem Auto auszusteigen und mit einem kleinen Mann, der direkt vor ihnen aussteigt und einen jungen kl\"affenden Hund dabeihat, vor dem sich Philip etwas f\"urchtet, ins Gespr\"ach zu kommen, damit sie das Auto wenden k\"onnte, aber es kommt anders: "`and somehow they were all on the route to the house"', sie waren also mehr oder weniger unfreiwillig in die Behausung gelangt. Nicht zu\-letzt k\"onnen aber die beiden Assoziationen, auf die Eve zur\"uckgreift, zwei Aspekte derselben Szene sein, die sie erlebt hat: n\"amlich w\"ahrend sie auf der B\"uhne eine der Rollen in einem Pinterst\"uck verk\"orperte ein abweisendes Klima auf Sei\-ten der Menschen im Zuschauerraum zu sp\"uren. Eves "`Or"' als gedankliche \"Uberleitung zur zwei\-ten Assoziation w\"are dann als eine Art Kippfigur zu denken, weil nicht entschieden werden kann, ob eine der beiden Szenen relevanter ist, aber sie v\"ollig in eins zu sehen auch nicht so recht gelingen mag. 

In einem der Romane von S\'{a}ndor M\'{a}rai, \textit{Wandlungen einer Ehe} (1941), wird unter umgekehrten Vorzeichen von einem \textit{Erlesnis} berichtet, im Vergleich zu dem Munros Variante beleuchtet werden kann. Bei M\'{a}rai erz\"ahlt die Protagonistin\textsuperscript{\tiny *} einer Freun\-din, die sie einige Jahre nicht gesehen hat, aus ihrem Leben und berichtet ihr von einer gedanklichen Begebenheit am Tag nach einem wichtigen Gespr\"ach mit dem Freund ihres Ehemannes: "`Jetzt fiel es mir wieder ein und es kam mir so vor, als h\"atte ich es irgendwo gelesen"' (\cite{Marai1941}:94). Eve h\"atte in einem Gespr\"ach etwa sagen k\"onnen: "`Gestern bin ich in eine Situation geraten, sag mal, kennst du St\"ucke von Pinter? Ja, genauso war es da. Oder wie wenn von Zuschauern nur Sturheit, Schweigen und Feinseligkeit bei mir ankommt."' Bei Munro blendet sich Eve Gelesenes bzw. Erlebtes ein, um eine Si\-tu\-a\-tion zu bew\"altigen, bei M\'{a}rai hat die Protagonistin\textsuperscript{\tiny *} ihrer Freundin zuvor detailliert das Gespr\"ach nacherz\"ahlt, von dem sie am Tag darauf gemeint hatte, sie habe dar\"uber irgendwo gelesen anstatt es \mbox{selbst} erlebt zu haben. Die Protagonistin\textsuperscript{\tiny *} f\"uhrt ihrer Freundin vor, wie sie die Fiktionalisierung von Erlebtem als Strategie eingesetzt hat, um den Inhalt jenes Gespr\"achs verarbei\-ten zu k\"onnen, in Bezug auf die n\"achsten Schritte im eigenen Leben -- \"ahnlich wie Eve, bis sie in den Raum hinein sagt: "`You see I can't move until somebody moves the truck"'. Sowohl bei Munro als auch bei M\'{a}rai wird eine komplizierte Situation mittels \textit{Erlesnis} gespiegelt, um eine L\"osung zu fin\-den. In \textit{Wandlungen einer Ehe} handelt es sich um ein imagin\"ares \textit{Erlesnis}, das einer anderen Person in direkter Rede erz\"ahlt wird, in \textit{Save the Reaper} wird eine Situation mit Hilfe eines \textit{Erlesnisses} beleuchtet, dessen Wirkung in zweifacher Hinsicht fiktional ist. Erstens bleibt offen, wie sich dasjenige ange\-f\"uhlt hat, was Eve "`in the middle of a Pinter play"' gesp\"urt hat, weil Eves Gedanken nicht expliziter wiedergegeben werden und Leser\textsuperscript{\tiny *} je nach Emp\-fin\-den der zuvor geschil\-derten aktuellen Szene sowie nach Kenntnissen (oder nicht) von Pinter-St\"ucken unterschiedliche Vorstellungen davon haben, welche Aspekte es sind, die Eve auf Assoziationen zu Pinter bringen. Zwei\-tens ist das \textit{Erlesnis} bei Munro auf der Ebene des Ereignisses fiktional, insofern Eve lediglich imaginiert, wie sie es erz\"ahlen w\"urde. 

Auf der Ebene des Textes hingegen wird Eves \textit{Erlesnis} durch Munros Erz\"ahlweise Teil der Szene. Dies f\"uhrt zu einem weiteren Aspekt: Dass wir als Leser\textsuperscript{\tiny *} in dem Augenblick \mbox{selbst} "`audience"' sind. Kurz nachdem Eve sich ihre beiden Assoziationen in narrativer Gestalt imaginiert hat, wird Philip von einem der M\"anner direkt an\-ge\-spro\-chen werden, ob er nicht Klavier spielen k\"onne und ihnen vielleicht etwas vorspielen wolle, woaufhin das Verhalten von Philip als "`mute, pressed against her side"' beschrieben wird, nicht un\"ahnlich dem, wie sich die als feindselig, stumm und stur empfundene Menschenmenge im Publikum auf relativ gr\"o{\ss}ere Ferne von der B\"uhne f\"ur die Schauspielerin Eve ange\-f\"uhlt haben k\"onnte, deren Angst sie von ihrer Warte eventuell nur nicht wahrgenommen hat, weil sie \mbox{selbst} in dem Augenblick andere Bed\"urfnisse hatte, nach Kontakt -- so wie auf der Ereignisebene im Text der t\"atowierte Mann in der Behausung, der Philip auffordert, ihnen etwas vorzuspielen: entweder ohne zu bedenken, dass sich die Situation f\"ur die drei unerwarteten Besucher\textsuperscript{\tiny *} v\"ollig anders darstellt als f\"ur ihn \mbox{selbst} -- ihm fehlt noch ein bisschen Musik f\"ur seinen Abend in der Bar -- oder in Schadenfreude provozierend -- so wird die auffordernde Frage an Philip sei\-tens des Greybeard aufgefasst, da er Einspruch erhebt und sagt, er solle den Jungen in Ruhe lassen -- oder beides.

Enden wird die Szene mit einem Tramper, der sich als eine junge Prostituierte he\-rausstellt, die Eves Auftauchen als Gelegenheit zur Flucht aus der Behausung zu nutzen gewusst hat. Als Eve abends ihrer Tochter von dem Ausflug erz\"ahlt -- "`Eve told her story"' -- l\"asst sie die andere, zensierte Version entstehen, in der sie bestimmte Dinge aus vermeintlicher R\"ucksicht auf die Emp\-fin\-dun\-gen der beiden Eheleute, die weniger alternativ eingestellt sind als die Gro{\ss}mutter, nicht erz\"ahlt. Auch ihr \textit{Erles\-nis} l\"a{\ss}t Eve darin nicht vorkommen, denn \"uber ihre Angst hat sie in der Abendversion nicht sprechen wollen. Wahrscheinlich hatte sie sich andere Adressaten\textsuperscript{\tiny *} daf\"ur vorgestellt -- oder zur Bew\"altigung ihrer Angst ohnehin keine anderen Adressaten\textsuperscript{\tiny *} als sich \mbox{selbst}.
 
Magdalene Redekop stellt fest, dass Munros Kurzgeschichten ber\"uhmt-ber\"uchtigt daf\"ur sind, sich Interpretation zu widersetzen. Als Begr\"undung sieht sie Munros "`profound connection with orality"': die Texte w\"urden wie Tratsch funktionieren, so dass die beste Antwort eine weitere Story sei. Im Zuge dieser \"Uberlegungen \"au{\ss}ert sich Redekop -- scheinbar aus eigener Lekt\"ureerfah\-rung -- dar\"uber, wie nicht-pro\-po\-si\-ti\-o\-na\-les Wissen aus Munro-Lekt\"ure sp\"urbar werden kann: "`that feeling you have when reading a Munro story that you are reading about your own life and, conversely, for those uncanny moments, while you are busy leading your life, when you become convinced that your are inside a Munro story"' (\cite{Redekop1999}:26). Mein \textit{Erlesnis} klingt zur H\"alfte wie das von Alluvial\_Fan in einem Blog-Kommentar vom 8. Dezember 2014, als Antwort auf eine Liste mit ty\-pischen Munro-Situationen: \href{http://the-toast.net/2014/12/08/tell-alice-munro-story/\#IDComment926099751}{"`And I am very glad not to be in an Alice Munro story"'} (\cite{AlluvialFan2014}). Die andere H\"alfte meines \textit{Erlesnisses} macht aus, dass ich w\"ahrend des Lesens versuche he\-rauszubekommen, ob ich in der Erz\"ahlweise ebenfalls Distanzierungsvorg\"ange wahrnehme, als bewusstes D\'{e}j\`{a}-vu von Neu\-tra\-li\-sie\-rungs\-ma{\ss}\-nahmen wie denjenigen, eigene Formulierungen in Wi\-ki\-pe\-dia-Eintr\"agen auf sogenannten -- unerw\"unschten -- "`POV"' (point of view) hin zu pr\"ufen, indem ich sie wie fremden Text lese.

So gesehen k\"onnte im Abschnitt "`Leseerlebnis"' des Wi\-ki\-pe\-dia-Eintrags zu der Story, die Eve erz\"ahlt, wiedergegeben werden, wie sie \mbox{selbst} die Szene in der Behausung mithilfe ihrer Pinter-Erfah\-rung "`gelesen"' hat. K\"onnte Eve als Rezensentin dieser Situation -- "`she'd say it was"' -- in demjenigen Text agieren, dessen Protagonistin\textsuperscript{\tiny *} sie ist? Ja, aber nach den geltenden Konventionen nicht des Textes \mbox{selbst}, denn Eve h\"atte ihre Rezension noch nicht mit b\"urgerlichem Namen an einem relevanten Ort in relevanter Sprache pu\-bli\-ziert. Eves Pinter-\textit{Erlesnis} kann zwei\-tens deswegen nicht Teil des "`Leseerlebnis"'-Abschnitts im "`Save the Reaper"'-Eintrag sein, weil eigentlich die Story \mbox{selbst} -- beziehungsweise bestimmte Szenen daraus -- als \textit{Erlesnis} in den "`Leseerlebnis"'-Abschnitt im Artikel "`Harold Pinter"' geh\"ort. Quasi eine neue Geschichte als Antwort auf ein Werk. Hat Pinter "`Save the Reaper"' gekannt? Las er \textit{The New Yorker}?

Das Versionierungsverfahren, das Munro ihre Protagonistin\textsuperscript{\tiny *} Eve anwenden l\"asst, wird von Munro \mbox{selbst} eingesetzt, indem sie im Juni und im Oktober 1998 zwei Fassungen einer Story ver\"of\-fent\-licht -- unter demselben Titel.

\pagebreak

\subsubsection{\flq For future editors\frq: \flq The London Encyclop{\ae}dia\frq \,\,und Doris Lessings \flq Alfred and Emily\frq}
\label{subsubsec:5.4.2}

\singlespacing
\begin{quote}
{\footnotesize Ist die Welt in die H\"ande der R\"ustungsindustrie gefallen? -- Ingeborg Maus, \textit{Menschenrechte,\,Demokratie und Frieden.\,Perspektiven globaler\,Organisation} (2015, S. 11)}\newline
\end{quote}
\onehalfspacing

Doris Lessings letztes Werk, \textit{Alfred and Emily} (2008), basiert auf den Biografien ihrer Eltern und weist einen intertextuellen Bezug zu \textit{The London Encyclop{\ae}dia} auf, erstmals he\-rausgegeben 1983, von Ben Weinreb und Christopher Hibbert. Ich schlage \mbox{hier} eine Interpretation vor, wonach es sich bei \textit{Alfred and Emily} um eine li\-te\-ra\-risch-fotografische Kommentierung eines Eintrags sowie des Vorworts zur \textit{Encyclop{\ae}dia} handelt und umgekehrt: Lessing nutzt \textit{The London Encyclop{\ae}dia} zur "`Illustrierung"' ihrer Aussagen und ihres Werks.
\href{http://www.theguardian.com/books/2008/may/17/fiction.dorislessing}{"`Lessing has never done the expected thing"'}, meint Rezensent Blake Morrison (\cite{Morrison2008}). Ich n\"ahere mich im Folgenden etwas Unerwartetem. Als ich diesen Beitrag begann, "`wusste"' ich davon noch nichts. Und mein \textit{Erlesnis} hat mich nicht nur \"uberrascht, sondern auch sehr ber\"uhrt.

\singlespacing
\begin{quote} 
"`It is, of course, possible that in spite of infinite care, we ourselves have lent credibility to further errors for future editors to discover."'\\
("`Introduction to the First Edition"', \textit{The London Encyclop{\ae}dia}, 1983, \cite{Weinreb1983}:xii)
\end{quote} 
\onehalfspacing

Wortw\"ortlich \"ubernommen wird in \textit{Alfred and Emily} der \textit{London Encyclop{\ae}dia}-Eintrag "`Royal Free Hospital, Pond Street, Hampstead, NW 3"', und dieser wird um ein Krankenhauszimmer-Foto erg\"anzt. Text und Foto geh\"oren ab 2008 in dieser Weise zusammen. Keine der mir bekannt gewordenen Kritiken geht darauf ein, dass das Foto in der \textit{London Encyclop{\ae}dia} nicht vorkommt, sondern erst in \textit{Alfred and Emily} angef\"ugt worden ist. Ich halte die Platzierung des Fotos f\"ur ein Hinweisschild, einen Wegweiser. 

Wie ich darauf komme? Ich habe mir vorgestellt, wie diese Text-Bild-Kombination zustandekam und wie ich es im Wi\-ki\-pe\-dia-Jargon formulieren w\"urde, h\"atte Lessing das Foto in einen Wi\-ki\-pe\-dia-Eintrag eingef\"ugt: \mbox{User}:Doris Lessing hat den Eintrag \flq Royal Free Hospital\frq \,\,als er\-g\"an\-zungs\-be\-d\"urf\-tig angesehen und hatte schon die Idee, ein privates Foto zu verwenden. Nun editiert \mbox{User}:Doris Lessing den Eintrag, indem das per\-s\"onli\-che Foto zuerst \flq bei Commons\frq, der Mediendatenbank der Wikimedia Foundation, hochgeladen wird. Im Wi\-ki\-pe\-dia-Eintrag \flq Royal Free Hospital\frq \,\,ist \mbox{User}:Doris Lessing in den Bearbeitungsmodus gewechselt. Am Ende des Artikeltextes wird nun der Link zum Bild eingebaut. Indem die \"Anderung an den Server gesendet wurde, ist die Abbildung Teil des Artikels geworden. Das Foto dient der Illustration dessen, was im Eintrag zu lesen ist. Der Versionsgeschichte ist zu entnehmen, dass der Eintrag nun 170 KB mehr hat und dass diese \"Anderung 25 Jahre nach der ersten Version von \flq Royal Free Hospital\frq \,\,get\"atigt worden ist. Sofern die Abbildung auch in der aktuellen Version des Eintrags noch zu sehen ist, kann davon ausgegangen werden, dass niemandem das erg\"anzte Foto als unpassend erschienen ist und dass demnach das Anf\"ugen dieser Datei zur Verbesserung des Artikels beigetragen hat. 

Bei Wi\-ki\-pe\-dia w\"are es die Commons-Datenbank, auf die verwiesen wird, weil die Abbildung dort "`entnommen"' worden w\"are. Im Fall von Lessings \textit{Alfred and Emily} ist quasi das Buch \mbox{selbst} die "`Datenbank"', denn die Abbildung ist genau dem Zusammenhang entnommen, den Lessing schildert. Als Verbindungshinweis (Link) zwischen \textit{Alfred and Emily} und \textit{The London Encyclop{\ae}dia} fungiert meiner Auffassung nach sowohl die Geste des Platzierens als auch das, was auf dem Foto zu sehen ist: ein m\"annlicher Patient, halbsitzend im Bett, der in die Kamera schaut, sowie eine Frau in Berufstracht, die auf einem Stuhl an der anderen Seite des Bettes sitzt, m\"oglicherweise Handarbei\-ten vor sich hat und den Kopf gesenkt h\"alt. Eine Bildunterschrift gibt es nicht. Das hei{\ss}t: Es wird davon ausgegangen, dass ohnehin alle wissen, woher das Foto stammt und wer darauf zu sehen ist. \mbox{Hier} wurde -- auch Lessings Danksagung nach zu schlie{\ss}en: "`My thanks to the photographer Francesco Guidicini who helped me with some very old and sometimes dilapidated photographs"' -- ein Familienalbum-Foto aus seinem Kontext gel\"ost und an einen Enzyklop\"adie-Eintrag angeh\"angt. Also sind aller Wahrscheinlichkeit nach Alfred und Emily auf dem Foto zu sehen. Der Eintrag ist bis dahin nur Teil einer renommierten Londoner Enzyklop\"adie gewesen, eines einb\"andigen Lexikons \"uber die Stadt, mit der die beiden Abgebildeten biografisch verbunden sind. Die Geste, das Foto zu erg\"anzen, verbindet die beiden Per\-so\-nen visuell mit dem Text des wieder abgedruckten Eintrags ebenso wie das Erg\"anzen des Fotos den Eintrag der Enzyklop\"adie mit demjenigen in Zusammenhang bringt, was Lessing in \textit{Alfred and Emily} pu\-bli\-ziert hat. 

Judith Kegan Gardiner merkt an, dass Lessing die "`alternative reality"' ihrer fiktionalen Werke ent\-wickelt, indem sie von "`familiar people in a transformed social context"' erz\"ahlt (\cite{Gardiner2009}:161). Aber in \textit{Alfred and Emily} ist der ver\"anderte Kontext dieses Fotos erstens kein an\-de\-rer, im Gegenteil, die Geste scheint zu sagen "`diese Leute geh\"oren dahin."' Zwei\-tens sind die Per\-so\-nen vor dem Ende von Part One nur scheinbar "`fami\-liar"', vertraut, denn ihnen war gerade ein anderes, fiktives, Leben zugedacht worden. Drittens wird \mbox{hier} der Kontext transformiert, indem f\"ur das Foto eine andere Art von Text (enzyklop\"adischer Eintrag) sowie ein anderes Publikationsformat (Lexikon) gew\"ahlt wurde. Viertens wird in diesem Fall keine Geschichte mit Worten erz\"ahlt, sondern durch die Platzierung eines Fotos bei einem Text strukturell auf einer Bild-Text-Ebene agiert, also in einem Vorgang, der sozusagen ohne Worte ist. 

Meiner Lesart zufolge dient das angeh\"angte Foto in \textit{Alfred and Emily} vor allem als Illustrationsbeispiel f\"ur die folgende redaktionelle Aussage in Weinrebs Vorwort zur \textit{London Encyclop{\ae}dia}: 

\singlespacing
\begin{quote}
"`[I]nterspersed between the lists of names and dates there sometimes springs to life a man or woman, long since dead, whose wit or wisdom, triumph or tragedy, transcends the sequence of events and imprints upon our mind the vivid picture of their person or the haunting echo of their voice."' (\cite{Weinreb1983}:xii)
\end{quote} 
\onehalfspacing

Im Vergleich dazu lautet, 25 Jahre sp\"ater, eine der redaktionellen Aussagen in \textit{Alfred and Emily} folgenderma{\ss}en: 

\singlespacing
\begin{quote}
"`Writing about my father's imagined life, my mother's, I have relied not only on traits of character that may be extrapolated, or extended, but on tones of voice, sighs, wistful looks, signs as slight as those used by skilful trackers."' (\cite{Lessing2008}:139)
\end{quote} 
\onehalfspacing

Sowohl Weinreb als auch Lessing sprechen \mbox{hier} in der Rolle von He\-rausgebern\textsuperscript{\tiny *} und die \"Ahnlichkeit in der Darstellung ihrer Absichten ist gro{\ss}. Al\-ler\-dings ist sie nicht weiter verbl\"uffend, wenn man davon ausgeht, dass Lessing sich mit Weinrebs Vorwort befasst hat, was ich f\"ur naheliegend halte, da ich meine, dass Lessing in \textit{Alfred and Emily} das Vorwort sowie einen der Eintr\"age mit ihren (f\"ur mich unerwarteten) Mitteln kommentiert. In welcher Weise dies meinem Verst\"andnis nach unternommen wird, ist der Punkt, auf den ich hinauswill -- neben einer Schilderung meines \textit{Erlesnisses}, das ich als mit dem Einsatz dieser Mittel in Zusammenhang stehend auffasse. 

Wie in \textit{Alfred and Emily} nachzulesen ist, haben sich die beiden abgebildeten Per\-so\-nen in einem Krankenhaus kennengelernt, weil die Krankenschwester Emily Dienst hatte, als dem Soldaten Alfred in den Jahren des Ersten Weltkriegs ein Bein amputiert werden musste. Nicht nur das Foto von ihnen, sondern auch die Biografien von Alfred und Emily \mbox{selbst} sind aufgrund der Aspekte ihres Lebens, die sie demzufolge gemeinsamen haben, als Erg\"anzung zum Eintrag aufzufassen. Denn auch diese Per\-so\-nen \"uberdauern -- ganz im Sinne der Aussage von Weinreb -- die Zeit, al\-ler\-dings nicht als Anekdotengeber\textsuperscript{\tiny *} im Rahmen der \textit{London Encyclop{\ae}dia}, sondern zum Leben erweckt in \textit{Alfred and Emily}, verfasst von deren Tochter Doris Lessing. Auf diesen Punkt komme ich sp\"ater nochmals zu sprechen, wenn ich einen weiteren Aus\-schnitt aus Weinrebs Vorwort heranziehe, um ihn aus der Perspektive meiner Lekt\"ure von \textit{Alfred and Emily} zu beleuchten.

\textit{The London Encyclop{\ae}dia} wurde im selben Jahr wie \textit{Alfred and Emily} in der dritten Auflage auf den Markt gebracht, "`with new photographs by Matthew Weinreb."' Aber bei diesem Eintrag findet sich weiterhin kein Foto -- wie ich anhand des Pr\"asenz-Exemplars im Lesesaal der W\"urttembergischen Landesbibliothek sehen konnte. Eigentlich aber hatte ich he\-rausfinden wollen, ob seit der Ausgabe von 1983 am Text des Eintrags etwas ge\"andert worden ist: auch nicht. Ich erl\"autere im Folgenden, warum es mir wichtig war, dies nach 25 Jahren zu pr\"ufen. Denn nach meiner Auffassung ist nicht nur die Platzierung des Fotos ein Hinweis, sondern auch, dass der gesamte Eintrag aus \textit{The London Encyclop{\ae}dia} von 1983 wortw\"ortlich in \textit{Alfred and Emily} wiedergegeben wird, und das ziemlich genau in der Mitte des Buches. Man k\"onnte das Krankenhauszimmerfoto als "`an zentraler Stelle platziert"' be\-zeichnen. Wie sieht der Gliederungsrahmen von \textit{Alfred and Emily} aus, in dem sich diese Position als zentral erweist? 

\textit{Alfred and Emily} setzt sich zusammen aus Texten verschiedener Art sowie "`the Tayler family album photos"' (\cite{Tiger2009}) von Emily und Alfred Tayler (\href{https://de.wikipedia.org/w/index.php?title=Alfred_und_Emily&oldid=138028858#Gliederung_.28Originalausgabe.29}{Gliederung}). Oft kommentiert worden ist bis\-her, dass und wie sich die beiden Teile des Buches unterscheiden. In seiner Rezension f\"ur \textit{The Guardian} formuliert Tim Adams es zum Beispiel so: 

\singlespacing
\begin{quote}
\href{http://www.theguardian.com/books/2008/may/11/fiction.biography}{"`Alfred Tayler and Emily McVeagh are the writer's parents and this is a book of two halves -- the first section is a novelist's game of might-have-beens: Lessing removes all the frustrations that circumscribed her growing up in Rhodesia, and gives Alfred and Emily the lives they wanted for themselves. The second section is another honest excavation of the lives they were all actually dealt. \textbf{The gap} is the one in which the writer has always lived."'} (\cite{Adams2008}) \textit{[Meine Hervorhebung]}
\end{quote} 
\onehalfspacing

Mit dem gap, der L\"ucke, beschreibt Adams, an welchen Stellen sich f\"ur Schriftsteller\textsuperscript{\tiny *} in thematischer Hinsicht erfah\-rungsgem\"a{\ss} Potenzial sch\"opfen l\"a{\ss}t f\"ur kreative Bearbeitungen eines Stoffes. Blake Morrison findet \textit{Alfred und Emily} ungew\"ohnlich: Als ob W\"urfel zweimal geworfen werden ("`a double throw of the dice"'), sei es gleich\-zei\-tig Belletristik und Sachtext, und dieselbe Geschichte werde auf zweierlei Art erz\"ahlt (\cite{Morrison2008}). Auch Denis Scheck meint, dass die Geschichte von Alfred und Emily ein zweites Mal erz\"ahlt wird -- und dann autobiografisch. Mitten im Buch wird Stil- und Tonlage gewechselt, was er eine k\"uhne Idee findet, mit der Lessing \href{http://www.deutschlandradiokultur.de/nachgetragene-tochterliebe.950.de.html?dram:article_id=136885}{"`aufs Wundervollste \"uber die Str\"ange"'} schl\"agt (\cite{Scheck2008}). Es handelt sich um ein \href{http://www.oxonianreview.org/wp/stories-to-herself/}{"`twinning of what never actually occurred with what did occur"'}, formuliert es David Sergeant und erg\"anzt: \href{http://www.oxonianreview.org/wp/stories-to-herself/}{"`it requires a slight retuning of the fictional ear, until it can register the cumulative effect of such incidents and their sequencing"'} (\cite{Sergeant2008}). Frank Kermode beschreibt die beiden Teile des Buches als "`two separate narratives"', "`the first fictional"' ("`a counterfactual imagining"'), "`the second autobiographical"', sieht aber ein gemeinsames Thema: "`a preoccuation with the destructive impact of war on ordinary happiness"' (\cite{Kermode2008}). Gewalt klingt auch in manchen metaphorischen Beschreibungen an, etwa bei Molly Pulda: "`Lessing splits her book down the middle, separating counterfactual history from autobiographical memory"' (\cite{Pulda2010}:3) oder Alexandra Schwartz in \textit{The New Yorker}: \href{http://www.newyorker.com/books/page-turner/on-doris-lessing-and-not-saying-thank-you}{"`she split [the story of her parents] in two, pairing the real account of their miseries and privations with an imagined counter-history of what their lives might have been if the First World War had never happened"'} (\cite{Schwartz2013}). Was an dieser Stelle in der Mitte des Buches passiert, wird auch gut auf den Punkt gebracht in einem Lesebericht mit Kommentar bei \textit{goodreads.com}: \href{https://www.goodreads.com/review/show/31104294}{"`Then, abruptly, the beautiful fiction ends and some seemingly random nonfiction begins [...] I don't know of many readers who can switch gears like that or who would want to even if they could"'} (\cite{Lara2008}), und der Kommentar darunter lautet: \href{https://www.goodreads.com/review/show/31104294?page=1\#comment_77325756}{"`I am so... to use her word... "`shell-shocked"' at the abrupt change that I had to get on Goodreads to find out what everyone else was saying, to see if I should even keep going. Glad I'm not the only one who feels this way!"'} (\cite{Haley2013}). 

Virginia Tiger sieht \mbox{hier} eine neue hybride Form eines Triptychons aus Erz\"ahlung, Notizbuch und Memoiren. Dem Triptychon voran stehe ein Vorwort, in dem die Autorin\textsuperscript{\tiny *} die Absicht der Dreiteilung erl\"autere. (Dies trifft f\"ur die Paperback-Ausga\-be zu, nicht aber f\"ur das Hardcover, wo das Vorwort als integraler Teil von Part One gelayoutet worden ist.) Als mittleren Teil des Triptychons macht Tiger den Abschnitt "`Explanation"' aus (\cite{Tiger2009}:22), der nach zwei Portr\"atfotos von Alfred Tayler auf den Text der "`novella"' folgt.

Zwar halte ich eine Perspektive, die entgegen einer offensichtlichen Part-One-/Part-Two-Strukturierung von einer Dreiteilung ausgeht, f\"ur interessant, aber eher noch \"uberzeugt mich die weitergehende Idee, \mbox{hier} ingesamt von einem Ensemble-Konzept auszugehen. Immerhin eine Position habe ich gefunden, die das Vorhaben \href{http://www.nytimes.com/2008/08/10/books/review/James-t.html}{im Ganzen} f\"ur aufschlussreicher h\"alt als es einer der Teile f\"ur sich genommen ist (\cite{James2008}). Selten ist \textit{Alfred and Emily} bis\-her als \href{http://www.oxonianreview.org/wp/stories-to-herself/}{"`Amalgam"'}\,be\-zeichnet\,worden (\cite{Sergeant2008}) oder als \href{http://www.babelio.com/livres/Lessing-Alfred-et-Emily/84706/critiques/621555}{"`Ensemble"'} (\cite{maevedefrance2014}). Der Lesebericht von maevedefrance beschreibt das Ensemble des Buches inklusive Fotos und en\-zy\-klo\-p\"a\-di\-schem Eintrag: Das Ensemble werde durch Fotos be\-rei\-chert (agr\'{e}ment\'{e}). Sie scheinen demnach "`Zutaten"' zu sein, keine integralen Bestandteile des Werkes, des Buches: 

\singlespacing
\begin{quote}
\href{http://www.babelio.com/livres/Lessing-Alfred-et-Emily/84706/critiques/621555}{"`Le livre, constitu\'{e} de deux parties, est organis\'{e} de mani\`{e}re originale sinon surprenante : la premi\`{e}re partie est une fiction : \textit{Le roman d'Alfred et Emily}; la deuxi\`{e}me (\textit{Alfred et Emily : deux vies}) est une r\'{e}flexion de Doris Lessing qui \'{e}voque l'impact qu'a eu la vie de ses parents sur la sienne, en tant que personne, \'{e}crivain et femme engag\'{e}e. L'ensemble du livre est agr\'{e}ment\'{e} de photos et d'un extrait du London Encyclopaedia qui relate l'histoire du Royal Free Hospital, premier h\^{o}pital public et gratuit de Londres, o\`{u} travaillait la m\`{e}re de Doris Lessing."'}\\ (\cite{maevedefrance2014}) 
\end{quote} 
\onehalfspacing

David Sergeant beschreibt zun\"achst die beiden Teile des Buches als "`novella-length fictionalised account"' und "`account"' und dass diese miteinander kombiniert werden, wodurch sie ein Amalgam werden. Danach listet er auf, welche weiteren Bestandteile in dieses Amalgam aufgenommen worden sind, darunter auch der enzyklop\"adische Eintrag:

\singlespacing
\begin{quote}
\href{http://www.oxonianreview.org/wp/stories-to-herself/}{"`It is a work that combines a novella-length fictionalised account of the lives of Lessing's parents, the Alfred and Emily of the title, as they might have been had the war not occurred, with an account of their lives as they actually turned out, raising two children on a scrappy farm in the Rhodesian bush. This amalgam also manages to assimilate, along the way, a foreword and coda to the fiction, an authorial explanation of its basis in fact, a long extract from an encyclopaedia about London, an assortment of grainy photographs, and zigzaggings through time and space touching on everything from prehistoric paintings to African insects, to Mugabe, to the colonial diet and a list of Lessing's childhood reading. And it works. The novel -- or is it biography? -- or is it history? -- is held together by the unabashed singularity of Lessing's voice: a voice that, in its flight, sheds the taxonomic shells that normally encase literary works."'}\\ (\cite{Sergeant2008})
\end{quote} 
\onehalfspacing

Sergeant und maevedefrance nehmen beide die formalen Besonderhei\-ten wahr und suchen nach angemessenen Beschreibungen, mit denen sie weder neue Genregrenzen erfinden noch das Werk innerhalb dieser Grenzen festzurren. Zu Tigers Idee des Triptychons hingegen ist zu fragen: Warum sollte das "`Notebook"' nicht alle Teile zwischen dem eigentlichen "`novella"'-Text von Part One und dem Beginn des ersten "`Essays"' in Part Two umfassen? Das w\"aren -- wie Tiger es fast vollst\"andig aufz\"ahlt: "`a notebook-like authorial Explanation, an encyclopedia entry on the Royal Free Hospital from The \textit{London Encyclop{\ae}dia}; the D. H. Lawrence epigraph to the work's final section, Part Two Alfred and Emily: Two Lives"' (\cite{Tiger2009}:22). Aus meiner Sicht fehlt \mbox{hier} die Nennung der beiden Epitaphe, auf Alfred und auf Emily, die zwischen der "`novella"' und der "`Explanation"' auf einer eigenen Seite platziert sind, dem Layout nach eigene Kapitel. Sie werden bei Tiger als "`the novella's ending"' be\-zeichnet. Sergeant sieht die beiden Epitaphe als eine Coda an und \mbox{nennt} sie im Zuge der Formulierung "`a foreword and coda to the fiction"', was ich einleuchtend finde. Im Gegensatz zu maevedefrance und Sergeant sieht Tiger die Fotos in der N\"ahe des Notizbuchs nicht als nennenswert an, mir aber scheinen sie bedeutsam zu sein. Die Ensemble-Idee lie{\ss}e zutage treten, dass nicht allein das Krankenzimmer-Foto als Illustration eines nebenstehenden Textes angesehen werden k\"onnte. Der Text des Epigraphs ist der erste Text nach dem Titelblatt von Part Two. Auf derselben Doppelseite links davon findet sich das Foto eines Ehepaares in jungen Jahren. Sie stehen leicht einander zugewandt und schauen in die Kamera. Das Zitat auf der rechten H\"alfte der Doppelseite ist mittig platziert und lautet:

\singlespacing
\begin{quote}
"`And dimly she realised one of the great laws of the human soul: that when the emotional soul receives a wounding shock, which does not kill the body, the soul seems to recover as the body recovers. But this is only appearance. It is, really, only the mechanism of reassumed habit. Slowly, slowly the wound to the soul begins to make itself felt, like a bruise which only slowly deepens its terrible ache, till it fills all the psyche. And when we think we have recovered and forgotten, it is then that the terrible after-effects have to be encountered at their worst."'\\(D. H. Lawrence, \textit{Lady Chatterly's Lover}, \cite{Lessing2008}:151)
\end{quote} 
\onehalfspacing

Dieses Zitat fungiert wie eine Kapitel\"uberschrift der folgenden ersten Episode, denn jene ist ohne Titel. Mit fast 40 Sei\-ten Umfang ist sie die l\"angste des zwei\-ten Teils. In ihr geht es um das, was im Zitat von D. H. Lawrence beschrieben wird: posttraumatische Belastungsst\"orungen durch die Erlebnisse w\"ahrend des Ersten Weltkriregs, und zwar bei Emily ebenso wie bei Alfred. 

Roberta Rubenstein hebt hervor, dass die letzten beiden Episoden von Part Two besondere Essays sind. Sie versteht "`Getting-off-the-Farm"' und "`Servant Pro\-blems"' als Anh\"ange zum "`process of filial reconciliation"', der zuvor abgeschlossen worden sei (\cite{Rubenstein2014}:193). Ich gebe Rubenstein insofern recht, als vor diesen beiden letzten Episoden f\"ur Lessing das Thema "`Bruder"' und "`My son John"' abgeschlossen worden zu sein scheint, mit der Aussage, die ich als Fazit lese: "`some men thoroughly enjoy war"' (\cite{Lessing2008}:252-253). \"Uber ihren Vater und ihre Mutter scheint Lessing danach tats\"achlich abgekl\"arter zu schrei\-ben. Aufgrund von Rubensteins Anmerkung aber entstand bei mir eine Vermutung, die mich auf einen interessanten Pfad gef\"uhrt hat: dass Lessing mit zwei Anh\"angen auf eine \"ahnliche Weise etwas "`illustriert"' wie mit dem angeh\"angten Krankenzimmer-Foto. Pl\"otzlich schien es mir so, als ob inbesondere diese beiden Episoden als ein Verm\"achtnis zu verstehen sein k\"onnten, gerade weil sie aus en\-zy\-klo\-p\"adischer Perspektive gesehen relativ unauff\"allig betitelt sind, also als ein gewisserma{\ss}en kaschiertes Verm\"achtnis -- \"ahnlich wie Alfreds Beinstumpf nach der Amputation auf dem Krankenzimmer-Foto unter der Bettdecke verborgen und daher nicht zu sehen ist. Im letzten Kapitel schildert Lessing neben der Uneinigkeit zwischen ihr und der Mutter bez\"uglich der Notwendigkeit schwarzer Hausangestellter die in S\"udrhodesien sp\"urbaren Folgen des Zwei\-ten Weltkriegs und wie die Mutter ihr Leben als Witwe f\"uhrt. Das vorletzte Kapitel "`Getting-off-the-Farm"' endet mit der Beerdigung des Vaters, und Lessing trifft in markanter Weise Feststellungen dar\"uber, wie sie das Sprechen \"uber Krieg in der Generation ihrer Eltern erinnert, wie sie aktuelle Kriege in afrikanischen L\"andern einsch\"atzt, sie erw\"ahnt, welcher K\"unstler welche Sujets hatte, und schlie{\ss}t die Passage mit einer Vermutung \"uber die aktuelle Situation:

\singlespacing
\begin{quote}
"`When those survivors of the First World War met, they would talk in a way that has fallen out of fashion. [...] The German small-mine worker down the hill, my father's friend [...] talks about the armament-makers, Krupps and the profiteers.\\ 
How strange that the words -- and the idea -- have dropped out of our minds. The \flq military industrial complex\frq \,\,does not have the same ring, does not remind us, or make us think. When a war starts up in Africa, a pointless war, apparently [...] what has been achieved at the end of it? A few hundred dead, but millons of pounds, spent on weapons, safely lodged in somebody's pockets.\\
Grosz' pictures were of the profiteers and armament-makers, who did well out of that war.\\
Profiteers and armament-makers -- gone from our speech and, so it seems, from our minds."'\\
(\cite{Lessing2008}:259-260)
\end{quote}

\onehalfspacing
Auch \mbox{hier} wird Bildliches im Text platziert, diesmal in Worten, indem von den Sujets der Werke des Malers Grosz die Rede ist. Lessing verf\"ahrt mit dieser Geste \"ahnlich wie beim Krankenzimmerfoto. "`Geschichte"' wird illustriert. Grosz malt R\"ustungsindus\-trielle -- Lessing spricht dar\"uber, dass Grosz R\"ustungsindustrielle portr\"atiert (auch wenn sie nichts dar\"uber sagt, wie). In Weinrebs Vorwort hei{\ss}t es: "`there sometimes springs to life a man or woman"' -- und Lessing stiftet ein Fa\-mi\-lien\-fo\-to. In mei\-ner Lesart sagt die Geste in beiden F\"allen so etwas wie: "`Und jetzt stellt euch vor, wie sie aussehen und wer sie sind."' Indem Rubenstein trotz ihres Augenmerks f\"ur diese letzten beiden Essays al\-ler\-dings auch 2014 kein Wort \"uber deren Inhalt verliert, scheint sich die Einsch\"atzung von Lessing zu best\"atigen, dass \"uber die Profiteure\textsuperscript{\tiny *} von Kriegen kaum gesprochen wird -- und anscheinend gibt es im Fall von Rubenstein -- im Unterschied zu Lessing -- nicht einmal eine Sprache daf\"ur, zu berichten, dass und wie andere \"uber die Profiteure\textsuperscript{\tiny *} von Kriegen schrei\-ben. Al\-ler\-dings sind Rubensteins Anmerkungen zu \textit{Alfred and Emily} insgesamt knapp gehalten -- in nahezu enzyklop\"adischem Stil, lie{\ss}e sich in diesem Zusammenhang anf\"ugen. 

Lessings formale Experimente lohnen der Aufmerksamkeit. Dies best\"atigen einige ihrer Interpretatoren\textsuperscript{\tiny *}. Kermode schreibt, Lessing habe sich \"uber Rezensenten\textsuperscript{\tiny *} be\-schwert, die "`stupidly ignored the technical inventiveness of \textit{The Golden Notebook}"' (\cite{Kermode2008}). Was jene "`hostile critics"' zur ihrer eigenen Verteidigung h\"atten sagen k\"onnen, aber nicht gesagt haben, so Kermode: sie seien von li\-te\-ra\-tur\-kritischen Er\-w\"a\-gungen bez\"uglich struktureller Aspekte abgelenkt worden durch "`the origina\-li\-ty of the book's messages"' oder "`the truth and force of what the book accessibly said."' Er erg\"anzt seine Vermutung zu Doris Lessings Standpunkt: "`She herself would argue, reasonably, that her technical innovations are of importance only in so far as they serve the story and its truths"' (\cite{Kermode2008}). Pulda best\"atigt, dass Lessing gehofft habe, \textit{The Golden Notebook} "`would talk through the way it was shaped"' und sie f\"ahrt fort: "`so does \textit{Alfred and Emily} communicate through its dual form"' (\cite{Pulda2010}:3). \"Ahnliches formuliert Phyllis Sternberg Perrakis bez\"uglich \textit{Shikasta}: "`Thus the narrative structure of the book contributes to the reader's experience of its meaning"' (\cite{Perrakis2007}); zitiert in \cite{Pulda2010}:3). Susan Watkins stellt in ihrer Lessing-Studie von 2010 fest, Lessings Arbei\-ten seit 2000 w\"urden implizieren, dass fiktionales Schrei\-ben fruchtbarer sein k\"onne als faktisches Schrei\-ben, insofern es Individuen und Nationen die M\"oglichkeit er\"offne, die Vergangenheit zu verstehen (\cite{Watkins2008}:161). Diese Idee werde in \textit{Alfred and Emily} verk\"orpert, indem der erste Teil spekulativ sei und der zweite "`a memoir."' F\"ur Lessing seien Genre-Zuordnungen fragil und sie arbeite an gegen Klassifizierungen aufgrund von Genre (Watkins 2010:162), etwa gegen Pauschalisierungen wie: Auto\-bio\-gra\-fie sei Wahrheit, ein Roman Fiktion und ein Essay Meinung (\cite{Watkins2008}:161). Auf Basis ihrer Interpretation von \textit{The Story of General Dann and Mara's Daughter, Griot and the Snow Dog} (2006) gelangt Watkins zu dem Schluss: "`Lessing's work in this period strongly suggests [...] that official histories are flawed and imperfect"' (\cite{Watkins2008}:141). Was bedeutet dies in Bezug auf Lessings letztes Werk konkret?

Lessings Buch scheint sich als Kommentar zur \textit{London Encyclop{\ae}dia} und ihrem Anspruch zu eignen. Lesen wir nach, wie die eingangs zitierte Passage des Vorworts von Weinreb in ihrem Kontext lautet und nehmen sie aus der Perspektive einiger Lekt\"uren von \textit{Alfred and Emily} unter die Lupe: 

\singlespacing
\begin{quote}
"`We are less inhibited [than H. B. Wheatley who, when editing the Diary of Samuel Pepys, in conformity with the proprieties of his time, omitted all that was indelicate], and when an incident or anecdote illustrates or enhances an entry it is told, we hope, with the smell and gusto of the period in which it occurred. The intervening years have also uncovered lost records which shed new light on old tales, enabling us to correct attributions, clarify identities and amend errors which repetition had hardened into history. It is, of course, possible that in spite of infinite care, we ourselves have lent credibility to further errors for future editors to discover. We have been chroniclers engaged in the recording of facts rather than historians concerned with their cause and consequence. It has been our task to compress within a paragraph that which has sometimes been the subject of a book and [...] to set down concisely and in plain words what is generally known and held to be true."' (\cite{Weinreb1983}:xii)
\end{quote}
\onehalfspacing

Im Eintrag zum Royal Free Hospital steht in dieser Enzyklop\"adie von 1983 zwar etwas \"uber eine Epidemie (eine weitere Krisenphase in der Geschichte des renommierten Hospitals), aber sie ist lange her: 1832. \mbox{Vier} Jahre nach der Er\"offnung seien \"uber 700 Cholera-Patienten\textsuperscript{\tiny *} behandelt worden und: "`A matron and nurse were employed while the epidemic lasted"' (\cite{LondonEncyclopaedia1983}:672). Danach gibt es Be\-rich\-tenswertes zu diesem Krankenhaus f\"ur die Jahre 1837, 1839, 1843, 1856, 1877, 1878, 1894, 1895, 1901, 1921, 1926-1930, und zum Zwei\-ten Weltkrieg ist zu lesen: "`The Royal Free suffered severe damage in the 2nd World War, with considerable loss of beds"' (\cite{LondonEncyclopaedia1983}:672). Aber an nennenswerten Auswirkungen des Ersten Weltkriegs scheint es keine gegeben zu haben, denn zwischen 1901 und 1921 steht nicht nur ein Abschnittswechsel, sondern pl\"otzlich scheint f\"ur eine Spanne von 20 Jahren kein "`recording of facts"' stattgefunden zu haben, wo in den Jahren zuvor die Spanne zwischen einem und sechs Jahren gelegen hat. Aus Sicht der beiden He\-rausgeber mussten 1983 im Eintrag zum Royal Free zwischen 1901 und 1921 keine "`errors"' korrigiert werden "`which repetition had hardened into history."' Wie lassen sie den Eintrag "`Royal Free Hospital"' also \"uber den Ersten Weltkrieg sprechen? Gar nicht. Der Erste Weltkrieg kommt im historisch erz\"ahlenden Eintrag zu dieser Institution nicht vor. Vielleicht hatten Weinreb und sein j\"ungerer Kollege Christopher Hibbert keine Informationen zur Hand, auf die ihre redaktionelle \"Uberzeugung anwendbar w\"are: "`when an incident or anecdote illustrates or enhances an entry it is told, we hope, with the smell and gusto of the period in which it occurred"' (\cite{LondonEncyclopaedia1983}:672). Folgt man der Darstellung in Lessings vorletztem Essay von Part Two, sprechen Weinreb und Hibbert eher im Ton ihrer eigenen Zeit, wenn der Erste Weltkrieg im Eintrag nicht vorkommt: F\"ur das Benennen von dessen Profiteuren\textsuperscript{\tiny *} g\"abe es demnach keine Sprache, und auch f\"ur die Nennung der Auswirkungen des Ersten Weltkriegs auf ein Londoner Krankenhaus nicht. Meiner Lesart zufolge ist der im Vorwort der \textit{London Encyclop{\ae}dia} prognostizierte Fall eingetreten, dass die beiden He\-rausgeber\textsuperscript{\tiny *} \mbox{selbst} daran beteiligt sein w\"urden, f\"ur k\"unftige Editors\textsuperscript{\tiny *} Aufgaben zu hinterlassen: \textit{Alfred and Emily} ist Lessings li\-te\-ra\-rische Bearbeitung des Eintrags "`Royal Free Hospital"' und stellt nicht zu\-letzt mit bildlichen Mitteln eine Kommentierung des Vorworts dar. 

Die Funktion einer bestimmten Abbildung soll abschlie{\ss}end genauer betrachtet werden, da es um etwas Undarstellbares gehen k\"onnte, im Sinne von Monika Schmitz-Emans:

\singlespacing
\begin{quote}
\href{http://www.actalitterarum.de/theorie/mse/enz/enzz01.html}{"`Durch Si­mu­la­ti­on oder Imi­ta­ti­on wis­sens­be­zo­ge­ner, wis­sen­schaft­li­cher, sach­in­for­ma­ti­ver Schreib­wei­sen re­flek­tie­ren li­te­ra­ri­sche Texte \"uber For­men des Wis­sens und sei­ner Ver­mitt­lung, \"uber die Ord­nun­gen des Wis­sens (Al­pha­bet, Le­xi­ko­gra­phik, en­zy­klo­p\"a­di­sche For­men... \mbox{Hier} geht es um die Prin­zi­pi­en und Im­pli­ka­tio­nen von Er­kennt­nis und Wis­sen, um die Dar­stell­bar­keit von Wis­sen -- und manch\-­mal auch um das Un­dar­stell­ba­re."'}\\ (\cite{Schmitz-Emans2010})
\end{quote}
\onehalfspacing

Dass auch nach 25 Jahren eine bestimmte enzyklop\"adische Auslassung weiterhin besteht, und in welcher Form, ist mit \textit{Alfred and Emily} darstellbar gemacht worden. Auch diese Generation also w\"urde in das Verschweigen der Auswirkungen des Ersten Weltkriegs eingeweiht: Dass und wie es geht, in einem enzyklop\"adischen Eintrag einen neuen Absatz zu machen und 20 Jahre zu \"uberspringen. \textit{Alfred und Emily} l\"asst sich demnach als Rebellion gegen diese fortgesetzte Auslassung lesen, neue Version: Beim enzyklop\"adischen Eintrag in der Mitte des Buches nun der Wegweiser zu einer Lesart von \textit{Alfred und Emily} als Kritik der \textit{The London Encyclop{\ae}dia}, in li\-te\-ra\-risch-illustrativer Form. Umgekehrt k\"onnte die Platzierung des Eintrags im Zentrum von \textit{Alfred und Emily} besagen: "`Jener Eintrag illustriert mein Thema punktgenau."' 

\singlespacing
\begin{quote}
"`He was suffering that need of the old: he needed to explain something before it was too late. Tell somebody, anybody, as if what he had to tell could have no reality unless it was in somebody else's mind, too."' \\(\cite{Lessing2008}:252, "`\"Uber Harry Tayler"')
\end{quote}
\onehalfspacing

In genau diesem Sinne ist \textit{Alfred und Emily} -- auch strukturell -- Doris Lessings Verm\"achtnis, mit dem sie gegen Geltungsanspr\"uche von "`official histories"' angeht. Am Beispiel eines enzyklop\"adischen Eintrags kritisiert sie fortgesetztes Schweigen dar\"uber, in wessen Taschen die Profite von Kriegen landen.
Durch das Platzieren und "`Illustrieren"' eines enzyklop\"adischen Eintrags im Zentrum von \textit{Alfred und Emily} macht Lessing deutlich, was "`in der Geschichte"', in Publikationen, allzu h\"aufig verschwiegen wird, n\"amlich so wie Lessing \mbox{hier} ironisch kunstvoll aufzeigt, dass Weinreb es gerade in diesem Fall besonders richtig zu formulieren scheint: als "`errors which repetition had hardened into history."'

Eine Freundin\textsuperscript{\tiny *}, die ich gebeten hatte, sich beide Ausgaben (hardcover und paperback) mit ihrem bildwissenschaftlich versierten Blick anzusehen, kommentierte unter anderem, dass zu den Abbildungen die Nachweise fehlen. Mir war nur aufgefallen, dass es keine Bildunterschriften gibt. Bei Durchsicht von etwa 10 \"ahnlichen Produkten desselben Verlagshauses aus den Jahren 2007-2009 war festzustellen: Alle Abbildungen haben Bildunterschriften und Nachweise. Und in Doris Lessings beiden Autobiografie-B\"anden? Ja: Fotos mit Bildunterschriften. Aus der Autobiografie Band 1 (\textit{Under my Skin}, 1994) nutzt Lessing f\"ur \textit{Alfred and Emily} drei der Familienfotos erneut. Anzunehmen ist also, dass sie nicht ihres Neuigkeitswerts wegen wieder aufgenommen wurden, denn sie wurden 2008 nicht zum ersten Mal gezeigt. Also wird Lessing diesen Fotos neben dem Illustrieren von biografischen Fakten in \textit{Alfred and Emily} knapp 15 Jahre sp\"ater eine weitere Funktion zugedacht haben. Von einem dieser Fotos ist ein an\-de\-rer Ausschnitt zu sehen: die Per\-so\-nen sind weniger deutlich zu erkennen, daf\"ur we\-sent\-lich mehr von der Umgebung: das Krankenhauszimmerfoto. Im Unterschied zur Autobiografie, wo die Abbildungen separat zusammengestellt auf Fotodruckpapier gedruckt sind, wurden die Abbildungen in \textit{Alfred and Emily} auf derselben Papiersorte, demselben Untergrund platziert wie der Text. 

Die Materalit\"at von Papier, Text und Abbildungen zog meine Aufmerksamkeit auf sich, ich lie{\ss} dies auf mich wirken, w\"ahrend ich erneut in dem Buch bl\"atterte, vorw\"arts, r\"uckw\"arts, hin und her, und mich daran zu erinnern versuchte, was mir beim ersten und zwei\-ten Lesen durch den Kopf gegangen war. Dass Lessing einen Enzyklop\"adie-Eintrag einbaut, hatte ich registriert, aber wie insbesondere an dieser Stelle eine Abbildung auf derselben Papiersorte "`Teil des Textes"' zu werden scheint, diese Idee kam mir erst jetzt. Als wir zu zweit in den beiden Ausgaben, Paperback und Hardcover, gebl\"attert hatten, war meiner Freundin\textsuperscript{\tiny *} bereits aufgefallen, dass das Foto des Ehepaares auf der R\"uckseite des Titelblattes von Part Two etwas durchscheint, so dass es dem Krankenhauszimmerfoto gegen\"ubersteht. Die beiden Teile sind nur durch das Titelblatt voneinander getrennt. Moniert worden ist, dass die Schreibung der Titelzeilen der beiden Teile inkonsistent sei: bei Teil 1 hei{\ss}e es, mit Doppelpunkt: "`Alfred and Emily: a novella"', bei Teil 2 sei hingegen ein Semikolon verwendet worden: "`Alfred and Emily; Two Lives"', was auf ungenaues Arbei\-ten zur\"uckzuf\"uhren sei ("`how shoddy was the book's publication"' (\cite{Tiger2009}:Fn. 4). Ich denke, es handelt sich bei "`Two Lives"' eben um keinen Untertitel, sondern um einen zwei\-ten Titel, der gleichranging ist, und lese das Semikolon als einen Kommentar zur Zweiteilung des Werks. Der erste Teil des Titels bezieht sich noch auf den ersten Teil des Buches, und ab "`Two Lives"' handelt es sich um eine neue Geschichte, in der die Traumata der beiden Per\-so\-nen \mbox{selbst} Oberhand gewinnen: Jede der beiden Per\-so\-nen hat im Buch ein Leben vor dem Krankenhauszimmerfoto und eines danach. Nur in der Mitte des Buches sind beide zusammen auf demselben Foto zu sehen: Da teilen sie etwas.

Als ich einige Wochen sp\"ater nach einer Pause zum Tisch zur\"uckkehre, wo das Buch liegt, f\"allt mir zum ersten Mal auf, dass auch der Text des Eintrags "`Royal Free Hospital"' durchscheint, auf der linken H\"alfte dieser Doppelseite in der Mitte des Buches. Eine zentral platzierte Aussage, denke ich: Rechts scheint ein Foto durch, links scheint ein Text durch. (Gerade hatte ich etwas \"uber mediale Besonderhei\-ten von li\-te\-ra\-rischen Texten gelesen, \cite{Hillenbach2012}:23.) Wenn es \mbox{hier} tats\"achlich um so etwas wie Transparenz geht ..., dachte ich, als ich eine Seite zur\"uckbl\"atterte, in den Text des "`Royal Free Hospital"'-Eintrags -- und in diesem Augenblick fiel mir auf, dass das Kranken\-hauszimmer\-foto auf der vorletzten Seite des Textes \"uber das Royal Free genau hinter der Stelle durchscheint, wo in \textit{The London Encyclop{\ae}dia} zwischen 1902 und 1921 vom Ersten Weltkrieg keine Rede ist. Mir kam in den Sinn, dass die Erz\"ahlung in Part One 1902 beginnt, Lessing setzt also an dem Punkt ein, wo \textit{The London Encyclop{\ae}dia} schweigt. Ein Foto aus dem ersten Band ihrer Autobiografie wird an der entsprechenden Stelle auf der R\"uckseite des Blattes platziert -- und anstatt auf Fotodruckpapier auf "`Textpapier"'. An dieser Stelle wird der Eintrag wortlos erg\"anzt (eine Abbildung wird "`hochgeladen"') -- und, damit wir es nicht zuf\"allig \"ubersehen, in der Mitte des Buches. Lessing editiert die Enzyklop\"adie insgesamt durch ihr Erz\"ahlen, also mit \textit{Alfred and Emily} als Werk, in dem auch zwei weitere Abbildungen aus dem Familienalbum ihren Ort finden, die in den Autobiografien noch nicht auf Fotodruckpapier zu sehen waren. Lessing, Li\-te\-ra\-tur\-nobelpreis 2007, beendet ihr schriftstellerisches Leben mit einem Bild an zentraler Stelle, auf unerwartete Weise wortlos: Wie zeigt man Undarstellbares kunstvoller und sagt im selben Atemzug, schweigend, mit einem Foto, wie Entscheidendes nicht verschwiegen wird? In wessen H\"ande war meine Wahrnehmung gefallen, dass ich so lange schon mit dem Buch gewohnt hatte, bis ich zu dieser Erkenntnis gelangt war?

Ich sehe \textit{Alfred and Emily} im Ganzen als ein \textit{Erlesnis} an, denn Lessing pr\"asentiert mit dem Werk allem Anschein nach gleichzeitig ihr Nachsinnen \"uber einen Prozess, dem sie das Entstehen bestimmter Emp\-fin\-dun\-gen zuschreibt, und das Ergebnis des Nachsinnens. Sie hatte bei einer Lekt\"ure darauf reagiert, dass in einem zeitgen\"ossischen enzyklop\"adischen Eintrag etwas fehlte: Informationen zu den Jahren 1902 bis 1920 hatte sie vermisst. Diese erachtete sie aber f\"ur so we\-sent\-lich, dass sie -- meiner \mbox{hier} vorgelegten Interpretation zufolge -- dem erworbenen nicht-propositio\-na\-len Wissen in k\"unstlerischer Gestaltung in Form von \textit{Alfred and Emily} Ausdruck verlieh. 

\newpage

\subsubsection{\flq B\"ose, das w\"ar gut\frq . Fliegende auf einem Bein (M\"uller)}
\label{subsubsec:5.4.3}

\"Uber Pflastersteine aus Buchstaben, W\"ortern und S\"atzen bewegen sich meine Augen bei der Lekt\"ure von Zeile zu Zeile, ich bl\"attere nur m\"uhsam um. Die sprachlichen Bilder wirken auf mich bedrohlich. Sie verst\"oren mich. Ich f\"uhle mich merkw\"urdig richtungslos und rechne jederzeit mit weiteren Informationen \"uber unangenehme Zust\"ande. Vermutlich werden sie lange andauern, ich wappne mich und stelle mich darauf ein. Nach einigen Sei\-ten Lekt\"ure fast keiner Aufnahme mehr f\"ahig, halte ich st\"orrisch durch, in der Hoffnung, he\-rauszubekommen, was es mit der Allgegenw\"artigkeit dieser Pflastersteine in diesem Text auf sich haben k\"onnte. Ich hole mir einen weiteren Pullover, versuche meine Verst\"ortheit zu \"uberwinden. W\"armer angezogen intensiviere ich beim Lesen meine Beobachtungen \"uber meine Reaktionen und zeichne gedanklich auf, wann ich das Buch weglege, wann ich weiterlese und warum ich nicht aufh\"oren kann, mich mit den Pflastersteinen zu befassen, die jetzt wie Kl\"otze in meinen Augen liegen, auf meinen Schultern, auch bei der zwei\-ten Lekt\"ure noch, bei der ich gerade zum letzten Drittel gelangt bin. Ende der 1980er Jahre in Westberlin, vor dem Mauerfall: Die Protagonistin\textsuperscript{\tiny *} Irene besucht einen Bekannten in dessen Wohnung. Er zeigt auf einige V\"ogel und sagt: "`Die fliegen zur Mauer"' und: "`Schau, wie sie gl\"anzen, wie fliegende Bl\"atter"' (S. 109) (\cite{Mueller1989}). Ich stutze, weil ich eine Dissonanz empfinde, eine Wechselwirkung zwischen Wortbedeutung und Kontext. Es scheint \mbox{hier} eine falsche oder irref\"uhrende Aussage zu geben, denn die tats\"achliche Bedeutung der \"Au{\ss}erung liegt nicht allein in der Wortbedeutung. Ich bin also miss\-trau\-isch geworden und be\"auge den Zusammenhang, in dem die Aussage steht. Wird \mbox{hier} etwas anderes gesagt als gemeint ist? Gl\"anzen tats\"achlich nur ein paar V\"ogel? 

Pl\"otzlich lese ich \textit{Fliegende Bl\"atter}, den Titel der Satirezeitschrift -- und werfe meine Pflastersteine ab. \textit{Reisende auf einem Bein} voller Karikaturen? Nicht allein ein Kontrast zwischen Wortbedeutungen, sondern zwischen verschiedenen Interpretationen dessen, was genannt wird: einander entgegengesetzte Bewertungen kommen ins Spiel, die in diesem Kontext beide m\"oglich sind. Ich empfinde einen Kontrast zwischen der Haltung des Sprechers\textsuperscript{\tiny *} und derjenigen Haltung, die die w\"ortliche Bedeutung der \"Au{\ss}erung impliziert. Ich lese weiter, erinnere mich an zuvor Gelesenes. Es tauchen m\"ogliche kontextuelle Signale als Drehmomente auf, meine Pflastersteine geraten in Bewegung, sie beginnen sich zu drehen, werden freundlich warm. Die Sonne kommt durch und grinst verbissen \"uber den Rand des Buches.

Drehmoment 1: Die Phrase "`fliegende Bl\"atter"' taucht in M\"ullers Text in einer Weise auf, die an Hannah H\"ochs Collagen erinnert. In "`Schnitt mit dem K\"uchenmesser Dada durch die letzte weimarer Bierbauchkulturepoche Deutschlands"' (1919/1920) ver\-wen\-det H\"och als auff\"alligste Buchstabenelemente Dada-Slogans. In Irenes Wahr\-neh\-mung sind es Wer\-be\-spots, die in der von ihr erlebten Antiutopie als satirisches Moment fungieren und die Utopie des Westens auf destruktive Weise entlarven. F\"ur mich scheint Irene die kleine tanzende Figur auf einem Bein in der Mitte von H\"ochs Collage zu sein -- ein Harlekin\textsuperscript{\tiny *} -- deren Kopf vom Hals getrennt nach links wegfliegt. \textit{Fliegende auf einem Bein}, das ist wie die \textit{Bl\"atter} durchsetzt von Satire, denn Irenes Blick be\-steht aus Brechungen und Unentscheidbarkei\-ten: 

\singlespacing
\begin{quote}
"`Bl\"atter wie Laub, oder Bl\"atter wie Papier, fragte sie."' (S. 110)
\end{quote}
\onehalfspacing 

Und einige Zeilen weiter: 

\singlespacing
\begin{quote}
"`In dem anderen Land gibt es zwei verschiedene W\"orter f\"ur Bl\"atter. Ein Wort f\"ur Laub und ein Wort f\"ur Papier. Dort muss man sich entscheiden, was man meint."' (S. 110)
\end{quote}
\onehalfspacing 

Drehmoment 2: M\"ullers teils monstr\"ose sprachliche Bilder funktionieren \"ahnlich wie bei Karikaturen, wo Bedrohlichkeit und Grauen als rei{\ss}erisches Mittel eingesetzt werden. M\"ullers Bildeffekte wirken auf mich an vielen Stellen so \"ubersteigert wie es die karikaturistische \"Uberzeichnung des Abgebildeten im Sinn hat, womit mir als Betrachter\textsuperscript{\tiny *} zugesetzt werden soll. Das Unwohlsein r\"uckt einem bei der Lekt\"ure sp\"urbar zu Leibe, wenn man zum Beispiel liest: 

\singlespacing
\begin{quote}
"`Es war eine Stille wie zwischen Hand und Messer gleich nach der Tat."' (S. 35) 

"`Irene f\"uhlte sich wie begraben.\\
Ihre Lider wurden l\"anger. Reichten f\"ur das ganze Gesicht.\\
F\"ur das ganze Zimmer reichten Irenes Lider."' (S. 44) 
\end{quote}
\onehalfspacing

Drehmoment 3: Satire ist darauf angelegt, eine Parallele zwischen gegens\"atzlichen Standpunkten herzustellen und angebliche Unterschiede in einem solchen Ma{\ss}e zu verfremden, dass pl\"otzlich eine N\"ahe zwischen bis\-her entfernt Gedachtem entsteht, Irene denkt zum Beispiel: 

\singlespacing
\begin{quote}
"`B\"ose, das w\"ar gut."' (S. 111)
\end{quote}
\onehalfspacing

V\"ollig neu liest sich aus diesem Blickwinkel die Bewegung, die Irene zwischen den Teilen ihrer eigenen Collage wahrnimmt: 

\singlespacing
\begin{quote}
"`Die Verbindungen, die sich einstellten, waren Gegens\"atze. Sie machten aus allen Photos ein einziges fremdes Gebilde. So fremd war das Gebilde, da{\ss} es auf alles zutraf."' (S. 50)
\end{quote}
\onehalfspacing 

Drehmoment 4: In dieser Paradoxie wird ein Scheitern von Strategien dargestellt, ebenfalls ein satirisches Mittel. H\"aufig bezieht sich die sprechende Position mit ein, so auch Irene, die sich mit dem Anfertigen einer Collage und dem Absenden von Karten und Briefen \"uber etwas Klarheit verschaffen will und dabei Vorhandenes umwandelt f\"ur neue Blickwinkel. Sie versucht es mit dem Rekombinieren von Elementen. Die Elemente machen aber nicht mit, denn in Irenes Selbst\-verst\"andigungsprozess spielen zwar Gedanken, Bilder, Gegenst\"ande und S\"atze eine glei\-cherma\-{\ss}en wichtige Rolle, aber alle scheinen auf alles zuzutreffen. Irene t\"auscht sich. So geht es nicht. Dass sie sp\"ater eine Person auf der Collage als den Diktator zu erkennen meint, mit dem R\"ucken zu ihr, tr\"agt zur Befreiung aus ihrer Lage nichts bei.

Drehmoment 5: Satire be\-steht nicht zu\-letzt in einer punktgenauen Dosierung von Wahnsinn im Inneren einer vern\"unftigen Argumentation. \textit{Reisende auf einem Bein} ist als sch\"ussige Darstellung eines \mbox{selbst}geleiteten Prozesses zur \"Uberwindung ei\-ner Traumatisierung interpretiert worden. Eine Entwicklung scheint sich immerhin darin anzudeuten, dass Irene gegen Ende der Erz\"ahlung eine Taubenfeder und ein beschriebenes Blatt Papier als Postsendung an sich \mbox{selbst} addressiert, dass sie sich -- statt anderen Karten zu schrei\-ben -- nunmehr sich \mbox{selbst} zuwendet. Al\-ler\-dings war Irene zur Taubenfeder das Wort "`Taubenm\"order"' eingefallen, als sie "`sich mit der Feder \"uber den Hals"' strich (S. 174), die sie am Morgen im Bad gefunden hatte. Auf eine Karte schreibt sie "`Taubenm\"orderin"', was den Eindruck ei\-ner Selbst\-\-an\-rede verst\"arkt, da Irene f\"ur die Anrede ein Femininum verwendet. Als Assoziation zu Taubenfeder und M\"orderin kommt mir: dass Tauben auch als Brieftauben eingesetzt werden, also als Kommunikationsmedium, und dass eine Taubenfeder bei Irene landete, die zwar von einer Taube ist, abgeworfen wurde, also tot ist, aber ein Schreibger\"at: eine leichte "`Feder zwischen Daumen und Zeigefinger."' (S. 174) Mit dem Wort "`Zwischen"' hatte der Text begonnen: "`Zwischen den kleinen D\"orfern unter Radarschirmen, die sich in den Himmel drehten, standen Soldaten. \mbox{Hier} war die Grenze des anderen Landes gewesen."' (S. 7) Als Irene begann, ihre Collage zu verfertigten (S. 50), um sich vernunftgesteuert eine Gelegenheit zu verschaffen, der eigenen Traumatisierung in einem kreativen Prozess beizukommen, hatte sie Abbildungen ausgeschnitten, aus Zeitungen. Pl\"otzlich war ihr wie im Wahn neben einem der Bilder ihr Daumennagel in den Blick gekommen (S. 49), aber jetzt, beim Halten der Taubenfeder, ist ein Zeigefinger mit im Spiel. Ich f\"uhle mich zu eigenen radikal umgest\"ulpten Denk\-ans\"at\-zen aufgefordert (\cite{Renneke2008}:313) und sp\"ure Irenes Zeigefinger im R\"ucken, als mein Blick sich in einer punktgenauen Dosis erneut Hannah H\"ochs satirischem Schnitt mit dem K\"uchenmesser zuwendet, aber nicht in der Stille nach der Tat, sondern in der Tat, mit einem so lauten Gel\"achter, dass der Kopf nach links wegfliegt.

W\"ahrend meiner Lekt\"ure der ersten beiden Drittel des Buches hatte ich das Spiel wohl nicht verstanden und dennoch mitgespielt. Der Humor des Textes, so tiefgr\"undig, war f\"ur mich fast nicht erkennbar gewesen. Erst nach meinem Staunen hatte ich aus einem anderen Anlass einen Beitrag von J\"urgen Wertheimer wieder zur Hand genommen, in dem 2002 ein \textit{Erlesnis} zu M\"ullers Collagen folgenderma{\ss}en geschildert wird: 

\singlespacing
\begin{quote}
"`Und so kommt es, dass man, w\"ahrend man die Serie der Katastrophenbilder durchgeht, \"uber die Spr\"unge stolpert und in die Gruben f\"allt, in den L\"ucken stecken bleibt und sich an den Wortr\"andern aufschrammt, pl\"otzlich und immer st\"arker zu lachen beginnt; und die Illustrationen, die den Text umtanzen und in ihm und mit ihm zu spielen scheinen, tragen in ihrer anti-illustrativen Unernsthaftigkeit nicht wenig dazu bei, den Trag\"odien ihre tragische Arroganz zu nehmen, hinter der doch meistens nur Dummheit steckt."' (\cite{Wertheimer2002}:82-83)
\end{quote}
\onehalfspacing

Aufgrund meiner Reaktion auf \textit{Fliegende auf einem Bein} scheint mir, dass ein solcher Gel\"achtereffekt auch mit ausschlie{\ss}lich sprachlichen Mitteln herbeizuf\"uhren war. In ihrer Komplexit\"at geh\"oren die intrikaten Ambiguit\"aten von M\"ullers Sprache nun unmittelbar zu meinem \textit{Erlesnis}: Wie es sich anf\"uhlt, gerade erkannt zu haben, auf welchem Wege es gelingen k\"onnte, mit etwas Aussichtslosem fertigzuwerden. Durch Satire. 

Lese ich erneut "`Reisende auf einem Bein"', den von mir vor einem Jahr begonnenen Wi\-ki\-pe\-dia-Eintrag, sehe die Referenzierungen aus 20 For\-schungsbeitr\"agen und Re\-zen\-si\-o\-nen, alle in bitterem Ernst, und kann ich mich auch \mbox{hier} nicht halten vor Lachen. Ich bewundere die fliegende Satire auf einem Bein, die Machtf\"ulle in Herta M\"ullers doppelb\"odigem Sprechen, und ich reagiere auf mein Empfinden von Bewunderung ausgelassen und befreit. Eine \"ahnliche Reaktion beschreibt A. L. Kennedy: \href{http://www.theguardian.com/books/booksblog/2013/feb/05/al-kennedy-home-thoughts-creative-writing}{"`Here is the volume of Raymond Carver I threw across the room when I was a student because it was so amazing, so tender with broken people"'} (\cite{Kennedy2013}). 

In \textit{Reisende auf einem Bein} arbeitet M\"uller also mit dem Dada-Prinzip einer gestaltenden text-bildlichen Rekombination von Vorhandenem, aber hier al\-lein mit sprach\-li\-chen Mit\-teln und in einer Variante gedoppelten satirischen Ernstes. 

Ein Remix-Prinzip liegt auch dem Arbei\-ten mit Texten und Bildern bei Wi\-ki\-pe\-dia zugrunde, wo offene Lizenzen es erm\"oglichen, ohne Nachfrage in vollem Umfang -- und nicht nur im Rahmen des Zitatrechts f\"ur Werke, die eine eigene sogenannte Sch\"opfungsh\"ohe aufweisen -- die Arbei\-ten an\-de\-rer nach Belieben abzuwandeln, mit Namensnennung -- und so gesehen ebenfalls fliegend auf einem Bein.

\newpage

\subsubsection{Gezanke in \flq rein GOLD\frq, Debatten im Kopf (Jelinek)}
\label{subsubsec:5.4.4}

\singlespacing
\begin{quote}
{\footnotesize "`Die Stimmung im Schlo{\ss} unertr\"aglich. Gez\"ank."' Elfriede Jelinek, \textit{rein GOLD}}
\end{quote}

\onehalfspacing

Meine Lekt\"ure von \textit{rein GOLD} begann mit Debatte, in meinem Kopf, und gelegentlich unterbrach ich deswegen das Lesen -- oder auch nicht. Sie klang etwa folgenderma{\ss}en: "`Strei\-ten ist wichtig, aber muss es in diesem Ton sein?"', -- und daraufhin: "`Ja, muss es, denn so kommt es dir vor. Du liest es doch als Gezanke, f\"ur dich klingt es so"', -- die erste Stimme weiter: "`... und das sei\-tenlang ohne Pause!"' -- ... dann wieder die andere: "`Wenigstens schrei\-ben die Wi\-ki\-pe\-dia-Streith\"ahne ihre Beitr\"age nicht sei\-tenlang."' -- "`Naja, die w\"urde wohl niemand lesen."' -- "`\mbox{Hier} ist es nur ein Dialog, dort ein Multilog."' -- "`Aber macht es das ertr\"aglicher?"' -- "`Nee, nur auf andere Weise aufreibender: Dort wechseln mal die Angriffspunkte, mal die Identit\"at, mal die Spielwiese innerhalb der Plattform, und dadurch auch diejenigen, die zuh\"oren und mitmachen."' -- "'Aber sie zeigen ebenfalls enorme Ausdauer."' -- "`... und mich musst du da mit einrechnen."' -- "`Wie kommst du eigentlich darauf, diese beiden Szenarien zu vergleichen?"' -- "`Das habe ich mich auch gerade gefragt. Scheint jedenfalls zu passen: Du hast dich hineinziehen lassen."' -- "`Sind wir uns jetzt tats\"achlich mal einig?"' -- Beim Lesen hatte ich manches Mal M\"uhe, meine Aufmerksamkeit zum Text zur\"uckkehren zu lassen. Auf faszinierende Weise traf das, was ich in theo\-retischem Diskurs bei Chantal Mouffe bez\"uglich an\-ta\-go\-nis\-ti\-schem Agonismus gelesen hatte, mit meiner Lekt\"ure z\"ankischer Aushandlungen in Jelineks sprachk\"unstlerischer Gestaltung zusammen. Das Lesen von Jelineks Stimmen brachte in meinem Kopf auf einer parallelen Tonspur mit erinnerten Kostproben der Diskurse von Wi\-ki\-pe\-dia-Streith\"ahnen einen polyphonen Satz zum Klingen. Das Erlebnis dieser Dreierkombination aus Mouffes \textit{Agonistics}, Jelineks \textit{rein GOLD} und den we\-ni\-ger freund\-lichen Aus\-handlungen bei Wi\-ki\-pe\-dia zusammen mit der in meinem Kopf laufenden Zweierdebatte hat mich auf eine h\"ochst merkw\"urdige Weise be\-rei\-chert: Bei der Lekt\"ure erschien es mir wahrscheinlicher zu werden, dass ich eine l\"angerere Debatte bei Wi\-ki\-pe\-dia w\"urde durchstehen k\"onnen -- so gut wie Jelineks Stimmen w\"urde wohl kein Kollektiv zanken k\"onnen. Ich traute mir also zu, \mbox{selbst} eine Debatte anzuzetteln. Mein Empfinden war, dass ich f\"ur etwaige Streitigkei\-ten gen\"ugend an "`Wie es sich anf\"uhlt"'-Wissen aus Jelinek und Mouffe (\cite{Mouffe2013}) erworben hatte, nicht zu\-letzt durch das in Gang gesetzte Konzert in meinem Kopf.

Im Folgenden gehe ich "`auf Konzertreise"' und h\"ore mir auf anderen B\"uhnen weitere Szenarien an, die ich mit denjenigen Gestaltungsaspekten vergleiche, die ich in \textit{rein GOLD} wahrnehme. Ich begebe mich nach Mesopotamien sowie zu den Propheten\textsuperscript{\tiny *}, in die Kulturen des Mittelalters und zu Wagners Tetralogie, in Utrecht und Berkeley zum essayistischen Schrei\-ben \"uber die Phase der westlichen Philosophie zwischen 1970 und 1985 sowie an meinen Schreibtisch in S\"udwestdeutschland, wo eine be\-stimmte Tonspur in \textit{rein GOLD} abschlie{\ss}end erneut angeh\"ort wird.

"`Zanken"' kommt vermutlich von "`zerrei{\ss}en"' und k\"onnte ein Ableitung von "`Zahn"' sein, womit das Verb urspr\"ung\-lich m\"oglicherweise beschreibt, wie Hunde mit ihren Z\"ahnen etwas in verschiedene Richtungen zerren. Abgeleitet davon gibt es die Abstrakta "`Zank"' und "`Gez\"ank"' sowie das Adjektiv "`z\"ankisch"' (\cite{Seebold2002}:1003). F\"ur die Auseinandersetzung in \textit{rein GOLD} k\"ame eine Beschreibung mit den beiden Re\-dens\-ar\-ten "`jemandem auf den Zahn f\"uhlen"' und "`jemandem den Zahn ziehen"' in Frage, weil B versucht, von W etwas he\-rauszufinden und weil sie ihn von etwas abbringen will. He\-rausfinden will sie, ob er so gehandelt hat wie sie das vermutet, und abbringen will sie ihn davon, zu meinen, dass er auch weiterhin ohne eigenes Risiko so handeln k\"onne wie bis\-her. Bei dem von mir im Titel verwendeten Begriff "`Gezanke"' handelt es sich um eine Abwandlung von "`Gez\"ank"', die mein Genervtsein zum Ausdruck bringt. Ich beschreibe die Redeweise in \textit{rein GOLD}: Es wird eine Auseinandersetzung von l\"angerer Dauer in z\"ankerischer Art gef\"uhrt. 

Zu einem Streitgespr\"ach kann es kommen, wenn zwei Sprecher\textsuperscript{\tiny *} zusammentreffen und die andere Person jeweils nur als H\"orer\textsuperscript{\tiny *} gesehen wird, nicht aber als Dialogpartner\textsuperscript{\tiny *}. Argumentiert wird dann im Grunde monologisch, das Argumentieren ist aber als adressatenspezifisches Handeln anzusehen (\cite{Rudolf1998}). \textit{rein GOLD} enth\"alt ausgedehnte Redebeitr\"age zweier Beteiligter, in denen es zwischendurch Signale gibt, aus denen zu schlie{\ss}en ist, dass die jeweils sprechende Person noch merkt, dass die andere Person da ist und m\"oglicherweise zuh\"ort. Al\-ler\-dings wird \"uber Anwesende gelegentlich auch in der dritten Person gesprochen, was als Signal in die entgegengesetzte Richtung zu deuten w\"are: Jemand redet dann nicht mit der anwesenden anderen Person, sondern spricht vor sich hin oder zum Publikum oder beides.

Wettstreit mittels abwechselnder Rede zwischen zwei Standpunkten wurde auch in der sumerischen Kultur schon als relevant genug angesehen, um schriftlich fixiert zu werden. Vom Hof der K\"onige der dritten Dynastie von Urim Ende des dritten Jahrtausends vor christlicher Zeitrechnung sind professionelle formale Debatten wie "`The debate between Sheep and Grain"' oder "`The debate between Bird and Fish"' in sumerischer Sprache \"uberliefert. Jeremy A. Black, Graham Cunningham, Eleanor Robson and G\'{a}bor Z\'{o}lyomi heben hervor, dass sich die Beteiligten scharfe und bissige Wortgefechte lieferten, an deren Ende eine Siegesentscheidung durch einen Gott oder den K\"onig erwartet wurde (\cite{Blacketal2004}:225-235). In \textit{rein GOLD} ist ein Gott \mbox{selbst} am Gezanke beteiligt. Wer gewonnen hat, wird \mbox{hier} nicht von einer dritten Instanz entschieden, sondern scheinbar gar nicht. Eventuell handelt es sich bei \textit{rein GOLD} nicht um einen Wettstreit: B und W zanken m\"oglicherweise, ohne dass einer\textsuperscript{\tiny *} gewinnen will. Das zweitgenannte sumerische Rededuell, zwischen Vogel und Fisch, weist in der Mitte einen narrativen Teil auf, in dem geschildert wird, wie Fisch das Nest von Vogel umdreht und die Eier zerst\"ort und wie Vogel daraufhin Fisch den Laich wegschnappt. Sie einigen sich, zur Kl\"arung bei Gott Enki ein Gerichtsverfahren anzustrengen. In der Entscheidung durch einen als besonders gerecht geltenden K\"onig wird Vogel zum Gewinner ernannt. Ein Gro{\ss}teil des Wettstreits be\-steht darin, sich gegensei\-tig aufgrund k\"orperlicher Unterschiede und an\-de\-rer Eigenhei\-ten zu beleidigen: Fisch habe keinen echten K\"orper mit Gliedma{\ss}en und stinke, Vogel sei eitel, hinterlasse \"uberall seinen Kot und nerve die Menschen mit seinem Kreischen und Qu\"aken. Zum letztgenannten Aspekt l\"asst sich f\"ur \textit{rein GOLD} beobachten, dass pers\"onliche Angriffe sei\-tens W gegen B erfolgen. Er verh\"ohnt zum Beispiel ihre Beschwerden und Beleidigungen in Bezug auf die ihr zugewiesene Rolle als Frau und er \"au{\ss}ert sich geringsch\"atzig gegen\"uber der beruflichen T\"atigkeit von B, dem Schrei\-ben. B hingegen nutzt gegen\"uber W andere Formen von Beleidigung, etwa wenn sie W's Burg herabsetzend als Einfamilienhaus be\-zeichnet (\cite{Vill2013}:76). Bei der "`Debatte von Vogel und Fisch"' handelt es sich um h\"ofische Unterhaltungskultur. Letzt\-lich ist eine G\"otterburg auch ein Hof, aber in \textit{rein GOLD} wird die Art und Weise des Wortgefechts aus Sicht von W implizit gerechtfertigt als im Rahmen der Familie traditionell \"ublich: Als W auf die Dis\-kus\-si\-onslose Hinrichtung seines Onkels Hermann zu sprechen kommt, vermerkt er, es sei innerhalb der Familie ohne interessante Dis\-kus\-si\-on nie gegangen (\cite{Jelinek2013}:92). W's Be\-zeichnung "`interessante Dis\-kus\-si\-on"' k\"onnte man al\-ler\-dings als eine euphemistische Beschreibung dessen ansehen, was B eingangs "`Gez\"ank"' (8) \mbox{nennt} und was ich als Teil der sprachk\"unstlerischen Gestaltung des Wortwechsels ansehe.

N\"achste Station der Konzertreise. Betrachtet man \textit{rein GOLD} wiederum aus dem Blickwinkel der gelehrten Praxis und schulischen \"Ubung des dialektischen Disputs im Mittelalter, l\"asst sich die Dreierkonstruktion aus Prophet\textsuperscript{\tiny *}, B und W mit einem weiteren Rollenaspekt beschrei\-ben. Die mittelalterliche Disputier-\"Ubung beginnt, indem ein Lehrer\textsuperscript{\tiny *} zwei Sch\"ulern\textsuperscript{\tiny *} eine Aufgabe in Frageform stellt. Oft spricht der Opponent\textsuperscript{\tiny *} zuerst und der Proponent\textsuperscript{\tiny *} antwortet ihm. Abschlie{\ss}end formuliert der Lehrer\textsuperscript{\tiny *} einen Kompromiss, entl\"asst die Disputanten\textsuperscript{\tiny *} und antwortet abschlie{\ss}end dann \mbox{selbst} auf die Einw\"ande, die durch den Opponenten\textsuperscript{\tiny *} vorgebracht worden sind (\cite{Perigot2014}). Bei \textit{rein GOLD} findet sich zu Beginn keine erkennbare Frage, sondern nur ein Bericht \"uber Gezanke, und zwar aus Sicht einer der beiden Beteiligten, durch B. Deren Rolle stellt sich -- der mittelalterlichen Form nach -- als die des Opponenten\textsuperscript{\tiny *} he\-raus, denn sie beginnt. Wer argumentiert in diesem Fall auf einen Kompromiss hin und antwortet auf die Einw\"ande, die durch den Opponenten\textsuperscript{\tiny *} vorgebracht worden sind? Der Leser\textsuperscript{\tiny *}, die \"of\-fent\-liche Meinung, die Rezensenten\textsuperscript{\tiny *} und Theater/Opern\-kri\-ti\-ker\textsuperscript{\tiny *}?

Es gibt einen weiteren, \"ahnlichen und ebenfalls streitkulturellen Aspekt aus der prophetischen Li\-te\-ra\-tur: In \textit{1. K\"onige 22} stellt sich ein Einzelner mit einer hintergr\"undigen Vision, die einen katastrophalen Ausgang sig\-na\-li\-siert, gegen die Mehrheit, die der positiven Version ihres Anf\"uhrers Beifall klatscht: "`Eine Dis\-kus\-si\-on gibt es nicht. Die Standpunkte stehen sich Dis\-kus\-si\-onslos gegen\"uber. Die Ereignisse \mbox{selbst} entscheiden dann \"uber die echte Prophetie"' (\cite{SeyboldUngern-Sternberg2011}:291-292). Jelinek l\"asst \textit{rein GOLD} enden mit W's pseudo-prophetischer Worth\"ulse: "`Mal sehn, was draus wird."' Auf der n\"achsten Seite des Buches folgt ein Abspann mit mehr oder minder genauen Angaben zu verwendeten Quellen und zur Anregerin des Werks, der Bayerischen Staatsoper M\"unchen. Sind die Informationen auf dieser Seite als ein Teil der Antwort zu lesen, als Andeutung von Ereignissen, zum Beispiel einer Auff\"uh\-rung? Jelinek hat im Titel ihres Werks \textit{rein GOLD. ein b\"uhnenessay} als Genre vermerkt, dass es sich um einen Essay handelt, genauer: um einen "`B\"uhnen"'essay, also f\"ur gesprochene Form, als Auff\"uh\-rung verstanden. Es wird ein Tochter-Vater-Verh\"altnis zur Auff\"uh\-rung gebracht und zwar in Form eines Wortgefechts mit Ge\-zan\-ke. Soll es eine Prophetie werden, die sich fl\"uchtig und transitorisch, aber in einer Art Verk\"orperung inszenieren l\"asst, in der "`Doppelheit von ph\"anomenalem Leib und se\-mio\-tischem K\"orper"' (\cite{Fischer-Lichte2012}:62)?

Unabh\"angig von der Genrebe\-zeichnung im Untertitel -- und ein Genre im Titel des Werks anzugeben (m\"oglicherweise mit der Intention, Skepsis hervorzurufen) kann ebenfalls in einer langen Tradition gelesen werden (\cite{Perigot2014}) -- ist \textit{rein GOLD} ein li\-te\-ra\-risches Werk in Buchform. Der Schutzumschlag zeigt einen Goldbarren auf schwarzem Grund (\textit{rein GOLD} 2013), das Buch \mbox{selbst} hat einen goldfarbenen Einband. Selten wird die Signifikanz der Materialit\"at eines gedruckten Werkes schon mittels des Designs f\"ur den Schutzumschlag der Hardcoverausgabe so ausdr\"ucklich vor Augen gef\"uhrt wie in diesem Fall: in und mit diesem Werk, f\"ur und durch dieses Werk. Darauf brachte mich eine Beschreibung, wonach die letzte Sammlung im \textit{Zw\"olfprophetenbuch}, unter dem Namen "`Maleachi"' \"uberliefert, "`aus so\-ge\-nann\-ten Dis\-kus\-si\-onsworten be\-steht, wobei der Prophet einen Dialog zwischen Gott und seiner Gemeinde simuliert"' (\cite{SeyboldUngern-Sternberg2011}:292). Wenn in \textit{rein GOLD} dieser Gott W hei{\ss}t und seine Gemeinde w\"are anwesend in Gestalt von B, dann h\"atte sich Autor\textsuperscript{\tiny *} Jelinek mit diesem Werk die Rolle des Propheten\textsuperscript{\tiny *} erschrieben. Durch und f\"ur dieses Werk und in diesem Werk. Und das k\"unstlerische Artefakt \textit{rein GOLD} w\"are gerade in seiner Materialit\"at ein prophetisches Buch -- oder soll sogar als eines der bekannten prophetischen B\"ucher angesehen werden. Denn "`das Buch"' scheint jenes Gold zu sein, um das auf verschiedenen se\-mio\-tischen, diskursiven und thematischen Ebenen gezankt wird. 

\textit{rein GOLD} ist ein Essay mit Sprecher\textsuperscript{\tiny *}wechsel, \"ahnlich wie der Beitrag, der im Folgenden herangezogen wird, um weitere Aspekte von Jelineks Werk beschreibbar zu machen. In "`Out of Bounds: Philosophy in an Age of Transition"' von Judith Butler und Rosi Braidotti stehen JB und RB f\"ur die beiden nicht-fiktiven Per\-so\-nen, die als Autoren\textsuperscript{\tiny *} des Essays firmieren (\cite{ButlerBraidotti2010}). B und W hingegen firmieren nicht als Autoren\textsuperscript{\tiny *}, sondern als Figuren, mit denen trans- und intertextuelle Bezugnahmen denkbar werden, zum Beispiel zu Br\"unnhilde und Wotan in Richard Wagners Tetralogie oder zu Elfriede und Friedrich, Tochter und Vater in der Bio\-gra\-fie Elfriede Jelineks, oder zu weiteren vergleichbaren Konstellationen. Im Fall des Vergleichs\-essays leben zum Zeitpunkt der Publikation beide Sprecher\textsuperscript{\tiny *} noch -- und wie ist es im Fall von \textit{rein GOLD}? Haben B und W je gelebt?

In beiden Essays wird abwechselnd gesprochen und die Sprecher\textsuperscript{\tiny *} haben dieselbe Anzahl an Beitr\"agen. Es beginnen jeweils andere als diejenigen, die enden. Prozentual macht Bs Anteil in \textit{rein GOLD} etwa 45\% aus, W redet also 10\% mehr als B. Im Essay "`Out of Bounds"' be\-steht Zeilengleichheit, al\-ler\-dings sind in RB's pers\"onlichen Passagen 24 Zeilen an Fu{\ss}noten zu finden und in den Passagen von JB keine. "`Out of Bounds"' hat etwa ein Neuntel der L\"ange von \textit{rein GOLD}. Bei \textit{rein GOLD} zeigt das Abwechseln die Abschnitte an, bei "`Out of Bounds"' entstehen Abschnitte beim Wechsel zwischen akademischem Ton und pers\"onlichen Passagen und umgekehrt. Diese Unterscheidung wird im Vorspann angemerkt mit: "`[T]his chapter will mix personal voices and individual accounts with the more standard academic tone. The sections related to the former will appear in italics"' (307). "`Out of Bounds"' beginnt im "`more standard academic tone"', also mit einer Passage, in der keine der beiden Autoren\textsuperscript{\tiny *} pers\"onlich schreibt, und der Essay endet mit einem Beitrag von RB, also mit Rosi Braidotti als He\-rausgeber\textsuperscript{\tiny *} eines Bandes, den dieser Essay beschlie{\ss}t. RB ist also Gastgeber\textsuperscript{\tiny *}. Teilen sich diese Rolle in \textit{rein GOLD} beide, liegt die Aufgabe des Gastgebers\textsuperscript{\tiny *} bei eine\textsuperscript{\tiny *} Dritten -- oder gibt es diese Funktion gar nicht? 

Wo befinden sich B und W, w\"ahrend sie sprechen? Stehen sie auf einer B\"uhne? Oder chatten sie im Netz? In \textit{rein GOLD} gibt es einen Vorspann von zwei Zeilen im Stile der "`Dramatis personae"' eines Dramentextes. In "`Out of Bounds"' wird diese Funktion durch die Angabe der Namen der Autoren\textsuperscript{\tiny *} erf\"ullt. In \textit{rein GOLD} er\"offnet B, erl\"autert ihre Haltung und den Anlass. W am Schluss ist nurmehr mit sich \mbox{selbst} befasst,\,will "`Mal sehn,\,was draus wird."'\,Die Hoffnung, die am Ende des Vergleichs­-essays von RB ge\"au{\ss}ert wird, mischt von der Redehaltung her "`wir"' und "`ich."' Da es im Vergleichsessay gemeinsam verfasste Teile gibt, wechseln sich pers\"onliche Passagen von RB und JB mit gemeinsam verfassten Passagen ab. W und B sprechen nirgends "`denselben Text"', wohl aber gemeinsam gegeneinander. Zwar schildern auch Butler und Braidotti Auseinandersetzungen, aber aus r\"uckblickender Perspektive, und es geht um Philosophie. Gezanke wird also in einen Bericht gekleidet. Ob die Autoren\textsuperscript{\tiny *} untereinander gezankt haben, wird nicht thematisiert. B und W hingegen zanken \mbox{selbst}, auch dann, wenn sie \"uber andere sprechen, sie scheinen aber verkleidet auf einer B\"uhne zu stehen. Wessen Gezanke f\"uhren sie auf?

RB und JB f\"uhren herk\"ommlicherweise verstanden einen Dialog, denn sie sind mit einem Thema befasst, zu dem jede\textsuperscript{\tiny *} von beiden etwas beisteuert: Wie sich die Wirkung von philosophisch einflussreichen Institutionen und Akteuren\textsuperscript{\tiny *} sowie deren Positionen und Handlungen in der Zeit von 1980 bis 1995 beschrei\-ben lassen und wie RB und JB \mbox{selbst} daran teilhatten. Die Themen in \textit{rein GOLD} sind un\-gleich schwe\-rer auszumachen, sie sind vielf\"altig und verschr\"ankt und es gibt im Unterschied zum Vergleichsessay keine Zwischen\"uberschriften. In "`Out of Bounds: Philosophy in an Age of Transition"' stehen diejenigen, die sprechen, in einem fachkollegialen Verh\"altnis zueinander, w\"ahrend es sich bei B und W um Tochter und Vater handelt, wie aus \mbox{Selbst}\-aussagen zu schlie{\ss}en ist. B: "`und ich sage dir auf den Kopf zu, Papa, du h\"attest das bl\"ode Schloss nicht gebraucht, wenn du es nicht bezahlen willst"' (28), W: "`Kind, h\"or zu"' (124) und W: "`jetzt spreche ich zu meiner Tochter, und Sie halten solange die Schnauze!"' (202). \mbox{Selbst} wenn B und W einander anreden, sprechen sie eher gegeneinander. W: "`Du sollst schlafen, Kind!"' (56) oder B: "`das hast Du mir versprochen, Papa!"' (182), B: "`Du wei{\ss}t, was ich meine, Papa"' (34) oder W: "`Du bist eine Mi{\ss}geburt"' (82) oder B: "`Du, Vater, du musst"' (169). Als W sich in seinem zwei\-ten Redebeitrag an seinen Onkel Herrmann erinnert, der von einem freundlichen Henker "`nicht ohne jene vorherige interessante Dis\-kus\-si\-on, ohne die es in unserer Familie ja nie gegangen"' (92) umgebracht worden sei, ist im Nebensatz von einer Familienkultur die Rede, in der notwendig diskutiert wurde. Unfreundlich w\"are ein Henker, der nicht erst diskutiert h\"atte, r\"asonniert W. Al\-ler\-dings f\"uhren B und W ein Gezanke auf, in dem sich Tochter und Vater vom Redestil her kaum unterscheiden und das keineswegs freundlich wirkt. In den Dialogpassagen des Vergleichsessays reden RB und JB einander nicht an, das abwechselnde Sprechen ist nicht aufeinander gerichtet und doch geht es um Gemeinsames.

In \textit{rein GOLD} wird auf thematischer und stilistischer Ebene unter anderem Richard Wagner zitiert. Zum Beispiel wird im Titel \textit{rein GOLD} \textit{Das Rheingold} von Wagner persifliert. Wagner wird im Abspann explizit referenziert, aber Verlautbarungen des NSU zum Beispiel nicht explizit, sondern in der Reihung "`Sonst nichts. Ein paar Zeitungen. Alles nichts."' Im Vergleichsessay referenzieren Judith Butler und Rosi Braidotti eigene Schriften, in denen sie sich bereits ausf\"uhrlicher autobiografisch ge\"au{\ss}ert haben. An einer Stelle zitieren sie -- mit dem begr\"undenden Vorspann "`requires more cautious phrasing"' -- aus einem Gespr\"ach zwischen Michel Foucault und Gilles Deleuze, \"alteren einflussreichen Fachkollegen\textsuperscript{\tiny *}. In deren Beitrag von 1972 wird abwechselndes Sprechen in \"ahnlicher Weise pr\"asentiert: Foucault beginnt, Deleuze endet (Foucault et al. 1972). Bei Jelinek wird allem Anschein nach ein begonnenes Gespr\"ach fort\-ge\-setzt, in k\"unstlerischer Form, bei Butler und Braidotti handelt es sich um den konkreten Anlass, gemeinsam einen Beitrag als abschlie{\ss}endes Kapitel f\"ur eine Philosophiegeschichte von 1980 bis 1995 zu verfassen. JB und RB kennen einander schon, als Fachkollegen\textsuperscript{\tiny *}, die dar\"uberhinaus manche Bezugspunkte gemeinsam haben. An einer Stelle wird gesagt: "`Both authors participated in the Coll\`{e}ge activities: one in the very year of its foundation, the other in recent years."' (325) -- was sich auf das 1983 in Frankreich gegr\"undete Coll\`{e}ge International de Philosophie bezieht. An pers\"onlichen Angriffen gibt es keine. Jelineks B und W kannten sich zu Beginn bereits. Sie werden als Tochter und Vater vorgestellt, die sich nicht zum ersten Mal begegnen, gegenseitig einige Schw\"achen kennen und dies in der Aus\-ein\-andersetzung f\"ur pers\"onliche Angriffe nutzen. \textit{rein GOLD} ist in eine mythische Nibelungen-Gegenwart gekleidet, deren Zeitspanne als unbestimmt wahrgenommen werden kann. In diesem Horizont werden Vorw\"urfe zu anhaltender Ausbeutung und Missachtung ge\"au{\ss}ert. In "`Out of Bounds"' ist das Thema die Phase von 1980 bis 1995 und Butler und Braidotti legen ihr Verst\"andnis des eigenen Handelns eingangs dar: "`Thirdly, we argue that the emergence of new experimental modes and venues of thinking combine to form both creative tensions and contradictions that have left a pro\-ble\-ma\-tic legacy for future generations of philosophers. Writing about them is a way of making ourselves accountable for this legacy."' (307) Hingegen scheint es bei \textit{rein GOLD} kein Thema zu sein, zuk\"unftigen Dritten gegen\"uber Rechenschaft abzulegen.

\textit{rein GOLD. ein b\"uhnenessay} lese ich als den Versuch (Essay), etwas auf eine B\"uhne zu bringen, dessen Form und/oder Le\-gi\-ti\-mi\-t\"at als B\"uhnenwerk unklar zu sein scheint. Die Geste des Werks macht auf mich den Eindruck, in seinem Genre ambig bleiben zu wollen, \mbox{selbst} wenn die Textform schon im Medium des Buches fixiert worden ist. Dass von \textit{rein GOLD} eine Urlesung stattfand, und zwar vor der Ver\"of\-fent\-lichung des Buches, k\"onnte darauf hinweisen, dass "`das Werk"' formal und medial fluide sein soll. Aufzeichnungen davon sowie Auff\"uh\-rungen und Aufzeichnungen einer dritten Variante gibt es mit dem gleichnamigen stark gek\"urzten Musiktheaterst\"uck. Es w\"are interessant zu \"uberlegen, welche bis\-her als feststehend angenommenen Anteile von \textit{rein GOLD} in einer Hypertextvariante fluide w\"urden, mit dem Web als "`B\"uhne"'.

N\"achste Station. Welche Aspekte von \textit{rein GOLD} werden bei einem Vergleich zwi\-schen der Sachtexte-Plattform Wi\-ki\-pe\-dia und diesem li\-te\-ra\-rischen Werk beschreibbar? Bei Wi\-ki\-pe\-dia f\"uhlt es sich so an, als ob alle schon immer geredet h\"atten, und in \textit{rein GOLD}? \mbox{Hier} wird ebenfalls ein bereits laufendes Gespr\"ach fort\-ge\-setzt, denn B er\"offnet mit: "`Ich versuche also zu pr\"azisieren, das ist ein etwas delikates Gebiet, es f\"allt mir schwer"' (7). Wie lange diese Auseinandersetzung schon andauert, wird nicht gesagt, aber jetzt scheint der Zeitpunkt gekommen, zu dem wir als Rezipienten\textsuperscript{\tiny *} eingeweiht werden sollen: Vorhang auf! Wenn bei Wi\-ki\-pe\-dia Aus\-ein\-andersetzungen gef\"uhrt werden, passiert dies ebenfalls auf offener B\"uhne, frei im Web. Bei \textit{rein GOLD} stellt sich nach dem ersten Redewechsel die Szene so dar: Die Gattin Fricka macht Wotan Vorw\"urfe, weil er sich finanziell \"ubernommen hat, die Tochter Br\"unnhilde verh\"ort ihn, stellt ihn zur Rede, und Wotan versucht sich he\-rauszureden. Das Zur-Rede-Stellen klappt anscheinend (W redet), aber das Verh\"or klappt nicht (W redet sich raus). W hat das Schlusswort, an dessen Ende er zwischen der Ank\"undigung seines Abgangs und seiner Lust laviert, weiterhin in Experimertierlaune verschiedenen Reizen zu folgen: "`Mal sehn, was draus wird"' (222). Wolfgang Schmitt und Franziska Sch\"o{\ss}ler beschrei\-ben den Schluss intertextuell vergleichend, indem sie feststellen, dass auch Richard Wagners \textit{G\"otterd\"ammerung} "`unerl\"ost"' ende (\cite{SchmittSchossler2013}:103). Susanne Vill vergleicht ebenfalls, indem sie erw\"ahnt, dass Br\"unnhilde in der Oper in ihrem Schlussgesang Abschied von Wotan nimmt. Den Effekt der Gestaltung der Abfolge der Redebeitr\"age in \textit{rein GOLD} liest Vill so, dass Jelinek die Tochter lediglich verstummen l\"asst, der Vater aber spreche weiter. Vill interpretiert den Schluss dahingehend, dass Jelineks Gestaltung bedeutet, dass B ihre Redemacht aufgibt (\cite{Vill2013}). Sieht man B und W als zwei verschiedene Figuren an, ist Vills Deutung nicht ganz von der Hand zu weisen. Al\-ler\-dings betonen Schmitt und Sch\"o{\ss}ler, es handele sich bei \textit{rein GOLD} um einen postdramatischen Theatertext, Br\"unnhilde und Wotan tr\"aten nicht als Per\-so\-nen auf. In \textit{rein GOLD} w\"urden die Grenzen zwischen den Figuren aufgel\"ost, was durch das sich \mbox{selbst} sch\"opfende Geld der \"okonomischen Entdifferenzierung entspreche. In ihrer Interpretation des Werks gibt es nur eine dramatis persona: das Geld. Dieser Argumentation f\"ur aufgel\"oste Grenzen in Richtung eines Monologs statt eines Dialogs geht im Beitrag von Schmitt und Sch\"o{\ss}ler das Referieren einer Deutung des Br\"unnhilde-Wotan-Verh\"altnisses von George Bernard Shaw voraus, der in seinem \textit{Wagner-Brevier}, im Original \textit{The Perfect Wagnerite. A commentary on The Ring of the Nibelungs} (1898) das Verh\"altnis zwi\-schen Wotan und Br\"unnhilde so beschrieb: "`[W]enn er zu ihr spricht, spricht er nur mit sich \mbox{selbst}"' (Shaw, zitiert bei \cite{SchmittSchossler2013}:105, Fn. 41). Schmitt und Sch\"o{\ss}ler begr\"unden ihren Standpunkt zwei\-tens damit, dass die je\-wei\-li\-gen Textfl\"achen auch von meh\-reren Schauspielern\textsuperscript{\tiny *} gesprochen werden k\"onnten. Vill sieht stilistische \"Ahnlichkei\-ten zum Rapp und es scheint ihr, als ob Jelinek versuche, sich im pausenlosen Reden ihrer \mbox{selbst} zu vergewissern. Unterbrochen werden m\"ochte niemand, sondern "`dass keiner ihm die Macht der aktiven Rede entrei{\ss}e"', so formuliert es Vill mit Bezug auf einen Rapper. 

Keine Unterbrechung kann al\-ler\-dings nur dann gelingen, wenn sie gemeinsam spre\-chen. Vill schreibt "`Jelineks Gespr\"achspartner Br\"unnhilde und Wotan"' und meint damit zwar vermutlich, dass Br\"unnhilde und Wotan einander Gespr\"achspartner sind, aber eine interessante Konstellation wie sie eher f\"ur Wi\-ki\-pe\-dia ty\-pisch ist, k\"ame zu\-stan\-de, wenn Br\"unnhilde und Wotan Jelineks Gespr\"achspartner w\"aren. Zu dritt w\"urde aus ihnen eine Gruppe. Bei Aushandlungen zu Wi\-ki\-pe\-dia-Eintr\"agen scheint es ebenfalls nicht immer von Belang zu sein, wer welchen Redebeitrag beisteuert, und die Teams setzen sich pro Anlass ad-hoc zusammen. Die L\"ange der Dis\-kus\-si\-onsbeitr\"age ist bei Wi\-ki\-pe\-dia we\-sent\-lich k\"urzer als in \textit{rein GOLD}. Und B und W sprechen jeweils in einer Art Redeflu{\ss}, der nicht argumentativ strukturiert ist, sondern sich durch assoziative Aneinanderkettungen auszeichnet. Assoziative Dis\-kus\-si\-onsbeitr\"age gibt es auch bei Wi\-ki\-pe\-dia auf Projekt- oder Dis\-kus\-si\-onssei\-ten, aber sie sind nicht die Regel wie \mbox{hier}. In \textit{rein GOLD} handelt es sich weder um einen Dialog noch um einen Disput, sondern lediglich um ein Gezanke \"ahnlich dem Format einer Parlamentsdebatte, in der nur nebenbei auf Vorredner\textsuperscript{\tiny *} eingangen wird. 
Dabei unterscheiden sich W und B in ihrer Art zu reden kaum, \mbox{selbst} in der Menge von Fragezeichen oder Ausrufe\-zei\-chen nicht. Sowohl B als auch W sprechen mal \"uber sich \mbox{selbst}, mal \"uber die andere anwesende Person, mal \"uber abwesende Per\-so\-nen, mal \"uber das Geld als Akteur\textsuperscript{\tiny *}, das Kapital oder Jesus, \"uber "`Erl\"oser"', "`Helden"', "`Arbeiter"', "`Zwerge"', "`Riesen"', "`nichts"' -- und ohnehin meist \"uber andere. Gerede sozusagen, das mit Anfeindungen durchsetzt ist. Meist ist der Wechsel der Sprechenden von der Abfolge her nicht inhaltlich motiviert, nur am Beginn seines dritten Beitrags unterbricht W einmal B. 

Vill beschreibt Jelineks Stil in \textit{rein GOLD} unter Bezugnahme auf Richard Wagners musikalisches und dichterisches Werk. Als solle es ein \"Aquivalent zu Wagners kompositorischer Technik sein, der Kunst des \"Ubergangs, imitiere Jelinek musikalische Verfahren mittels ver\"astelter Intertexte, die r\"atselhaft verfremdet w\"urden. In Sprachspielen mit Kinderabz\"ahlversen, mit Assonanzen und Alliterationen, die Wagners Stabreime parodieren, w\"urden Zusammenh\"ange ausgeblendet und der Sinn einer Mitteilung unkenntlich gemacht, so Vill. Sie stellt zum anderen fest, dass Jelinek mittels Burleske und derben Sp\"a{\ss}en das Erhabene von Wagners Musik konter\-ka\-riert (\cite{Vill2013}). Auch in Aushandlungen bei Wi\-ki\-pe\-dia kommen gelegentlich Reaktionen vor, die man als Humor be\-zeichnen k\"onnte, \mbox{selbst} wenn das Thema des Eintrags, um den es geht, gar nichts Lustiges an sich hat.

Worum geht es B/W eigentlich? Handeln sie etwas aus? Falls ja, was? Und falls nein, gibt es ein anderes Ziel f\"ur die Auseinandersetzung? Welchen Zweck erf\"ullt das Gezanke? Das Team B/W ist befasst mit eigenen inneren Konflikten, mit Trag\"odien, News und Skandalen im "`sich repetierenden Dialog"' (\cite{SchmittSchossler2013}:90). "`Im Zusammenhang mit der Figur der Wiederholung, der Deutung gegenw\"artiger Skandale im Lichte fr\"uherer ge\-sell\-schaftlicher Ersch\"utterungen, spielt die Narration eine ganz besondere Rolle, als Archiv und Speicher jederzeit abrufbarer und performativ aktualisierbarer Verhaltens- und Deutungsmuster"' (\cite{Gelzetal2014}). B/Ws Re\-fle\-xion von Vergangenem und ihre Bearbeitung skandalf\"ormiger Themen ebenso wie ihr Stil einer z\"ankischen Mischung mit pers\"onlichen Konflikten strukturiert den Redefluss ebenso wie die intensiv verschachtelten Wortspiele. Al\-ler\-dings kann auch das Jelinek-Werk \mbox{selbst} als Skandal angesehen werden, weil es eine F\"ulle von Beispielen freim\"utiger Rede (\textit{parrhesia}) enth\"alt, die nicht selten erst der Anlass f\"ur Skandale sind. Auch bei Wi\-ki\-pe\-dia entsteht nicht selten eine umfangreiche Debatte erst, wenn es um Skandal\"oses geht. Lesen wir auch Sachtexte als Narrative, werden Aushandlungen bei Wi\-ki\-pe\-dia zwar nicht f\"ur li\-te\-ra\-rische Dialogana\-lyse interessant, wohl aber f\"ur die Ana\-lyse von Kommunikation und Konsensherstellung, was als eines der Ziele der Aushandlungen bei Wi\-ki\-pe\-dia postuliert worden ist. Yiftach Nagar hat sich mit der sogenannten NPOV-Policy befasst ("`Neutral Point of View"'), mit der erreicht werden soll, dass Artikelinhalte m\"oglichst neutral formuliert werden. Nagar hat unter anderem versucht he\-rauszufinden, wie sich die Strukturierungsprozesse der Debatte auf Dis\-kus\-si\-onssei\-ten zur NPOV-Policy w\"ahrend der Aushandlung von Interpretationen der Policy beschrei\-ben lassen. Er hat beobachtet, dass die Hauptform des Austauschs im Stellen von Fragen und Vorschlagen von Antworten be\-steht (\cite{Nagar2012}). Bei \textit{rein GOLD} hingegen sind die meisten Fragen lediglich rhetorischer Art und werden nicht im Dialog beantwortet -- wenn \"uberhaupt. Der B\"uhnenessay scheint sich eher als ein Rededuell zu entpuppen. Bei der aktuellen Auseinandersetzung sollen andere ihnen zuschauen und miterleben, was sie auff\"uhren und wie sie sich auff\"uhren. 

Wie kam es nach eigenen Aussagen zum Duell? B sagt, "`Papa"' habe sich f\"ur den Bau seiner Burg finanziell \"ubernommen und "`Mama"' (8), seine Gattin Fricka, "`macht Papa Vorw\"urfe wegen dem Kredit"' (8). B fasst zusammen: "`Die Stimmung im Schlo{\ss} unertr\"aglich. Gez\"ank"' (8). Statt Fricka ist f\"ur die Auff\"uh\-rung des Essays also Br\"unnhilde im Einsatz? Und wie geht es aus? W macht mehrfach Ank\"undigungen f\"ur seinen Abgang, aber der Schluss seines letzten Beitrags klingt eher nach "`weiter wie bis\-her"'. Und wer zur Auff\"uh\-rung gekommen und beim Duell zugegen ist, soll \mbox{selbst} beurteilen, ob es f\"ur eine\textsuperscript{\tiny *} der beiden t\"odlich endet -- oder f\"ur keine\textsuperscript{\tiny *} oder f\"ur beide? Der Begriff des Duells kommt im Text nicht vor, aber ein Duell, das sich herk\"ommlicherweise zwischen M\"annern abspielt, k\"onnte \mbox{hier} in eine Form der Aus\-ein\-andersetzung zwischen eine\textsuperscript{\tiny *} weiblichen und eine\textsuperscript{\tiny *} m\"annlichen Sprecher\textsuperscript{\tiny *} gebracht worden sein, auch wenn -- oder gerade weil -- sich die beiden Stimmen nicht we\-sent\-lich voneinander unterscheiden. An einer Stelle beansprucht B ein "`ich sagte es schon"' von etwas, dass sie zuvor wiederholt hatte, was aber der Textgestaltung nach zuerst von W ge\"au{\ss}ert worden ist: ein ty\-pischer Fall f\"ur Mi{\ss}verst\"andnisse, die zu einem Duell f\"uhren k\"onnen. 

Um Anschein von Rollenverteilung und deren Aushandlung im Ton von Vorw\"urfen geht es insbesondere im textlichen Umkreis einer wiederkehrenden Sentenz, bei der neue Geschlechterrollen thematisiert werden: wer (wie ein) Held und wer (wie) Jesus sein oder sprechen kann oder soll. Die folgenden Textstellen eignen sich als Beispiel daf\"ur, wie in \textit{rein GOLD} Themen bearbeitet werden, scheinbar assoziativ und doch stringent. Dabei weisen nur Anreden oder besitzanzeigende F\"urw\"orter darauf hin, wessen Redebeitrag es ist. Zuerst spricht eine Frau so, wie es der historischen Figur Jesus zugeschrieben wird, und zwar im Femininum ("`Ich bin die, die ihr sucht"'). Dann wird im Zusammenhang des Nicht-gesucht-Werdens die Aussage im Maskulinum verwendet, danach wieder im Femininum in Zusammenhang mit einer weiblichen Abwandlung von Jesus, und nochmals im Femininum die "`Worte Jesu"'. Dann geht es um das Einsperren und Einstellen einer Ungesuchten, erneut wird ein weiblicher Jesus benannt, mit scheinbarer Unbeholfenheit als Braut getarnt. Bald wird das Heldenm\"adchen schon Vorbild f\"ur den Helden, dann wird der Satz von Jesus in w\"ortlicher Rede im Femininum wiedergegeben und Wotan sagt voraus, dass sein Kind Jesus sein kann. Abschlie{\ss}end wird statt des Polizeinotrufs die Nation Deutschland der Adressat f\"ur die Aussage des Heldenm\"adchens.

\singlespacing
\begin{enumerate}
\item W: "`Ich bin die, die ihr sucht, sagt die Nazibraut, der man am Telefon vorhin, als sie den Polizeinotruf erw\"ahlte, nicht geglaubt und nicht getraut hat. Wer uns getraut, der geh\"ort gehaut. Sie sagt dasselbe wie Jesus zu seinen F\"angern. Genau: Es sind Jesu Worte!"' (58).
\item W: "`Und am Ende hat er das meiste: sich \mbox{selbst} gegeben, das gr\"o{\ss}te Opfer. Ein ganz spezieller Idiot, kein Zweifel. Man muss andere opfern, nie sich \mbox{selbst}! Ich bin der, den ihr sucht, das sagten schon viele, auch die, die gar nicht gesucht wurden"' (89).
\item B: "`Diese Frau, die Heldenfrau, mit der wurde niedlich umgegangen, Badvorleger und Gardinen, die brauchen wir Helden schon, die Frau? Ja, diese weibliche Jesus-Taschenbuchausgabe mit diesem \flq Ich bin die, die ihr sucht!\frq \,\,Eher Gutrune als Br\"unnhilde. Sanfter. Sanfter als ich auf jeden Fall"' (138). 
\item B: "`sie sagt: Ich bin die, die ihr sucht. Sie spricht die Worte Jesu, ich sagte es schon, aber ich finde das derma{\ss}en cool"' (167). 
\item B: "`sie sagt nichts, sie bestreitet nichts und wird eingesperrt, weil sie diejenige ist, die nicht gesucht wird"' (167).
\item B: "`h\"atten sie nie gefunden, wenn sie nicht von \mbox{selbst} gekommen w\"are, ein weiblicher Jesus, na ja, eine Abart von ihm, Jesu Braut, auch nicht schlecht. Sie ist die, die gesucht wird, und sie wird auch gleich eingestellt"' (168).
\item W: "`Wenn die klingeln, machst du sch\"on die T\"ur auf, Held. und dann sagst du: Was wollt Ihr? So wie das Heldenm\"adchen sagen wird: Ich bin die, die ihr sucht!"' (202).
\item W: "`Ich gebe dir die Worte ein, Kind, da{\ss} auch du Jesus sein kannst, wenn du willst. Wenn der Held durch das Feuer kommt, beziehungsweise wenn er f\"ur dich durchs Feuer gegangen sein wird, sagst du zu ihm einfach, was Jesus \mbox{selbst} damals sagte: Ich bin die, die Sie suchen"' (204).
\item W: "`Deutschland. Ich bin die, die ihr sucht, das wirst du sagen, das wird mein Heldenm\"adchen sagen, mit den Worten des Herrn Jesus, welche er sprach"' (207).
\end{enumerate}
\onehalfspacing

Aus Jesus wurde eine Nazibraut, weil er/sie in einer \"ahnlichen Situation so \"ahnlich sprach wie jener. Ihr Handeln wird ein Vorbild f\"ur Helden und sie stellt sich -- beziehungsweise duelliert sich -- mit dem Staat. B und W, wir Helden, f\"uhren in \textit{rein GOLD} demnach ein gemeinsames Duell gegen den Staat auf und zanken in einem Dialog, der wie bei Platon nur noch der Form nach dialogisch ist (\cite{SeyboldUngern-Sternberg2011}:290). Handelt es sich vielmehr um einen auf zwei Sprecher\textsuperscript{\tiny *} verteilten autoritativen Monolog? Die Debatte im Kopf ist um.

\pagebreak

\subsection{Zusammenfassung}
\label{subsubsec:5.5}

In den Abschnitten dieses Kapitels \textit{Metaebene I} habe ich die Konzepte er\"ortert, die ich in meiner Argumentation einsetze, habe zwei\-tens das Feld dargestellt, auf dessen Ver\"anderung mein Erkenntnisinteresse zielt, und drittens anhand eigener Interpretationen von \mbox{vier} zeitgen\"ossischen li\-te\-ra\-rischen Texten eigene \textit{Erlesnisse} in schriftlicher Form zur Sprache gebracht wie sie Teil eines Beitrags sein k\"onnten, der bei Wi\-ki\-pe\-dia als belegf\"ahig gilt. 

Zun\"achst wurde der neue Terminus \textit{Erlesnis} n\"aher definiert und seine kon\-zep\-tio\-nel\-le Funk\-tion in Zusammenhang mit "`Li\-te\-ra\-tur\textsuperscript{\~.\~.}lekt\"ure"', "`Leseerlebnis"' und "`Lesebericht"' vorgestellt. Wie Cunningham bei seiner Erfindung des Wiki \"ubernehme ich aus der Architektur zus\"atzlich das Konzept "`Entwurfsmuster"', um eine aus meiner Sicht li\-te\-ra\-tur\-vermittelnde Abfolge zu denken: von Entwurfsmuster 1 (li\-te\-ra\-rische Texte) zu Handlung 1 (Lesen) zu Entwurfsmuster 2 (\textit{Erlesnis}) zu Handlung 2 (Leseberichte pu\-bli\-zieren) zu Entwurfsmuster 3 (Wi\-ki\-pe\-dia-Eintr\"age) und Handlung 3 (\textit{Erlesnisse} in "`Leseerlebnis"'-Abschnitten zusammenfassend dar\-stel\-len).

Um Ergebnisse meiner \textit{Erlesnis}-Suche in Teasern f\"ur die Hauptsei\-ten-Rubrik "`Schon gewusst?"' der deutschspra\-chi\-gen Wi\-ki\-pe\-dia-Version ging es im zwei\-ten Abschnitt, der gleichzeitig als eine weitere Einf\"uh\-rung zu Wi\-ki\-pe\-dia konzipiert war, diesmal mit einem Schwerpunkt auf Li\-te\-ra\-tur\-artikeln. In meinem Befund, dass in etwa 15\,\% der Teaser zumindest eine \textit{Erlesnis}-Chance angedeutet wird, sehe ich einen ersten Hinweis darauf, dass f\"ur Leseerlebnis-Abschnitte bei Wi\-ki\-pe\-dia mit Akzeptanz zu rechnen w\"are.

In einem dritten Schritt ergab die Ana\-lyse von deutschspra\-chi\-gen \"Au{\ss}erungen in li\-te\-ra\-tur\-wissenschaftlichen Beitr\"agen inklusive der Li\-te\-ra\-tur\-ver\-mitt\-lungsfor\-schung, dass in diesen Beispielen weder Vertrautheit mit der Idee und Funktionsweise von Wi\-ki\-pe\-dia noch Wertsch\"atzung f\"ur Li\-te\-ra\-tur\-eintr\"age bei Wi\-ki\-pe\-dia zum Ausdruck kommt.

Viertens habe ich einige Texte von Li\-te\-ra\-tur\-nobelpreistr\"agern\textsuperscript{\tiny *} aus einer durch Wikipedia-Konzepte und -Realit\"aten inspirierten Perspektive interpretiert und dabei ei\-ge\-ne \textit{Erlesnisse} zur Sprache gebracht. (Diese Zusammenfassung konzentriert sich auf die Bezugnahmen zu Wi\-ki\-pe\-dia, eine zusammenfassende Schilderung der \textit{Erlesnisse} hingegen formuliere ich in Abschnitt \textit{~\ref{subsubsec:7.1.4} \textit{Erlesnis}-Darstellungen in den eigenen Interpretationen}. Im Beitrag zu Alice Munros \textit{Save the Reaper}, 1998 und 1998 in zwei Varianten erschienen, wurden zun\"achst andernorts be\-kann\-te Versionierungsverfahren demjenigen bei Wi\-ki\-pe\-dia gegen\"ubergestellt, um dann in eingehender Arbeit am Text zu zeigen, wie Munros fiktionalisierte Varianten in Kombination mit der Ver\"of\-fent\-lichung der zwei Fassungen gesehen werden k\"onnten. Zu Doris Lessings \textit{Alfred and Emily} von 2008 schlage ich eine Interpretation vor, die in den Blick r\"uckt, was ich f\"ur Lessings kunstvolle Auseinandersetzung mit Geltungsanspr\"uchen eines enzyklop\"adischen Beitrags halte. Bei meiner Interpretation von Herta M\"ullers \textit{Reisende auf einem Bein} von 1989 kommen aus Wi\-ki\-pe\-dia-Sicht zwei Punkte zur Sprache, erstens, dass Hannah H\"ochs bekannteste Collage, mit der ich M\"ullers Satire zusammendenke, mit Remix-Verfahren arbeitet wie es heute durch Freie Lizenzen der in Wi\-ki\-pe\-dia-Eintr\"agen verlinkten Abbildungen m\"oglich ist, zwei\-tens, welche Folge f\"ur den Wi\-ki\-pe\-dia-Eintrag "`Reisende auf einem Bein"' ich sehe, dass die mir bis\-her bekannt gewordene For\-schungsli\-te\-ra\-tur zu diesem Text auf dessen satirische Aspekte nicht eingeht. Meine Lekt\"ure von Elfriede Jelineks \textit{rein GOLD} (2013) ergibt, dass -- kennt man sich mit der Aushandlungskultur in der deutschspra\-chi\-gen Wi\-ki\-pe\-dia etwas aus und liest den Text in dessen Modulationen -- einige neue Aspekte des kunstvollen Jelinekschen Gezankes formulierbar werden, die sich unter anderem dazu eignen, die Geste des Dialoghaften in diesem B\"uhnenessay zu hinterfragen.

Im folgenden Teil, \textit{Objektebene II}, mache ich die B\"uhne frei f\"ur mein Leseerlebnis-Experiment bei Wi\-ki\-pe\-dia. 

\pagebreak

\section{Objektebene II}
\label{sec:6}

In diesem Kapitel stelle ich zun\"achst vor, wie der jeweilige Text meiner neuen "`Leseerlebnis"'-Abschnitte in verschiedenen Werkartikeln lautet, die zu meinem Experiment geh\"oren. Der darauffolgende Abschnitt hat zwei Teile. Im ersten berichte ich \"uber Reaktionen aus der deutschspra\-chi\-gen Wi\-ki\-pe\-dia-Community auf einzelne der "`Leseerlebnis"'-Abschnitte. Im zwei\-ten Teil fasse ich eine Debatte zusammen, die ich herbeigef\"uhrt habe, und in der es mir gegen Schluss gelungen ist, die "`Leseerlebnis"'-Abschnitte in einem allgemeineren Rahmen beurteilen zu lassen. Anhand der aktuellen Reaktionen versuche ich in der Zusammenfassung zu \textit{Objektebene II} eine Einsch\"atzung geben, inwiefern das Einbeziehen von Aussagen \"uber nicht-pro\-po\-si\-ti\-o\-na\-les Wissen aus Li\-te\-ra\-tur\-lekt\"ure auch k\"unftig mit Akzeptanz rechnen k\"onnte. 

\subsection{Leseerlebnis-Abschnitte in Wikipedia-Eintr\"agen}
\label{subsubsec:6.1}

Zwischen November 2014 und M\"arz 2015 sind f\"ur einige Wi\-ki\-pe\-dia-Eintr\"age zu li\-te\-ra\-rischen Werken neuartige Abschnitte mit dem Titel "`Leseerlebnis"' (oder "`Leseerlebnisse"' im Plural) verfasst und eingef\"ugt worden. Die Werke \mbox{selbst} z\"ahlen zur Gegenwarts\-li\-te\-ra\-tur\- und erfuhren in der Zeit seit dem Start der deutschspra\-chi\-gen Version von Wi\-ki\-pe\-dia im Mai 2001 Erst- beziehungsweise Neuauflagen. Demzufolge gab es Besprechungen in gro{\ss}en Tageszeitungen, die aus Sicht der Wi\-ki\-pe\-dia-Community relevanzstiftend sind, oder bereits For\-schungsli\-te\-ra\-tur, die ohne all\-zu\-gro{\ss}en Aufwand erreichbar war -- oder beides. Damit war den Relevanzkriterien Gen\"uge getan. 

In Leseerlebnis-Abschnitten fasse ich Leseberichte aus verschiedenen Arten von Quel\-len zusammen. Manchmal verwende ich zus\"atzlich Zitate, mit dem Ziel, die An\-schau\-lich\-keit der Darstellung zu erh\"ohen und dem Erlebnischarakter von Lekt\"ure durch direkte Aussagen mehr Gewicht zu geben. Meist vollst\"andig offline verfasst und sp\"ater im Artikel erg\"anzt, wurden diese Zusammenfassungen dann mit dem \"Anderungskommentar "`Neuer Abschnitt: Leseerlebnis"' platziert, eine Explizitheit, die Kritiker\textsuperscript{\tiny *} auf den Plan rufen sollte, so dass sich eventuell eine Dis\-kus\-si\-on erg\"abe. Dieser textliche K\"oder hat allem Anschein nach niemanden besonders interessiert.

In sechs F\"allen konnte dennoch eine Reaktion herbeigef\"uhrt werden. Meiner Vermutung nach ist dies allein durch die Pr\"asentation in der Hauptsei\-tenrubrik "`Schon gewusst?"' gelungen. Zum meinem ersten Artikel mit Leseerlebnis-Abschnitt, der von der Hauptseite verlinkt wurde, "`Pour que tu ne te perdes pas dans le quar\-tier"', gab es am fr\"uhen Morgen des Platzierungstages eine kritische Anmerkung auf der dortigen Dis\-kus\-si\-onsseite. Wenige Stunden sp\"ater wurde eine so geistreiche Gegenrede formuliert wie sie mir \mbox{selbst} wohl kaum in den Sinn gekommen w\"are. Es ent\-wickelte sich daraus leider keine weitere Debatte. Auf diesen und die anderen f\"unf F\"alle wird in der Auswertung n\"aher eingegangen, siehe \textit{Objektebene II}, Abschnitt \textit{~\ref{subsubsec:6.2.1} Reaktionen auf einzelne Leseerlebnis-Abschnitte}.

Die Reihung der Eintr\"age erfolgt \mbox{hier} chronologisch nach Einf\"ugung des Leseerlebnis-Abschnitts, da sich auf diese Weise im Lesefluss die Entwicklung der Schreibweisen innerhalb dieser neuen Art von Abschnitt besser zeigt. Erschienen sind die Werke in folgender Chronologie: \textit{Reisende auf einem Bein} (M\"uller 1989), \textit{La Vie commune} (Salvayre 1991), \textit{Save the Reaper} (Munro 1998), \textit{Die Lichter l\"osche ich} (Pirzad 2001), \textit{Im Caf\'{e} der verlorenen Jugend} (Modiano 2007), \textit{Alfred und Emily} (Lessing 2008), \textit{Atemschaukel} (M\"uller 2009), \textit{Stimmungen lesen. \"Uber eine verdeckte Wirklichkeit der Li\-te\-ra\-tur} (Gumbrecht 2011), \textit{Gr\"aser der Nacht} (Modiano 2012), \textit{Rein Gold: Ein B\"uhnenessay} (Jelinek 2013), \textit{Pas pleurer} (Salvayre 2014), \textit{Pour que tu ne te perdes pas dans le quartier} (Modiano 2014). 

In der Aufmerksamkeits\"okonomie f\"ur Eintr\"age zu li\-te\-ra\-rischen Werken spielt bei Wi\-ki\-pe\-dia al\-ler\-dings weniger die Publikationschronologie eine Rolle, sondern vor allem die Bekanntheit des einzelnen Autors\textsuperscript{\tiny *}, denn Platzierungen auf Bestenlisten oder Preisvergaben werden von Teilen der Wi\-ki\-pe\-dia-Community prinzipiell als re\-le\-vanz\-stei\-gernd angesehen. Nicht selten scheint daher die Vollst\"andigkeit der Liste der Preisvergaben in Eintr\"agen zu Autoren\textsuperscript{\tiny *} wichtiger zu sein als die Vollst\"andigkeit der Liste der Werke. Da es \"uberwiegend neu ver\"of\-fent\-lichte Werke sind, denen eine solche Aufmerksamkeit zukommt, wird in diesem Kontext das Datum der Erstpublikation, auf die es unter li\-te\-ra\-tur\-wissenschaftlichen Gesichtspunkten ankommt, als nachrangig angesehen. Das Datum der Erstpublikation wird in der Perspektive der aufmerksamkeits\"okonomischen Logik bei Wi\-ki\-pe\-dia einerseits sogar aufgewogen, falls das Werk wiederaufgelegt worden ist und aktuelle Rezensionen zu finden sind, durch die medien\"of\-fent\-liche Relevanz entsteht, die in dieser Logik der Bedeutsamkeit eines Lemmas zugutekommt. An\-de\-rerseits, da es bei Wi\-ki\-pe\-dia ohnehin keine redaktionelle T\"atigkeit einer Gruppe von Usern gibt, die sich vor dem Anlegen von Artikeln Gedanken dar\"uber machen, welche Auswahl getroffen werden soll, sondern alle \mbox{User} \mbox{selbst} ausw\"ahlen beziehungsweise sich eventuell zu zweit oder dritt absprechen, zu welchem Lemma sie einen neuen Artikel anlegen wollen, kommt es eher auf die pers\"onliche Interessenlage Einzelner an und nicht zuallererst auf ein vermutetes Interesse bei denjenigen, die Wi\-ki\-pe\-dia als Informationsquelle nutzen, und daher auch nicht un\-be\-dingt auf Preisvergaben, sondern auf die Entscheidung einzelner \mbox{User} dar\"uber, mit dem Verfassen welchen Lemmas sie als N\"achstes ihre Zeit verbringen m\"ochten. Drittens kann auch eine Mischung aus diesen beiden Faktoren bei der Wahl eines neuen Artikelthemas ausschlaggebend sein: Wenn manche Wi\-ki\-pe\-dia-Autoren\textsuperscript{\tiny *} sich durch Preisvergaben angespornt f\"uhlen, weil es sich dann um ein Thema handelt, das aktuell Aufmerksamkeit erregt und vom User, der einen Artikel verfasst, vermutet wird, dass viele Leser\textsuperscript{\tiny *} sich freuen k\"onnten, bei Wi\-ki\-pe\-dia schon einen Eintrag zu finden, spielt sowohl die Aufmerksamkeits\"okonomie als auch die pers\"onliche Entscheidung eine entscheidende Rolle beim Zustandekommen eines Eintrags. Diese \"Uberlegungen haben nicht zu\-letzt dazu gef\"uhrt, dass ich in einem Fall den neuen Artikel eines Mitautors\textsuperscript{\tiny *} f\"ur eine Pr\"asentation in der Hauptsei\-tenrubrik "`Schon gewusst?"' vorgeschlagen habe, um au{\ss}er in Artikeln zu Werken des aktuellen Nobelpreistr\"agers\textsuperscript{\tiny *} noch weitere Testballons zu haben, bei denen auch innerhalb der Wi\-ki\-pe\-dia-Community mit einem gewissen Ma{\ss} an aktueller Aufmerksamkeitsbe\-reit\-schaft zu rechnen ist; siehe diesbez\"uglich auch Abschnitt \textit{~\ref{subsec:2.6} Methodenwahl} in Kapitel \textit{~\ref{sec:2} Research Design}.

Aus diesen Gr\"unden ist in der Argumentation der vorliegenden Arbeit f\"ur das Vor\-stel\-len und Diskutieren der von mir angelegten Leseerlebnis-Abschnitte in Wi\-ki\-pe\-dia-Ein\-tr\"a\-gen eher eine Reihung nach Datum der Artikelerstellung von Belang, denn neu angelegte Artikel erhalten routinem\"a{\ss}ig in der Community etwas mehr Aufmerksamkeit als \"Anderungen an bereits bestehenen Eintr\"agen: Zum einen werden in der sogenannten Eingangskontrolle, bei der einige \mbox{User} alle j\"ungsten \"Anderungen systematisch durchsehen, intensiver als an an\-de\-rer Stelle bei Wi\-ki\-pe\-dia neue Eintr\"age auf deren Relevanz und Qualit\"at hin gepr\"uft. Zwei\-tens wird eine Auflistung der jeweils neuesten Artikel f\"ur einige der Themenportale automatisiert per bot erstellt, so auch f\"ur den Li\-te\-ra\-tur\-bereich, \mbox{hier} f\"ur deutschspra\-chi\-ge und internationale Li\-te\-ra\-tur, und zwar f\"ur Per\-so\-nen, die als Autoren\textsuperscript{\tiny *} relevant sind, und ebenso f\"ur Sachbegriffe wie f\"ur einzelne Werke. Die eigens geschriebenen Programme haben die Funktion, neue Artikel auf bestimmte Kategorien hin zu durchsuchen, so dass ein Lemma fachlich zuordnet wird, entweder von einem erfahrenen Artikelersteller\textsuperscript{\tiny *} \mbox{selbst} oder von einem User, der\textsuperscript{\tiny *} sich auf systematische Arbei\-ten spezialisiert hat und bei Artikeln Kategorien erg\"anzt. Wiederum andere User, die sich f\"ur ein be\-stimmtes Wissengebiet bei Wi\-ki\-pe\-dia besonders interessieren, schauen Themenportale nach den angezeigten Neuzug\"angen durch und erhalten auf diesem Weg Kenntnisse dazu, welche li\-te\-ra\-rischen Werke neuerdings ein eigenes Lemma erhalten haben, weil dezidiert zum Werk ein Artikel neu angelegt worden ist und es nicht mehr lediglich in der Liste der Werke des Autors\textsuperscript{\tiny *} zu finden ist. Es kann auch vorkommen, wie im Fall von Zoya Pirzad in der deutschspra\-chi\-gen Wi\-ki\-pe\-dia, dass jemand zur Autorin\textsuperscript{\tiny *} \mbox{selbst} noch keinen Eintrag verfasst hat, wohl aber zu ihrem zweiten Roman, der 2001 in Iran ein Besteller war, \textit{Die Lichter l\"osche ich}. Drittens erhalten manche Artikel durch das Vorschlagen f\"ur die Hauptsei\-tenrubrik "`Schon gewusst?"' erh\"ohte Aufmerksamkeit, und falls der Vorschlag nach einer Dis\-kus\-si\-onsphase angenommen wird, erst recht dann, wenn der Teaser mit Verlinkung zu diesem Eintrag f\"ur zwei Tage auf der Hauptseite zu lesen ist. Al\-ler\-dings erzielen Eintr\"age, deren Teaser an den Tagen Montag bis Freitag auf der Hauptseite platziert werden, meist h\"ohere Zugriffszahlen. Dass Li\-te\-ra\-tur\-artikel eher f\"ur einen Samstag oder Sonntag eingetragen wurden, hatte ich w\"ahrend einiger Monate beobachtet und mich dann entschieden, \mbox{selbst} im Adhoc-Team zur Rubrik "`Schon gewusst?"' mitzumachen, um in diesem aufmerksamkeitsgenerierenden Kontext die Relevanz von neuen Artikeln mit Li\-te\-ra\-tur\-bezug zu st\"arken, damit eines Tages auch f\"ur die Eintr\"age zu li\-te\-ra\-rischen Werken, die ich vorhatte zu verfassen, eine Pr\"asentation an einem Wochentag wahrscheinlicher w\"urde. Da auf der "`Schon gewusst?"'-Dis\-kus\-si\-onsseite zur Vorbereitung der Auswahl von Teasern ein Vier-Augen-Prinzip vereinbart worden ist, so dass niemand seinen eigenen Vorschlag aussuchen kann, musste \mbox{hier} ein gewisser Vorlauf einkalkuliert werden.

Am Beispiel des Eintrags "`Atemschaukel"' erl\"autere ich meine Arbeitsweise. Die Angaben zu den Belegen der Leseerlebnis-Abschnitte werden jeweils als Fu{\ss}noten direkt unter dem wiedergegebenen Text angezeigt.

\singlespacing
\subsubsection{"`Die Lichter l\"osche ich"' (Pirzad 2001)}
\label{subsubsec:6.1.1}

Eintrag "`Die Lichter l\"osche ich"' in der Version vom \href{https://de.wikipedia.org/w/index.php?title=Die_Lichter_l\%C3\%B6sche_ich\&oldid=136571720}{8. Dezember 2014, 9:24 CET}

(Dieser Eintrag dient als Beispiel f\"ur einen kurzen Artikel, bei dem ich Leseerlebnisse in den Abschnitt "`Rezeption"' integriert habe.)

\textbf{Artikelaufbau}, 1 Inhalt und Interpretation, 2 Rezeption, 3 Auszeichnungen, 4 Ausgaben, 5 Rezensionen in deutscher Sprache, 6 Einzelnachweise, 7 Weblinks.

\textbf{Rezeption}\\
Besch\"onigt wird die normale Absurdit\"at des All\-tagslebens nicht, aber es wird mit Fantasie und Spannung erz\"ahlt, atmosph\"arisch fein und voller Details. Der Leser k\"onne beim Lesen kaum vergessen, dass hinter dem fiktiven Land mit einer \"Ol\-industrie die Realit\"at der Islamischen Republik Iran aufd\"ammert, meint Sybill Mahl\-ke in ihrer Rezension f\"ur den Tagesspiegel.[3] In ihrer Rezension f\"ur Die Welt schreibt Tanja Langer, es werde ein Moment in der iranischen Geschichte beschrieben "`-- doch so, als ginge es um heute."'[1]

{\tiny[1] Tanja Langer: "`Die rote Linie \"uberschrei\-ten. Zoya Pirzad erz\"ahlt in ihrem Roman \textit{Die Lichter l\"osche ich} eine leise Geschichte aus dem iranischen All\-tag"', in: \textit{Die Welt}, 22. April 2006, LITERARISCHE-WELT, S. 4\par}
{\tiny[3] Sybill Mahlke: \href{http://www.tagesspiegel.de/kultur/fernes-land/693416.html}{"`Fernes Land"'}, in: \textit{Der Tagesspiegel}, 15. M\"arz 2006\par}

\subsubsection{"`Pour que tu ne te perdes pas dans le quartier"' (Modiano 2014)}
\label{subsubsec:6.1.2}

Eintrag "`Pour que tu ne te perdes pas dans le quartier"' in der Version vom \href{https://de.wikipedia.org/w/index.php?title=Pour_que_tu_ne_te_perdes_pas_dans_le_quartier\&oldid=137248287}{30. Dezember 2014, 19:31 CET}

In der Einleitung: "`An Leseerlebnissen wurde geschildert, dass Unausgesprochenes sich schlimmer anf\"uhlt als Fakten, dass Pro\-ble\-ma\-ti\-sches aus der Vergangenheit einen mysteri\"osen Modergeruch entfaltet und dass Modianos leichtf\"u{\ss}iger sprachlicher Stil das Empfinden eines Flie{\ss}ens oder sogar eines Flirtens oder Verzaubertseins hervorrufen kann."'

\textbf{Artikelaufbau}: 1 Titel und Motto, 2 Inhalt, 3 Erz\"ahlweise und Interpretation, 4\,Leseerlebnis, 5 Bekannte Namen, 6 Ausgaben, 7 Weblinks, 8 Einzelnachweise.

\textbf{Leseerlebnis}\\
In ihrer Rezension f\"ur \textit{Elle} kommt Olivia de Lamberterie zu dem Ergebnis, dass man nach der Lekt\"ure dieser Verweigerung einer Kindheit das Ge\-f\"uhl hat, dass Dinge, die weder ausgesprochen wurden noch aufgel\"ost worden sind, m\"oglicherweise noch schlimmer sind als zu wissen, dass Annie Astrand Akrobatin war, eine gewisse Zeit im Gef\"angnis verbracht hat und dass Daragane anschlie{\ss}end mit ihr in Mont\-mar\-tre lebte, als sie ihm den Zettel mit der Adresse ihrer Wohnung zusteckte.[6] Francis Richard nimmt Bezug auf das Stendhal-Motto des Romans, wenn er schreibt, dass es vorausdeute, wie die beiden Zei\-ten der Vergangenheit von Daragane nicht vollst\"andig aus dem Schat\-ten her\-vor\-kom\-men, einmal die vor et\-was mehr als 60 Jah\-ren und die andere f\"unfzehn Jahre sp\"ater. Modiano verbinde auf leichtf\"u{\ss}ige Art diese beiden Vergangenhei\-ten mit der Gegenwart, nicht ohne dabei einige Zonen mit einer Warnung zu versehen, um beim Leser ein Gef\"uhl des Mysteri\"osen wachzuhalten.[3] Bei Modiano sei nichts wie es scheint ("`l'apparence ne fait pas l'essence"') und die verschiedenen Erz\"ahlf\"aden bleiben in der Schwebe. Daher wolle man sich darin verlieren, als ob das Buch mit einem flirte, so der Leseeindruck bei Caroline Doudet.[9] F\"ur Bruno Corty bleibt nach der Lekt\"ure f\"ur erfahrene Modiano-Leser die Empfindung eines Flie{\ss}ens und einer merkw\"urdigen Atmosph\"are in Vermischung mit dem Modergeruch einer pro\-ble\-ma\-tischen Vergangenheit. Neuen Lesern im Universum von Patrick Modiano, das seinesgleichen suche, sagt Corty ein Ge\-f\"uhl des Verzaubertseins voraus.[2]

{\tiny[2] Bruno Corty, \href{http://www.lefigaro.fr/livres/2014/10/16/03005-20141016ARTFIG00019--pour-que-tu-ne-te-perdes-pas-dans-le-quartier-de-modiano-l-ombre-d-un-doute.php}{Pour que tu ne te perdes pas dans le quartier de Modiano : l'ombre d'un doute}, \textit{lefigaro.fr}, 16. Oktober 2014, in franz\"osischer Sprache\par}
{\tiny[3] Francis Richard, \href{http://www.contrepoints.org/2014/10/25/185863-pour-que-tu-ne-te-perdes-pas-dans-le-quartier-de-patrick-modiano}{\flq Pour que tu ne te perdes pas dans le quartier\frq\, de Patrick Modiano}, \textit{contrepoints.org}, 25. Oktober 2014\par}
{\tiny[6] Olivia de Lamberterie, \href{http://www.elle.fr/Loisirs/Livres/News/Le-roman-de-la-semaine-Pour-que-tu-ne-te-perdes-pas-dans-le-quartier-de-Patrick-Modiano-2846552}{Le Roman de la semaine? \flq Pour que tu ne te perdes pas dans le quartier\frq\,\,de Patrick Modiano}, \textit{elle.fr}, 10. Oktober 2014\par}
{\tiny[9] Caroline Doudet, \href{http://leschroniquesculturelles.com/2014/10/30/pour-que-tu-ne-me-perdes-pas-dans-le-quartier-de-patrick-modiano/}{Pour que tu ne te perdes pas dans le quartier, de Patrick Modiano}, \textit{leschroniquesculturelles.com}, 30. Oktober 2014\par}

\subsubsection{"`Im Caf\'{e} der verlorenen Jugend"' (Modiano 2007)}
\label{subsubsec:6.1.3}

Eintrag "`Im Caf\'{e} der verlorenen Jugend"' in der Version vom \href{https://de.wikipedia.org/w/index.php?title=Im_Caf\%C3\%A9_der_verlorenen_Jugend&oldid=137646716}{10. Januar 2015, \mbox{22:23 CET}}

In der Einleitung:\\ "`Die Schwerelosigkeit, nach der Louki sich gesehnt hat, sp\"urt der Leser bei der Lekt\"ure der sch\"onen Prosa, die manchmal \mbox{selbst} \flq irgend\-wie schwebt\frq.[1]"'

\textbf{Artikelaufbau}, 1 Inhalt, 2 Interpretation, 2.1 Guy Debord und weitere Phantome, 3 Rezeption, 4 Leseerlebnis, 5 Rezensionen, 6 For\-schungsli\-te\-ra\-tur, 7 Ausgaben, 7.1\,Print, 7.2 H\"orbuch, 8 Weblinks, 9 Einzelnachweise.

\textbf{Rezeption}\\
Die internationale Rezeption der franz\"osischen Originalfassung setzte im eng\-lisch\-spra\-chigen Raum auch ohne \"Ubersetzung bereits 2008 ein, und bis Ende 2009 lagen \"Uber\-set\-zun\-gen ins Katalanische, ins Kastilische, ins Persische, ins Arabische und ins Russische vor. 2010 und 2011 folgten Japanisch und Italienisch. Eine deutschspra\-chi\-ge Fassung wurde erst 2012 verlegt.

\textbf{Leseerlebnis}\\
Die Lekt\"ure des Romans hinterl\"asst glei\-cherma\-{\ss}en einen tiefen Eindruck von Dichtung wie den eines Unbehagens, das nicht lokalisierbar ist: eine merkw\"urdige Emp\-fin\-dung, die den Leser bei der Gurgel packt.[10] Die Schwerelosigkeit, nach der Louki sich sehnt, sp\"urt der Leser "`durch die sch\"one, manchmal beil\"aufige, manchmal \mbox{selbst} irgendwie schwebende Prosa"', so empfindet es Gerrit Bartels, in seiner Rezension im Tagesspiegel am 10. Juni 2012.[1] Pascal Gavillet beschreibt f\"ur Tribune de Gen\`{e}ve am 4. Oktober 2007, dass da eine Beklommenheit ist angesichts des Ge\-f\"uhls von Verlust, von Erinnerung, die weniger wird, und auch von Orientierung, die verloren geht.[10] In Les Echos meinte Denis Cosnard am 2. Oktober 2007, man k\"onne einfach nicht anders als der leicht absch\"ussigen Stra{\ss}e zu folgen und damit einer Louki, mit der es unentrinnbar bergab gehe.[10] Patrick K\'{e}chichian schrieb in Le Monde am 5. Oktober 2007, dass manche B\"ucher uns h\"arter machen und dass andere, die wertvoller sind und notwendiger, uns empfindlicher machen und uns die Waffen abnehmen -- so wie dieses Portrait einer Frau, die so nah ist und so verloren, von Modiano entlang der Grenze zwischen Licht und Schatten ge\-zeichnet, ersch\"utternd.[10]

{\tiny[1] Gerrit Bartels: \href{http://www.tagesspiegel.de/kultur/patrick-modianos-roman-cafe-der-verlorenen-jugend-das-glueck-der-schwerelosigkeit/6730142.html}{"`Patrick Modianos Roman \textit{Caf\'{e} der verlorenen Jugend}. Das Gl\"uck der Schwerelosigkeit"'}, \textit{Tagesspiegel}, 10. Juni 2012\par}
{\tiny[10] \href{http://www.alalettre.com/actualite-modiano-jeunesse-perdue.php}{Dans le caf\'{e} de la jeunesse perdue de Patrick Modiano}, \textit{alalettre.com}, ohne Datumsangabe\par}

\subsubsection{"`Pas pleurer"' (Salvayre 2014)}
\label{subsubsec:6.1.4}

Eintrag "`Lydie Salvayre"' in der Version vom \href{https://de.wikipedia.org/w/index.php?title=Lydie_Salvayre\&oldid=136929747}{19. Dezember 2014, 17:49 CET}

\textbf{Artikelaufbau}, 1 Leben, 2 Werk, 3 La Vie commune (1991) "`In dieser qu\"alenden Situation wird der Leser dazu gebracht, mit beiden Sei\-ten zu sympathisieren"', 4 Pas pleurer (2014), 4.1 Stil und Leseerlebnis, 5 Auszeichnungen, 6 Werke (Auswahl), 7 Li\-te\-ra\-tur, 8 Weblinks, 9 Einzelnachweise.

\textbf{Stil und Leseerlebnis}\\
Es sind zwei Perspektiven miteinander verflochten zu h\"oren: Den Zeitzeugen Ber\-na\-nos ekelt es und er prangert an, wie die Nationalisten "`die schlechten Armen"' ("`les mauvais pauvres"') terrorisieren, w\"ahrend die damals jugendliche Zeitzeugin Montse, \mbox{selbst} "`mauvaise pauvre"', sich kaum mehr erinnert, au{\ss}er dass diese begeisternden Erlebnisse am Beginn der Befreiungsbewegung zu den intensivsten ihres Lebens z\"ahlen. Zwei Redeweisen, zwei Visionen, die mit unserer Gegenwart auf merkw\"urdige Weise zusammenklingen, wodurch die Erz\"ahlkunst von Lydie Salvayre mit aller Kraft zum Ausdruck kommt: zwischen Heftigkeit und Leichtigkeit, zwischen Brutalit\"at und Feinge\-f\"uhl.[6] Die Figuren des Romans wirken stark \"uberzeichnet in ihrer Pers\"onlichkeit, was die Autorin damit begr\"undet, dass sich in einer Zeit des B\"urgerkriegs pers\"onliche Einstellungen radikalisieren, religi\"ose wie politische.[2] Salvayres K\"onnen ist getragen von einer Prosa, die einem einerseits makellos erscheint und einem an\-de\-rerseits in fr\"ohlichem Ton hart zuzusetzen vermag.[6]

{\tiny[2] Marianne Grosjean, \href{http://www.tdg.ch/culture/livres/Lydie-Salvayre-injecte-de-la-joie-libertaire-dans-l-horreur-franquiste/story/25099586}{Lydie Salvayre remporte le Prix Goncourt pour Pas pleurer} (Interview mit der Autorin), \textit{Tribune de Gen\`{e}ve}, zu\-letzt aktualisiert am 5. November 2014\par}
{\tiny[6] \href{http://www.babelio.com/livres/Salvayre-Pas-pleurer/633376}{Pas pleurer}, \textit{babelio.com}, Oktober/Dezember 2014\par}

\subsubsection{"`La Vie commune"' (Salvayre 1991)}
\label{subsubsec:6.1.5}

Eintrag "`La Vie commune"' in der Version vom \href{https://de.wikipedia.org/w/index.php?title=La_Vie_commune\&oldid=138022681}{21. Januar 2015, 18:28 CET}

In der Einleitung: "`Thematisiert wird in dieser Geschichte unter anderem, welche Folgen es haben kann, wenn es jemandem nicht gelingt, seine Unzufriedenheit zu \"au{\ss}ern. Salvayre beschreibt dies am Beispiel von zwei Frauen unterschiedlichen Alters, die sich neuerdings ein B\"uro teilen m\"ussen. Obwohl nur Suzanne, die \"altere, erz\"ahlt, beginnt man als Leser doch, mal mit der einen und mal mit der anderen Seite zu sympathisieren."'

\textbf{Artikelaufbau}, 1 Inhalt, 2 Interpretation, 3 Rezeption, 3.1 Leseerlebnis, 4 Li\-te\-ra\-tur, 5 Ausgaben, 6 Weblinks, 7 Einzelnachweise.

\textbf{Rezeption}\\
La Vie commune, Salvayres zweiter Roman, wird als der Beginn ihrer Karriere als Schriftstellerin angesehen.[8] Kaum hatte Salvayre nach La d\'{e}claration ihr zweites Buch, La Vie commune, ver\"of\-fent\-licht, da wusste man bereits, dass sie zur fran\-z\"o\-si\-schen Li\-te\-ra\-tur etwas was Starkes, J\"ahzorniges beitr\"agt.[9] Alle 8 Jahre ist bis\-her von La Vie commune eine neue Auflage in einem anderen Verlag erschienen, zu\-letzt im zweitgr\"o{\ss}ten Verlagshaus in Frankreich, Gallimard, bei dem auch die j\"ungsten beiden franz\"osischspra\-chi\-gen Li\-te\-ra\-tur\-nobelpreistr\"ager verlegt werden.

\begin{quote}
\textit{(Die Formulierungen in diesem Abschnitt wurden zwischen User:Ges\-tum\-blindi und User:C.Koltzenburg ausgehandelt)}\\
\textbf{Leseerlebnis}\\
Anfangs finde man das Buch am\"usant, schreibt Julia Scheeres in der New York Times, bald aber ist es mal unerquicklich, mal gruselig und voller genauer Beobachtungen, und dennoch gefallen einem die 119 Sei\-ten gut. Man sp\"urt einen Hass, der so leidenschaftlich ist, dass er bei Suzanne sexuelle Fantasien hervorruft. Man lebt beim Lesen im Kopf einer manischen Meg\"are, es ist auf delikate Art d\"uster und man wird nerv\"os. Der Leser sympathisiert mit der Gequ\"alten ebenso wie mit der Qu\"alenden, stellt Scheeres fest.[1] Beim Lesen dieses Romans lernt man in mikroskopi\-scher Auf\-l\"osung die Funktionsweise und Wirkung kleiner m\"orderischer Phrasen kennen, deren uneingestandene Ziele die Verachtung und Dem\"utigung des Gegen\"ubers sind.[5] Man sp\"urt fast \mbox{selbst} den Druck, der sich in Suzanne aufbaut, wo es schon vor der Begegnung mit der Neuen kaum noch Raum gab.[2] Verl\"asst man bei der Lekt\"ure das fiktionale Universum f\"ur einen Moment, kann es sein, dass man pl\"otzlich meint, dass zum Beispiel die Gro{\ss}buchstaben des NON! im elften Kapitel von Salvayre \mbox{selbst} kommen, tats\"achlich geschrieben, weil Suzanne es nicht kann.[3] Mit welchen Me\-tho\-den man unangenehme Begegnungen am Arbeitsplatz l\"ose, suche man sich \mbox{selbst} aus. Es gebe radikalere und andere. Jedenfalls betrachtet man nach dem Lesen dieses Romans seine Kollegen nicht mehr mit denselben Augen.[10]

{\tiny[1] Julia Scheeres, \href{http://query.nytimes.com/gst/fullpage.html?res=9E0CE2DF1731F934A25751C1A9609C8B63}{The Office}, \textit{The New York Times}, 17. Dezember 2006\par}
{\tiny[2] M. A. Orthofer, \href{http://www.complete-review.com/reviews/salvayre/everyday.htm}{Everyday Life by Lydie Salvayre}, \textit{Complete review}\par}
{\tiny[3] Warren Motte, Voices in her Head, in: \textit{SubStance} 33,2 (2004)/ Special Section: Contemporary Novelist Lydie Salvayre, S. 13-29, S. 18, 15\par}
{\tiny[5] admin, \href{http://ideesinfos.free.fr/?p=15}{La vie commune}, \textit{id\'{e}es infos : bibs, epn}, 25. Juli 1999\par}
{\tiny[10] Myrinna, \href{http://www.babelio.com/livres/Salvayre-La-vie-commune/8457/critiques/33911}{En lisant ce roman, vous ne regarderez plus votre coll\`{e}gue de la m\^{e}me fa\c{c}on...}, \textit{babelio.com}, 1. Mai 2010\par}
\end{quote}

\subsubsection{"`Atemschaukel"' (M\"uller 2009)}
\label{subsubsec:6.1.6}

Eintrag "`Atemschaukel"' in der Version vom \href{https://de.wikipedia.org/w/index.php?title=Atemschaukel\&oldid=137657441}{11. Januar 2015, 10:43 CET}

In der Einleitung: "`We\-sent\-liche Motive sind Hunger und Heimweh. Das Werk ist als ergreifend empfunden worden."'

\textbf{Artikelaufbau}, 1 Handlung, 2 Entstehung, 3 Erz\"ahlstruktur, 4 Motive, 5 Rezensionen, 6 Leseerlebnis, 7 Ausgaben, 8 For\-schungsli\-te\-ra\-tur, 9 Weblinks, 10 Einzelnachweise.

\textbf{Rezensionen}\\
Die Sprache dieses Romans, die altert\"umlich und mal klar, mal \"uberbordend sei, konserviere die untergegangene Welt des k.u.k-Sprachraums, in dem meh\-rere Sprachen nebeneinander existierten: Jiddisch, Russisch, Ungarisch, Rum\"anisch und Deutsch, so Ina Hartwig. M\"ullers Wortbildungen wie "`Eigenbrot"' und "`Herzschaufel"' ebenso wie "`Atemschaukel"' erinnern Hartwig an die fr\"uhe Lyrik Paul Celans.[6] Daniela Strigl hebt hervor, dass Herta M\"uller in Atemschaukel li\-te\-ra\-rische Bilder f\"ur das Au{\ss}ersprachliche gefunden hat, in einer zeitlosen Studie \"uber den Menschen in extremis, die mit den Erfah\-rungen des schrecklichen 20. Jahrhunderts ges\"attigt sei.[7] Bezeugt werde das komplexe Verh\"altnis von Erinnerung und Sprache, so Michael Lentz.[8]

\textbf{Leseerlebnis}\\
Michael Lentz findet Atemschaukel ergreifend.[8] Ina Hartwig berichtet, dass die Leserin durch Herta M\"ullers "`Beschw\"orung der poetischen Kraft im Ungl\"uck"' an ihre eigene Grenze gelangt. Diese poetische Kraft liegt unter anderem darin, dass M\"uller mit der Metaphorik des Hungerengels eine Gefahrenzone entstehen l\"asst, denn, so findet Hartwig: "`Den Hungerengel muss man sich wie einen Geist vorstellen, den der Hungernde sich schafft, um gegen ihn k\"ampfen zu k\"onnen"' -- nur dass der \"Uberlebende feststellen muss, dass der Hungerengel ihn f\"ur immer in Besitz genommen hat, ein W\"urgegriff, so Hartwig, der dazu f\"uhrt, dass Leo niemandem mehr sein Herz wird schenken k\"onnen. Hartwig empfindet Atemschaukel auch deswegen als eine He\-rausforderung und als ein "`schwierig sch\"ones Geschenk."'[6]

{\tiny[6]Ina Hartwig, "`Ein Held namens Hungerengel"', in: Ina Hartwig, \textit{Das Geheimfach ist offen. \"Uber Li\-te\-ra\-tur}, S. Fischer, Frankfurt am Main, 2012, ISBN 978-3-10-029103-5, S. 97-100, \"uberarbeitete Fassung der Rezension in: \textit{Frankfurter Rundschau}, 21. August 2009\par}
{\tiny[8]Michael Lentz, Herta M\"ullers \textit{Atemschaukel}, in: \textit{Textleben : \"uber Li\-te\-ra\-tur, woraus sie gemacht ist, was ihr vorausgeht und was aus ihr folgt}, S. Fischer, Frankfurt am Main 2011, ISBN 978-3-10-043934-5, S. 243-250, S. 250.\par}

\onehalfspacing
An diesem Beispiel l\"asst sich zeigen, wie unterschiedlich mit referenzierten Quellen umgegangen werden kann. In diesem Fall wurde in einer \"Au{\ss}erung von Michael Lentz etwas Nennenswertes sowohl f\"ur den Abschnitt "`Rezensionen"' als auch f\"ur "`Leseerlebnis"' gefunden. Die Aussagen sind dem letzten Abschnitt eines Essays entnommen und lauten urspr\"ung\-lich folgenderma{\ss}en: "`\textit{Atemschaukel} ist mit Herzblut geschrieben. Es ist ein Manifest der Erinnerung und der Sprache, deren komplexes Verh\"altnis es auf ergreifende Weise bezeugt"' (\cite{Lentz2011}:250). Im Abschnitt "`Rezensionen"' wurde als Beschreibung des Objekts vermerkt: "`Bezeugt werde das komplexe Verh\"altnis von Erinnerung und Sprache, so Michael Lentz"', und im Abschnitt "`Leseerlebnis"' das Adjektiv in einem lapidaren Satz verwertet, n\"amlich als: "`Michael Lentz fin\-det Atemschaukel ergreifend."' -- womit eine Empfindung auf eine Meinung reduziert wird, denn wenn die Aussage ein Erlebnis wiedergeben soll, h\"atte es in etwa hei{\ss}en m\"ussen: "`Michael Lentz empfindet Atemschaukel als ergreifend."' -- so wie es in der Einleitung zum Artikel zusammengefasst ist: "`Das Werk ist als ergreifend empfunden worden."' Von an\-de\-rer Seite war daran seit mehr als einem Monat nichts ge\"andert worden, obwohl die Seite seitdem etwa 1500 Mal abgerufen worden war. Es wurde in diesem Abschnitt dar\"uber hinaus auch die Regel nicht eingehalten, Rezeption sei im Pr\"ateritum wiederzugeben, was bei an\-de\-rer Gelegenheit aktuell angemerkt worden war. (Auf Beobachtungen dieser Art wird auf \textit{Metaebene II} in Abschnitt \textit{~\ref{subsubsec:6.2.1} Reaktionen auf einzelne Abschnitte} n\"aher einzugehen sein.)

Eine weitere Aussage, die der von Ina Hartwig \"ahnlich ist, wurde f\"ur eine Wiedergabe nicht ausgew\"ahlt. Lentz formulierte etwa in der Mitte desselben Essays: "`Den Terror des Hungers fasst Herta M\"uller in Bilder extrem beanspruchter Phy\-sis, die dem Leser zu Leibe r\"ucken"' (\cite{Lentz2011}:246-247). Im Abschnitt "`Leseerlebnis"' h\"atte man dies als Zitat wiedergeben k\"onnen, dennoch wurde eine Paraphrase von Hartwigs Aussage bevorzugt. F\"ur die zusammenfassenden S\"atze "`Ina Hartwig berichtet, dass die Leserin durch Herta M\"ullers \flq Beschw\"orung der poetischen Kraft im Ungl\"uck\frq \,\,an ihre eigene Grenze gelangt. Diese poetische Kraft liegt unter anderem darin, dass M\"uller mit der Metaphorik des Hungerengels eine Gefahrenzone entstehen l\"asst, denn, so findet Hartwig: ..."' lautete bei Hartwig die Stelle urspr\"ung\-lich so: "`Ganz klar begibt die Autorin sich mit dieser Metaphorik in eine Gefahrenzone. Nicht, dass sie sich die Geschichte eines anderen \flq leiht\frq \,\,(im Unterschied zur Lagerli\-te\-ra\-tur eines Imre Kert\'{e}sz oder eines Warlam Schalamow), ist das Pro\-blem. Das muss erlaubt sein. Aber die Beschw\"orung der poetischen Kraft im Ungl\"uck bringt die Leserin an eine, n\"amlich ihre eigene Grenze."' Der \textit{Erlesnis}-Anteil der \mbox{hier} berichteten Leseerfah\-rung wurde f\"ur den "`Leseerlebnis"'-Abschnitt zugespitzt auf den Effekt bei der Leserin, die \mbox{hier} rezensiert: ihre Schilderung, dass M\"uller sich mit der Hungerengel-Metaphorik in eine Gefahrenzone begeben habe. Im Wi\-ki\-pe\-dia-Eintrag formuliere ich den Grund daf\"ur erst nach dessen Effekt.

\singlespacing
\subsubsection{"`Alfred und Emily"' (Lessing 2008)}
\label{subsubsec:6.1.7}

Eintrag "`Alfred und Emily"' in der Version vom \href{https://de.wikipedia.org/w/index.php?title=Alfred_und_Emily&oldid=138028858}{21. Januar 2015, 21:48 CET}

In der Einleitung:\\ "`Das Werk ist international viel beachtet worden und hat in Rezensionen ebenso wie in der For\-schungsli\-te\-ra\-tur vielf\"altige Beschreibungen von Leseerlebnissen hervorgerufen."'

\textbf{Artikelaufbau}, 1 Inhalt, 2 Interpretationen, 2.1 Verh\"altnis der beiden Teile zueinander, 2.2 Hybride Form, 2.3 Weitere Aspekte, 3 Entstehung, 4 Rezeption, 5\,Lese\-erlebnis, 6 Gliederung (Originalausgabe), 7 Ausgaben, 8 For\-schungsli\-te\-ra\-tur, 9 Ein\-zel\-nach\-wei\-se.

\textbf{Rezeption}\\
F\"ur den New York Review of Books stellt Tim Parks Alfred and Emily in das Umfeld an\-de\-rer "`family memoirs"' und bespricht im selben Beitrag weitere \mbox{vier} Werke, von Marie Brenner (Apples and Oranges: My Brother and Me, Lost and Found, 2008), von Rachel Sontag (House Rules, 2008), von Miranda Seymour (Thrumpton Hall: A Memoir of Life in my Father's House, 2007) sowie von Isabel Allende (The Sum of Our Days, 2007). Parks res\"umiert, dass Lessings Alfred and Emily sich durch seine Art von Wut und politischem Engagement auszeichne. Lessings Hauptinteresse gelte \mbox{hier} weniger der Familie als vielmehr einer leidenschaftlichen Anti-Kriegs-Position. Sie k\"onne damit rechnen, von der modernen Leserschaft daf\"ur bewundert zu werden.[18]

\textbf{Leseerlebnis}\\
Dass der Text durchsetzt ist mit vielen Familienfotos, bringt einen als Leser durch\-ein\-ander, meint Tiger. \"Ahnlich wie bei Sebald, zum Beispiel in \textit{Austerlitz}, w\"urden die nebeneinandergestellten Abbildungen ebenso den Text erl\"autern wie der Text die Abbildungen. Tiger erg\"anzt, dass ihr das Lesen \"ahnlich vorgekommen sei wie das Betrachten eines Werks von Escher, etwa \textit{Zeichnen}, wo Illusionen von Perspektive und Dimension erzeugt w\"urden.[1][19] Lizzie kommentiert ebenfalls die Form und ihre Wirkung: "`Ich finde nicht, dass es stimmt, was bei Wi\-ki\-pe\-dia steht: dass es Lessings letzter Roman sei. Es ist ehrlichgesagt keiner, und es wird die Leute umso mehr verwirren, wenn sie \mbox{hier} einen Roman erwarten."'[20]\\
Sayers findet das Werk wegen seiner Kombination aus Fiktion und Sachtext bewegend.[7] F\"ur Conrad trifft dies nur auf den zwei\-ten Teil zu, den sie als wahrhaft bewegend empfunden hat, und sogar ergreifend fand sie den Versuch einer Vers\"ohnung mit der Mutter, weil dieser Versuch im nicht-fiktionalen Teil des Buches unternommen werde.[10] Sarah Norris meint: Beim Lesen \"uber Lessings Verh\"altnis zu ihren Eltern, vor allem zur Mutter, erschaudert es einen, weil man sich an die eigenen K\"ampfe um Autonomie erinnert sieht.[21] Bei Birkenhauer klingt bez\"uglich des dargestellten Themas zustimmende Emp\"orung an, wenn er schreibt: "`\mbox{Vier} Jahre lang hat die Mutter in einem der gr\"o{\ss}ten Londoner Krankenh\"auser als Schwes\-ter Sol\-da\-ten sterben sehen. Immer kurz nach den gro{\ss}en Schlachten auf dem Kontinent waren alle Londoner Krankenh\"auser in Alarmzustand. Welch perverse Kopplung an den geplanten Tod."'[9] Clodagh fand etwas anderes zum Heulen: "`Irgendwas ist da mit dem eng\-lischen Kolonialleben in Afrika, was mich zum Weinen bringt. Das halbe Buch war f\"ur mich schmerzlich zu lesen."'[22]\\
Scheck berichtet begeistert von Lessings Stil- und Tonlagenwechsel in der Mitte des Buches, denn damit schlage die Autorin "`aufs Wundervollste \"uber die Str\"ange."'[12] Andere schildern die Wirkung weniger erfreut: "`Und dann endet die sch\"one Erz\"ah\-lung abrupt und ein irgendie wahllos zusammengew\"urfelter Teil beginnt. Ich kenne nicht viele Leser, die so umschalten k\"onnen oder die es wollen w\"urden -- \mbox{selbst} wenn sie es k\"onnten"'[23], woraufhin dieser Kommentar das Erlebnis konkretisiert: "`Mich hat bei diesem abrupten Wechsel ein Sch\"utzengrabenschock erwischt -- um ihren Ausdruck zu verwenden (shell shock) --, so dass ich he\-rausfinden musste, was andere denken und ob ich weiterlesen soll. Bin froh, dass ich nicht der Einzige bin, der dieses Gef\"uhl hat!"'[24]\\
Sergeant hat sich \mbox{selbst} beim Lesen beobachtet und meint, w\"ahrend der Lekt\"ure betreibe man \mbox{selbst} R\"uckschau. Leser fragen sich eventuell, wo sie eigentlich gewesen sind und was das genau war. Wir haben beim Lesen von Alfred und Emily die starke Empfindung, so Sergeant, dass die Zeit knapp wird, und auch fragen wir uns, was in der verbleibenden Spanne wohl noch erreicht werden kann. Was von diesen beiden als schwerwiegender empfunden wird, balanciere sich aus, je nachdem, wie alt man \mbox{selbst} ist. Lessing findet nicht zu allen ihren Fragen eine L\"osung, aber darum geht es nicht allein, meint er, denn ihre unbeantworteten Fragen wirken f\"ur Leser wie etwas Lebendiges, das Echos produziert: Von \mbox{hier} aus k\"onnen wir uns weiterbewegen, weil wir eine genauere Vorstellung davon erhalten haben, wo wir waren und wo wir sind, so Sergeant.[13] Eine \"ahnliche Beobachtung macht auch Lizzie, wenn es hei{\ss}t: "`Die Fragen ihres Lebens sind nicht gel\"ost, aber immerhin stellt sie sie. Lessing stellt unsere Fragen und zeigt, ob sie beantwortbar sind"', und, eingangs sowie am Schluss der Rezension: "`Das Buch ergibt kaum Sinn, soviel kann ich sagen. Objektiv ist es bizarr zu lesen und wirklich fragmentiert und sogar innerhalb der einzelnen Fragmente wird hin- und hergesprungen wie verr\"uckt"', "`Dieses Buch wird immer wichtig f\"ur mich bleiben, und vermutlich muss es daf\"ur nicht einmal Sinn ergeben."'[20]

{\tiny[1] Virginia Tiger, "`Life Story: Doris, \textit{Alfred and Emily}"', in: \textit{Doris Lessing Studies}, Vol. 28, No. 1, 2009, S. 22-24.\par}
{\tiny[7] Valerie Sayers, \href{http://www.washingtonpost.com/wp-dyn/content/article/2008/07/31/AR2008073102585.html}{"`A Separate Peace"'}, in: \textit{The Washington Post}, 3. August 2008\par}
{\tiny[9] Franz Birkenhauer, \href{http://www.sf-magazin.de/doris-lessing-alfred-und-emily,117.html}{"`Erz\"ahl uns eine Geschichte!"'}, \textit{sf-magazin.de}, 28. November 2008\par}
{\tiny[10] Bernadette Conrad, \href{http://www.nzz.ch/aktuell/startseite/der-weite-weg-zurueck-nach-hause-1.2438185}{"`Lessing \"uber ihre Eltern. Der weite Weg zur\"uck nach Hause"'}, in: \textit{Neue Z\"urcher Zeitung}, 21. April 2009\par}
{\tiny[12] Dennis Scheck, \href{http://www.deutschlandradiokultur.de/nachgetragene-tochterliebe.950.de.html?dram:article_id=136885}{Nachgetragene Tochterliebe}, \textit{deutschlandradiokultur.de}, 2. Dezember 2008\par}
{\tiny[13] David Sergeant, \href{http://www.oxonianreview.org/wp/stories-to-herself/}{"`Stories to Herself"'}, in: \textit{The Oxonian Review of Books}, summer 2008: volume 7: issue 3\par}
{\tiny[19] M. C. Escher, \href{http://en.wikipedia.org/wiki/File:DrawingHands.jpg}{Abbildung Zeichnen} (1948)\par}
{\tiny[20] Lizzie, Alfred and Emily by Doris Lessing, \href{https://www.goodreads.com/review/show/445489394}{"`I've put off reviewing this one a bit, because I'm not entirely sure what to say. This book was really, really important to me -- but this book is wacko ..."'}, \textit{goodreads.com}, 6. August 2014\par}
{\tiny[21] Sarah Norris, \href{http://www.barnesandnoble.com/review/alfred-and-emily}{Alfred and Emily by Doris Lessing}, \textit{Barnes \& Noble review}, 25. September 2008\par}
{\tiny[22] Clodagh, Alfred and Emily by Doris Lessing, \href{https://www.goodreads.com/review/show/133541837}{"`more doris. yay ..."'}, \textit{goodreads.com}, 8. Januar 2011\par}
{\tiny[23] Lara, Alfred and Emily by Doris Lessing, \href{https://www.goodreads.com/review/show/31104294}{"`What a strange read! ..."'}, \textit{goodreads.com}, 22. September 2008\par}
{\tiny[24] Haley, Alfred and Emily by Doris Lessing, \href{https://www.goodreads.com/review/show/31104294?page=1\#comment_77325756}{"`I am so... to use her word... 'shell-shocked' at the abrupt change ..."'}, \textit{goodreads.com}, 22. September 2008\par}

\subsubsection{"`Save the Reaper"' (Munro 1998)}
\label{subsubsec:6.1.8}

Eintrag "`Save the Reaper"' in der Version vom \href{https://de.wikipedia.org/w/index.php?title=Save_the_Reaper\&oldid=137361185}{3. Januar 2015, 9:34 CET}

\textbf{Artikelaufbau}, 1 Handlung, 2 Interpretationen, 2.1 Intertextuelle Bez\"uge, 2.2 Le\-se\-er\-lebnis, 3 Ausgaben und Versionen, 4 Li\-te\-ra\-tur, 5 Einzelnachweise.

\textbf{Leseerlebnis}\\
Judith Maclean Millers Erinnerung nach gibt es bei Save the Reaper im Ein\-gangs\-raum des verwahrlosten Hauses einen Toten. Miller fragt sich, warum sie das erinnert. Der Schnitter (Reaper) suggeriere Tod als Thema, im Laufe der Erz\"ahlung sterbe aber niemand.[4] Duncan teilt das Unbehagen bez\"uglich der Behausung und erg\"anzt, dass in dieser Erz\"ahlung verschiedene Dinge r\"atselhaft bleiben: warum in der Beschreibung "`Massive disorder was what they had to make their way through -- the kind that takes years to accumulate"' das "`massive disorder"' am Beginn so m\"achtig sei, wer die Mosaikwand kreiert habe, welche Art von Tochter als Erwachsene noch Spiele wie "`Was hasst du am meisten an deiner Mutter?"' spiele. Manches davon erhelle sich, wenn man Eves Nostalgie als unglaub\-w\"urdig und skurril wahrnehme, und sich vorstelle, was so eine Nostalgie f\"ur Eve an Konsequenzen bedeute.[3] F\"ur Birkenstein entsteht der Eindruck, dass Eve nicht einschrei\-ten kann, weil sie wei{\ss}, dass die Eignerin des Hauses in ein Heim gebracht worden ist.[6] Miller hat sich unter stilistischen Gesichtspunkten mit der Stelle befasst, wo der kleine Mann, der die drei einl\"adt in die merkw\"urdige Behausung zu kommen, sagt "`Mary, she owns it, but Harold he put her in the Home, so now he does. It wasn't his fault, she had to go there."' Miller merkt in Klammern an, dass sie denkt, dass es \mbox{selbst}verst\"andlich Harolds Schuld war. Und diese vielen Kommas seien Anlass f\"ur ihr Gef\"uhl, dass da Dinge im Gange seien, die noch unheimlicher sind. Miller nimmt eine weitere Stelle n\"aher in Augenschein: "`\flq I told her maybe there was pictures in the front but she couldn't go in there you got that shut up,\frq\,\,the little man said. Harold said, \flq You shut up.\frq"' Da sei irgendwas mit den Pausen in diesem Dialog und mit der Wiederholung der Wendung "`shut up"'. Falls es nicht Mary ist, die in dem Haus verscharrt wurde, gebe es da irgendetwas oder irgendjemand Anzuklagenden, darin ist Miller sich sicher.[4] Aufgrund ihrer Kenntnis einer anderen Munro-Geschichte entsteht f\"ur de Papp Carrington beim Lesen von Save the Reaper die Ahnung, dass Eve nach dem Ende der Geschichte ermordet wird.[1]

{\tiny[1] Ildik\'{o} de Papp Carrington, \href{http://canlit.ca/site/getPDF/article/10625}{Where are you, mother? Alice Munro's Save the Reaper}, in: \textit{Canadian Literature / Litt\'{e}rature canadienne} (173) 2002, 34-51.\par}
{\tiny[3] Isla Duncan, \textit{Alice Munro's Narrative Art}, Palgrave Macmillan, New York 2011, ISBN 978-0-230-33857-9 (hardcover), ISBN 978-1-137-00068-2 (ebook), S. 17, 80, 83, 85-86, 106.\par}
{\tiny[4] Judith Maclean Miller, Deconstructing Silence: The Mystery of Alice Munro, in: \textit{Antigonish Review} 129 (Spring 2002), S. 43-52.\par}
{\tiny[6] Jeff Birkenstein, The Houses That Alice Munro Built: The Community of \textit{The Love of a Good Woman}, in: \textit{Critical Insights. Alice Munro}, edited by Charles E. May, Salem Press, Ipswich, Massachusetts, 2013, ISBN 978-1-4298-3722-4 (hardcover), ISBN 978-1-4298-3770-5 (ebook), S. 212-227, S. 223.\par}

\onehalfspacing

Im Eintrag "`Save the Reaper"' ist der Leseerlebnis-Abschnitt als einer von zwei Unterabschnitten zu "`Interpretationen"' platziert worden, nicht wie in manchen der anderen Eintr\"age als Unterabschnitt zu "`Rezeption"' oder als eigener Abschnitt nach "`Rezeption"'. In diesem Beispiel folgt der Unterabschnitt auf "`Intertextuelle Bez\"uge"' und steht vor einem ausf\"uhrlichen Abschnitt mit dem Titel "`Ausgaben und Versionen"'. Im Abschnitt "`Leseerlebnis"' werden ausschlie{\ss}lich Positionen referiert, die bereits an an\-de\-rer Stelle im Artikel referenziert wurden: Birkenstein, Miller, de Papp Carrington und Redekop.

\singlespacing
\subsubsection{"`Rein Gold: Ein B\"uhnenessay"' (Jelinek 2013)}
\label{subsubsec:6.1.9}

Eintrag "`Rein Gold: Ein B\"uhnenessay"', Version vom \href{https://de.wikipedia.org/w/index.php?title=Rein_Gold:_Ein_B\%C3\%BChnenessay\&oldid=137866703}{17. Januar 2015, 12:11 CET}

Artikelanfang: "`Rein Gold: Ein B\"uhnenessay (Eigenschreibweise rein GOLD) ist ein Prosawerk von Elfriede Jelinek. Die Urlesung fand am 1. Juli 2012 im Prinzregententheater in M\"unchen statt, pu\-bli\-ziert wurde es am 8. M\"arz 2013 im Rowohlt Verlag und hat Beschreibungen intensiver Leseerlebnisse hervorgerufen."'

\textbf{Artikelaufbau}, 1 Inhalt, 2 Bez\"uge zu anderen Werken, 2.1 Titel, 2.2 Weitergehende intertextuelle Bez\"uge, 3 Stil, 4 Entstehung und Quellen, 5 Rezensionen, 6 Leseerlebnisse, 7 Ausgabe, 8 Urlesung, 9 Inszenierungen, 10 Besprechungen der B\"uhnen\-versionen, 11 Li\-te\-ra\-tur zum Text, 12 Weblinks, 13 Einzelnachweise.

\textbf{Rezensionen}\\
In Rein Gold w\"urden die Mythen, die der Komponist Wagner sich ange\-eig\-net hatte, kapitalismuskritisch verbunden mit der aktuellen Gegenwart und ihren Realit\"aten, in der Wotan einen Streit austrage mit seiner marxistischen Tochter Br\"unnhilde, so Tim Caspar Boehme in seiner Rezension f\"ur die \textit{taz}. Jelinek unternehme in ihrer Lesart des 3. Akts der Walk\"ure einen Medienwechsel von Musik zu Text. Boehme hat Jelineks Beitrag zum Wagner-Jahr 2013 beeindruckt.[7]\\
Ina Hartwig hat Rein Gold f\"ur \textit{Die Zeit} rezensiert und schreibt, dass es in diesem Werk nicht nur um die Herrschaft des Geldes, sondern auch um die des Mannes gehe. Originell findet die Rezensentin, wie neue Hassges\"ange im Umfeld der Morde des NSU mit dem gegenseitigen Beschuldigen und Ankeifen von Br\"unnhilde und Wotan verflochten w\"urden, und regt an, das Buch, eine funkelnde Zumutung, zu lesen.[8]\\
Judith von Sternburg \"au{\ss}ert in ihrer Rezension "`Der Hort der Niegelungenen"', die in der \textit{Frankfurter Rundschau} pu\-bli\-ziert wurde, dass Jelinek in Rein Gold das Epos durch Details erweitert, die Wagners Ringzyklus karikiert. Die Autorin stehe dem Komponisten und dessen massiver Welterkl\"arungswucht in nichts nach, al\-ler\-dings fehle die Musik und dies mindere die Schlagkraft. In der grandiosesten der Hass\-tiraden nimmt der Text die deutsche Neigung zu Heldentum und Geiz aufs Korn.[5]\\
Dirk Pilz findet in der \textit{Neuen Z\"urcher Zeitung}, dass der B\"uhnenessay einer von Jelineks besten Texten ist, aufgrund seiner streng durchkomponierten Dichte und Dring\-lichkeit, die \"ubervoll mit Anspielungen ist, und weil der Gegenwart keinerlei Vers\"ohnungsangebot gemacht wird. Zum Untertitel \textit{Ein B\"uhnenessay} meint Pilz, dass Jelineks Werke sich herk\"ommlichen Kategorien entziehen und daher der Leser entscheide, was Rein Gold ist. Seine Idee ist, dass Ws Schlusssatz "`Mal sehn, was draus wird"' auch der erste sein k\"onnte.[9]

\textbf{Leseerlebnisse}\\
Pilz hat sich beim Lesen in einem wilden Gestr\"upp wiedergefunden, das einem kein Durchatmen erlaube und kein Wohlgef\"uhl. Er illustriert sein Empfinden mit einem Dialog: Man m\"ochte "`Aufh\"oren. Aufh\"oren!"' schreien, riefe da nicht schon Br\"unnhilde: "`Danke, Leute, ich versuch's ja, ehrlich."' Pilz gibt zu Beginn seiner Rezension eine Aussage des Regisseurs Steman von vor ein paar Jahren wieder: Als Leser von Jelineks Texten gerate man in eine Notwehr-Situation, weil die \"uber\-frach\-te\-ten Texte ungeheuer nervten. Deswegen k\"onne man bei Inszenierungen mit Jelineks Texten fast nur eigensinning umgehen, n\"amlich m\"oglichst frei.[9] Auf Tim Caspar Boehme wirkt der B\"uhnenessay wie ein Sprachexzess[7] und Christine Ammann, die von Jelineks Bildgewalt verbl\"ufft ist, meint, das Buch lese man wahrscheinlich nicht "`in einem Rutsch"', sondern \mbox{hier} und da, "`um Jelineks Sprachkunst und die ihr eigent\"umliche klammheimliche Freude in H\"appchen zu genie{\ss}en."' Amman haben die Assoziationsketten \"uberrascht, mit denen Jelinek Welten in einem Crash aufeinandertreffen l\"asst. Dazu komme Wortwitz, der so abgr\"undig sei, dass es einem manchmal geradezu die Sprache verschlage.[10] F\"ur Judith von Sternburg springt der Text neben der Ring-Tetralogie entlang und kommt mal \mbox{hier} mal da hinter ihr hervor. Den Text in seiner Menge und Masse empfindet sie wie ein Gedr\"angel, mit dem das b\"urgerlich-klassische Theatermaterial eingekreist wird. Sie hat den B\"uhnenessay als eine Ring-Umrundung gelesen, in die sich Assoziationen zu den sumpfigen NSU-Morden einmischen, die sich an das Geschehen im Ring anlehnen, das von Mord, Totschlag und brutaler Rechthaberei gepr\"agt ist. Auch scheint es ihr manchmal so, als ob die G\"otter bei Jelinek endlich die Wahrheit sagen: Wotan sieht es als Beweis seines Gottseins an, dass er ein Papier vorweisen kann -- von ihm \mbox{selbst} unterschrieben. Von Sternburg empfindet Verwunderung dar\"uber, wie es Jelineks Text gelingt, mit der Tetralogie verbunden zu sein und jenseits von ihr zu bestehen.[5] Bei der Lekt\"ure von rein Gold fallen nach Arno Widmanns Empfinden Wotans Gedanken \"ubereinander her "`wie die massigen Spieler im american football"' und sie seien "`endlich so schnell wie wir sie in unseren K\"opfen erleben."' Und sie w\"urden nicht kontrolliert von einer Vernunft, die das er\"ortert. Er hat beim Lesen Teile des Textes im Gaumentheater seines Mundes (Ginka Steinwachs) aufgef\"uhrt und dabei festgestellt, dass der Text nichts \"uberh\"oht, dass er "`redet wie wir alle es tun"' und dass man ihm deshalb "`leicht auf den Leim"' geht. W\"ahrend man Br\"unnhildes Text spreche, begreife man nicht, wie sie auf das Helden- und Stammtischgerede gekommen ist. Man lande pl\"otzlich in einer Passage, in der es hei{\ss}t, so formuliert es Widmann, "`die Deutschen seien Helden, aus Menschenmaterial zu Helden geschwei{\ss}t, wie Metall zu einer Dose."' Diese Aneinanderreihung "`im woll\"ustigen Gewebe der S\"atze der Elfriede Jelinek"', diese "`Gedanken, die gleichzeitig im Kopf umher irren"': "`Jeder Leser muss sie \mbox{selbst} auf die Stimmen verteilen."'[11] Hartwig stellt fest, dass die Lekt\"ure von Rein Gold zu produktivem Denken anregt. Von Gegen\-ein\-an\-der\-ge\-rede und Beschuldigungen wimmelt es nur so, die Dialoge empfindet sie als "`quirlige Gedankenmusik mit giftigen T\"onen,"' f\"ur die man eingangs Geduld braucht, bis man den Sprachfluss goutieren k\"onne. Man m\"usse daf\"ur "`die Schleusen des Bewusstseins \"offnen,"' so beschreibt Hartwig ihren Weg zum Leseerlebnis, bei dem sie erst dann "`die w\"usten Assoziationsstr\"ome in sich aufnehmen"' kann. Das Buch \mbox{selbst} sei wundersch\"on kompakt.[8]

{\tiny[5] Judith von Sternburg: \textit{Der Hort der Niegelungenen. Elfriede Jelinek liest in "`rein gold"' Wotan, Wagner, uns und allen die Leviten}, in \textit{Frankfurter Rundschau}, 16. Juli 2013, S. 31.\par}
{\tiny[7] Tim Caspar Boehme: \href{http://www.taz.de/1/archiv/digitaz/artikel/?ressort=ku\&dig=2013/05/18/a0031}{Wenn G\"otter vor sich hin d\"ammern. OHE! OHE! Auch an seinem 200. Geburtstag bleibt Richard Wagner ein so kontroverser wie anregender Komponist: Sein "`Ring des Nibelungen"' inspirierte \mbox{selbst} Schriftsteller von der Sp\"atromantik bis heute, wie Neuerscheinungen von Elfriede Jelinek, Eckhard Henscheid und \'{e}l\'{e}mir Bourges zeigen}, \textit{taz.de}, 18. Mai 2013.\par}
{\tiny[8] Ina Hartwig: \textit{Walk\"urenritt durchs Theoriegebirge. Elfriede Jelineks sarkastische hochaktuelle Wagner-Interpretation "`Rein Gold"'}, in \textit{Die Zeit}, 13. Juni 2013, S. 53.\par}
{\tiny[9] Dirk Pilz, "`Wenn die Menschen weg sein werden. Fortgesetzer Wutgesang: Elfriede Jelineks B\"uhnenessay Rein Gold"', in \textit{Neue Z\"urcher Zeitung}, 26. September 2013, S. 36.\par}
{\tiny[10] Christine Ammann, \href{http://www.belletristik-couch.de/elfriede-jelinek-rein-gold.html}{"`Elfriede Jelinek trifft Richard Wagner"'}, \textit{belletristik-couch.de}, Mai 2013\par}
{\tiny[11] Arno Widmann, \href{http://www.berliner-zeitung.de/vom-nachttisch-geraeumt/elfriede-jelinek--rein-gold--sonst-nichts--ein-paar-zeitungen-,22893728,22894736.html}{Elfriede Jelinek: Rein Gold. Sonst nichts. Ein paar Zeitungen.}, \textit{berliner-zeitung.de}, 29. Mai 2013\par}

\newpage
\subsubsection{"`Reisende auf einem Bein"' (M\"uller 1989)}
\label{subsubsec:6.1.10}

Eintrag "`Reisende auf einem Bein"' in der Version vom \href{https://de.wikipedia.org/w/index.php?title=Reisende_auf_einem_Bein\&oldid=137915300}{18. Januar 2015, 19:15 CET}

In der Einleitung: "`Bei der Lekt\"ure bekommt man \mbox{selbst} zu sp\"uren, wie sich Irenes d\"unne Haut anf\"uhlt."'

\textbf{Artikelaufbau}, 1 Inhalt, 2 Interpretationen, 2.1 Gro{\ss}stadtroman subvertiert, 3 Stil, 4 Rezeption, 5 Leseerlebnis, 6 Ausgaben, 7 \"Ubersetzungen, 8 Rezensionen, 9 For\-schungsli\-te\-ra\-tur, 10 Einzelnachweise.
 
\textbf{Rezeption}\\
2013 ist Reisende auf einem Bein im Fischer Taschenbuch-Verlag in der drit\-ten Auf\-lage erschienen. Zwischen 1990 und 2013 ist Reisende auf einem Bein in mindestens 9 Sprachen \"ubersetzt worden. Eine \"Ubersetzung ins Franz\"osische gibt es noch nicht.\\
Einen \"Uberblick zur fr\"uhesten Rezeption gibt Norbert Otto Eke 1991.[17] In den Rezensionen wurde M\"ullers Ton in Reisende aur [sic] einem Bein als knapp, spr\"od und ausdrucksstark beschrieben, aber auch als n\"orgelnd -- wo es um die bundesrepublikanische Wirklichkeit gehe --, was als st\"orend und kaum erhellend empfunden wurde. Einerseits beschrieb man M\"ullers Sprache als meisterhaft assoziativ, an\-de\-rerseits als ein teilweise unbeholfenes expressionistisches Krachen. M\"ullers Bilder wurden als zug\"ang\-lich wahrgenommen, aber auch als schief. Manchen der Rezensenten fehlte ein Handlungsfaden, an dem sie sich h\"atten orientieren k\"onnen. Auf andere hat M\"ullers Art und Weise, Details immer mit dem Ganzen korrespondieren zu lassen, im exemplarischen Erz\"ahlen \"uberanstrengt gewirkt. Eke zitiert aus G\"unther Franzens Besprechung, M\"ullers Wahrnehmungsoptik lasse "`das aufgebl\"ahte Exterieur der Ge\-sell\-schaft schruft\-lich\-keitn, bis das archaische Fundament sichtbar wird."'[21] Abschlie{\ss}end erw\"ahnt Eke aus einer Hochschulzeitung noch ein Rezensions-Beispiel, in dem sich ein "`Scheitern in der Ver\-mitt\-lung von pers\"onlicher und ge\-sell\-schaft\-li\-cher Thematik"' zeige, denn in der Rezension hei{\ss}e es, wo die Autorin ihrer inneren diffusen Zerrissenheit Ausdruck verleihe, ersticke sie in Metaphern und es w\"urden Nichtigkei\-ten dramatisiert und mystifiziert. Eke merkt an, dass M\"uller in der Kritik auf eine "`gesch\"arfte Aufmerksamkeit"' gesto{\ss}en sei, weil sie 1989 in Reisende auf einem Bein erstmals das Leben in der Bundesrepublik thematisierte, und zwar mit dem "`fremden"' Blick einer Ankommenden.[17]

\textbf{Leseerlebnis}\\
Reisende auf einem Bein ist nach Auffassung von Hans Ester kein Lesevergn\"ugen, denn alles von dem, wie sich Irenes d\"unne Haut anf\"uhlt, bekommt der Leser \mbox{selbst} zu sp\"uren. Wir schauen \"uber Irenes Schulter in einen tiefen Abgrund an Sinnlosigkeit. Irene ist keine Patientin, von der wir uns distanzieren k\"onnten, sondern sie sieht extrem genau, was wir \mbox{selbst} allzu gern negieren w\"urden.[1] Hineingezogen in Irenes Blick, wird dem Leser Bekanntes in Wahrnehmungsfragmente zerschnitten und als Fremdes und Trosloses gezeigt, formuliert es Peter Laudenbach.[22] Irene kann und will nicht an etwas Sch\"ones denken und wird von einer Trost- und Hoffnungslosigkeit gel\"ahmt, die sich am Ende auch auf den Leser \"ubertr\"agt, so hat es Ursula Homann bei der Lekt\"ure erlebt.[15] Karl Schulte hat das Werk mithilfe des Bildes von einem starren und dennoch kreisenden Gegenstand gelesen: "`Als Ganzes gleicht die Erz\"ahlung einem Mobile, an dem Fetzen der Wirklichkeit h\"angen, die sich st\"andig bewegen und doch auf der Stelle bleiben."'[2] F\"ur Susanne Schaber blieb nach der Lekt\"ure die schmerzliche Erfah\-rung zur\"uck, der Welt ausgesetzt zu sein, eine Irritation, die nicht so leicht abzusch\"utteln sei.[23] Reisende auf einem Bein ist nicht leicht zu lesen, res\"umiert Binder, weil mit verfremdeten Illusionen sowie Tr\"aumen und Visionen, die sich erst sp\"ater als solche he\-rausstellen, eine "`Art impliziter \flq Poetik der Verunsicherung\frq"' betrieben wird, die sich auf den Leser \"ubertr\"agt.[12]

{\tiny[1] Hans Ester: "`Reizigster op \'{e}\'{e}n been"', in: \textit{Trouw} 18 (1993), S. 4.\par}
{\tiny[2] Karl Schulte: "`\textit{Reisende auf einem Bein}. Ein Mobile"', in: \textit{Der Druck der Erfah\-rung treibt die Sprache in die Dichtung. Bildlichkeit in Texten Herta M\"ullers}, he\-rausgegeben von Ralph K\"ohnen, P. Lang, Frankfurt am Main 1997, ISBN 3-631-30662-8, S. 53-62.\par}
{\tiny[12] Karin Binder: "`Reisende auf einem Bein (1989)"', in: \textit{Handbuch der deutschspra\-chi\-gen Exilli\-te\-ra\-tur. Von Heinrich Heine bis Herta M\"uller}, he\-rausgegeben von Bettina Bannasch und Gerhild Rochus, De Gruyter, Berlin 2013, ISBN 978-3-11-025674-1, ISBN 978-3-11-025675-8, S. 464-471.\par}
{\tiny[15] Ursula Homann: "`Herta M\"uller. \textit{Reisende auf einem Bein}"', in: \textit{Deutsche B\"ucher}, Band 20 (1990), S. 109-110.\par}
{\tiny[22] Peter Laudenbach: "`Jeder f\"ur jeden ein Passant. Herta M\"ullers Reisende auf einem Bein"', in: \href{https://de.wikipedia.org/wiki/Die_Tageszeitung}{\textit{taz}}, 24. Oktober 1989; zitiert in Eke 1991, S. 124.\par}
{\tiny[23] Susanne Schaber: "`Mit einem Fuß im Osten, dem anderen im Westen. Reisende auf einem Bein, eine Erz\"ahlung der Rum\"aniendeut\-schen Herta M\"uller"', in: \href{https://de.wikipedia.org/wiki/Neue_Luzerner_Zeitung}{\textit{Luzerner Neuste Nachrichten}}, 29. Dezember 1989; zitiert in Eke 1991, S. 124.\par}

\subsubsection{"`Gr\"aser der Nacht"' (Modiano 2012)}
\label{subsubsec:6.1.11}

Eintrag "`Gr\"aser der Nacht"' (Hauptautor\textsuperscript{\tiny *} \mbox{User}:Magiers), Version vom \href{https://de.wikipedia.org/w/index.php?title=Gr\%C3\%A4ser_der_Nacht\&oldid=138052609}{22. Januar 2015, 17:04 CET}

\textbf{Artikelaufbau}, 1 Inhalt, 2 Hintergrund, 3 Ver\"of\-fent\-lichung und Rezeption, 3.1 Leseerlebnisse, 4 Ausgaben, 5 Rezensionen, 6 Einzelnachweise.

\begin{quote}
\textit{(Hauptautor\textsuperscript{\tiny *} des Eintrags sowie Autor\textsuperscript{\tiny *} des folgenden ersten Abschnitts ist \mbox{User}:Magiers)}\\
"`\textbf{Ver\"of\-fent\-lichung und Rezeption}\\
\textit{Gr\"aser der Nacht} erschien in seiner deutschen \"Ubersetzung rund einen Monat nach der Bekanntgabe des Nobelpreises f\"ur Li\-te\-ra\-tur 2014 an Patrick Modiano. Der Hanser Verlag hatte die urspr\"ung\-lich f\"ur Fr\"uhjahr 2015 vorgesehene Ver\"of\-fent\-lichung[5] auf den 10. November 2014 vorgezogen.[6] In der Folge wurde der Roman in zahlreichen deutschspra\-chi\-gen Feuilletons -- ausnahmslos positiv -- besprochen.[7] In der SWR-Bestenliste Dezember/Januar wurde \textit{Gr\"aser der Nacht} auf den dritten Platz gew\"ahlt.[8] Auch in der Verkaufs-Bestsellerliste von Buchreport platzierte sich der Roman mit dem h\"ochsten Rang 11 am 24. November 2014.[9]\\
F\"ur Gerrit Bartels eignet sich \textit{Gr\"aser der Nacht} \flq gut als Modiano-Einstiegslekt\"ure\frq.[4] J\"org Aufenanger hingegen, der \"Ubersetzer von Modianos Roman Ein so junger Hund, empfiehlt \textit{Gr\"aser der Nacht} eher \flq f\"ur fortgeschrittene Modianoleser, f\"ur die Initiierten\frq.[2] Tilman Krause sieht den Nobelpreistr\"ager jedenfalls im Roman \flq auf der H\"ohe seines K\"onnens\frq \,\,und in einem f\"ur den Autor eher ungew\"ohnlichen \flq Zustand einer sti\-lis\-ti\-schen und thematischen Fokussiertheit\frq.[10] Sebastian Hammel\-eh\-le \mbox{liest} am En\-de gar einen Kriminalroman, der \flq f\"ur Modianos Verh\"altnisse ungewohnt geradlinig\frq sei.[3] Joseph Hanimann stellt al\-ler\-dings klar, dass \textit{Gr\"a\-ser der Nacht} jederzeit \flq subtile und in den Mittellagen reichhaltige Li\-te\-ra\-tur\frq bleibe, die hohe Anforderungen an Elisabeth Edls \"Uberset\-zungs\-kunst stelle.[11] Hans-Jost Weyandt zieht das Fazit: \flq Nie ist es einfacher gewesen, Zugang zu finden zum Werk eines Nobelpreistr\"agers, und kaum einmal gelangt eine sentimental motivierte Prosa zu einer erz\"ahlerischen Klarheit wie in diesem Roman.\frq[12]"'
\end{quote}

\textbf{Leseerlebnisse}\\
Rezensenten thematisierten unter anderem, was bei \textit{Gr\"aser der Nacht} als Thriller gewirkt hat und welche Assoziationen zu vorherigen Kunsterlebnissen ihnen bei der Lekt\"ure kamen. Dar\"uber hinaus wurde das Charakteristische von Modianos Stil an sich als ein Leseerlebnis beschrieben.\\
Tilman Krause schrieb in der Welt: "`Bei Modiano kommen einem immer franz\"osische Chansons in den Sinn."' Er empfand die Lekt\"ure wie einen Ausflug, bei dem man Gesch\"opfen begegne, die verzaubert sind und im Wind schwanken, aber auch Schatten von politischen Verbrechen: "`Wir haben Figuren in Umrissen kennengelernt, die Kinofans an Truffauts Filme erinnern werden"', sie seien "`Nachtgeweihte wie von Novalis"'.[10] Joseph Hanimann berichtete in der S\"uddeutschen Zeitung, dass die durcheinanderwirbelnden Erinnerungen wie ein unterbelichteter Schwarz-Wei{\ss}-Film von Georges Franju anmuteten[11] und Hans-Jost Weyandt in der taz waren beim Lesen von \textit{Gr\"aser der Nacht} "`Mantelgestalten wie aus einem Melville-Film"' begegnet, sowie "`eine junge Frau wie aus einem Piaf-Chanson."'[12]
Judith von Sternburg schrieb in ihrer Rezension in der Frankfurter Rundschau, in \textit{Gr\"aser der Nacht} gebe es eine Atmosph\"are der Beunruhigung. Die Handlung sei konstruiert wie in einem Thriller, mit einem Mordfall, \"uber den alles im Nebel bleibe. Aber Modiano mache es einem m\"oglich, sich beim Lesen v\"ollig auf die Atmosph\"are zu konzentrieren. Dies l\"age vor allem an der Passivit\"at von Jean in seiner beunruhigten, aber schweigsamen Ergebenheit gegen\"uber Dannie.[13] Weyandt sah in Jean eine nostalgische Figur, "`die sich zum teilnahmslosen Beobachten verurteilt versteht"' und meinte, Modiano mache in \textit{Gr\"aser der Nacht} dieser Figur den Prozess: Leute, die "`v\"ollig zu ver\-schwin\-den scheinen hinter ihren Beobachtermasken."' Kalte Schattensei\-ten von Modianos ber\"uhmter sensibler Distanz in der Beschreibung hat Weyandt in diesem Roman entdeckt, und die flie{\ss}enden "`\"Uberg\"ange von der Diskretion zum Des\-inte\-resse"' in der Gestalt von Jean als pro\-ble\-ma\-tisch empfunden.[12]\\
F\"ur Gerrit Bartels im Tagesspiegel sorgte vor allem Modianos Stil f\"ur Vergn\"ugen bei der Lekt\"ure. Er fand, man m\"usse daher nicht alle Details entr\"atseln k\"onnen. Die Prosa schwebe sanft in einem Ton, der zwischen Gl\"ucklichsein und Traurigsein changiere, je nachdem, ob es ums Erinnern geht oder um Zei\-ten, die vergangen und daher verloren sind: "`Wer einen Roman von Patrick Modiano liest, macht oft dieselbe Erfah\-rung wie viele seiner Helden: Die Zeit verschwimmt, Vergangenheit und Gegenwart gehen ineinander \"uber, und ob dieser Modiano-Roman nun aus den siebziger Jahren stammt oder ein ganz aktueller ist, ger\"at bei der Lekt\"ure schnell in Vergessenheit."'[4] F\"ur Sebastian Hammelehle im Spiegel war es "`der ty\-pische, der magische Modiano-Sog"', in den Jean ger\"at, "`und der Leser mit ihm."' F\"ur diese Situation gab der Rezensent einen Tipp: Man habe "`bei der Lekt\"ure von \textit{Gr\"aser der Nacht} am besten den Stadtplan von Paris zur Hand."'[3] J\"org Aufenanger, gleicher Jahrgang wie Modiano, braucht vermutlich nicht einmal einen Stadtplan, denn, wie er in sei\-ner Rezension in der Berliner Zeitung berichtet, hat er ab den Endsechzigern \mbox{selbst} in Paris gelebt. Er hat den Roman doppelt autobiografisch gelesen: F\"ur Modiano und f\"ur sich \mbox{selbst}, mit sportlichen Assoziationen bei den Zeitspr\"ungen, die man versuche mitzumachen, um der Vergangenheit auf die Spur zu kommen.[2] Weyandt aller\-dings beschrieb den Text als so offen, dass man mit dem Lesen auch irgendwo in der Mitte beginnen k\"onne, und: "`Es wirkt alles leicht, fast magisch, doch zugleich ist es transparent."'[12]

{\tiny[2] \href{https://de.wikipedia.org/wiki/J\%C3\%B6rg_Aufenanger}{J\"org Aufenanger}: \href{http://www.berliner-zeitung.de/magazin/patrick-modiano-roman--graeser-der-nacht--buchtipps-fuer-den-herbst,10809156,29041850.html}{Buchtipps f\"ur den Herbst}. In: \textit{Berliner Zeitung} vom 14. November 2014.\par}
{\tiny[3] Sebastian Hammelehle: \href{http://www.spiegel.de/kultur/literatur/patrick-modiano-graeser-der-nacht-neuer-roman-des-nobelpreis-traegers-a-1002172.html}{Paris in den Sechzigern: Die st\"arkste Droge ist, auf ein M\"adchen zu warten}. In: \textit{Der Spiegel} vom 12. November 2014.\par}
{\tiny[4] Gerrit Bartels: \href{http://www.tagesspiegel.de/kultur/roman-graeser-der-nacht-von-patrick-modiano-bald-schon-bin-ich-alt/10944754.html}{Bald schon bin ich alt}. In: \textit{Der Tagesspiegel} vom 7. November 2014.\par}
{\tiny[10] \href{https://de.wikipedia.org/wiki/Tilman_Krause}{Tilman Krause}: \href{http://www.welt.de/kultur/literarischewelt/article134135284/In-Paris-waehrt-die-Liebe-nur-ein-Vierteljahr.html}{In Paris w\"ahrt die Liebe nur ein Vierteljahr}. In: \textit{Die Welt} vom 8. November 2014.\par}
{\tiny[11] Joseph Hanimann: \href{http://www.sueddeutsche.de/kultur/neuer-roman-von-patrick-modiano-paris-wie-es-flimmert-und-wirbelt-1.2209555}{Paris, wie es flimmert und wirbelt}. In: \textit{S\"uddeutsche Zeitung} vom 9. November 2014.\par}
{\tiny[12] Hans-Jost Weyandt: \href{http://www.taz.de/1/archiv/digitaz/artikel/?ressort=ku\&dig=2014/11/08/a0032\&cHash=d4db2f45eddbe2484f5ecdbbacc3c469}{Subtile Schuldgef\"uhle}. In: \textit{die tageszeitung} vom 8. November 2014.\par}
{\tiny[13] Judith von Sternburg: \href{http://www.fr-online.de/literatur/patrick-modiano--graeser-der-nacht--die-zeit-der-beunruhigung,1472266,29017966.html}{Die Zeit der Beunruhigung}. In: \textit{Frankfurter Rundschau} vom 11. November 2014.\par}

\subsubsection{"`Stimmungen lesen. \"Uber eine verdeckte Wirklichkeit der Literatur"' (Gumbrecht 2011)}
\label{subsubsec:6.1.12}

Eintrag "`Stimmungen lesen. \"Uber eine verdeckte Wirklichkeit der Li\-te\-ra\-tur"'\\ in der Version vom \href{https://de.wikipedia.org/w/index.php?title=Stimmungen_lesen._\%C3\%9Cber_eine_verdeckte_Wirklichkeit_der_Literatur\&oldid=139943406}{19. M\"arz 2015, 19:25 CET}

In der Einleitung: "`Der Essayband ist viel beachtet worden und hat widerspr\"uchliche Leseerlebnisse hervorgerufen."'

\textbf{Artikelaufbau}, 1 Inhalt, 1.1 Zu Liedern Walthers von der Vogelweide (um 1200), 1.2 Zum ersten Schelmenroman Lazarillo de Tormes (anonym, 1554), 1.3 Zu William Shakespeares Sonnett XVIII (1609), 1.4 Zu Mar\'{i}a de Zayas' Novellen (1635-1650), 1.5 Weitere Essays, 2 Stil, 3 Einsch\"atzungen zum Band, 3.1 Mark-Georg Dehrmann (\textit{SZ}, 2011), 3.2 Manfed Koch (\textit{NZZ}, 2011), 3.3 Detlev Sch\"ottker (\textit{FAZ}, 2011), 3.4 Wolf\-gang Schnei\-der (\textit{DR Kultur}, 2011), 3.5 Steffen Martus (\textit{Die Zeit}, 2012), 3.6 Stefan Hajduk (\textit{KulturPoetik}, 2012), 3.7 Andrei Corbea-Hoi{\c s}ie (\textit{Arcadia}, 2013), 4 Rezeption, 4.1 Leseerlebnis, 5 Ausgaben, 6 Einzelnachweise, 7 Weblinks.

\textbf{Rezeption}\\
\textit{Stimmungen lesen. \"Uber eine verdeckte Wirklichkeit der Li\-te\-ra\-tur} hat in den gro{\ss}en Zeitungen sowie dem bundeswei\-ten \"of\-fent\-lich-recht\-li\-chen Rundfunk ebenso wie in der Fach\-li\-te\-ra\-tur Beachtung gefunden. Rezensionen erschienen in \textit{S\"uddeutsche Zeitung}, \textit{Neue Z\"urcher Zeitung}, \textit{Frankfurter Allgemeine Zeitung}, \textit{Deutschlandradio Kultur} und \textit{Die Zeit} sowie in \textit{KulturPoetik} und in \textit{Arcadia: international journal of literary culture}.

\textbf{Leseerlebnis}\\
Gumbrechts Buch mache Mut zum sinnlichen Lesen, berichtet Manfred Koch, und er w\"urde gern mehr dar\"uber erfahren, was dessen Autor zu sagen hat, wenn er "`das unmittelbare \"asthetische Erlebnis festzuhalten versucht"', f\"ur das jener Hinweise von seiner Intuition erwartet.[6] Detlev Sch\"ottker stellt fest, das Buch k\"onne seinen Lesern viele bereichende Einsichten vermitteln, so etwa diejenige, dass "`\mbox{selbst} der Meisterdenker unter den zeitgen\"ossischen Philologen die sentimentalische Stimmung nicht aus seinem Gef\"uhlshaushalt verbannt"' habe.[7] Die "`Stimmung"' des Buches \mbox{selbst} sei irritierend, berichtet Steffen Martus, denn "`nicht wirklich zuversichtlich"' f\"ur die Zukunft einer \flq stimmungsorientierten\frq\,\,Li\-te\-ra\-tur\-wissenschaft erscheine ihm der Hauch von Wehmut, der die Lekt\"uren umgebe.[1]

{\tiny[1] Steffen Martus, \textit{Sehnsucht entziffern, Sommer beschw\"oren. Der Romanist Hans Ulrich Gumbrecht fordert eine neue, \flq stimmungsorientiertere\frq \,\,Li\-te\-ra\-tur\-wissenschaft}, in: \textit{Die Zeit}, 9. Juni 2011, S. 50\par}
{\tiny[6] Manfred Koch, \textit{Zeigen, was uns ergreift. Hans Ulrich Gumbrechts Wiederentdeckung der \"asthetischen Stimmung}, in: \textit{Neue Z\"urcher Zeitung}, 9. April 2011, S. 66\par}
{\tiny[7] Detlev Sch\"ottker, \textit{Heidegger in der Tasche. Hans Ulrich Gumbrecht liest mit Gef\"uhl}, in: \textit{Frankfurter Allgemeine Zeitung}, 16. April 2011, Nr. 90, S. Z5\par}

\subsubsection{Zusammenfassung}
\label{subsubsec:6.1.13}

Bei meiner Darstellung der in Leseerlebnis-Abschnitten referierten \"Au{\ss}erungen bin ich nach inhaltlichen Gesichtspunkten verfahren und habe entgegen allgemein postulierter Handhabung bei Wi\-ki\-pe\-dia nicht danach unterschieden, wer die Aussagen getroffen hat (Namhaftigkeit war kein ausschlie{\ss}endes Kriterium) oder wo eine Aussage pu\-bli\-ziert worden ist (Renommiertheit des Fundortes war kein ausschlie{\ss}endes Kriterium). Al\-ler\-dings war ich bei meiner Suche nach Belegen in der Reihenfolge so verfahren, dass ich in den meisten F\"allen nur bei Mangel an \textit{Erlesnis}-nahen Funden in der Namhaftigkeits-Kategorie auf unbekannteren Websites oder in Leseforen nach Beitr\"agen recherchiert habe, deren \"Au{\ss}erungen zu Leseerlebnissen mir \textit{Erlesnis}-nah genug erschienen. Die Tendenz meiner Auswahl geht also dahin, aus Renomm\'{e}e-Quellen auch Aussagen in Leseerlebnis-Abschnitten zu pr\"asentieren, die weniger \textit{Erlesnis}-nah sind (was auch damit zu tun hat, dass dort weniger oder keine \textit{Erlesnisse} zu finden waren), w\"ahrend ich bei der Auswahl in den weitere Quellen vor allem das Kriterium "`\textit{Erlesnis}-nah"' habe gelten lassen. Waren die Belege als Ensemble beisammen, bin ich so verfahren wie eingangs geschildert.

\pagebreak
\subsection{Experimentieren mit Leseberichten. Konzeption, Hergang, Auswertung}
\label{subsec:6.2}

Ein Experiment durchzuf\"uhren zu wollen basierte auf zwei Annahmen: Erstens ist bei Wi\-ki\-pe\-dia nicht alles machbar, was technisch machbar w\"are, denn die Community hat \"uber die Jahre einige Regeln f\"ur ihre Verfahrensweisen und Inhalte ausgehandelt, deren aktueller Stand bei Neuerungen erkundet werden muss. Zwei\-tens w\"are es unter diesem Gesichtspunkt f\"ur mein Erkenntnisinteresse, welche Bedingungen es auf Sei\-ten Wi\-ki\-pe\-dias f\"ur die Darstellbarkeit von nicht-pro\-po\-si\-ti\-o\-na\-lem Wissen in Eintr\"agen zu li\-te\-ra\-rischen Werken gibt, nicht ausreichend, neue Artikel anzulegen, denn es gibt derzeit nach allgemeiner Einsch\"atzung keine gro{\ss}e Anzahl an Usern, die bereit w\"aren, sich intensiv mit der Qualit\"at der Li\-te\-ra\-tur-Artikel an\-de\-rer zu befassen. Demzufolge schien es ratsam, \mbox{selbst} daf\"ur zu sorgen, dass bez\"uglich der von mir zu diesem Zweck verfassten Eintr\"age eine gewisse Aufmerksamkeit erzielt w\"urde.

Ziel des Experiments war, he\-rauszufinden, ob aktuell Vorbehalte gegen\"uber expliziten Ausf\"uh\-rungen zu "`Leseerlebnissen"' ge\"au{\ss}ert w\"urden, und falls ja, welche Vorbehalte dies w\"aren und mit welchem Ma{\ss} an Stichhaltigkeit in der Argumentation bzw. mit welcher Vehemenz. Die Bedeutsamkeit der Einw\"ande w\"urde neben den genannten inhaltlichen Faktoren unter anderem daran messbar sein, von wem die Einw\"ande ge\"au{\ss}ert werden und ob diejenige Person eventuell andere \"uber diese Vorg\"ange informieren w\"urde, um Neuerungen dieser Art gegebenenfalls mit einer Abstimmung zu verhindern. Die Dis\-kus\-si\-onsseite "`Grillenwaage"', auf der im Januar 2015 die \mbox{hier} ausgewertete Debatte stattfand, war in der Feiertagsferienzeit um den Jahreswechsel stark frequentiert worden, weil jemand ein Thema aufbrachte, zu dem viele sich \"au{\ss}erten, so auch ich. Im Zuge dieser belebten Dis\-kus\-si\-on entspann sich tats\"achlich auch ein Li\-te\-ra\-tur-Zweig, bei dem verschiedene Anliegen zur Sprache gebracht und nach meiner Einsch\"atzung sogar au{\ss}ergew\"ohnlich eingehend debattiert wurden.

Im Folgenden bespreche ich im ersten Abschnitt Reaktionen auf einzelne Lese\-erleb\-nis-Abschnitte und im zwei\-ten Abschnitt Ergebnisse einer zu Teilen von mir in Gang gebrachten Debatte bei dieser "`Grillenwaage"', einem Be\-nut\-zeraccount in der deutsch\-spra\-chi\-gen Wi\-ki\-pe\-dia, der von \mbox{vier} Leuten gemeinsam angelegt wurde, allein mit dem Ziel, auf der Dis\-kus\-si\-onsseite dieses Accounts eine Art Debattierclub in eigener Regie f\"uhren zu k\"onnen. \mbox{Hier} vorab ein kurzer \"Uberblick dazu, f\"ur welche der Artikel Reaktionen erzielt werden konnten. Die Ausflistung erfolgt wiederum in der Reihenfolge meines Artikelerstellens bzw. Erg\"anzens von Leseerlebnis-Abschnitten, fett markiere ich dabei die Artikel, auf die ich anschlie{\ss}end eingehe: 

¸* "`Die Lichter l\"osche ich"'. Der Artikel wurde angelegt von \mbox{User}:C.Kolt\-zen\-burg, meine Lesebericht-Auswertungen habe ich am 5. November 2014 und 13. November 2014 in den Artikel eingef\"ugt, al\-ler\-dings ohne sie (angesichts des knappen Eintrags) in einem eigenen Leseerlebnis-Abschnitt zu platzieren. Der "`Schon gewusst?"'-Vorschlag erfolgte am 5. November 2014, die "`Schon gewusst?" Pr\"asentation am 8. Dezember 2014, keinerlei Reaktionen (Stand: 15. M\"arz 2015).

¸* \textbf{"`Pour que tu ne te perdes pas dans le quartier"'}. Der Artikel wurde angelegt von \mbox{User}:C.Kolt\-zen\-burg, bei "`Schon gewusst?"'-vorgeschlagen am 23. November 2014, der Leseerlebnis-Abschnitt eingef\"ugt am 26. November 2014 (w\"ahrend der Vorschlagsphase), auf die "`Schon gewusst?"'-Pr\"asentation am 4. Dezember 2014 kam binnen einer Stunde eine Reaktion auf der Hauptsei\-ten-Dis\-kus\-si\-onsseite sowie eine Antwort darauf, danach keine weiteren Reaktionen (Stand: 15. M\"arz 2015).

¸* "`Im Caf\'{e} der verlorenen Jugend"'. Angelegt wurde dieser Eintrag von \mbox{User}:C.Kolt\-zenburg, der "`Schon gewusst?"'-Vorschlag war am 24. Oktober 2014, die Pr\"asentation am 15. November 2014. Den Leseerlebnis-Abschnitt habe ich erst danach eingef\"ugt, am 26. November 2014. Obwohl es zum Artikel zuvor Anmerkungen auf der Dis\-kus\-si\-onsseite gegeben hatte, gab es keine Reaktionen auf den neu eingef\"ugten Abschnitt (Stand: 15. M\"arz 2015).

¸* "`Pas pleurer"'. Der Artikel (zu Lydie Salvayre) wurde am 6. November 2014 angelegt von \mbox{User}:Goesseln, mit "`Schon gewusst?"'-Vorschlag am 13. November, am 5. Dezember 2014 (w\"ahrend der Vorschlagsphase) habe ich in diesem biografischen Eintrag einen Abschnitt zu \textit{Pas pleurer} eingef\"ugt, der aus einer Inhaltsangabe und einem Unterabschnitt "`Stil und Leseerlebnis"' be\-steht. Der Artikelersteller\textsuperscript{\tiny *} liess mich gew\"ahren und kommentierte nichts. Die "`Schon gewusst?"'-Pr\"asentation erfolgte am 10. Dezember 2014, keine Reaktionen in diesem Artikel (Stand: 15. M\"arz 2015) -- wohl aber im darin verlinkten Eintrag "`La Vie commune"'.

¸* \textbf{"`La Vie commune"'}. Dieser Artikel wurde am 9. Dezember 2014 von \mbox{User}:C.Kolt\-zenburg angelegt, und zwar gleich mit Leseerlebnis-Abschnitt (anstelle eines Ab\-schnitts "`Re\-zep\-tion"'). Bei "`Schon gewusst?"' habe ich den Artikel am 7. Januar 2015 vorgeschlagen, die Pr\"asentation erfolgte am 13. Januar 2015. Es hatte auf der Artikel-Dis\-kus\-si\-ons\-sei\-te am ersten Tag, nachdem der Artikel neu war, eine Reaktion zum Leseerlebnis-Abschnitt gegeben und die Dis\-kus\-si\-on wurde von diesem \mbox{User} teilweise in der Debatte bei "`Grillenwaage"' fortgesetzt. Auf der Artikeldis\-kus\-si\-onsseite kam noch \mbox{User}:Olag hinzu, er \"au{\ss}erte sich eigentlich dagegen (sieht es als Provokation), agiert im Artikeltext aber nicht. Keine weiteren Reaktionen (Stand: 15. M\"arz 2015).

¸* "`Atemschaukel"'. Erster Autor\textsuperscript{\tiny *} dieses Eintrags ist \mbox{User}:Tweebak. Der Artikel wurde 8. Oktober 2009‎ angelegt und ist nicht bei "`Schon gewusst?"' vorgeschlagen worden. Ich habe am 26. Dezember 2014 einen Leseerlebnis-Abschnitt eingef\"ugt, und diesen am 28. Dezember 2014 und am 11. Januar 2015 erweitert, um vielleicht doch noch Reaktionen zu erzielen, aber es erfolgten keine (Stand: 15. M\"arz 2015).

¸* \textbf{"`Alfred und Emily"'}. Der Artikel wurde in der eng\-lischspra\-chi\-gen Wi\-ki\-pe\-dia von \mbox{User}:Teatreez angelegt, in der deutschspra\-chi\-gen Wi\-ki\-pe\-dia wurde f\"ur die \"Uberset\-zung die dortige Versionsgeschichte importiert, nach erheblicher Erweiterung durch \mbox{User}:C.Kolt\-zen\-burg. F\"ur die deutschspra\-chi\-ge Wi\-ki\-pe\-dia hat \mbox{User}:C.Kolt\-zen\-burg den Eintrag \"ubersetzt und wiederum erweitert, den Artikel am 4. Dezember 2014 bei "`Schon gewusst?"' vorgeschlagen und w\"ahrend der Vor\-schlags\-pha\-se einen Le\-se\-erleb\-nis-Ab\-schnitt eingef\"ugt, und zwar am 27. Dezember 2014. Die "`Schon gewusst?"'-Pr\"a\-sen\-ta\-tion erfolgte am 7. Januar 2015. Es gab am ersten Pr\"asentationstag auf der Artikel-Dis\-kus\-si\-onsseite eine Reaktion speziell zum Leseerlebis-Abschnitt, auf die ich geantwortet habe, danach aber keine weiteren Reaktionen (Stand: 15. M\"arz 2015).

¸* "`Save the Reaper"'. Angelegt wurde der Artikel durch \mbox{User}:Jackentasche, eine "`Schon gewusst?"'-Pr\"a\-sen\-ta\-tion erfolgte am 7. Februar 2014. Den Leseerlebnis-Ab\-schnitt eingef\"ugt habe ich am 31. Dezember 2014, keine Reaktionen (Stand: 15. M\"arz 2015).

¸* \textbf{"`rein GOLD"'}. Der Artikel war schon 2013 von \mbox{User}:C.Kolt\-zen\-burg angelegt worden und es gab keine "`Schon gewusst?"'-Pr\"asentation. Den Leseerlebnis-Abschnitt habe ich am 5. Januar 2015 eingef\"ugt und innerhalb der Debatte bei "`Grillenwaage"' verlinkt am 19. Januar 2015. Trotz Debatten-Zugriffszahlen zwischen dem 18.\,\,und dem 20.\,\,Januar in H\"ohe von 162, 100 und 158 wurden beim Eintrag \mbox{selbst} keine markant h\"oheren Zugriffszahlen verzeichnet. In der Debatte gab es am selben Tag eine (positive) Reaktion. Es erfolgten keine \"Anderungen am Artikeltext und auch keine weiteren Reaktionen (Stand: 15. M\"arz 2015).

¸* "`Reisende auf einem Bein"'. Der Artikel wurde bereits im Fr\"uhjahr 2014 angelegt (von \mbox{User}:C.Kolt\-zen\-burg) und dessen "`Schon gewusst?"'-Pr\"asentation war am 10. Mai 2014. Einen Leseerlebnis-Abschnitt habe ich am 8. Januar 2015 eingef\"ugt und darauf gab es keine Reaktionen (Stand: 15. M\"arz 2015).

¸* \textbf{"`Gr\"aser der Nacht"'}. Angelegt wurde der Eintrag durch \mbox{User}:Magiers und es gab keine "`Schon gewusst?"'-Pr\"asentation. In Absprache mit \mbox{User}:Magiers habe ich (zu einem Zeitpunkt, als die Debatte bei der "`Grillenwaage"' beendet zu sein schien), einen Leseerlebnis-Abschnitt verfasst und gegenlesen lassen. \mbox{User}:Magiers gab eine positive Reaktion im Sinne eines OK f\"ur das Einf\"ugen, steht der Idee \mbox{selbst} eher ablehnend gegen\"uber, l\"asst mich aber gew\"ahren. Ich habe den Abschnitt am Tag darauf in den Artikel eingef\"ugt (22. Januar 2015), keine weiteren Reaktionen (Stand: 15. M\"arz 2015).

¸* \textbf{"`Stimmungen lesen. \"Uber eine verdeckte Wirklichkeit der Li\-te\-ra\-tur"'}. Erstelldatum des Artikels ist der 15. M\"arz 2015, angelegt von \mbox{User}:C.Kolt\-zen\-burg, gleich mit Leseerlebnis-Abschnitt. Am selben Tag schlug ich den Artikel bei "`Schon gewusst?"' vor. Ebenfalls am 15. M\"arz gab es auf der Dis\-kus\-si\-onsseite des Artikels eine Reaktion, insofern positiv, als meine Erl\"auterung des Sinns des Abschnitts akzeptiert wurde, im Rahmen eines Austauschs \"uber meh\-rere Abschnitte des Artikels. Seither keine weiteren Reaktionen (Stand: 27. M\"arz 2015).

\pagebreak
\subsubsection{Reaktionen auf einzelne Leseerlebnis-Abschnitte}
\label{subsubsec:6.2.1}

Auf sechs Artikel bzw. deren Leseerlebnis-Abschnitte gab es konkrete Reaktionen, wobei ich vermute, dass sie alle erst infolge meines Bestrebens, mehr Aufmerksamkeit auf die Artikel zu lenken, herbeigef\"uhrt werden konnten. Die Reaktionen auf \textbf{"`Pour que tu ne te perdes pas dans le quartier"'}, \textbf{"`La Vie commune"'} und \textbf{"`Alfred und Emily"'} und \textbf{"`Stimmungen lesen. \"Uber eine verdeckte Wirklichkeit der Li\-te\-ra\-tur"'} begannen h\"ochstwahrscheinlich durch deren Pr\"asentation bzw., im j\"ungsten Fall, durch deren Vorgeschlagenwerden bei "`Schon gewusst?"', bei den Eintr\"agen \textbf{"`rein GOLD"'} und \textbf{"`Gr\"aser der Nacht"'} bin ich so vorgegangen, dass ich diese Beispiele f\"ur Leseerlebnis-Abschnitte jeweils in einem Rahmen, der mir ge\-eig\-net erschien, an\-ge\-spro\-chen habe. In letztgenannten Fall habe ich dies mit einer direkten Kooperationsanfrage beim Artikelersteller\textsuperscript{\tiny *} unternommen. Mein Verfahren, das sich erst im Zuge des durchgef\"uhrten Experiments und zusammen mit dem Experiment weiter hat ausbauen lassen, k\"onnte f\"ur weitere allgemeine For\-schung zum einem methodisch weiterverwendbar sein, zum anderen k\"onnte das Vorgehen im Lichte der Reaktionen und der Ergebnisse f\"ur Arbei\-ten im Rahmen der Wikipedistik interessant sein, etwa f\"ur Variet\"aten in der Aushandlungskultur der deutschspra\-chi\-gen Wi\-ki\-pe\-dia im Vergleich zu anderen Sprachversionen.

Bez\"uglich der Frage, ob die Gr\"o{\ss}e des Artikels ein Einflussfaktor ist f\"ur die Be\-reit\-schaft, mit einem Einspruch t\"atig zu werden, l\"asst sich mit nur sechs F\"allen zwar keine Aussage treffen, aber da "`Leseerlebnis"' in Abschnitts\"uberschriften des Inhaltsverzeichnisses zu finden war, konnte leicht dorthin gesprungen werden, ohne einen langen Artikel durchlesen oder durchsuchen zu m\"ussen. Der Eintrag "`Pour que tu ne te perdes pas dans le quartier"' hatte am ersten "`Schon gewusst?"'-Pr\"asentationstag bei 6 Abbildungen einen Umfang von etwa 4 Sei\-ten pdf, "`Lydie Salvayre"'/ Pas Pleurer etwa 3 Sei\-ten pdf bei einer Abbildung, "`Alfred und Emily"' etwa 6 Sei\-ten pdf mit 4 Abbildungen und "`La Vie commune"' 2 Sei\-ten pdf mit einer Abbildung.\newline

\textbf{"`Pour que tu ne te perdes pas dans le quartier"'}

\singlespacing
\begin{quote}
\href{https://de.wikipedia.org/w/index.php?title=Wikipedia_Diskussion:Hauptseite\&diff=prev\&oldid=136447261}{"`\textbf{Pour que tu ne te perdes pas dans le quartier}\\
Ich empfinde den auf schon gewusst pr\"asentierten Artikel als Werbeeintrag ... schon in der Einleitung des Artikels steht \flq An Leseerlebnissen wurde geschildert [...] und dass Modianos leichtf\"u{\ss}iger sprachlicher Stil das Empfinden eines Flie{\ss}ens oder sogar eines Flirtens oder Verzaubertseins hervorrufen kann\frq ... das ist f\"ur mich kein neutraler Artikel und mich wundert gerade sehr, dass er auf der Hauptseite pr\"asentiert wird. Auf das \flq Leseerlebnis\frq \,wird sp\"ater sogar noch ausf\"uhrlich eingegangen."'} (\cite{UserAnghy2014})\\
\\
\href{https://de.wikipedia.org/w/index.php?title=Wikipedia_Diskussion:Hauptseite\&diff=prev\&oldid=136449008}{"`Interessante Frage.\\
In der Artikeldis\-kus\-si\-on \mbox{selbst} steht ja keinerlei Kritik am Artikel.\\
Die Werbegefahr ist wohl gering - kaum ein Artikelgegenstand braucht so wenig Werbung wie das neuste Werk des neuen Literaturnobelpreis\-tr\"a\-gers...\\
Meist ist ja die WP heillos \"uberfordert, wenn es um Belletristik geht, wie oft genug auch bei Werken der bildenden K\"unste. Da liest man dann auch \"uber literarische Klassiker Nacherz\"ahlungen wie von einem Grundsch\"uler. Dazu oft genug lange Listen von Personen.\\
Da ist dieser Artikel eine l\"obliche Ausnahme. Da geht es wirklich um das Buch."'} (\cite{UserHummelhum2014}) 
\end{quote}
\onehalfspacing

Im Einwand auf den Eintrag "`Pour que tu ne te perdes pas dans le quartier"', dem neuesten Werk von Patrick Modiano, wird ein Empfinden geltend gemacht: dass es sich um einen Werbeeintrag handele. Aufgrund der Beschreibung von Leseerlebnissen sei der Artikel nicht neutral und es sei verwunderlich, dass er dennoch auf der Hauptseite pr\"asentiert werde. Schon in der Einleitung stehe etwas mit Flie{\ss}en und Flirten und Verzaubertsein, und im Artikel werde "`sogar noch ausf\"uhrlich"' auf das Leseerlebnis eingangen. Am selben fr\"uhen Morgen etwa 3 Stunden sp\"ater wird entgegnet, das sei eine interessante Frage und in der Artikeldis\-kus\-si\-on sei keine Kritik am Artikel zu finden. Es wird die Einsch\"atzung ge\"au{\ss}ert, dass \mbox{hier} keine Werbungsabsicht bestehe. Als Begr\"undung wird genannt, ohnehin ben\"otige kaum ein Artikelgegenstand so wenig Werbung wie das neueste Werk eines aktuellen Li\-te\-ra\-tur\-nobelpreistr\"agers. In der Entgegnung wird ausgef\"uhrt, dass es sich bei diesem Artikel im \"Ubrigen um eine positive Ausnahme handele, denn meist sei die Wi\-ki\-pe\-dia "`heillos \"uberfordert, wenn es um Belletristik geht, wie oft genug auch bei Werken der bildenden K\"unste. Da liest man dann auch \"uber li\-te\-ra\-rische Klassiker Nacherz\"ahlungen wie von einem Grundsch\"uler. Dazu oft genug lange Listen von Per\-so\-nen."' Dieser Artikel sei gut. Eine Entgegnung darauf gab es leider nicht und auch ich bin nicht eingestiegen, denn ich war zu diesem Zeitpunkt noch zu \"uberrascht. Damit, dass "`mein Thema"' derart prominent Anlass f\"ur einen Austausch w\"urde, hatte ich nicht gerechnet. Ich f\"uhlte mich vor allem durch die geistreiche Entgegnung angespornt, in dieser Art weiter vorzugehen, in der Hoffnung, weitere Reaktionen zu erzielen, um he\-rauszufinden, ob meine Idee in einer Form Akzeptanz finden w\"urde, auf die ich bei sp\"ateren Aushandlungen gegebebenfalls verweisen k\"onnte, um meine Position zu st\"arken.\newline

\textbf{"`La Vie commune"'}

In diesen Eintrag hatte ich kaum eine Woche nach der Hauptsei\-tenverlinkung f\"ur "`Pour que tu ne te perdes pas dans le quartier"' einen Leseerlebnis-Abschnitt anstelle des \"ublichen Abschnitts "`Rezeption"' eingebaut, und zwar bereits am Tag der Artikelerstellung. Gleichzeitig war der Eintrag zur Autorin\textsuperscript{\tiny *} in der Rubrik "`Schon gewusst?"' auf der Hauptseite zu sehen und ich nehme an, dass es die Verlinkung im Per\-so\-nenartikel war, der \mbox{User}:Ges\-tum\-blin\-di folgte, um dann etwas im Artikeltext zu korrigieren: Ein von mir "`distanzlos"' wiedergegebenes Leseerlebnis wurde mit dem Namen der Rezensentin versehen, die Tageszeitung genannt und verlinkt, die Aussagen in indirekte Rede gesetzt und die Abschnitts\"uberschrift von "`Leseerlebnis"' zu "`Rezeption"' ge\"andert. Es ergab sich ein l\"angerer Austausch, weil ich eine im Eintrag vorgenommene \"Anderung auf der Artikeldis\-kus\-si\-onsseite an\-ge\-spro\-chen habe, ein weiterer \mbox{User} dazukam (w\"ahrend die Debatte bei "`Grillenwaage"' angelaufen war und nunmehr dieser Eintrag als neuer Artikel auf der Hauptseite pr\"asentiert wurde) und weil ich mich im Folgenden dann etwas st\"orrisch verhielt. Zum Abschluss des Austauschs auf der Artikeldis\-kus\-si\-onsseite begr\"undete ich inhaltlich, warum ich \mbox{selbst} die beiden nicht-namhaften Quellen nicht entfernen wolle. Dabei blieb es. \mbox{User}:Ges\-tum\-blin\-di und \mbox{User}:C.Kolt\-zen\-burg nahmen in der Dis\-kus\-si\-on auf verschiedene Wi\-ki\-pe\-dia-Spielregeln Bezug und Differenzen wurden nicht weiter ausgetragen, vor allem, weil mir an dem Artikel we\-sent\-lich mehr lag als dem Kollegen\textsuperscript{\tiny *}. Auch \mbox{hier} also ein Gew\"ahren-Lassen.\newline

\textbf{"`Alfred und Emily"'}

Am ersten Hauptsei\-ten-Pr\"asentationstag dieses Eintrags (8. Januar 2015) unternahm jemand den Schritt, nach dem Lesen von mindestens der Intro und einem der Abschnitte direkt zur Dis\-kus\-si\-onsseite des Artikels zu klicken und eine skeptische Anmerkung zu hinterlassen. 

\singlespacing
\begin{quote}
\href{https://de.wikipedia.org/w/index.php?title=Diskussion:Alfred_und_Emily\&oldid=137566068}{\textbf{"'Stil im Abschnitt \flq Leseerlebnis\frq}\\
Der Abschnitt wirkt stilistisch, als sei er aus einer umfassenderen Rezension \,\,\"uber \,\,das \,\,Werk \,\,he\-rauskopiert. Formulierungen wie \flq ...meint Ti\-ger\frq \,\,oder \flq Bewegend findet Sayers...\frq \,\,werden benutzt, ohne dass dem Leser vorher die Damen und Herren Tiger / Sayers mit vollem Namen vor\-ge\-stellt werden. Auch wenn ich keine entsprechende Textreferenz in der Such\-ma\-schi\-ne finden konnte, habe ich deshalb den starken Verdacht, dass hier abgeschrieben wurde."'} (\cite{User80.156.174.352015})
\end{quote}
\onehalfspacing

Auf meine Entgegnung, dass ich es \mbox{selbst} so verfasst habe, kam kein weiterer Beitrag. Meine Vermutung, dass jemand nach dem Lesen der Intro gleich im Inhaltsver\-zeich\-nis auf den Abschnitt "`Leseerlebnis"' geklickt hatte, habe ich als Hinweis auf ein eventuelles Leser\textsuperscript{\tiny *}interesse auch in der Debatte ge\"au{\ss}ert: 

\singlespacing
\begin{quote}
\href{https://de.wikipedia.org/w/index.php?title=Benutzer_Diskussion:Grillenwaage\&diff=next\&oldid=138021151}{"`Und dies hat mich best\"arkt: Neulich wurde mal ein von mir angelegter Artikel bei \flq Schon gewusst?\frq \,\,pr\"asentiert. Auf der Artikeldisk landete eine Anmerkung, die den Stil des Abschnitts Leseerlebnis betraf. Ich vermute, der Leser hat die Intro gelesen und danach direkt auf diesen Abschnitt geklickt: Das hatte ihn am meisten interessiert."'}\\ (\cite{UserC.Koltzenburg2015})
\end{quote}
\onehalfspacing

Konkret auf diese Reaktion Bezug nehmend gab es im Laufe der Debatte jedoch keine weitere \"Au{\ss}erung.\newline

\textbf{"`Rein Gold: Ein B\"uhnenessay"'}

Den Leseerlebnis-Abschnitt dieses Eintrags verlinkte ich im Laufe der Debatte bei "`Grillenwaage"', als bei einem in Li\-te\-ra\-tur und Sprache augenscheinlich bewanderten \mbox{User} schon eine positive Tendenz auszumachen war (19. Januar). Die Reaktion l\"asst sich in gewisser Hinsicht fast \mbox{selbst} als ein \textit{Erlesnis} meines Leseerlebnis-Abschnitts ansehen. Zudem wird \"uberlegt, was akademische Li\-te\-ra\-tur\-interpretation zu leisten vermag und ein Wi\-ki\-pe\-dia-Eintrag erscheint in dieser Hinsicht als ein Ort, der viele M\"oglichkei\-ten bietet. In Sachen Leseerlebnis solle mit den Richtlinien gro{\ss}z\"ugig verfahren werden, um Informationen dieser Art in Artikeln bieten zu k\"onnen.

\singlespacing
\begin{quote}
\href{https://de.wikipedia.org/w/index.php?title=Benutzer_Diskussion:Grillenwaage\&diff=137961506\&oldid=137943802}{"`Ja, genau, und nur so k\"onnen wohl die von mir oben (\"ubrigens v\"ollig spontan) zusammengestellten \flq Anforderungen\frq \,\,erf\"ullt werden: Dem Artikelleser, der das Werk noch nicht kennt, zumindest eine Ahnung vermitteln, \flq worum es geht\frq \,\,bzw. welches \flq Feeling\frq \,\,in dem Werk weht, also eben das \flq Wesen\frq. Dem anderen, der es zwar schon gelesen aber viel\-leicht mit einem schnellen Urteil (\flq ziemlich abstrus!\frq ) zur Seite gelegt hat, Eindr\"ucke anzubieten, die ihm entweder vielleicht ganz direkt die Augen \"offnen, zumindest aber helfen (egal ob er sie teilt oder nicht), seinen Blick dahingehend zu sch\"arfen.\\
Nehmen wir als Beispiel das Schlo{\ss} oder ein paar andere Sachen von Kafka: V\"ollig unm\"oglich, diese alptraumhafte Atmosph\"are einer offenbar naturgesetzlichen Vergeblichkeit mit einer reinen Inhaltszusammenfassung r\"uberzubringen, und auch die akademischen Ausdeutungen ver\-m\"o\-gen das nicht zu leisten. Wie sollten sie auch? Denn es sind ja nicht die gelehrten Deutungen und Schlussfolgerungen, sondern eben diese gef\"uhlte Atmosph\"are, die den Reiz und das Wesen des Werks ausmachen.\\
Was aber, wenn es keine entsprechenden Quellen gibt und man sich diese sozusagen \flq atmosph\"arische\frq \,\,Charakterisierung \mbox{selbst} aus den Fingern saugen mu{\ss}? Ich finde, da{\ss} man auf diesem speziellen Gebiet die Ma{\ss}gaben von TF und POV zumindest au{\ss}ergew\"ohnlich kulant handhaben sollte."'} (\cite{UserEpipactis2015b})
\end{quote}
\onehalfspacing

Ich entgegne, dass es zu Kafka sehr wohl eine "`akademische Ausdeutung"' gibt, die sich zu diesem Punkt als Beleg eignen w\"urde, und verlinke das Inhaltsverzeichnis des Bandes, mit Nina Orts Studie zu \textit{Josefine, die S\"angerin oder Das Volk der M\"ause}. Siehe dazu auch meine Ausf\"uh\-rungen zum Zusammenhang der Debatte, im folgenden Abschnitt \textit{~\ref{subsubsec:6.2.2} Debatte zu Leseerlebnis-Abschnitten}.\newline

\textbf{"`Gr\"aser der Nacht"'}

Am Beispiel des Eintrags zu diesem Modiano-Werk, den \mbox{User}:Magiers verfasst hat, wollte ich gern ausprobieren, ob jemand einverstanden ist, wenn ich einen Leseerleb\-nis-Ab\-schnitt in einem bereits bestehenden Artikel erg\"anze -- w\"ahrend die Debatte lief, an der sich dieser \mbox{User} zum gegebenen Zeitpunkt nicht mehr beteiligte. \mbox{User}:Magiers hatte mitbekommen, dass ich Verfasser\textsuperscript{\tiny *} der beiden anderen Werkartikel zu Modiano bin, zwei umfangreiche Eintr\"age, die ausf\"uhrliche Leseerlebnis-Abschnitte aufweisen. Da mir bekannt ist, dass dem Kollegen\textsuperscript{\tiny *} eine Wertsch\"atzung gegen\"uber Artikel\-erstellern\textsuperscript{\tiny *} wichtig ist (wir hatten schon vorher gelegentlich ko\-ope\-riert), rich\-te\-te ich sozusagen meine Kooperationsanfrage an den Hauptautor\textsuperscript{\tiny *}, und zwar auf dessen Be\-nut\-zerdis\-kus\-si\-onsseite. Ich verfasste den Abschnitt auf einer Unterseite in meinem Be\-nut\-zernamensraum und verlinkte die Seite, so dass sich \mbox{User}:Magiers meinen Vorschlag vorab w\"urde ansehen k\"onnen: 
\href{https://de.wikipedia.org/w/index.php?title=Benutzer_Diskussion:Magiers&diff=next&oldid=138016279}{"`Finde ich prima. [...] Deine Le\-se\-er\-fah\-r\-un\-gen bringen erst viele Details in den Artikel ein, die ihm bis\-her fehlen."'} Dieser \mbox{User} sieht es also \"ahnlich wie in der Debatte zuvor \mbox{User}:Epipactis, insofern Leseerlebnisse als Bereicherung des Artikels empfunden werden. Zwei Tage sp\"ater frage ich zu einer bestimmten Rezension etwas nach und daraufhin bringt \mbox{User}:Magiers noch zum Ausdruck, dass Leseerlebnisse (der von mir eingef\"uhrte Begriff wird in Anf\"uh\-rungsstrichen verwendet) bis\-her in Rezeptionsabschnitten Platz gefunden haben und dass ihm\textsuperscript{\tiny *} \href{https://de.wikipedia.org/w/index.php?title=Benutzer_Diskussion:Magiers&diff=next&oldid=138104304}{zwei eigene Beispiele} einfallen. Die Inhalte sind also willkommen, auch wenn ein separater Abschnitt als nicht so einleuchtend empfunden wird.

In diesem Zweiergespr\"ach, das zeitgleich mit der Debatte zustandekam, \"uberwiegt eine positive Einsch\"atzung. \mbox{User}:Magiers hat die Rolle eines Administrators\textsuperscript{\tiny *} inne und schreibt Artikel zu fiktionalen Texten, \"uberwiegend zu Krimis, ist nach eigenen Aussagen kein Li\-te\-ra\-tur\-wissenschaftler\textsuperscript{\tiny *} und wird in der Community wegen seines auf Ausgleich bedachten Umgangs sehr gesch\"atzt, bei dem unter anderem Gew\"ahren-Lassen als die richtige Art angesehen wird, Wertsch\"atzung zum Ausdruck zu bringen.\newline 

\textbf{"`Stimmungen lesen. \"Uber eine verdeckte Wirklichkeit der Li\-te\-ra\-tur"'}

Diese Reaktion auf den Rezeptionsabschnitt eines Eintrags stammt von Mitte M\"arz, der \href{https://de.wikipedia.org/w/index.php?title=Diskussion:Stimmungen_lesen._\%C3\%9Cber_eine_verdeckte_Wirklichkeit_der_Literatur&oldid=139806486}{Austausch} fand also knapp zwei Monate sp\"ater als die anderen statt, zu einem Zeitpunkt, als ich den neuen Artikel gerade bei "`Schon gewusst?"' vorgeschlagen hatte. Auch mit diesem Kollegen\textsuperscript{\tiny *} hatte ich bei an\-de\-rer Gelegenheit bereits kooperiert. Auf die Frage \href{https://de.wikipedia.org/w/index.php?title=Diskussion:Stimmungen_lesen._\%C3\%9Cber_eine_verdeckte_Wirklichkeit_der_Literatur&diff=next&oldid=139796600}{"`ist diese Rezeptionsaufspaltung \mbox{hier} li\-te\-ra\-tur\-wissenschaftl. bedingt?"'} antwortete ich, etwas popularisierend: \href{https://de.wikipedia.org/w/index.php?title=Diskussion:Stimmungen_lesen._\%C3\%9Cber_eine_verdeckte_Wirklichkeit_der_Literatur&diff=next&oldid=139796600}{"`Ja: 3. sind die fachlichen Einsch\"atzungen zu den Aussagen und der Machart des Pu\-bli\-zierten (also die Li\-te\-ra\-tur\-kritik, in diesem Fall zu einem li\-te\-ra\-tur\-theo\-retischen Thema). Das ist also das, was in einem Artikel \"uber einen Roman im Abschnitt \flq Interpretationen\frq\, stehen w\"urde; 4. ist Wirkung im messbaren Sinne, man k\"onnte noch Verkaufszahlen erg\"anzen; 5. sind Berichte \"uber pers\"onliche Reaktionen auf den Band, aus der Sicht von \flq Lesern\frq, nicht in ihrer Eigenschaft als Fachkollegen, sondern eben als Leser, quasi nicht \flq bei der Arbeit\frq\, am Schreibtisch, sondern zuhause im Sessel"'}, was vom Kollegen\textsuperscript{\tiny *} als eine sinnvolle Trennung angesehen wird. Der Leseerlebnis-Abschnitt \mbox{selbst} wird weder inhaltlich noch stilistisch infrage gestellt.

\subsubsection{Debatte zu Leseerlebnis-Abschnitten}
\label{subsubsec:6.2.2}

In einem der Club-\"ahnlichen Zusammenh\"ange, die allen offenstehen, ent\-wickelte sich bei der deutschspra\-chi\-gen Wi\-ki\-pe\-dia im Dezember 2014 reger Dis\-kus\-si\-onsbedarf zu einem bestimmten Verhalten von Usern, das vom Initiator des Themas nicht gutgehei{\ss}en wurde, womit er Zustimmung ebenso wie Gegenreden erhielt. Mit welchen Mitteln bei dem von ihm ge\"au{\ss}erten Anliegen welche Art von \"Anderung erreicht werden k\"onnte, wurde sehr unterschiedlich eingesch\"atzt. Die Beteiligung an der Debatte war vor allem in den Tagen um den Jahreswechsel sehr rege und vielseitig, weshalb ich auf die Idee kam, es auch mal mit einen Li\-te\-ra\-tur\-thema zu probieren.

In der Debatte hie{\ss} es am 18. Januar:

\singlespacing
\begin{quote}
\href{https://de.wikipedia.org/w/index.php?title=Benutzer_Diskussion:Grillenwaage\&diff=next\&oldid=137891471}{"`einen Abschnitt wie das dortige Leseerlebnis ["`La Vie commune"'] finde ich in seiner mangelnden Distanziertheit verfehlt. Aber nunja... vielleicht verstehe ich einfach den grunds\"atzlich anderen, neuen Ansatz f\"ur das Erstellen von Li\-te\-ra\-tur-Artikeln, den du \mbox{hier} einbringen m\"ochtest, noch nicht."'} (\cite{UserGestumblindi2015})
\end{quote}
\onehalfspacing

Dieser Wi\-ki\-pe\-dia-Kollege\textsuperscript{\tiny *} formulierte das Ziel meiner Initiative so, dass es sich um einen grunds\"atzlich anderen Ansatz handeln w\"urde. Ablehnung wird einlei\-tend nicht bez\"uglich der Inhalte des Abschnitts ge\"au{\ss}ert, sondern aufgrund mangelnder Distanziertheit. Demnach w\"aren die Themen der Leseerlebnis-Abschnitte akzeptabel, solange der Darstellungsstil distanziert genug ist. 

Tags darauf, am 19. Januar, hatte ich den Leseerlebnis-Abschnitt f\"ur den relativ umfangreichen Eintrag zu \textit{rein GOLD} fertig und verlinkte ihn in der Debatte, woraufhin \mbox{User}:Epipactis zustimmend schrieb: 

\singlespacing
\begin{quote}
\href{https://de.wikipedia.org/w/index.php?title=Benutzer_Diskussion:Grillenwaage\&diff=137961506\&oldid=137943802}{"`Ja, genau, und nur so k\"onnen wohl die von mir oben (\"ubrigens v\"ollig spontan) zusammengestellten \flq Anforderungen\frq \,\,erf\"ullt werden: Dem Artikelleser, der das Werk noch nicht kennt, zumindest eine Ahnung vermitteln, \flq worum es geht\frq \,\,bzw. welches \flq Feeling\frq \,\,in dem Werk weht, also eben das \flq Wesen\frq .\,\,Dem anderen, der es zwar schon gelesen aber viel\-leicht mit einem schnellen Urteil (\flq ziemlich abstrus!\frq ) zur Seite gelegt hat, Eindr\"ucke anzubieten, die ihm ent\-weder vielleicht ganz direkt die Augen \"offnen, zumindest aber helfen (egal ob er sie teilt oder nicht), seinen Blick dahingehend zu sch\"arfen."'} (\cite{UserEpipactis2015b})
\end{quote}
\onehalfspacing

Hier wird eine eigene Lese-Wirkung beschrieben und im Zuge dessen eine Reihe von Fragen formuliert, die man als eine Funktionsbeschreibung der Leseerlebnis-Abschnitte ansehen kann. Diese Reaktion \mbox{selbst} w\"urde ich aus meiner Warte als \textit{Erlesnis} ansehen. Sie hat mich sehr angespornt, weiter mit neuartigen Aspekten und abweichenden Formulierungen in Wi\-ki\-pe\-dia-Eintr\"agen zu experimentieren. Am 21. Januar lautete die Einsch\"atzung desselben Users so: 

\singlespacing
\begin{quote}
\href{https://de.wikipedia.org/w/index.php?title=Benutzer_Diskussion:Grillenwaage\&diff=next\&oldid=138032479}{"`Die Idee eines Abschnitts \flq Leseerlebnis\frq \,\,(oder wie auch immer man ihn betiteln will) halte ich auf jeden Fall f\"ur fundamental, um die \flq wei{\ss}en Flecken\frq \,\,abzudecken, die neben der zwangsl\"aufig n\"uchternen Ana\-lyse und der (sofern der Artikelschreiber nicht \mbox{selbst} \"uber li\-te\-ra\-risches Ta\-lent ver\-f\"ugt) meist ebenso trockenen (im schlimmsten Fall geradezu nichts\-sa\-gen\-den) Inhaltszusammenfassung mMn oft und deutlich sp\"urbar sind. Na\-t\"ur\-lich dabei nicht \"ubertreiben -- wenn es zwanzig Stimmen gibt, mu{\ss} man nicht un\-be\-dingt alle bringen, und vielleicht auch nicht gerade diejenigen, die zum Verst\"andnis tiefgr\"undige Vorkenntnisse oder intensive Be\-sch\"af\-ti\-gung mit dem Thema erfordern."'} (\cite{UserEpipactis2015c}
\end{quote}
\onehalfspacing

Hier wird mit einer Kombination aus Stil und Umfassenheit argumentiert, und zwar f\"ur Leseerlebnis-Abschnitte, die auch anders genannt werden k\"onnten. Aber die Idee an sich wird f\"ur grundlegend gehalten, um Artikel zu verbessern, vor allem um sie besser lesbar zu machen. 

User:Kr\"achz schrieb am 22. Januar: 

\singlespacing
\begin{quote}
\href{https://de.wikipedia.org/w/index.php?title=Benutzer_Diskussion:Grillenwaage\&diff=next\&oldid=138039032}{"`Ich sympatisiere mit Ges\-tum\-blin\-dis Erwiderung, dass der g\"angige \flq Re\-zep\-tions\-ab\-schnitt\frq \,\,f\"ur das \flq Leseerlebnis\frq \,\,ausreichend Raum gibt. L\"asst sich Kritik/Rezeption/Einordnung allgemein von Leseerlebnis trennen? Und ist \flq Leseerlebnis\frq \,\,in der Li\-te\-ra\-tur\-wissenschaft eine etablierte Gr\"o\-{\ss}e? Ich wei{\ss} es nicht, deinen Abschnitt in "`Alfred und Emily"' finde ich aber trotz aller Bedenken gelungen."'} (\cite{UserKrachz2015}) 
\end{quote}
\onehalfspacing 

Auch \mbox{hier} also wird gewisses Interesse sig\-na\-li\-siert, aber ebenso Zur\"uckhaltung bezogen auf formale Aspekte. Ein bereits vorhandenes Beispiel wird aber dennoch gut gefunden. 

Am Tag darauf formulierte \mbox{User}:Epipactis Unterst\"utzung in dieser Art: 

\singlespacing
\begin{quote}
\href{https://de.wikipedia.org/w/index.php?title=Benutzer_Diskussion:Grillenwaage\&diff=next\&oldid=138067900}{"`MMn ist die Umsetzung des bis\-her Besprochenen in der derzeitigen Version von La Vie commune schon sehr gut gelungen. Alle Abschnitte sind erfreulich knapp, das ist das erste Plus. [...] Auch das \flq Leseerlebnis\frq \,\,h\"alt sich innerhalb des (f\"ur meinem Geschmack) zumutbaren Umfangs."'} (\cite{UserEpipactis2015d})
\end{quote}
\onehalfspacing

Es gab noch zwei weitere Beitr\"age von bis\-her Beteiligten, in denen man sich einig war, dass "`anonyme Stimmen"' -- wie sie genannt wurden -- als nicht ge\-eig\-net anzusehen sind, einerseits, weil sie unbekannt seien, an\-de\-rerseits, weil sich der Ar\-ti\-kel\-schrei\-ber\textsuperscript{\tiny *} \mbox{selbst} dahinter verstecken k\"onne. Ich hatte zuvor inhaltlich begr\"undet, warum ich sie als relevant ansehe, dass sie n\"amlich beschrei\-ben,
\href{https://de.wikipedia.org/w/index.php?title=Benutzer_Diskussion:Grillenwaage\&type=revision\&diff=138095833\&oldid=138073134}{"`was sich durchs Lesen dieses Buches f\"ur sie ver\"andert hat."'}

In dieser Zeit hatte diese "`Grillenwaage"' etwa 100 Abrufe am Tag. Zwar ist den statistischen Angaben nicht zu entnehmen, welche Abschnitte der Debatte gelesen wurden, aber man kann davon ausgehen, dass einige dieses Thema registriert haben, \mbox{selbst} wenn sie sich nicht veranlasst f\"uhlten, sich zu beteiligen. Ge\"au{\ss}ert haben sich au{\ss}er mir drei andere mehr als einmal. An den diskutierten Abschnitten der Artikel sind seither (Stand: 15. M\"arz) keine \"Anderungen vorgenommen worden, man hat mich also gew\"ahren lassen. 

Da ich f\"ur diese Werke noch nicht gen\"ugend \textit{Erlesnisse} gefunden hatte, bestehen die Abschnitte zwar weit \"uberwiegend aus allgemeineren Leseerlebnissen, aber nach dieser Debatte sehe ich die Idee eines eigenen Abschnitts f\"ur Leseerlebnisse als etabliert an -- soweit sich das schon jetzt -- nach erst zwei Monaten -- sagen l\"asst. Andersherum formuliert k\"onnte man sagen, dass die eloquenteste Stimme in der Debatte die meiste Unterst\"utzung der Idee formuliert hat.

Etwa zehn Tage nachdem die Debatte abgeklungen war, habe ich auf Nachfrage bei \mbox{User}:Edith Wahr eine weitere Einsch\"atzung erfahren, und diesmal handelt es sich um einen Verfasser\textsuperscript{\tiny *} von Li\-te\-ra\-tur\-artikeln. \mbox{User}:Edith Wahr ist der Auffassung, dass Leseerleb\-nis-Abschnitte in Wi\-ki\-pe\-dia-Eintr\"agen nicht notwendig sind, da es \textit{per\-len\-tau\-cher.de} bereits gebe. Al\-ler\-dings wird im folgenden Dis\-kus\-si\-onsbeitrag nicht konkreter gesagt, welche "`\flq enzy\-klop\"a\-di\-schen\frq \,\,Zwecke"' gemeint sind, f\"ur die sich Erlebnisse "`tats\"achlich nicht so gut"' eignen, die im Feuilleton zur Sprache kommen. Im Wortlaut: 

\singlespacing
\begin{quote}
\href{https://de.wikipedia.org/w/index.php?title=Benutzer_Diskussion:Edith_Wahr\&diff=prev\&oldid=138494233}{"`ich lese ja dieser Tage Daniel Woodrell, ganz zauberhaft; aber auch erstaunlich, wieviel man in li\-te\-ra\-tur\-\flq wissenschaftlichen\frq \,\,Zeitschriften zu diesem Autor findet, n\"amlich gar nix. In F\"ulletons [sic], gerade auch den deutschen, hingegen eine ganze Menge, nur so wirklich eignet sich das dort gerne verbreitete \flq Leseerlebnis\frq \,\, tats\"achlich nicht so gut f\"ur \flq enzy\-klop\"adische\frq \,\, Zwecke, und nach einiger \"Uberlegung bin ich zu dem Schluss gekommen, dass ein Digest dieser \flq Erlebnisse\frq \,\, beim Per\-len\-tau\-cher besser aufgehoben sind, und der hat uns/mir diese Arbeit schon abgenommen."'} (\cite{UserEdithWahr2015b})
\end{quote}
\onehalfspacing

In einem weiterem Beitrag wird meines Erachtens angedeutet, was nach Ansicht dieses Users einen Wi\-ki\-pe\-dia-Eintrag von einer Rezensionsnotiz unterscheiden sollte: ein S\"attigungsgrad. Dass Leseerlebnisse einen Artikel in diesem Sinne abrunden k\"onnten (wie \mbox{User}:Epipactis deren Funktion in der Debatte sah), wird in der folgenden Standpunktschilderung nicht in Betracht gezogen.

\singlespacing
\begin{quote}
\href{https://de.wikipedia.org/w/index.php?title=Benutzer_Diskussion:Edith_Wahr\&diff=next\&oldid=138508487}{"`So ziemlich jede Zeitung hat "`Winters Knochen"' rezensiert, alle fanden es super, nur wird es da eben nicht viel spezifischer als z.B. "`Daniel Woodrell erz\"ahlt knapp und doch stimmungsvoll, in oft \"uberraschenden Bildern. Es ist ein \mbox{selbst} gleichsam bis auf die Knochen reduzierter, ein w\"ahlerischer Text, der aber das, was ihm wichtig ist, wie im Scheinwerferlicht aufleuchten l\"asst."' (so etwa die FR). Und wenn man genug solcher Schnipsel aneinanderreiht, ob als Zitat oder Paraphrase, dann kommt sowas wie beim Per\-len\-tau\-cher dabei he\-raus. Was ja nicht das schlimmste w\"are, aber so wirklich s\"attigend auch nicht, abgesehen davon, dass es den Per\-len\-tau\-cher nunmal schon gibt. Nunja. [...]"'} (\cite{UserEdithWahr2015c})
\end{quote}
\onehalfspacing

Nach meiner Auffassung eignen sich weder das Konzept noch das knappe Format der Rezensionsnotizen bei Per\-len\-tau\-cher daf\"ur, \textit{Erlesnisse} angemessen zur Geltung zu bringen. Al\-ler\-dings habe ich einige Beispiele gefunden, in denen Formulierungen in den Rezensionsnotizen recht nah an \textit{Erlesnisse} herankamen, vgl. Abschnitt \textit{~\ref{subsubsec:7.1.2} Rezensionsnotizen bei per\-len\-tau\-cher.de} 

\subsection{Zusammenfassung}
\label{subsec:6.3}

Auf \textit{Objektebene II} befasste ich mich mit dem Experiment, bestehend aus dem Platzieren neuartiger Abschnitte, in denen \textit{Erlesnisse} enthalten sein k\"onnen, und dem Anregen von Austausch dazu in einer Debatte sowie in einigen Zweier/Dreier-Gespr\"achen auf Artikeldis\-kus\-si\-onssei\-ten. Ziel des Experiments war, he\-raus\-zu\-fin\-den, ob aktuell Vorbehalte gegen\"uber expliziten Ausf\"uh\-rungen zu Leseerlebnissen ge\"au{\ss}ert w\"urden. Die Idee wurde \"uberwiegend positiv aufgenommen, einschr\"ankende \"Ande\-rungen an den Leseerlebnis-Abschnitten der diskutierten Artikel wurden keine mehr vorgenommen. Als Ergebnis festhalten l\"asst sich, dass das Einbeziehen von Aussagen \"uber nicht-pro\-po\-si\-ti\-o\-na\-les Wissen aus Li\-te\-ra\-tur\-lekt\"ure auch k\"unftig mit Akzeptanz rechnen k\"onnte, sofern in ausreichend distanzierter Weise formuliert w\"urde und, wo dies m\"oglich ist, Belege zu finden sind.

\pagebreak

\section{Metaebene II}
\label{sec:7}

In diesem letzten Teil meiner Arbeit komme ich auf das doppelseitige Pro\-blem der Bedingungen der Darstellbarkeit nicht-pro\-po\-si\-ti\-o\-na\-len Wissens in Werkartikeln bei Wi\-ki\-pe\-dia zur\"uck und widme mich erneut der Literatur-Seite dieser Bedingungen: Gibt es f\"ur das Verfassen von Wikipedia-Eintr\"agen nicht gen\"ugend zitierbare Quellen, kommt der Prozess, Erkenntnisse aus Li\-te\-ra\-tur\textsuperscript{\~.\~.}lekt\"ure web\"offentlich zu vermitteln, ins Stocken. Werden hingegen Leseberichte, die \textit{Erlesnisse} enthalten, namhaft pu\-bli\-ziert, gelangen wertvolle Aussagen in den \"of\-fent\-lichen Raum, die geeignet sind, Wertsch\"atzung f\"ur Literaturlekt\"ure zum Ausdruck zu bringen. Allem Anschein nach -- so die Hypothese, die ich mithilfe meines Formalobjekts (\textit{Erlesnis}) erarbeiten konnte, siehe auch \textit{Kapitel ~\ref{sec:2}, Research Design} -- wird der Schritt, \textit{Erlesnisse} jenseits von Freund\textsuperscript{\tiny *}eskreisen, also Fremden gegen\"uber, zu \"au{\ss}ern, im deutschsprachigen Raum weitgehend als eine Preisgabe mit gro{\ss}en "`Frei\-m\"u\-tig\-keits\-an\-tei\-len"' empfunden, zumal dann, wenn ein solcher "`parrhesiatischer"' Schritt in den web\"offentlichen Raum stattfinden soll. Diesbez\"uglich scheint es gesellschaftliche Konventionen zu geben, mit denen gebrochen werden muss, will man trotz eines Risikos, das weitgehend als nicht kalkulierbar wahrgenommen wird, zum Ziel gelangen.

Im ersten Teil entwickle ich Ans\"atze einer Typologie von Schreibweisen \"uber \textit{Erlesnisse} und im zwei\-ten Teil er\"ortere ich, was meine Erkenntnisse in der Li\-te\-ra\-tur\-ver\-mitt\-lungsfor\-schung zu theo\-retischen Debatten beitragen, die im weitesten Sinne von \textit{Erlesnissen} und Fragen ihrer Ver\-mitt\-lung au{\ss}erhalb eines Freund\textsuperscript{\tiny *}eskreises inspiriert sind. 

\pagebreak
\subsection{\textit{Parrhesia} im Schreiben \"uber Literatur\textsuperscript{\~.\~.}: Ans\"atze einer Typologie}
\label{subsec:7.1}

F\"ur die Ans\"atze einer Typologie gehe ich in \mbox{vier} Schritten vor. Im ersten Schritt ana\-lysiere ich Beispiele aus Essayb\"anden von Ulrike Draesner, Hans Ulrich Gumbrecht und Ina Hartwig. In Schritt 2 bespreche ich Beispiele aus Rezensionsnotizen der Plattform \textit{per\-len\-tau\-cher.de}. Weil Verfasser\textsuperscript{\tiny *} der Rezensionsnotizen ebenso wie Wi\-ki\-pe\-dia-Autoren\textsuperscript{\tiny *} \"uber Leseberichte an\-de\-rer schrei\-ben, sind auch sie f\"ur die Darstellbarkeit nicht-pro\-po\-si\-ti\-o\-na\-len Wissens in ihrer Zusammenfassung abh\"angig davon, dass wiederum in den Rezensionen, \"uber die sie berichten, etwas dar\"uber gesagt wird, wie es sich angef\"uhlt hat, den Text zu lesen. Im dritten Schritt thematisiere ich \textit{Erlesnis}-Formulierungen in von mir verfassten Lesererlebnis-Abschnitten in Werkartikeln bei Wi\-ki\-pe\-dia. Schritt 4 geht auf meine \textit{Erlesnis}-Schilderungen in den \mbox{vier} eigenen Interpretationen ein. Die beiden Schritte, in denen direkte Interpretationen von li\-te\-ra\-rischen Texten auf \textit{Erlesnisse} hin besprochen werden, bilden also den Rahmen f\"ur die Schritte 2 und 3, in denen Schreibweisen \"uber Leseberichte an\-de\-rer im Mittelpunkt stehen. Meine Ergebnisse formuliere ich im f\"unften Unterabschnitt in einer Zusammenfassung in typologischer Perspektive, unter anderem, indem ich Aussagen und Schreibweisen, die ich in dieser Arbeit zuvor besprochen habe, miteinander vergleiche.

Das Verfassen und Mitarbei\-ten an enzy\-klop\"adischen Eintr\"agen bei Wi\-ki\-pe\-dia hat mich we\-sent\-lich skeptischer werden lassen gegen\"uber verallgemeinernden Aussagen ("`man liest"'), die von Einzelnen pu\-bli\-ziert werden: Solange nicht gesagt wird, wer au{\ss}er einem \mbox{selbst} beim Lesen eines Textes bestimmte Emp\-fin\-dun\-gen hatte, tendiere ich inzwischen dazu, Beschreibungen als Eigenaussagen zu verstehen, die lediglich "`un-parrhesiastisch"' verpackt wurden. Auch mir f\"allt es nicht immer leicht, bestehenden Konventionen auszuweichen.

F\"ur \"of\-fent\-liches Schrei\-ben \"uber das Entstehen von Emp\-fin\-dun\-gen im Lese- und/oder Zuh\"orkontakt mit li\-te\-ra\-rischen Texten und \"uber nicht-pro\-po\-si\-ti\-o\-na\-les Wissen, das beim Lesen erworben wurde (\textit{Erlesnis}) unternehme ich in diesem Abschnitt erste Ans\"atze zur einer Typologie der \textit{parrhesia} im Schrei\-ben \"uber Li\-te\-ra\-tur und Li\-te\-ra\-tur\textsuperscript{\~.\~.}. Als Beweggrund f\"ur das Erarbei\-ten einer solchen Typologie -- so ist aus der Argumentation in den vorigen Kapiteln deutlich geworden -- gilt \mbox{hier}, dass mehr Aufmerksamkeit notwendig ist, wenn das Schrei\-ben \"uber Li\-te\-ra\-tur -- und vor allem \"uber Li\-te\-ra\-tur\textsuperscript{\~.\~.} -- nicht zu\-letzt f\"ur einen Verwertungsrahmen Relevanz erhalten soll, in dem nur Aussagen namhafter Per\-so\-nen willkommen sind, zum Beispiel als Belege in Eintr\"agen zu li\-te\-ra\-rischen Werken bei Wi\-ki\-pe\-dia. 

\subsubsection{Schreibweisen bei Ulrike Draesner, Hans Ulrich Gumbrecht und Ina Hartwig}
\label{subsubsec:7.1.1}

Konventionsgem\"a{\ss} werden eigene Leseerlebnisse als intersubjektive Aussagen formuliert, auch wenn naheliegt, dass die G\"ultigkeit einer Aussage nicht bereits mit anderen besprochen worden ist. Im Folgenden stelle ich stark gek\"urzt einige Textstellen aus Essayb\"anden von Hans Ulrich Gumbrecht, \textit{Stimmungen lesen. \"Uber eine verdeckte Wirklichkeit der Li\-te\-ra\-tur} (2011) (\cite{Gumbrecht2011}), Ina Hartwig, \textit{Das Geheimfach ist offen. \"Uber Li\-te\-ra\-tur} (2012) (\cite{Hartwig2012}), und Ulrike Draesner, \textit{Heimliche Helden. \"Uber Heinrich von Kleist, Jean-Henri Fabre, James Joyce, Thomas Mann, Gottfried Benn, Karl Valentin u.v.a.: Essays} (2013) (\cite{Draesner2013}) vor, die ich mehr oder minder als Beschreibungen von nicht-pro\-po\-si\-ti\-o\-na\-lem Wissen empfinde. Diese Stellen w\"urde ich f\"ur Leseerlebnis-Abschnitte in Wi\-ki\-pe\-dia-Eintr\"agen zusammenfassend dar\-stel\-len oder sogar zitieren -- je nachdem, welche weiteren Belege mir zur Verf\"ugung st\"unden, um einem solcherma{\ss}en dezidierten Leseerlebnis-Abschnitt im Rahmen des Artikelganzen durch seine Stoffmenge gen\"ugend Gewicht zu geben. 

Ich typologisiere die folgenden Textstellen nicht inhaltlich, sondern danach, welche Ausdrucksweise in ihnen gew\"ahlt worden ist: "`ich"'; "`wir"' / "'uns"' (und die empha\-ti\-schere Variante "`Wir alle"' / "`jedem von uns"'); "`man"' sowie eine Art vom implizitem "`man"'; "'der Leser\textsuperscript{\tiny *}"' / "`beim Lesen"'; implizite Frage oder explizite Frage. 

\textbf{"'ich"'}

\singlespacing
\begin{quote}
"`Thomas Manns Worte [...] jagten mir Schauer \"uber die Haut und versetzten mich in die Lage, mir etwas vorzustellen, das ich nicht kannte. Nur eines war mir daran vertraut: das wunderbare \flq in die Lage versetzt werden\frq."'\\
"`[...] 
dass ich fast erschrecke: dar\"uber, wie viel sich seit 1974 ver\"andert hat; dar\"uber, wie alt ich \mbox{selbst} bin. Die Lekt\"ure l\"asst mich f\"uhlen, bis in welche Schichten des Herkommens ich als Leserin reiche [...]"'\\
"`[...] und wie ich, die etwas lernte, etwas f\"uhlte und den Erwachsenen fort\-an ein aus Worten abgeleitetes, aber kaum in Worte zu fassendes Wissen verheimlichte."'\\
"`Die Szene st\"urzt dennoch nicht ab, weil Krulls Lust im Halbklaren bleibt f\"ur den Leser, untergr\"undig aber durch den Text wandert als Begehren des Erz\"ahlers, eines alten Felix, nach sich und der Welt. \\ Eben dies war, worauf die 12-J\"ahrige reagierte. Es ist der Kern des Bu\-ches, wo es nicht gealtert ist, wo ich es nicht bin. Sein lebendiger Punkt, in dem etwas spricht, das mit einzelnen Vokabeln zu be\-zeichnen notorisch misslingt."' (\cite{Draesner2013}:340, 341, 345, 350)
\end{quote}
\onehalfspacing



Ulrike Draesners \textit{Erlesnis} stellt insofern einen Idealfall dar, als \"uber ein Erlesis explizit aufgrund der Literarizit\"at eines Textes berichtet wird.

Die Ausdrucksweise in der ersten Person Singular erscheint mir als die unmittelbarste Art und Weise, um den Erlebnisfaktor einer Lekt\"ure in einem so schnellg\"angigen und web\"of\-fent\-lichen Umfeld wie Wi\-ki\-pe\-dia in Worten wiederzugeben. Auch wenn man diese direkte Art der Beschreibung eher in Leseforen erwarten w\"urde: In zitierter Form geben sie nach meinem Empfinden dem li\-te\-ra\-rischen Text besonderes Gewicht. Ich stelle mir vor, es landet jemand bei diesem Eintrag, der am liebsten jemand anderen kennen w\"urde, um danach zu fragen, wie das Buch denn ist. Dann liest sich meiner Vermutung nach ein Zitat in erster Person singular fast wie ein "`Beweis"' daf\"ur, wie nah mindestens einer anderen Person diese Lekt\"ure schon gegangen ist. Und wenn sie sogar so namhaft ist, einen eigenen Wi\-ki\-pe\-dia-Eintrag zu haben (der Name ist dann mit einer Verlinkung hinterlegt), macht eine solche direkte Aussage sogar Leute neugierig, die ansonsten auch weniger direkte Aussagen sch\"atzen, wenn nicht sogar bevorzugen -- so meine Vermutung, auf deren Basis \textit{Erlesnisse} in erster Person Singular typologisch gesehen als die we\-sent\-lichsten zu betrachten sind.

\textbf{"`wir"'/"'uns"'}

\singlespacing
\begin{quote}
"`Die ersten Notizen in der Nacht vom 12. auf den 13. Juli 1960 entstehen auf einer Reise nach Marokko und st\"urzen uns sofort in eine schw\"ul-intime Gedankenwelt."' (\cite{Hartwig2012}:53)
\end{quote}
\onehalfspacing


\singlespacing
\begin{quote}
"`Aber nie beschreibt er die erregende Landschaft der Stadt, und kein einziges Worte erm\"oglicht es uns, eine Vorstellung von dem Haus zu haben, wo all seine Notizen geschrieben werden."' (\cite{Gumbrecht2011}:117)
\end{quote}
\onehalfspacing

\singlespacing
\begin{quote}
"`Jeden Schritt gehen wir mit Selma, wir sp\"uren den Riemen ihres Rucksacks, den sie \"uber die Schulter streift, sehen sie zaghaft l\"acheln, weil sie in Gedanken versinkt, ihr Leben rekonstruiert, sich bitterb\"ose Klarheit \"uber ihre Situation verschafft."' (\cite{Hartwig2012}:69)
\end{quote}
\onehalfspacing

\singlespacing
\begin{quote}
"`Wie erz\"ahlt man ein Leben? Sprache bedeutet Ordnung, Muster, Phantasie. Joyce testet sie aus, dehnt sie, um unsere Freiheit im Wahrnehmen der Welt zu erweitern."' (\cite{Draesner2013}:289)
\end{quote}
\onehalfspacing

\textbf{"`Wir (alle)"' / "`jedem von uns"'}

\singlespacing
\begin{quote}
"`[B]ei ihm liest man die bisher \"uberzeugendsten, kontrolliertesten Schil\-derungen jener absurd-ohnm\"achtigen Situation, in die wir alle vor dem Fernseher hineingeraten sind."' (\cite{Hartwig2012}:77)
\end{quote}
\onehalfspacing

\singlespacing
\begin{quote}
"`In kaleidoskopisch angeordneten Geschichten zeigt \textit{Anders} auf, zu wel\-chen L\"ugen und Identit\"atsfindungen Leben nach einem zerst\"orerischen Krieg \flq inspiriert\frq, wie lange die Situation des Nachkrieges andauern kann -- nicht (nur) auf der politischen B\"uhne, sondern mitten in der Ge\-sell\-schaft, bei jedem von uns [...]"' (\cite{Draesner2013}:292)
\end{quote}
\onehalfspacing

Peter Lamarque hinterfragt diese Konvention der ersten Person Plural bei Gelegenheit einer Aussage zu emotionalen Reaktionen auf Tolstois \textit{Anna Karenina} und ich halte seine Fragen insofern f\"ur berechtigt, als sich der Raum, in dem \"uber Li\-te\-ra\-tur geschrieben wird, seit der "`Erfindung"' dieser Schreibweise we\-sent\-lich vergr\"o{\ss}ert hat, auch wenn sich die Grenzen zwischen Angeh\"origen verschiedener ge\-sell\-schaftlicher Gruppen dadurch nicht flie{\ss}ender gestalten. Lamarque schreibt:

\singlespacing
\begin{quote}
"`But who is the \flq we\frq\,\,referred to in this quotation, the \flq we\frq\,\,who feels intense urges, intense stress, intense desires and hopes? What do we say of readers who do not respond with such intensity? Is it taken to be just a fact of the matter that readers will respond in this way? Or true of sensitive readers? Or readers properly informed about Li\-te\-ra\-ture?"' \cite{Lamarque2014}:196)
\end{quote}
\onehalfspacing

\textbf{"`man"'}

\singlespacing
\begin{quote}
"`Die Lekt\"ure besonders des zwei\-ten Teils des Tagebuchs wird man nicht ohne Tr\"anen bew\"altigen, so nah bringt diese junge, stolze, lebenshungrige und zugleich bescheidene, einf\"uhlsame, kluge Intellektuelle uns die Situation der j\"udischen Bev\"olkerung im okkupierten [...] Paris.
"'\\ (\cite{Hartwig2012}:61)
\end{quote}
\onehalfspacing

\singlespacing
\begin{quote}
"`
Man f\"uhlt sich zur\"uckversetzt in die fr\"uhe Zeit Jelineks, als sie eine junge Popliteratin war, \"atzend und flink; man erinnert sich an \textit{Die Liebhaberinnen} [...]
"' (\cite{Hartwig2012}:177-178)
\end{quote}
\onehalfspacing

\singlespacing
\begin{quote}
"`Eigner erzeugt einen Sprachsirup, durch den das Erleben doppelt gezogen wird. Man sieht im gro{\ss}en Kleinen, wie Sprache Raum \"offnet, um Figuren entstehen zu lassen [...]"' (\cite{Draesner2013}:318)
\end{quote}
\onehalfspacing


\singlespacing
\begin{quote}
"`
\textit{To the Lighthouse} [...] be\-steht aus drei Teilen, deren mittlerer in einem bildintensiven Prosafluss das Vergehen der Zeit in einem verlassen stehenden Haus so schildert, dass man versteht, dass inzwischen anderswo Schlachten geschlagen werden und N\"achte nicht mehr enden."' (\cite{Draesner2013}:348)
\end{quote}
\onehalfspacing

\singlespacing
\begin{quote}
"`Die Worte werden eingerammt wie Eisenpf\"ahle, an denen man sich weh tun kann, obwohl sie auch eine Art Schutzfunktion erf\"ullen.
"'\\ (\cite{Hartwig2012}:32-33)
\end{quote}
\onehalfspacing

\singlespacing
\begin{quote}
"`Schrei\-ben, was nicht ist, schrei\-ben, was ist, das durch Schrei\-ben erst Hergestellte: Man h\"ort das Beckett'sche Lachen im Hintergrund, das man von Kert\'{e}sz bis\-her nicht kannte."' (\cite{Hartwig2012}:107)
\end{quote}
\onehalfspacing


\singlespacing
\begin{quote}
"`Diesem Buch kann man sich nicht entziehen. Es ent\-wickelt erhebliche Sogwirkung und l\"asst einen dennoch in seltsam ungekl\"arter Gef\"uhlslage zur\"uck."' (\cite{Hartwig2012}:157)
\end{quote}
\onehalfspacing

\textbf{"`man"' implizit}


\singlespacing
\begin{quote}
"`Das absichtlich stressige Sprachgequirl, in dem keiner und alle sprechen d\"urfen (die Per\-so\-nen, die Berge, das Wasser, die K\"orperteile), ist virtuos und sollte un\-be\-dingt genossen werden."' (\cite{Hartwig2012}:179)\end{quote}
\onehalfspacing

\singlespacing
\begin{quote}
"`Es ist schon ziemlich komisch, wie der westdeutsche Ingenieur mit seiner Delegation pl\"otzlich in das schwitzige Magma aus Kumpelhaftigkeit und sozialistischem Jargon hineinplatzt."' (\cite{Hartwig2012}:141)
\end{quote}
\onehalfspacing


\textbf{"'Leser"' / "`beim Lesen"'}

\singlespacing
\begin{quote}
"`Gelegentlich mischt sich ein mysteri\"oses Ich ein, das den Leser anspricht, um ihn vor falschen Erwartungen zu warnen."' (\cite{Hartwig2012}:142)
\end{quote}
\onehalfspacing

\singlespacing
\begin{quote}
"`
Gedichte
 [...] von Luis de G\'{o}ngora. Auch sie stimulieren ein ums andere Mal die Imagination der Leser, doch all diese starken Impulse sind in Bewegung und verweisen in verschiedene Richtungen der Vorstellung, ohne sich je zum koh\"arenten Bild einer Wirklichkeit zusammenzuf\"ugen."' (\cite{Gumbrecht2011}:79)
\end{quote}
\onehalfspacing


\singlespacing
\begin{quote}
"`[...] K\"unstlerroman auf der (Gebirgs-)H\"ohe der Zeit geschrieben. Beim wiederholten Lesen gewinnt er die Leichtigkeit, die ihm geb\"uhrt."'\\ (\cite{Hartwig2012}:243)
\end{quote}
\onehalfspacing


\singlespacing
\begin{quote}
"`Beim Lesen wird zunehmend sp\"urbar, wie sehr Lars Brandts Buch um einen tieftraurigen Kern kreist, angetrieben von einer unerf\"ullten Sehnsucht."' (\cite{Hartwig2012}:159)\\
\end{quote}
\onehalfspacing

\singlespacing
\begin{quote}
"`Aber die Beschw\"orung der poetischen Kraft im Ungl\"uck bringt die Le\-se\-rin an eine, n\"amlich ihre eigene Grenze."' (\cite{Hartwig2012}:100)
\end{quote}
\onehalfspacing

\textbf{implizite Frage}

\singlespacing
\begin{quote}
"`Wie eine nichterz\"ahlbare Geschichte zu erz\"ahlen sei, bleibt offen."'\\ (\cite{Hartwig2012}:110)
\end{quote}
\onehalfspacing

\singlespacing
\begin{quote}
"`[...] und einmal steigt Katja einem von ihnen hinterher, aus Verzweiflung oder aus Provokation, man wei{\ss} es nicht, wie so vieles in diesem Buch sich nicht mit letzter Sicherheit beantworten l\"asst, obwohl doch st\"andig nach Antworten gesucht wird."' (\cite{Hartwig2012}:141)
\end{quote}
\onehalfspacing

\textbf{explizite Frage}

\singlespacing
\begin{quote}
"`Was musste raus? Und in welcher Sprache? Eine Sprache der Schuld, der Not, der Bew\"altigung? Das m\"ochte man wissen, unabh\"angig davon, ob wom\"oglich \"au{\ss}ere Gr\"unde dem Nobelpreistr\"ager das sp\"ate Bekenntnis abgen\"otigt haben."' (\cite{Hartwig2012}:153-154)
\end{quote}
\onehalfspacing

\singlespacing
\begin{quote}
"`[...] ist sie wirklich die r\"atselhafteste Erscheinung in diesem so mehrdeutigen Roman? [...] Wer kein Vergn\"ugen an Entschl\"usselungsarbeit hat, dem ist die Lekt\"ure dieses Buches nicht zu emp\-fehlen."'\\ (\cite{Hartwig2012}:239)
\end{quote}
\onehalfspacing

\singlespacing
\begin{quote}
"`So handelt der Roman in aller K\"urze auf \mbox{vier} Ebenen zugleich von Rede\-akten: Warum haben die Figuren, von denen erz\"ahlt wird, ihre erlogenen Geschichten in die Welt gesetzt? Wie lange haben sie sie \mbox{selbst} geglaubt? Warum und wie erz\"ahlen sie die beiden heutigen Figuren einander? Und warum und wie bietet der Autor Sch\"adlich sie uns schlie{\ss}lich an?"' (\cite{Draesner2013}:294)
\end{quote}
\onehalfspacing


\singlespacing
\begin{quote}
"`Ich dachte an Schlachten. Kleist kannte das. Und wie es ist, wenn man einmal an diesem Haken h\"angt? Die Gesetze am eigenen Leib erfahren hat, die das T\"oten erlauben?"' (\cite{Draesner2013}:101)
\end{quote}
\onehalfspacing

\singlespacing
\begin{quote}
"`L\"ugt er? Warum sagt er nicht, was er will? \\ Er l\"ugt nicht. Er sagt, was er will.\\ Hat er vergessen, dass er Kriemhilds wegen kam?
"' (\cite{Draesner2013}:59)
\end{quote}
\onehalfspacing

\singlespacing
\begin{quote}
"`Da ist sie doch, die psychische Obession. Aber wovon wird eigentlich gesprochen?"' (\cite{Hartwig2012}:151-152)
\end{quote}
\onehalfspacing

\singlespacing
\begin{quote}
"`Mittendrin: Kleist. Als Romantiker? Ich h\"ore ihn lachen."'\\ (\cite{Draesner2013}:101)
\end{quote}
\onehalfspacing

Das Stellen einer Frage sehe ich in typologischer Hinsicht als eine besondere Form an, so dass ich entsprechende Beispiele nach diesem Kriterium gruppiert habe und nicht nach einem der anderen. Die besondere Rolle ergibt sich aus meiner Sicht dadurch, dass mit einer Frage inhaltlich zum Ausdruck gebracht wird, was einen bewegt hat, das hei{\ss}t: was man beim Schrei\-ben erinnert (bzw. formulieren will f\"ur einen bestimmten Kontext oder Adressaten\textsuperscript{\tiny *}kreis). Und da einen unl\"osbar scheinende Fragen -- auch von diesen gibt es \mbox{hier} Beispiele -- nach der Lekt\"ure manchmal noch lange besch\"aftigen, halte ich diese Ausdrucksweise f\"ur besonders interessant, wenn es um die Beschreibung von \textit{Erlesnissen} geht, \mbox{selbst} wenn sie implizit ist und einige Interpretationsarbeit erfordert, will man sie anders als mit einem Zitat wiedergeben.

Auch wenn die Typologisierung \mbox{hier} nur aufgrund sehr knapper Textstellen vor\-ge\-nom\-men wurde, konnte ersichtlich werden, wie \textit{Erlesnisse} durch konventionelle Distanzierungsweisen indirekter werden. Insbesondere die Art und Weise, wie eine Aussage gerade nicht als \textit{Erlesnis} in deutlicheren Worten geschildert wird, k\"onnen al\-ler\-dings als besonders interessant gelten. Meiner Beispielauswahl nach zu urteilen sind sie sehr zahlreich. Auf Leser\textsuperscript{\tiny *}seite w\"are \mbox{hier} \"ahnlich wie bei li\-te\-ra\-rischen Texten mit \"asthetischem Anspruch einiges an Interpretation n\"otig, wollte man auf den Punkt bringen, worin genau das \textit{Erlesnis} in diesem oder jenem Fall bestanden hat. Es wurde nach Ausdrucksweise typologisiert ("`ich"'; "`wir"' / "'uns"'; "'der \mbox{Leser\textsuperscript{\tiny *}"' /} "`beim Lesen"'), sowie danach, ob Fragen gestellt werden. Wollte man spezifischer typologisieren, d\"urften inhaltliche Gesichtspunkte nicht au{\ss}en vor bleiben, denn \textit{Erlesnisse} sind ephemere Ph\"anomene, die sich wohl eher in Zwi\-schen\-r\"au\-men ansiedeln. 

\pagebreak
\subsubsection{Rezensionsnotizen bei ‹perlentaucher.de›}
\label{subsubsec:7.1.2}

Im Rahmen einer Er\"orterung von Bedingungen der Darstellbarkeit nicht-proposi\-tionalen Wissens in Werkartikeln bei Wi\-ki\-pe\-dia z\"ahlt \textit{per\-len\-tau\-cher.de -- Das Kulturmagazin} insofern zu den Einflussgr\"o{\ss}en, als Wi\-ki\-pe\-dia-Autoren\textsuperscript{\tiny *} vermutlich dieses Angebot in einem ersten Schritt nutzen, um Anhaltspunkte zu erhalten, ob das Werk gen\"ugend Relevanz f\"ur einen eigenen Eintrag besitzt, gemessen an der Anzahl an Rezensionen in gro{\ss}en Tageszeitungen. Der andere Gesichtspunkt, unter dem \textit{per\-len\-tau\-cher.de} indirekt zu den Bedingungen z\"ahlt, ist die Einsch\"atzung, die Stefan Neuhaus gibt: "`Wer sich \"uber die Li\-te\-ra\-tur\-kritische Aufnahme eines Buches informieren m\"ochte, nutzt den \flq Per\-len\-tau\-cher\frq"' (\cite{Neuhaus2010}:20), denn wenn deren Nutzungs\-brei\-te in der For\-schung als Faktum konstatiert worden ist, wird die Seite aus Sicht von Wi\-ki\-pe\-dia-Richtlinien als eine meinungsbildende Gr\"o{\ss}e angesehen. In Hinsicht auf die Arbeit des Verfassens eines Eintrags ist \textit{per\-len\-tau\-cher.de} vor allem dann relevant, wenn nicht allzuviel Zeit investiert werden soll, aber dennoch etwas Ge\-eig\-netes f\"ur den Abschnitt "`Rezeption"' gesucht wird. Im Idealfall werden Wi\-ki\-pe\-dia-Autoren\textsuperscript{\tiny *} aber versuchen, an die Rezensionen \mbox{selbst} heranzukommen. Denn Schrei\-ben \"uber Li\-te\-ra\-tur be\-steht bei \textit{per\-len\-tau\-cher.de} pro Werk aus dem Klappentext und aus Rezensionsnotizen, die zusammenfassende Darstellungen von Rezensionen in gro{\ss}en deutschspra\-chi\-gen Tageszeitungen bieten, ohne Verfasser\textsuperscript{\tiny *}angabe und mit \"Uber\-schrif\-ten in knapper Form, zum Beispiel "`Rezensionsnotiz zu Neue Z\"urcher Zeitung, 09.04.2011"'.
 
Dass bei \textit{per\-len\-tau\-cher.de} aus Sicht der Verlags- und Vertriebsbranche, also ver\-kaufs\-ori\-en\-tiert, ausgew\"ahlt wird, ist nicht allein an Werbefeldern zu sehen, die jeden Eintrag rahmen, sondern auch daran, dass Buchtitel, die nicht ins Deutsche \"ubersetzt vorliegen, keine Aufmerksamkeit erhalten. Zu den zw\"olf Beispielen der vorliegenden Arbeit gibt es also keine Rezensionsnotizen f\"ur \textit{Pour que tu ne te perdes par dans le quartier}, \textit{Pas pleurer} und \textit{La Vie commune}. Relevanzstiftend waren f\"ur den Wi\-ki\-pe\-dia-Eintrag also gen\"ugend weitere Kriterien als nur die deutschspra\-chi\-ge Rezeption in Tagszeitungen (n\"amlich renommierte Li\-te\-ra\-tur\-preise). Zu den anderen neun Werken sind bei \textit{per\-len\-tau\-cher.de} zwischen einer und f\"unf Rezensionsnotizen aufgef\"uhrt, die ich im Folgenden daraufhin ana\-lysiere, inwiefern in ihnen nicht-pro\-po\-si\-ti\-o\-na\-les Wissen aus Li\-te\-ra\-tur\-lekt\"ure zur Sprache kommt.
Ich beginne mit dem Beispiel des Essaybandes von Gumbrecht. Dass in Rezensionsnotizen zu diesem Titel nicht-pro\-po\-si\-ti\-o\-na\-les Wissen zur Sprache kommen k\"onnte, liegt insofern auf der Hand, als diese Art des Wissens \mbox{selbst} im Band thematisiert wird, auch wenn ich es nirgends explizit als nicht-pro\-po\-si\-ti\-o\-na\-les Wissen benannt gefunden habe. Bez\"uglich der Rezension von Steffen Martus wird in der entsprechenden Notiz als dessen Leseerlebnis berichtet, jener habe gestaunt: 

\singlespacing 
\begin{quote}
\href{http://www.perlentaucher.de/buch/hans-ulrich-gumbrecht/stimmungen-lesen.html}{"`Doch auch wenn der Rezensent unter einigen \flq Handkantenschl\"agen\frq \,\,zusammenzuckt, die Gumbrecht seinem Thema zuf\"ugt, muss er doch auch staunen: \flq \"uber zar\-te, mit\-f\"uh\-len\-de, fast liebevolle Lekt\"uren\frq."'}\\ (\cite{Anonym*ohneDatum1})
\end{quote}
\onehalfspacing

Bei Martus \mbox{selbst} ist dieser Eindruck verhaltener formuliert, denn der Zielpunkt des Satzes ist ein an\-de\-rer, n\"amlich eine Betrachtung dar\"uber, was Gumbrechts Schreibweise aus Sicht von Martus f\"ur im Fach ansonsten \"ubliche Vorgehensweisen bedeutet, \mbox{hier} das Original der Aussage, die in der Rezensionsnotiz teils paraphrasiert wiedergegeben wird: 

\singlespacing 
\begin{quote}
"`Gumbrecht haut seine Themen mit zwei, drei Handkantenschl\"agen zu\-recht -- die oftmals journalistisch knappen und in den Grundbotschaften bis\-wei\-len redundanten Texte sind aus einer Artikelserie f\"ur die \textit{Frankfurter Allgemeine Zeitung} entstanden. Zugleich aber bietet er zar\-te, mit\-f\"uh\-len\-de, fast liebevolle Lekt\"uren an, die mit leichter Ber\"uh\-rung die Werke umschmeicheln, f\"ur sie werben und eben in dieser Leichtf\"u{\ss}igkeit die Li\-te\-ra\-tur\-wissenschaft provozieren."' (\cite{Martus2011}:50)
\end{quote}
\onehalfspacing

Meinem Eindruck nach hatte in diesem Fall der Verfasser\textsuperscript{\tiny *} der Rezensionsnotiz dem eigenen Empfinden nach bei Martus ein Leseerlebnis gefunden und es in der Formulierung "`muss er doch auch staunen"' auf eine Weise expliziert, die in der Rezension so nicht zu finden ist. 

Wenn es in weiteren Rezensionsnotizen zu Gumbrecht 2011 hei{\ss}t, der Band werde \href{http://www.perlentaucher.de/buch/hans-ulrich-gumbrecht/stimmungen-lesen.html}{"`[p]ositiv, wenn auch nicht gerade emphatisch"'}\,besprochen\,(\cite{Anonym*ohneDatum2}) oder \href{http://www.perlentaucher.de/buch/hans-ulrich-gumbrecht/stimmungen-lesen.html}{"`Angetan zeigt sich Rezensent [xy] von"'} oder \href{http://www.perlentaucher.de/buch/hans-ulrich-gumbrecht/stimmungen-lesen.html}{"`[n]ichtsdestoweniger findet er das Werk sehr erhellend"'} (\cite{Anonym*ohneDatum5}), ist dies aus dem Blickwinkel meiner Fragestellung nicht nah genug an der Beschreibung eines \textit{Erlesnisses}, denn Wertungen dieser Art werden in vielen Wi\-ki\-pe\-dia-Eintr\"agen bereits im Abschnitt "`Rezeption"' wiedergegeben (Beispiel: Eintrag "`Gr\"aser der Nacht"', Abschnitt \href{https://de.wikipedia.org/w/index.php?title=Gr\%C3\%A4ser_der_Nacht\&oldid=136750896\#Ver.C3.B6ffentlichung_und_Rezeption}{Ver\"offentlichung\,und\,Rezeption},\,Text\,von\,\mbox{User}:Magiers, siehe\,auch\,Abschnitt\,\textit{~\ref{subsubsec:6.1.11}}).

Wenn in einer Rezensionsnotiz hingegen das Berichten einer Wertung kombiniert wird mit einer Beschreibung bez\"uglich der Arbeitsweise des Autors\textsuperscript{\tiny *} des rezensierten Werks, kommt dies aus meiner Sicht der Andeutung einer \textit{Erlesnis}-Chance gleich wie ich sie in Abschnitt \textit{5.2 Wissen \"uber Li\-te\-ra\-tur bei Wi\-ki\-pe\-dia} etwa in einigen der Teaser zu li\-te\-ra\-rischen Werken bei "`Schon gewusst?"' diagnostiziert habe. Dies trifft auf die Notiz \"uber Bernadette Conrads Rezension zu Lessings \textit{Alfred und Emily} zu, weil in dieser Aussage eine Frage des "`Wie"' in den Mittelpunkt ger\"uckt wird, \mbox{selbst} wenn es nicht konkreter wird als zu schrei\-ben, durch Conrad werde das Wie kommentiert:

\singlespacing
\begin{quote}
\href{http://www.perlentaucher.de/buch/doris-lessing/alfred-und-emily.html}{"`Sehr beeindruckend findet Conrad, wie die Autorin ihre unbew\"altigte Familiengeschichte angeht und sie preist dieses Buch nicht nur als be\-r\"uh\-ren\-de, sp\"ate Liebestat, sondern als li\-te\-ra\-rische Kostbarkeit."'}\\ (\cite{Anonym*ohneDatum4})
\end{quote}
\onehalfspacing

Auch wenn in einer Kombination aus Leseerlebnis und sprachlichen Eigenhei\-ten \"uber den Inhalt einer Rezension berichtet wird, sch\"atze ich dies als relativ nah an einem \textit{Erlesnis} ein, nicht hingegen die Beschreibung sprachlicher Eigenhei\-ten allein. \mbox{Hier} ein interessantes Beispiel wie Aspekte der Literarizit\"at mit Leser\textsuperscript{\tiny *}reaktionen verbunden werden:

\singlespacing 
\begin{quote}
\href{http://www.perlentaucher.de/buch/zoya-pirzad/die-lichter-loesche-ich.html}{"`Aber gerade diese vermeintliche Leichtigkeit verh\"ulle eine Substanz, die sich in Nebens\"atzen und Sei\-tenblicken manifestiert und die Rezensentin immer wieder innehalten l\"asst."'} (\cite{Anonym*ohneDatum3})
\end{quote}
\onehalfspacing

Als \"ahnlich nah an einem \textit{Erlesnis} sch\"atze ich es ein, wenn in einer Kombination aus Thema und pers\"onlichen Schlussfolgerungen des Rezensenten\textsuperscript{\tiny *} \"uber den Inhalt einer Rezension berichtet wird, wie in diesem Fall zu Patrick Modianos \textit{Im Caf\'{e} der verlorenen Jugend} (2007):

\singlespacing 
\begin{quote}
\href{http://www.perlentaucher.de/buch/patrick-modiano/im-cafe-der-verlorenen-jugend.html}{"`Damit gelinge Modiano ein \flq Chanson triste\frq \,\,auf das alte Paris, seufzt die Rezensentin, die nun ihrerseits die Welt ein St\"uck weit mehr als im stetigen Vergehen begriffen sieht."'} (\cite{Anonym*ohneDatum6})
\end{quote}
\onehalfspacing

\pagebreak
\subsubsection{Leseerlebnis-Abschnitte in Eintr\"agen zu literarischen Werken}
\label{subsubsec:7.1.3}

F\"ur eine Wiedergabe in Leseerlebnis-Abschnitten von Werkartikeln bei Wi\-ki\-pe\-dia habe ich aus meiner Sicht nur drei Beispiele gefunden, in denen tats\"achlich etwas \"uber erworbenes nicht-pro\-po\-si\-ti\-o\-na\-les Wissen berichtet wird. Zwei davon sind Leseforen entnommen, eines einer Hochschulzeitschrift im Web. Ob es zu den von mir ausgew\"ahlten Werken und zu diesem Zeitpunkt \textit{Erlesnisse} in Print gegeben h\"atte, kann ich nicht sagen.

Das erste Beispiel ist dem Artikel "`La Vie commune"' entnommen, die anderen beiden finden sich in "`Alfred und Emily"'. Meine Formulierungen der \textit{Erlesnisse} lauten in den Eintr\"agen folgenderma{\ss}en: 

\singlespacing 
\begin{quote}
\href{https://de.wikipedia.org/w/index.php?title=La_Vie_commune\&oldid=138094789\#Leseerlebnis}{Mit welchen Methoden man unangenehme Begegnungen am Arbeitsplatz l\"ose, suche man sich \mbox{selbst} aus. Es gebe radikalere und andere. Jedenfalls betrachtet man nach dem Lesen dieses Romans seine Kollegen nicht mehr mit denselben Augen.} (User:C.Koltzenburg, 23. Januar 2015, 22:15 CET)
\end{quote}
\onehalfspacing

Die ersten beiden S\"atze verfasse ich in indirekter Rede, das \textit{Erlesnis} \mbox{selbst} gebe ich im Indikativ mit einer "`man"'-Konstruktion wieder, im Original war es als Prognose in zweiter Person Plural formuliert worden: "`En lisant ce roman, vous ne regarderez plus votre coll\`{e}gue de la m\^{e}me fa\c{c}on"'.

In den folgenden beiden Beispielen gebe ich die Aussagen meiner Quellen als Beobachtungen derjenigen wieder, die ich zitiere. Dies in einem so empirischen Duktus zu rahmen, ist der Orientiertheit an Fakten geschuldet, die bei Wi\-ki\-pe\-dia vor\-herr\-schend ist. Ich habe also nach einer M\"oglichkeit gesucht, den Aussagen, wo schon kein objektives, so wenigstens ein faktischeres Gewand zu geben.

\singlespacing 
\begin{quote}
\href{https://de.wikipedia.org/w/index.php?title=Alfred_und_Emily&oldid=138028858#Leseerlebnis}{Sergeant hat sich \mbox{selbst} beim Lesen beobachtet und meint, w\"ahrend der Lekt\"ure betreibe man \mbox{selbst} R\"uckschau. Leser fragen sich eventuell, wo sie eigentlich gewesen sind und was das genau war. Wir haben beim Lesen von Alfred und Emily die starke Empfindung, so Sergeant, dass die Zeit knapp wird, und auch fragen wir uns, was in der verbleibenden Spanne wohl noch erreicht werden kann. Was von diesen beiden als schwerwiegender empfunden wird, balanciere sich aus, je nachdem, wie alt man \mbox{selbst} ist. Lessing findet nicht zu allen ihren Fragen eine L\"osung, aber darum geht es nicht allein, meint er, denn ihre unbeantworteten Fragen wirken f\"ur Leser wie etwas Lebendiges, das Echos produziert: Von \mbox{hier} aus k\"onnen wir uns weiterbewegen, weil wir eine genauere Vorstellung davon erhalten haben, wo wir waren und wo wir sind, so Sergeant.} (User:C.Koltzenburg, 21. Januar 2015, 21:48 CET)
\end{quote}
\onehalfspacing

Hier verwende ich eine Mischung an Ausdrucksweisen, mit denen ich versucht habe, den gew\"ahlten Stil des eng\-lischspra\-chi\-gen Originals in etwa wiederzugeben. Zu fin\-den sind: "`man"', "`Leser fragen sich"', "`wir"', indirektes "`man"' und direktere Aussagen werden sozusagen abgefedert mit "'so Sergeant"' oder "`meint er"'. 

\singlespacing 
\begin{quote}
\href{https://de.wikipedia.org/w/index.php?title=Alfred_und_Emily&oldid=138028858#Leseerlebnis}{Eine \"ahnliche Beobachtung macht auch Lizzie, wenn es hei{\ss}t: "`Die Fragen ihres Lebens sind nicht gel\"ost, aber immerhin stellt sie sie. Lessing stellt unsere Fragen und zeigt, ob sie beantwortbar sind"', und, eingangs sowie am Schluss der Rezension: "`Das Buch ergibt kaum Sinn, soviel kann ich sagen. Objektiv ist es bizarr zu lesen und wirklich fragmentiert und sogar innerhalb der einzelnen Fragmente wird hin- und hergesprungen wie verr\"uckt"', "`Dieses Buch wird immer wichtig f\"ur mich bleiben, und vermutlich muss es daf\"ur nicht einmal Sinn ergeben."'}\\ (User:C.Koltzenburg, 21. Januar 2015, 21:48 CET)
\end{quote}
\onehalfspacing

In diesem dritten \textit{Erlesnis}-Beispiel w\"ahle ich als Darstellungsform direkte Zitate in \"Ubersetzung. Ich kn\"upfe abfedernd an die vorigen Aussagen an, indem ich sage, dass es sich auch \mbox{hier} um Beobachtungen handelt.

\pagebreak
\subsubsection{\textit{Erlesnis}-Darstellungen in den eigenen Interpretationen}
\label{subsubsec:7.1.4}

Durchg\"angig in erster Person Singular formuliert werden \textit{Erlesnisse} in den eigenen Interpretationen zu Elfriede Jelinek (\textit{rein GOLD}, 2013), Doris Lessing (\textit{Alfred and Emily}, 2008), Herta M\"uller (\textit{Reisende auf einem Bein}, 1989) und Alice Munro (\textit{Save the Reaper}, 1998 und 1998) auf \textit{Metaebene I}. Die Interpretationen finden sich in den Abschnitten \textit{~\ref{subsubsec:5.4.1}} (Munro), \textit{~\ref{subsubsec:5.4.2}} (Lessing), \textit{~\ref{subsubsec:5.4.3}} (M\"uller) und \textit{~\ref{subsubsec:5.4.4}} (Jelinek).

Vorausgeschickt sei, dass ich zu diesem Zeitpunkt Ulrike Draesners Essay von 2013 noch nicht kannte, denn Draesners Beitrag h\"atte mich vermutlich stark beeinflusst, weil ich ihre Darstellungsweise der \textit{Erlesnisse} sehr ansprechend finde. Als Wi\-ki\-pe\-dia-Autor\textsuperscript{\tiny *} w\"urde ich mir von dieser Art we\-sent\-lich mehr belegf\"ahige Aussagen namhafter Autoren\textsuperscript{\tiny *} w\"unschen.

Der Beitrag "`Fiktionalisierte Varianten von Versionierungsverfahren in Alice Munros \flq Save the Reaper\frq"' bespricht auf Textebene ein doppelt fiktionalisiertes \textit{Erlesnis}. Dieses wird von Munro in einem fiktionalen Text zur Sprache gebracht und findet nur in der Vorstellung der Protagonistin\textsuperscript{\tiny *} statt, die sich in einer bestimmten Si\-tu\-a\-tion an Ge\-le\-se\-nes erinnert. Mein eigenes \textit{Erlesnis} ist ebenfalls doppelter Natur und bezieht sich zum einem auf mein Gef\"uhl, Erleichterung dar\"uber zu empfinden, nicht \mbox{selbst} Teil einer Munro-Story zu sein, und zum anderen darauf, w\"ahrend des Lesens \"uber Distanzierungsvorg\"ange in Munros Erz"ahlweise nachzusinnen und aus diesem Anlass dar\"uber reflektiert zu haben, wie ich mit dieser Art von Vorg\"angen im eigenen sprachlichen Ausdruck verfahre.

In "`\flq For future editors\frq: \flq The London Encyclop{\ae}dia\frq \,\,und Doris Lessings \flq Alfred and Emily\frq"' wird \textit{Alfred and Emily} insgesamt als ein \textit{Erlesnis} aufgefasst, das ich als eine k\"unstlerisch gestaltete Reaktion auf einen enzy\-klop\"adischen Eintrag lese. Das von mir geschilderte eigene \textit{Erlesnis} speist sich aus Erstaunen \"uber die Wahl der Mittel (indem sie als Schriftstellerin ein Bild wortlos an zentraler Stelle platziert). Es hat mit meinem Staunen dar\"uber zu tun, dass ich so lange gebraucht habe, um zu der Einsicht zu gelangen, wie Lessing Undarstellbares zur Darstellung bringt und dass sie im Zuge dessen vormacht, wie Kunst offiziell Verschwiegenes sichtbar machen kann.

Im Essay zu Herta M\"ullers \textit{Reisende auf einem Bein}, den ich betitele mit "`\flq B\"ose, das w\"ar gut\frq . Fliegende auf einem Bein (M\"uller)"', fu{\ss}t meine Interpretation auf einem \textit{Erlesnis}, das mich um eine besondere Sorte von "`Wissen wie"' bereichert hat und mit meinem Lachen in zweifacher Hinsicht ein Gef\"uhl des Befreitseins zum Ausdruck bringt, einmal in Bezug auf mein Lesen des Textes, bei dem ich erst bei der zwei\-ten Lekt\"ure die Satire-Signale entdecke, und zwei\-tens bezogen auf die mir bis\-her bekannt gewordenen For\-schungsergebnisse, die ich im Wi\-ki\-pe\-dia-Eintrag referenziere.

Die Konzertreise unter dem Motto "`Gezanke in \flq rein GOLD\frq, Debatten im Kopf (Jelinek)"' stellt als \textit{Erlesnis} dar, wie beim Lesen eine doppelte Tonspur entstanden ist, die mich zu Aushandlungen und Streitgespr\"achen an anderen Orten und zu anderen Zei\-ten begleitet hat, und welchen Effekt die Lekt\"ure aus diesem Grund f\"ur die im Rahmen dieser Arbeit geplanten Schritte hatte: Varianten von Gezanke und seinen m\"oglichen Diskursen als solche zu reflektieren und dieser Vielfalt im Verlauf meines Debatten-Experiments in der deutschspra\-chi\-gen Wi\-ki\-pe\-dia-Version gewahr zu bleiben und es f\"ur die eigene Vorgehensweise w\"ahrend der Aushandlungen zu nutzen. 

\subsubsection{Zusammenfassung in typologischer Perspektive}
\label{subsubsec:7.1.5}

Bevor ich das vorige zusammenfasse, m\"ochte ich noch kurz zwei Punkte ansprechen. Ein Beispiel ist mir aufgefallen, anhand dessen ich eine Aussage in einer Rezension mit deren Zusammenfassung in einer Rezensionsnotiz und mit meiner Handhabung im Wi\-ki\-pe\-dia-Eintrag vergleiche sowie eine Spekulation dazu wage, wie die ursp\"ung\-liche Aussage als direktes \textit{Erlesnis} formuliert werden k\"onnte. Ein weiteres Beispiel bezieht sich auf die Deutbarkeit von ge\"au{\ss}erten Assoziationen als \textit{Erlesnisse}.

In der Rezensionsnotiz zur Rezension von Gumbrechts Essayband in \textit{Die Zeit} (von Steffen Martus) wird auf eine Empfindung fokussiert, die in der Rezension \mbox{selbst} nicht als solche expliziert ist -- m\"oglicherweise, weil Martus kurz darauf, am Ende seiner Rezension, auf eine andere, ihm eventuell wichtigere, zu sprechen kommt:

\singlespacing 
\begin{quote}
"`Bei aller polemischen Verve gegen die akademisch-sterile Deutungsroutine irritiert jedoch die \flq Stimmung\frq \,\,des Buchs \mbox{selbst}. An einer Stelle z\"ahlt Gumbrecht sich zu einer \flq Generation von etwas infantilen Alten\frq. Ein Hauch von Wehmut umgibt diese Lekt\"uren, in denen es oft um ver\-geb\-li\-che Hoffnungen, Traurigkeit und ungestillte Sehnsucht geht. Das alles stimmt nicht wirklich zuversichtlich f\"ur die Zukunft einer \flq stimmungsorientierten\frq \,\,Li\-te\-ra\-tur\-wissenschaft."' (\cite{Martus2011}:50)
\end{quote}
\onehalfspacing

Im Wi\-ki\-pe\-dia-Eintrag steht die in der Rezensionsnotiz als Martus' Leseerlebnis des Staunens formulierte Passage im Abschnitt "`Stil"' und das explizit von Martus als Leseerlebnis Benannte wurde f\"ur den Abschnitt "`Leseerlebnis"' genutzt. In der \mbox{hier} vorgeschlagenen Typologie hat diese Schilderung von Martus bereits etwas vom Charakter eines \textit{Erlesnisses}, auch wenn nichts \"uber ein eventuell erworbenes nicht-pro\-po\-si\-ti\-o\-na\-len Wissen berichtet wird. Dieser Schritt bis zu einem \textit{Erlesnis} h\"atte nach meiner Einsch\"atzung etwa so lauten k\"onnen:

\singlespacing
\begin{quote}
Ich habe mich gefragt, ob ich mir eine solche Li\-te\-ra\-tur\-wissenschaft w\"un\-sche und festgestellt, dass mir Schreibweisen zukunftsweisender er\-schei\-nen, die sich nicht vorrangig aus wehm\"utigen Emp\-fin\-dun\-gen speisen, sondern [\mbox{hier} einf\"ugen, was das entsprechend positiv Formulierte w\"are].\\ (Spekulatives \textit{Erlesnis} auf Basis meiner Lesart der Rezension von Steffen Martus)
\end{quote}
\onehalfspacing

Im Rahmen einer Rezension wird eine solche Art der pers\"onlichen Aussage eventuell als un\"ublich angesehen und meist vermieden. Aus anderen der oben besprochenen Rezensionen l\"asst sich dies ebenfalls herlei\-ten. 

Bei dem folgenden Beispiel habe ich mich gefragt, in welchem Ma{\ss}e Assoziationen als \textit{Erlesnisse} deutbar sein k\"onnten -- und von wem.

\singlespacing
\begin{quote}
"`Family album photographs are interspersed throughout \textit{Alfred and Emily}, the experience of the reader being one of dislocation, rather like that experienced with a Sebald novel -- say \textit{Austerlitz} -- where the juxtaposed visuals no more explicate the text than the text explicates the visuals. Reading the work, for me, was akin to viewing a graphic by Escher, say \flq Drawing hands\frq, with its perspectival and dimensional illusions."' (\cite{Tiger2009}:24, Fn. 3)
\end{quote}
\onehalfspacing

Assoziationen dieser Art k\"onnen zwar in einem Wi\-ki\-pe\-dia-Eintrag \href{https://de.wikipedia.org/wiki/Alfred_und_Emily\#Leseerlebnis}{referiert} werden. In beiden F\"allen wird neben einem Vergleich auch deren Effekt beschrie\-ben. Mir stellt sich in Bezug auf \textit{Erlesnisse} aber die Frage, was unbedarftere Leser\textsuperscript{\tiny *} mit diesen Informationen w\"urden anfangen k\"onnen, wenn nicht gesagt wird, was an nicht-pro\-po\-si\-ti\-o\-na\-lem Wissen erworben wurde -- au{\ss}er diesen beiden Assoziationen -- so interessant sie auch sein m\"ogen. Und \mbox{selbst} Leser\textsuperscript{\tiny *}, die beide Texte aus eigener Lekt\"ure kennen und Eschers \textit{Zeichnen} ebenfalls, w\"urden schwerlich ahnen k\"onnen, ob sich f\"ur Tiger ein \textit{Erlesnis} ent\-wickelt hat und falls ja, welches.

Zusammenfassend l\"asst sich sagen, dass die Bandbreite an Ausdrucksweisen f\"ur \textit{Erlesnisse} im Deutschen recht hoch ist. Es lie{\ss}e sich zum einen textstilistisch he\-rausarbei\-ten, bei welchem Typ von Aussage es sich um eine direktere Darstellung handelt und bei welchem eher das Indirekte \"uberwiegt. Zum anderen w\"are es interessant, eine solche Typologie eventuell f\"ur eine empirische Studie zu \textit{Erlesnis}-Wahrnehmungen in verschiedensten ge\-sell\-schaftlichen Gruppen im Rahmen der Li\-te\-ra\-tur\-ver\-mitt\-lungsfor\-schung als ersten Schritt zu verwenden. 

\pagebreak

\subsection{Update f\"ur die Theorie der Literaturvermittlung}
\label{subsec:7.2}

F\"ur Debatten der Li\-te\-ra\-tur\-ver\-mitt\-lungsfor\-schung sind mit den Ergebnissen der vorliegenden Arbeit einige Punkte zutage gef\"ordert worden, die in diesem Abschnitt umrissen werden sollen.

Neu in Betracht gezogen werden sollte die Bedeutsamkeit von Akteuren\textsuperscript{\tiny *}, die Li\-te\-ra\-tur\-ver\-mitt\-lung im Rahmen gemeinn\"utziger Ak\-ti\-vi\-t\"aten leisten ("`unbezahlbar"'): in Verkaufskatalogen, in Leseforen und bei Wi\-ki\-pe\-dia (\mbox{hier} inklusive der Be\-nut\-zerdis\-kus\-si\-onssei\-ten). Als Voraussetzung f\"ur eine fruchtbare Debatte sehe ich an, multivariante Umgebungen wie Wi\-ki\-pe\-dia wertsch\"atzen zu lernen und sich \mbox{selbst} darauf einzulassen, auszuprobieren, in welcher Weise deren Paradigmen f\"ur Li\-te\-ra\-tur\-ver\-mitt\-lung zu nutzen w\"aren. Neben dem Aspekt, dass unentgeltlich t\"atige Autoren\textsuperscript{\tiny *} bereits heute einen we\-sent\-lichen Beitrag zur Li\-te\-ra\-tur\-ver\-mitt\-lung leisten, hat die Tatsache, dass sie an Beitr\"agen mit freier Zeiteinteilung arbei\-ten und vor allem allein aufgrund pers\"onlicher Interessen die Werke ausw\"ahlen in der Li\-te\-ra\-tur\-ver\-mitt\-lungsfor\-schung aus meiner Sicht bis\-her nicht ausreichend Beachtung gefunden, zumindest allem Anschein nach nicht unter Einbeziehung von Wi\-ki\-pe\-dia als einer Plattform, auf der Informationen \"uber Li\-te\-ra\-tur gesucht -- und oft auch gefunden -- werden. Mein Vorschlag ist daher, in theo\-retische Debatten die genannten drei Faktoren einzubeziehen, denn man kann von Verkaufskatalogen oder Wikpedia halten, was man will: Die Eintr\"age werden bei Google vorzugsweise gerankt und schon allein dieses Faktum erfordert eine Auseinandersetzung mit den frei im Web verf\"ugbaren Inhalten. Was dort zu finden ist, hat Einfluss auf potenzielle Leser\textsuperscript{\tiny *}, die eine erste schnelle Websuche unternehmen, um sich \"uber ein Werk zu informieren. In etwas geringerem Ma{\ss}e gilt dies auch f\"ur gr\"o{\ss}ere mehrspra\-chi\-ge Leseforen wie \textit{goodreads.com} oder, etwa im franz\"osischspra\-chi\-gen Raum, f\"ur \textit{babelio.com}, denn in meiner Suche nach ge\-eig\-netem Material f\"ur Wi\-ki\-pe\-dia-Eintr\"age bin ich \mbox{hier} per Websuche mit dem Titel des Werks f\"undig geworden, was mich immerhin dazu bewogen hat, die von Bourdieu angedachte Gleichwertigkeit von Rezeptionsarten (vgl. \cite{Zahner2010}:65) in Wi\-ki\-pe\-dia-Eintr\"agen in die Tat umzusetzen. Ich komme damit zum n\"achsten Punkt.

Als weiteres Ergebnis der vorliegenden Arbeit kann festgehalten werden, dass auf der Plattform Wi\-ki\-pe\-dia einiger Spielraum be\-steht, der f\"ur Li\-te\-ra\-tur\-ver\-mitt\-lung genutzt werden kann. Dieser Spielraum kann der allgemeinen \"Of\-fent\-lichkeit als Anre\-gung die\-nen, das eigene Verh\"altnis zu li\-te\-ra\-rischen Werken umzustellen von Li\-te\-ra\-tur\-genuss als Besch\"aftigung im privaten oder semi-\"of\-fent\-lichen Raum hin auf eine web\"of\-fent\-lich und kostenlos zug\"ang\-liche Bereicherung des kulturellen Wis\-sens\-schat\-zes. F\"ur die Li\-te\-ra\-tur\-ver\-mitt\-lungsfor\-schung w\"are infolgedessen gefordert, eine eventuell gegen\"uber dem "`Produkt"' Wi\-ki\-pe\-dia bestehende und gehegte pauschale Ablehnung daraufhin mit einem frischen Blick zu inspizieren, was durch eine aktivere Nutzung der Platt\-form f\"ur die Li\-te\-ra\-tur\-ver\-mitt\-lung zu gewinnen w\"are. Falls bis\-her gewisse Nach\-rich\-ten \"uber ein angeblich durchgehend unfreundliches Arbeitsklima in der Wi\-ki\-pe\-dia-Community zu einer ablehnenden Haltung gef\"uhrt haben, w\"are sich klar zu machen, dass Millionen von Informationssuchenden zu den Eintr\"agen gelangen, ohne sich an Streitigkei\-ten zu beteiligen oder sonstwie "`zu Wi\-ki\-pe\-dia"' zu z\"ahlen. Sich darauf zur\"uckzuziehen, dass die Artikel schlecht seien, ist angesichts des hohen Rankings dieser Artikel bei Google eine Haltung, die meines Erachtens f\"ur eine zeitgem\"a{\ss}e Li\-te\-ra\-tur\-ver\-mitt\-lung nicht vertr\"aglich ist. Denn nicht zu\-letzt kann durch das gemeinsame Verfassen und Editieren von Eintr\"agen zu li\-te\-ra\-rischen Werken auf einer anerkannten Plattform erreicht werden, dass die ge\-sell\-schaftliche Relevanz von offener Debatte \"uber Li\-te\-ra\-tur\textsuperscript{\~.\~.} nachvollziehbar und weithin augenscheinlich wird, wenn die Informationen auch inhaltlich ansprechend gestaltet sind. Womit ich beim konzeptionellen Hauptaugenmerk dieser Arbeit angelangt bin.

Neben Li\-te\-ra\-tur\textsuperscript{\~.\~.} als dem Entstehen von Emp\-fin\-dun\-gen im Lese- und/oder Zuh\"or\-kontakt mit li\-te\-ra\-rischen Texten und Li\-te\-ra\-tur\textsuperscript{\~.\~.}lekt\"ure als dem Nachsinnen \"uber einen Prozess, dem das das Entstehen bestimmter Emp\-fin\-dun\-gen zugeschrieben wird, stelle ich mit dieser Arbeit \textit{Erlesnis} als ein neues Arbeits\-konzept zur Dis\-kus\-si\-on. Verbunden ist dies mit dem Vorschlag, in sozialpsychologisch fundierten Debatten vor allem die Voraussetzungen zu er\"ortern, unter denen mehr Leser\textsuperscript{\tiny *} sich bereitf\"anden, freim\"utig \"uber nicht-pro\-po\-si\-ti\-o\-na\-les Wissen aus Li\-te\-ra\-tur\-lekt\"ure zu sprechen (und zu schrei\-ben), um mit diesen Neuigkei\-ten die \"of\-fent\-liche Debatte \"uber einzelne li\-te\-ra\-rische Texte zu bereichern. Vielleicht k\"onnen hierzu als Erstes Anregungen aus der Art gewonnen werden, in der Ulrike Draesner 2013 mit ihrem Beitrag zur eigenen Re-Lekt\"ure des \textit{Felix Krull} ein \textit{Erlesnis} vorgelegt hat, das aus meiner Sicht brilliant ist (\cite{Draesner2013}:339-354).

Allgemein kann ich als meine Einsch\"atzung festhalten, dass die Grundlagen des Tuns neu gedacht werden sollten. Ich meine zum Beispiel, dass Li\-te\-ra\-tur\-ver\-mitt\-lung nicht \"uberwiegend bezogen auf: "`an welchem Ort"', "`in was f\"ur einer Art von Veranstaltung"', "`wessen Werke"', "`welche Werke"', "`wie vermitteln"' und "`wen er\-reichen"' theo\-retisch diskutiert werden sollte, sondern auch bezogen auf das Erlebnis der eigenen Lekt\"ure, das WAS des Gelernten, um ein Sich-Verf\"uhren-Lassen-zum-Lesen anzuregen durch das \"Au{\ss}ern eigener \textit{Erlesnisse}, in subjektiver Form ohne Allgemeing\"ultigkeitsanspruch, als Geschenke f\"ur im (web)\"of\-fent\-lichen Raum gef\"uhrte Debatten. Da dieser Raum vielf\"altig und kaum \"uberschaubar ist, soll \textit{Tabelle 2} (siehe unten) einen ersten \"Uberblick anhand der \mbox{hier} vorgestellten neuen Erkenntnisse zu schriftlicher Li\-te\-ra\-tur\-ver\-mitt\-lung bieten.

In \textit{Tabelle 2} sind zum Einen diejenigen Bereiche fett markiert, die in allgemein verf\"ugbares Wissen verwandelt werden k\"onnen, indem es k\"unftig in Artikeln bei Wi\-ki\-pe\-dia zu finden ist. Ein zweiter Aspekt, der zuvor in textueller Hinsicht weit gefasste Begriff "`Lesebericht"', der sowohl wissenschaftliche Interpretationen als auch Kritiken und Berichte in Leseforen beinhaltete, wird \mbox{hier} tentativ wieder auf zwei Bereiche aufgeteilt, womit der eher kultursoziologischen Konzeption in \textit{Tabelle 1} Rechnung getragen wird (siehe Kapitel \textit{~\ref{sec:3} For\-schungsstand}, Abschnitt \textit{~\ref{tab:1} Theo\-rie der Li\-te\-ra\-tur\-ver\-mit\-tlung}). F\"ur die Br\"uc\-ken\-funk\-tion des Entwurfsmusters \textit{Erlesnis} bleibt der weiter gefasste Begriff passender, diese ist aber aufgrund des Tabellenkonzepts so nicht darstellbar, daran w\"are noch weiterzuarbei\-ten. Hilfsweise werden in dieser Tabelle die Titel der Bereiche, in denen ich einzelne \textit{Erlesnisse} f\"ur formulierbar halte, kursiv dargestellt: nahezu \"uberall. Meines Erachtens k\"onnen \textit{Erlesnisse} nicht als "`gesichertes Wissen"' gelten, also sind sie allein in didaktisch ausgerichteten Kontexten nicht formulierbar, sie w\"aren aber erst recht f\"ur die ge\-nann\-ten Zielgruppen interessant. \mbox{Hier} entsteht nach meiner Auffassung ein Dilemma, das vielleicht als Erstes von der Li\-te\-ra\-tur\-ver\-mitt\-lungsfor\-schung zu diskutieren w\"are, noch vor der Li\-te\-ra\-tur\-didaktik, um eine umfassende Perspektive zu gew\"ahrleisten, denn auch Studierende und Sch\"uler\textsuperscript{\tiny *} suchen au{\ss}erhalb des unmittelbaren Kontextes didaktischer Institutionen im Web nach Informationen, ohne dass ich sie damit schon zur Gruppe "`allgemeine \"Of\-fent\-lichkeit"' z\"ahlen w\"urde, denn -- so meine Argumentation -- das per Web gefundene Wissen bleibt in ihrem All\-tag abforderbar durch sie nach "`Leistung"' beurteilende Akteure\textsuperscript{\tiny *}. 

In Kapitel \textit{~\ref{sec:3} Forschungsstand} im Abschnitt \textit{~\ref{subsec:3.2} Theorie der Literaturvermittlung} stellte ich ein Flowchart vor, in dem eine Abfolge von drei Entwurfsmustern und drei Handlungen konzipiert ist:

\singlespacing
¸* Entwurfsmuster 1 = publizierte literarische Texte\\
¸* Handlung 1 = Lesen\\
¸* Entwurfsmuster 2 = \textit{Erlesnis}\\ 
¸* Handlung 2 = als Leseberichte publizieren\\
¸* Entwurfsmuster 3 = Wikipedia-Eintr\"age\\
¸* Handlung 3 = \textit{Erlesnisse} in "`Leseerlebnis"'-Abschnitten dar\-stel\-len
\onehalfspacing

Zusammen mit diesem Flowchart l\"asst sich \textit{Tabelle 2} nunmehr als ein Er\-kun\-dungs\-werk\-zeug f\"ur diejenigen R\"aume nutzen, in denen \textit{Erlesnisse} in schriftlicher Form an die \"Of\-fent\-lichkeit gebracht werden k\"onnen. 

\singlespacing
\label{tab:2}
\textbf{Tabelle 2: Das Entwurfsmuster \textit{Erlesnis} in seinem Umfeld (Kolt\-zen\-burg)}\\\newline
\begin{tabular}{p{1,6cm}|p{2,1cm}p{2,1cm}p{2,3cm}p{2,5cm}p{2,5cm}}
& Sprachkunst & Forschung, \newline Inter- \newline pretation & Kritiken, \newline Leseberichte & \textbf{Lehre} & \textbf{Popula-}\newline \textbf{risierung} \\\hline
&&& \\
Funktion & \textit{Gestaltung} & \textit{Wissens-} \newline \textit{produktion} & \textit{Wissens-} \newline \textit{produktion} & \textbf{Wissens-}\newline \textbf{transfer} & \textit{\textbf{Wissens-}} \newline \textit{\textbf{transfer}} \\
&&& \\
Aus- \newline richtung & \textit{k\"unstlerisch} & \textit{\textbf{wissen-}} \newline \textit{\textbf{schaftlich}}\textsuperscript{1} & \textit{essayistisch} & \textbf{didak-} \textbf{tisch}\textsuperscript{2} & \textit{\textbf{journalis-}} \textit{\textbf{tisch}}\textsuperscript{1} \\
&&& \\
Diskurs- \newline form & \textit{frei}\textsuperscript{3} & \textit{\"uberwiegend} \newline \textit{argumen-} \newline \textit{tierend} & \textit{essayistisch} & \textbf{erkl\"arend, \newline narrativ} & \textit{\textbf{berichtend}}, \newline \textit{\textbf{narrativ}} \\
&&& \\
Inhalt & \textit{\textbf{Neues}}\textsuperscript{4}\textsuperscript{,}\textsuperscript{5} & \textit{\textbf{Neues}}\textsuperscript{5} & \textit{\textbf{Neues}}\textsuperscript{5} & \textbf{sogenanntes} \textbf{Gesichertes} & \textit{\textbf{Interes-}} \textit{\textbf{santes}} \\
&&& \\
Ziel- \newline gruppe & \textit{\textbf{allgemeine}} \newline \textit{\textbf{\"Offent-}} \textit{\textbf{lichkeit}} & \textit{\textbf{Fachleute}} & \textit{\textbf{allgemeine}} \newline \textit{\textbf{\"Offent-}} \textit{\textbf{lichkeit}} & \textit{\textbf{Studie-}} \newline \textit{\textbf{rende}}, \newline \textit{\textbf{Sch\"uler\textsuperscript{\tiny *}}} & \textit{\textbf{allgemeine}} \newline \textit{\textbf{\"Offent-}} \textit{\textbf{lichkeit}} \\
&&& \\\hline
&&& \\
Text- \newline sorte & \textit{frei} & \textit{wissen-} \newline \textit{schaft-} \newline \textit{licher} \newline \textit{Beitrag} & \textit{Essay} & \textbf{enzyklo-} \textbf{p\"adischer Eintrag} & \textit{\textbf{enzyklo-}} \textit{\textbf{p\"adischer Eintrag}} \\
&&& \\
\end{tabular}
\footnotesize{¸[1] mit Nachweisen f\"ur einzelne Aussagen (Fu{\ss}noten)}\\
\footnotesize{¸[2] mit Literaturangaben\\
\footnotesize{¸[3] \textit{Save the Reaper} enth\"alt ein \textit{Erlesnis}, vgl. Abschnitt \textit{~\ref{subsubsec:5.4.1} Fiktionalisierte Varianten von Versionierungsverfahren in Alice Munros \flq Save the Reaper\frq}, und \textit{Alfred and Emily} ist im Ganzen ein \textit{Erlesnis}, vgl. Abschnitt \textit{~\ref{subsubsec:5.4.2} \flq For future editors\frq:\,\,\textit{The London Encyclop{\ae}dia} und Doris Lessings \textit{Alfred and Emily}}.\\
\footnotesize{¸[4] kann zitiert werden, ein Gedicht zum Beispiel dann in voller L\"ange, wenn es mit einer Freien Lizenz pu\-bli\-ziert worden ist\\ 
\footnotesize{¸[5] kann zitiert werden im Rahmen des Zitatrechts\footnote{Wie ein Gedicht, dessen Text noch nicht gemeinfrei ist, dennoch im Detail enzy\-klop\"adisch dargestellt werden kann, siehe \href{http://de.wikipedia.org/wiki/Nur_zwei_Dinge}{"`Nur zwei Dinge"'} (Gottfried Benn).}\\
\onehalfspacing

\normalsize
F\"ur die Theorie der Li\-te\-ra\-tur\-ver\-mitt\-lung bedeutet dies aus meiner Sicht, dass auf der Basis der Ergebnisse dieser Arbeit neue Parameter erwogen werden k\"onnen, um Perspektiven einzubeziehen, die gemein\"utzigen Ak\-ti\-vi\-t\"aten nicht zu\-letzt im fachlichen \mbox{Selbst}verst\"andnis mehr Gewicht geben. Denn, um es knapp auf den Punkt zu bringen: \textit{Erlesnisse} m\"ussen niemandem besondere Kosten verursachen, sondern sind in den Commons ansiedelbar.

\subsection{Zusammenfassung}
\label{subsec:7.3}

Auf \textit{Metaebene II} wurde zun\"achst anhand eines ersten typologischen Entwurfs die Bedeutung von \textit{parrhesia} im Schrei\-ben \"uber Li\-te\-ra\-tur\textsuperscript{\~.\~.} dargestellt. Ich habe Schreib\-weisen \"uber \textit{Erlesnisse} miteinander verglichen, um eine erste Einsch\"atzung bez\"uglich distanzierender und weniger distanzierender Aussagen zu gewinnen. Es war zu sehen, dass es allein im Deutschen eine gro{\ss}e Bandbreite gibt, wie \textit{Erlesnisse} formuliert werden k\"onnen. Offen blieb, in welchem Ma{\ss}e bei eher indirekten Aussagen eine Interpretationsleistung auf Sei\-ten von Lesern\textsuperscript{\tiny *} erwartbar w\"are.

Im zwei\-ten Abschnitt wurden Bedingungen der Darstellbarkeit nicht-pro\-po\-si\-ti\-o\-na\-len Wissens in Wi\-ki\-pe\-dia-Artikeln mit Aspekten g\"angiger Theorien der Li\-te\-ra\-tur\-ver\-mitt\-lung zu\-sam\-men\-gedacht und festgestellt, welche neuen Punkte aus den Er\-kennt\-nis\-sen dieser Arbeit eine Bereicherung f\"ur k\"unftige Debatten dar\-stel\-len k\"onnten.

\newpage
\section{Fazit}
\label{sec:8}

\textit{Nicht-pro\-po\-si\-ti\-o\-na\-les Wissen aus Li\-te\-ra\-tur\-lekt\"ure und Bedingungen seiner Darstellbarkeit in Wi\-ki\-pe\-dia-Eintr\"agen zu li\-te\-ra\-rischen Werken} befasste sich mit der Frage, unter welchen Bedingungen \textit{Erlesnisse} Dritter in Li\-te\-ra\-tur\-artikeln bei Wi\-ki\-pe\-dia darstellbar sind, ob \textit{Erlesnis} ge\-eig\-net ist, in der Li\-te\-ra\-tur\-ver\-mitt\-lungsfor\-schung in weitergehende Debatten anzuregen und ob die gew\"ahlte Methode die gestellten Fragen l\"osen kann.

Zu den Bedingungen der Darstellbarkeit nicht-pro\-po\-si\-ti\-o\-na\-len Wissens aus Li\-te\-ra\-tur\-lekt\"ure bei Wi\-ki\-pe\-dia z\"ahlt, dass gen\"ugend Per\-so\-nen der Li\-te\-ra\-tur\-szene an publizistisch namhaften Orten \textit{Erlesnisse} verfassen, die in enzy\-klop\"adischen Eintr\"agen als Beleg verwendet werden k\"onnen. An Schreibweisen f\"ur Leseberichte, die \textit{Erlesnisse} zur Sprache bringen, gibt es im Deutschen eine gro{\ss}e Bandbreite. Aufsei\-ten der deutschspra\-chi\-gen Wi\-ki\-pe\-dia-Community ist mit Akzeptanz f\"ur dezidierte Leseerlebnis-Abschnitte zu rechnen, da deren Inhalte nicht als Gegenentw\"urfe zur bestehenden Ordnung des Sagbaren aufgefasst werden. Nicht-pro\-po\-si\-ti\-o\-na\-les Wissen war bis\-her in Li\-te\-ra\-tur\-artikeln bei Wi\-ki\-pe\-dia keine sp\"urbare Gr\"o{\ss}e und sein Anteil f\"ur die Aussagekraft eines Eintrags unerprobt. Durch Ergebnisse der vorliegenden Arbeit konnte aufgezeigt werden, dass die Spielr\"aume, in denen \textit{Erlesnisse} in Wi\-ki\-pe\-dia-Eintr\"agen zur Spache gebracht werden k\"onnen, zumindest in der aktuellen Phase des Projekts bezogen auf die deutschspra\-chi\-ge Version recht gro{\ss} sind, so dass die Ordnung des Sagbaren nunmehr die Artikulation von nicht-pro\-po\-si\-ti\-o\-na\-lem Wissen aus Li\-te\-ra\-tur\-lekt\"ure zu beinhalten scheint. Subversiv bleibt dabei allein der Aspekt, dass das Eigentliche, die \textit{Erlesnisse}, in Leseerlebnis-Abschnitten eher nebenbei untergemischt werden. Dies kann sich \"andern, sobald auf Li\-te\-ra\-tur\-seite mehr Zitierbares zur Verf\"ugung steht. Es w\"are ein Beitrag zur F\"orderung der Wahrnehmung der spezifischen St\"arken von Li\-te\-ra\-tur.

Literarische Texte treffen in der \"Of\-fent\-lichkeit stets auf ein Umfeld, in dem au{\ss}erli\-te\-ra\-rische Diskurse dominant sind und nicht-pro\-po\-si\-ti\-o\-na\-les Wissen als Quelle des Erlebens daher kaum wahrgenommen wird. Mit \textit{Erlesnis} als Formalobjekt konnte in dieser Arbeit ein gr\"o{\ss}erer Zusammenhang sichtbar gemacht werden. Es zeigt sich ein Pro\-blem, dass zun\"achst nur in Form einer Hypothese formulierbar ist, n\"amlich, dass aufsei\-ten Li\-te\-ra\-tur\-be\-wan\-der\-ter im deutschspra\-chi\-gen Raum eine gewisse Scheu be\-steht, sich \"uber eigene \textit{Erlesnisse} im \"of\-fent\-lichen Raum in direkter Ausdrucksweise schriftlich zu \"au{\ss}ern. Ansatzweise konnten f\"ur den multidisziplin\"aren Aus\-hand\-lungs\-raum Wi\-ki\-pe\-dia nachteilige Folgen einer solchen angenommenen Zur\"uckhaltung auf\-ge\-zeigt werden. Al\-ler\-dings ist vermutbar, dass nicht allein gemeinn\"utzige Li\-te\-ra\-tur\-ver\-mitt\-lung davon betroffen ist, sondern dass auch erwerbsorientierte Ver\-mitt\-lungsinitiativen in einer medial gepr\"agten Erlebnisge\-sell\-schaft einen Gewinn daran h\"atten, \textit{Erlesnisse} bei der Ver\-mitt\-lung von Li\-te\-ra\-tur ins Rampenlicht zu holen. Die Li\-te\-ra\-tur\-ver\-mitt\-lungsfor\-schung ist gefordert, sich eingehender und auch auf theo\-retischer Ebene mit web\"of\-fent\-lichen Inhalten sowie gemeinn\"utzigen Aspekten aus\-ein\-an\-der\-zu\-set\-zen.

Die vorliegende Arbeit hat nicht nur in der Hinsicht Pilot\textsuperscript{\tiny *}charakter, dass \"uber Li\-te\-ra\-tur\-eintr\"age bei Wi\-ki\-pe\-dia noch keine For\-schung pu\-bli\-ziert worden ist, sondern ist auch methodisch innovativ verfahren, indem Verfahren der Li\-te\-ra\-tur\-wissenschaft (Interpretation) mit denen der Wikipedistik (Experiment) kombiniert wurden. Die gestellten Fragen zu einem vorwiegend von gegenseitiger Ignoranz gepr\"agten Ver\-h\"alt\-nis konnten -- wenn auch in typologischer Hinsicht ansatzweise -- auf diesem Wege gel\"ost werden.

\newpage
\section{Verzeichnis zitierter Texte}
\label{sec:9}

\singlespacing

Beitr\"age in Buchformaten werden allein mit Jahr referenziert (unabh\"angig davon, in wel\-chem Medium sie zug\"ang\-lich gemacht wurden), Beitr\"age in Zeitschriften mit Ausgabe und Jahr (auch bei Periodika, die nur im Web erscheinen). Nicht bei allen Zeitschriftenbeitr\"agen steht hinter dem Autor\textsuperscript{\tiny *}namen eine Monatsangabe und bei manchen, z.B. \textit{The New Yorker}, wird eine Ausgabe mit Datum angegeben. Native Webpublikationen werden mit dem Datum verzeichnet, sofern eines zu sehen oder zu eruieren war, bei Texten aus \textit{Wi\-ki\-pe\-dia} inklusive der Versionsangabe mit Uhrzeit.

W\"urde ich die Plattform Wi\-ki\-pe\-dia einem herk\"ommlichen Pu\-bli\-kationsorgan ent\-spre\-chend behandeln, m\"ussten als "`Zitate daraus"' oder "`pu\-bli\-ziert in"' auch einzelne S\"atze oder Teile davon mit einer "`Li\-te\-ra\-tur\-angabe"' referenziert werden. Allzu kleinteilige Text(teil)e sind in diesem Verzeichnis aber nicht aufgef\"uhrt. Al\-ler\-dings erscheint mir ein Platz\-hal\-ter-Ein\-trag wie \href{https://de.wikipedia.org}{Wi\-ki\-pe\-dia, deutschspra\-chi\-ge Version} auch nicht angemessen zu sein, denn in einer For\-schungs\-arbeit \"uber einen Print-Brockhaus w\"urde ich auch die einzelnen verwendeten Eintr\"age in einem solchen Verzeichnis auff\"uhren und nicht das Gesamtwerk. Einzelne Dis\-kus\-si\-onsbeitr\"age werden mit Au\-tor\textsuperscript{\tiny *}\-na\-men angegeben, denn Wi\-ki\-pe\-dia wird von Individuen ge\-macht, die sich am Verfassen von Artikeln beteiligen und sich in Debatten konstruktiv \"au{\ss}ern (auf ei\-ner Skala bis zur jeweils negativsten Variante davon, und dies tun sie ebenfalls als zitierbare Individuen). Ist eine bestimmte Version bzw. \"Anderung einer Wi\-ki\-pe\-dia-Seite gemeint, werden Datum und Uhrzeit hinter dem Sei\-tentitel (Lemma) in Klammern angegeben, unabh\"angig davon, ob es sich um die Seite eines Eintrags handelt, um eine Dis\-kus\-si\-onsseite oder um eine andere Seite "`bei Wi\-ki\-pe\-dia"'.

\section{Verzeichnis der Abbildungen}
\label{sec:10}

\begin{enumerate}
\item Schematische Darstellung zum \href{https://commons.wikimedia.org/wiki/Category:Alice_Munro#/media/File:Munro_Save_the_Reaper_1998_percentages.png}{Vergleich von L\"ange und Position einzelner Abschnitte relativ zur Gesamt\-l\"ange in "`Save the Reaper"'} (Alice Munro), Zeitschriftenversion 1998 (1255 Zeilen), Buchversion 1998 (34,5 Sei\-ten), siehe \textit{Abbildung ~\ref{fig:1}} in Abschnitt \textit{~\ref{subsubsec:5.4.1} Fiktionalisierte Varianten von Versionierungsverfahren in Alice Munros \flq Save the Reaper\frq}
\item \textit{Tabelle 1 in Abschnitt ~\ref{tab:1}} "`Textsorten der wissenschaftlichen Li\-te\-ra\-tur\-ver\-mitt\-lung"' (Huter)
\item \textit{Tabelle 2 in Abschnitt ~\ref{tab:2}} "`Das Entwurfsmuster \textit{Erlesnis} in seinem Umfeld"' (Kolt\-zen\-burg)  
\end{enumerate}

\vspace{1cm}
[Ende von: Claudia Kolt\-zen\-burg (September 2015), \textit{Nicht-pro\-po\-si\-ti\-o\-na\-les Wissen aus Li\-te\-ra\-tur\-lekt\"ure und Bedingungen seiner Darstellbarkeit in Wikipedia-Eintr\"agen zu li\-te\-ra\-rischen Werken}, \"uberarbeitete und gek\"urzte Web-Fassung (pdf) einer Dissertation \href{https://publikationen.uni-tuebingen.de/xmlui/handle/10900/64678}{an der Universit\"at T\"ubingen, Philo\-so\-phi\-sche Fakult\"at}]

\end{document}